\documentclass[preprint,12pt]{elsarticle}
\usepackage[top=1in, bottom=1in, left=1in, right=1in]{geometry}
\usepackage{bm}
\usepackage{amsmath}
\usepackage{setspace}
\usepackage{array}
\usepackage{multirow}
\usepackage{colortbl}
\usepackage{graphicx}
\usepackage{subfigure}
\usepackage[table,xcdraw]{xcolor}
\usepackage[outdir=./]{epstopdf}
\usepackage{booktabs}
\usepackage{tabu}
\usepackage{bbm}
\usepackage{float}
\usepackage[singlelinecheck=false]{caption}
\DeclareMathAlphabet      {\mathbf}{OT1}{cmr}{bx}{n}
\usepackage[utf8]{inputenc}
\usepackage[english]{babel}
\usepackage{natbib}
\usepackage{bookmark}
\usepackage{mathrsfs}
\usepackage{color}
\usepackage{mathtools} 
\newcommand{\comment}[1]{}
\usepackage{amssymb}
\usepackage{xfrac}
\usepackage{algorithm}
\usepackage{algorithmic}
\usepackage{lscape}
\usepackage{pdflscape}
\usepackage{lineno}
\usepackage{soul,xcolor}

\begin{document}
\begin{spacing}{1.15}
\begin{frontmatter}

%\title{The effect of prior knowledge on the uncertainty quantification and efficient propagation from limited data}
\title{The effect of prior probabilities on quantification and propagation of imprecise probabilities resulting from small datasets}
\author{ Jiaxin Zhang, Michael D. Shields$^{*}$ \footnote{*Corresponding author. email: jiaxin.zhang@jhu.edu (J Zhang), michael.shields@jhu.edu (MD Shields)} }
\address{Department of Civil Engineering, Johns Hopkins University
	Baltimore, MD 21218, USA.}

\begin{abstract}
This paper outlines a methodology for Bayesian multimodel uncertainty quantification (UQ) and propagation and presents an investigation into the effect of prior probabilities on the resulting uncertainties. The UQ methodology is adapted from the information-theoretic method previously presented by the authors (Zhang and Shields, 2018) to a fully Bayesian construction that enables greater flexibility in quantifying uncertainty in probability model form. Being Bayesian in nature and rooted in UQ from small datasets, prior probabilities in both probability model form and model parameters are shown to have a significant impact on quantified uncertainties and, consequently, on the uncertainties propagated through a physics-based model. These effects are specifically investigated for a simplified plate buckling problem with uncertainties in material properties derived from a small number of experiments using noninformative priors and priors derived from past studies of varying appropriateness. It is illustrated that prior probabilities can have a significant impact on multimodel UQ for small datasets and inappropriate (but seemingly reasonable) priors may even have lingering effects that bias probabilities even for large datasets. When applied to uncertainty propagation, this may result in probability bounds on response quantities that do not include the true probabilities. 
\end{abstract}

\end{frontmatter}
%%%%%%%%%%%%%%%%%%%
%             INTRODUCATION            %
%%%%%%%%%%%%%%%%%%%
\section{Introduction}

Uncertainty quantification (UQ) is the science of quantitatively characterizing and, if possible, reducing uncertainties in the computational evaluation of engineering/mathematical systems. It is used to determine how likely certain outcomes are if some parts of the system are not exactly known. Practically speaking, UQ is playing an increasingly important role in performance prediction, reliability analysis, risk evaluation and decision making. Uncertainty can be broadly classified into two categories \cite{DerKiureghian2009}: \emph{epistemic}, resulting from a lack of complete knowledge \textcolor{black}{and modeling approximations},  and \emph{aleatory}, resulting from inherent randomness.  It is widely accepted that probability theory provides an appropriate framework for the treatment of aleatory uncertainty although it is open to debate as to what mathematical treatment is most appropriate for epistemic uncertainty. 

It is often considered preferable to view all uncertainty probabilistically - given the well-understood and intuitive nature of \textcolor{black}{probability theory} . This desire has given rise to the field of so-called imprecise probabilities wherein aleatory uncertainties are modeled using standard probability theory and epistemic uncertainties provide a level of ``imprecision.'' Despite efforts to develop a unified theory of imprecise probability \cite{Walley1991,Walley2000}, there remain numerous approaches to model this imprecision that include the use of random sets \cite{Molchanov2005,Fetz2004,fetz2016}, intervals and probability boxes \cite{Moore1979,Weichselberger2000,ferson2002,ferson2004,schobi2017}, Bayesian \cite{raftery1995,sankararaman2013} and frequentist \textcolor{black}{\cite{walley1982,cattaneo2017}} methods, and combinations of these theories \cite{dubois1991,dubois2005} among many others (e.g.\ \cite{Dubois2012}). Additionally, Dempster-Shafer theory \cite{Dempster1967,Shafer1976} and fuzzy set theory \cite{Zadeh1965} aim to relax the constraints on probability measures to account for this imprecision. For the interested reader, an extensive review of many of these approaches for engineering applications can be found in \cite{beer2013imprecise}.

In this work, we apply a multimodel Bayesian probabilistic approach, extended from \cite{zhang2018}, to quantify and propagate combined aleatory and epistemic uncertainty. Specifically, epistemic uncertainties manifest as uncertainties in the form of probability models characterizing a dataset (referred to as \emph{model-form uncertainties} {\color{black} or \emph{structural uncertainties} \cite{draper1995}}) and uncertainties in the parameters of the probability models (referred to as \emph{parameter uncertainties}). Both model-form and parameter uncertainties are quantified from the given data using Bayesian inference. The result of this multimodel approach is a set of candidate probability models (each having associated probability) and the joint parameter probability densities for each model and provides a near comprehensive description of the uncertainties in the system rooted in probabilities (or more precisely probabilities of probabilities). 

The imprecise probabilities quantified using the proposed approach are propagated through a model of a physical system using a Monte Carlo approach with importance sampling reweighting for simultaneous propagation of the full set of probability models \cite{zhang2018}. The procedure is critical in reducing the computational expense of Monte Carlo-based imprecise probability propagation by reducing a multi-loop Monte Carlo to a single-loop Monte Carlo. 

Given that the approach employed herein is fully Bayesian in its construction and the datasets used for inference are necessarily small, prior probabilities are likely to influence the quantified uncertainties in important ways. The primary objective of this work is to improve our understanding of the influence of prior probabilities in both model-form and parameter uncertainties on the quantified uncertainties and, ultimately, on propagated uncertainties. We employ informative and noninformative priors for both model-form and parameter uncertainties under realistic data availability constraints. More specifically, priors are formulated based on either assumed ignorance (noninformative) or using \textcolor{black}{historical} data available from literature (informative) that may or may not be entirely appropriate for the present analysis. In other words, the informative prior may be incorrect in seemingly subtle ways. This is combined with the reality of small datasets for Bayesian inference used for quantifying imprecise probabilities and, an example of a simple plate buckling problem with uncertainty in material properties is used to illustrate how prior probabilities come to have a strong influence (good or bad) on multimodel uncertainty quantification and propagation from small datasets. Moreover, the effect of the prior is studied {\color{black} to observe, for this application,} the rate of convergence of imprecise probabilities to the ``true'' probabilities for increasing dataset size. Even in these large data case, it is shown that seemingly rational (but ultimately incorrect) priors can create biases that preclude convergence to the true probabilities. 

The paper is structured as follows. Sections 2 and 3 provide the basic theory for multimodel Bayesian uncertainty quantification for model-form (Section 2) and parameter uncertainties (Section 3). Section 4 reviews a new Monte Carlo-based method for multimodel uncertainty propagation proposed in a recent work by the authors \cite{zhang2018}. Some minor modifications are included. Section 5 discusses the formulation of model and parameter prior probabilities. The influence of these prior probabilities on multimodel uncertainty quantification and propagation is then studied in the context of a simplified plate buckling problem in Section 6 where priors are considered to be either noninformative or rooted in \textcolor{black}{historical} data from literature. Several such priors are considered for both model-form and parameter uncertainties and their influence studied systematically. Finally, some concluding remarks are provided in Section 7.

\section{Bayesian multimodel inference and model-form uncertainty}
\label{sec:multimodel}

Among the most important problems in computational science and engineering are the quantification of model-form uncertainty and its use for model selection, with widespread recognition of these challenges dating back more than 30 years \cite{draper1987,dijkstra1988}. Probabilistic model selection has taken considerable strides forward with the development of approaches based on Bayes' theory and {\textcolor{black}information theory}. Bayesian approaches involving the notion of posterior model probabilities follow the work of Raftery \cite{raftery1995} and have been revived in the more recent work of Beck \cite{beck2004, beck2010, cheung2010} and Oden \cite{farrell2015, oden2016,prudencio2015}. Meanwhile the information-theoretic approach is derived from the work of Akaike \cite{akaike1974new, akaike1976} and its further generalizations \cite{schwarz1978, hurvich1989regression, konishi2008}. 

The issue of model selection is fundamental to the quantification of input uncertainties for physics-based calculations given limited data. In such cases, multiple probability models may reasonably fit the data.  Generally, both the Bayesian and information-theoretic model selection methods are employed to select a single ``best" model based on the given data and a set of candidate models. That model is then the sole model used for uncertainty propagation without any further consideration for \textcolor{black}{the assessment and propagation of model-form uncertainty \cite{draper1987,draper1995}. This approach is known to potentially misrepresent the true uncertainties in statistical quantities of interest (typically underestimating them) \cite{draper1995}, while also implicitly asserting that a ``true" model exists, which is counter to the definition of a model \cite{Chatfield1995}.} In certain cases, model averaging is performed {\color{black}(see \cite{hoeting1999bayesian} and the associated commentary)} and, while this may be preferable, it still ignores much of the true model-form uncertainty by propagating only averaged quantities rather than their full probability structure. Moreover, such selection needs very large datasets. Given \textcolor{black}{scarcity of data}, it is often impossible to identify a unique best model so that we need to quantify model-form uncertainty and retain multiple candidate models and their associated probabilities - a method referred as to multimodel inference \cite{burnham2004multimodel}. The previous work of the authors \cite{zhang2018} presented an information-theoretic approach for multimodel inference to quantify and propagate these model-form uncertainties. This work seeks to generalize this in a fully Bayesian framework.

\subsection{Bayesian multimodel inference}

{\color{black}In this work, we consider the specific case of probability model selection from sparse data. In all subsequent discussion, a model $M_i$ refers to a parametric probability model for a random variable $X$ (typically expressed in the form of a probability density function, $p(x)$) having parameters $\boldsymbol{\theta}_i$. Given a collection of $m$ candidate models $\bm{\mathcal{M}}=\{M_j\}$ with model parameters $\bm{\theta}_j \hspace{3pt} j=1,\dots,m$, our objective in this section is to assess the ``goodness-of-fit'' of each model given a dataset $\bm{d}$ of independent observation of $X$ and ultimately infer probabilities that each model is the ``best'' model for the data.}

In the Bayesian setting, selection between two models $M_i$ and $M_j$ given data $\bm{d}$ is often performed by estimating the ratio of posterior odds as 
\begin{equation}
\underbrace { \frac{p(M_i | \bm{d})}{p(M_j | \bm{d})} }_{ \textup{Posterior \ odds} } = \underbrace { \frac{p(M_i )}{p(M_j  )} }_{ \textup{Prior \ odds} } \times \underbrace { \frac{p(\bm{d} | M_i)}{p(\bm{d} | M_j)} }_{ \textup{Bayes' factor } } \label{eq: BF}
\end{equation}
where Bayes' factor is defined as the ratio of the evidence of $M_i$ and $M_j$, and the prior odds is the ratio of model prior of $M_i$ and $M_j$. If the posterior odds are greater than one, then model $M_i$ is selected while if the posterior odds are less than one, model $M_j$ is selected.

Intuitively, Bayes' factor can be easily generalized for comparison of multiple candidate models. {\color{black}Consider the aforementioned collection of $m$ parametric models $\bm{\mathcal{M}}$, with each model $M_j$ having an associated prior probability $\pi_j = p(M_j)$ with $\sum_{j=1}^m\pi_j=1$.} Bayes' rule relates posterior model probabilities $\hat{\pi}_j$ to prior model probabilities $\pi_j$ via the formula 
\begin{equation}
\hat{\pi}_j = p(M_j| \bm{d}) = \frac{p(\bm{d} | M_j)p(M_j )}{\sum_{k=1}^{m} p(\bm{d} | M_k)p(M_k )},  \quad j=1,\dots, m \label{eq: evidence}
\end{equation}
having $\sum_{j=1}^m\hat{\pi}_j=1$ and where
\begin{equation}
p(\bm{d}| M_j)={ \int_{\Theta_j} {p({ \bm{d}}|\bm{\theta}_j,M_j)p(\bm{\theta}_j|M_j)d\bm{\theta}_j }  }, \quad j=1,\dots, m \label{eq:evidence0}
\end{equation}
is the marginal likelihood or evidence of model $M_j$. Typically, the model $M_k\in\bm{\mathcal{M}}$ with highest probability $p(M_k| \bm{d})$ is deemed the most plausible in the set $\bm{\mathcal{M}}$ for the given data $\bm{d}$. In the multimodel inference context, rather than selecting the model with the highest probability, the models are ranked according to their probabilities given by Eq.\ (\ref{eq: evidence}) and all models with non-negligible probability are retained. 

In Bayesian parameter estimation the evidence $p(\bm{d}| M_j)$ is just a normalization factor that does not need to be evaluated explicitly using Markov Chain Monte Carlo (MCMC). However, as evident from Eq.\ (\ref{eq: evidence}), the evidence $p(\bm{d}| M_j)$ is critical in Bayesian multimodel inference and consequently needs to be calculated with caution. The following section discusses evidence calculation.

\subsection{Bayesian evidence calculation}
\label{sec:evidence_calc}

The evidence in Eq.\ \eqref{eq:evidence0} can be computed in a number of different ways. In rare cases, the integral can be evaluated analytically. Usually, approximate or statistically exact (i.e. Monte Carlo) methods are necessary.

% Because of the computational advantages of having closed form expression for evidence $p(\bm{d}| M_j)$, it may be preferable to use a computable approximation for $p(\bm{d}| M_j)$ if exact analytical expression is not available. 

One efficient approximation uses Laplace's approach \cite{konishi2008} to approximate the evidence $p(\bm{d}| M_j)$ as 
\begin{equation}
p(\bm{d}| M_j) \approx \exp\left\{\log(p(\bm{d} | \bm{\theta}_j^*, M_j))\right\} p(\bm{\theta}_j^* | M_j)(2\pi)^{K_j/2}n^{-K_j/2} |H^*(\bm{\theta}_j^{*})|^{-1/2}
\label{eq: bic_approx1}
\end{equation}
Taking the logarithm of this expression and multiplying it by $-2$, we obtain 
\begin{equation}
-2\log(p(\bm{d}| M_j)) \approx  -2\log(p(\bm{d} | \bm{\theta}_j^*, M_j)) + K_j\log(n) + \log{|H^*(\bm{\theta}_j^{*})|} - K_j\log(2\pi) - 2\log(p(\bm{\theta}_j^* | M_j)) \label{eq: bic_approx2}
\end{equation}
where $K_j$ is the dimension of the parameter vector $\bm{\theta}$, $\bm{\theta}_j^{*}$ is the maximum likelihood estimate and $H^*$ is the inverse Hessian of the negative log likelihood (Fisher information matrix). Ignoring the terms in Eq.\ \eqref{eq: bic_approx2} with order less than $O(1)$ with respect to the large sample size $n$ yields the Bayesian Information Criteria (BIC) \cite{schwarz1978}
\begin{equation}
\textup{BIC}_j = \textup{BIC}(M_j) = -2\log(p(\bm{d} | \bm{\theta}_j^*, M_j)) + K_j \log(n) \label{eq: bic}
\end{equation}
where $n$ is the dataset size. This quantity can be used to construct an asymptotic approximation to Bayes' factor,  namely $\textup{BF}_{i,j} \approx \exp(-(\textup{BIC}_i-\textup{BIC}_j))/2$ \cite{raftery1995}. Combined with the model prior  $\pi_j = p(M_j)$, posterior model probabilities from Eq.\ \eqref{eq: evidence} can be expressed as
\begin{equation}
\hat{\pi}_j^{\textup{BIC}} \approx \frac{\exp(-\frac{1}{2} (\textup{BIC}_j-\textup{BIC}_{\min} )) \pi_j}{\sum_{k=1}^{m} \exp(-\frac{1}{2} (\textup{BIC}_k-\textup{BIC}_{\min})) \pi_k}  \label{eq: bic_weights}
\end{equation}
where $\textup{BIC}_{\min} = \min_{j}({\textup{BIC}_j})$. Assigning uniform prior model probabilities to the set $\bm{\mathcal{M}}$, $\pi_j \equiv 1/m$, yields what are referred to as BIC model weights. In fact, Eq.\ \eqref{eq: bic_weights} can be considered generalized BIC model weights for arbitrary prior model probabilities.  Notice also that Eq.\ \eqref{eq: bic} may be thought of as an implicit approximation to evidence $p(\bm{d}| M_j)$ under a \emph{noninformative parameter prior} (or \emph{Jeffreys parameter prior}) even though it does not explicitly depend on a parameter prior. 

% It also should be emphasized that the asymptotic justification for this approximation, dependent on the amount of dataset size relative to the dimension of parameter space of the model, may not perform well in small dataset size. 

% As similar as BIC, Akaike Information Criterion (AIC) established by Akaike \cite{akaike1974new} also uses the maximum likelihood estimate 
% \begin{equation}
% \textup{AIC}(M_j) = -2\log(p(\bm{d} | \bm{\theta}_k^*, M_j)) + 2K \label{eq: aic}
% \end{equation}
% Different from the BIC, AIC is referred to as information-theoretic since it provides a simple estimation of the K-L information. Although AIC is proposed based on the information-theoretic setting, it can also be conceived in a Bayesian context through the use of a class of \emph{savvy priors} (see \cite{burnham2004multimodel} for  explanation).

The information-theoretic multimodel selection (introduced in \cite{burnham2004multimodel} and employed in the authors' previous work \cite{zhang2018}) can be shown as a special case of the Bayesian evidence-based multimodel selection used herein. Akaike \cite{akaike1974new} showed that the maximized log-likelihood is a biased estimator of the K-L information and that the bias is approximately equal to $K_j$. Hence, the Akaike Information Criterion (AIC) is defined 
\begin{equation}
\textup{AIC}_j=\textup{AIC}(M_j) = -2\log(p(\bm{d} | \bm{\theta}_j^*, M_j)) + 2K_j \label{eq: aic}
\end{equation}
as an approximation of the K-L information. By rescaling the AIC as
\begin{equation}
\Delta_j=\textup{AIC}_j-\textup{AIC}_{\text{min}}
\end{equation}
the marginal likelihood of the model $M_j$ given the data can be expressed as $\exp{-(\frac{\Delta_j}{2})},\hspace{3pt}j=1,\dots,m$ \cite{akaike1981}. By normalizing these likelihoods to sum to 1, they are treated as model probabilities (as in Eq.\ \eqref{eq: bic_weights}) with
\begin{equation}
\hat{\pi}_j^{\text{AIC}}=\dfrac{\exp{(-\frac{\Delta_j}{2})}}{\sum_{k=1}^m\exp{(-\frac{\Delta_k}{2})}}
\label{eq:AIC_weights}
\end{equation}
As shown by \cite{burnham2004multimodel}, Eq.\ \eqref{eq:AIC_weights} is in fact a special case of Eq.\ \eqref{eq: bic_weights} in which the prior model probabilities $\pi_j$ take the form
\begin{equation}
\pi_j=\dfrac{\exp{(\frac{1}{2}K_j\log(n)-K_j)}}{\sum_{k=1}^m\exp{(\frac{1}{2}K_k\log(n)-K_k)}}
\end{equation}
This form of priors are referred to as {\it savvy} (shrewdly informed) priors because they depend on $n$ and $K_k$. 

The BIC and AIC based results are important because they illustrate directly the influence of priors in the asymptotic case. While the model probabilities in Eq.\ \eqref{eq: evidence} are general, they can be approximated (in large data cases) by Eq.\ \eqref{eq: bic_weights} and the AIC derived model probabilities are an instance of this approximation under certain prior information.

Because both the BIC and AIC are asymptotic quantities that require large dataset size, they are of limited practical use here. Although a small data correction of the AIC, denoted AICc, has been derived \cite{hurvich1989regression,hurvich1995model} and used in our previous work \cite{zhang2018}, this again implies a certain prior form and our objective here is to investigate the effect of the prior. Consequently, we must rely on other estimators for Eq.\ \eqref{eq:evidence0} that do not have asymptotic conditions or assume a prior form. We favor a Monte Carlo-based statistical estimator given by
\begin{equation}
\hat{p}(\bm{d}| M_j)=\frac{1}{N_k}\sum_{k=1}^{N_k}{p({ \bm{d}}|\bm{\theta}_j^k,M_j)},\quad \bm{\theta}_j^k \sim p(\bm{\theta}_j | M_j), \quad j=1,\dots, m \label{eq:evidence_mc}
\end{equation}
in which samples $\bm{\theta}_j^k$ are drawn from the parameter prior distribution and $N_k$ is the number of samples. The computational cost of the Monte Carlo-based algorithm for probability models used in this paper is moderate, and its efficiency can be improved with parallel computing. For complex or high dimensional model evidence calculation, MCMC-based algorithms, including Chib and Jeliazkov  \cite{chib2001} and nested sampling \cite{skilling2004nested} may be preferable as discussed in the recent review literature \cite{bos2002, friel2012, zhao2016}.

% Nevertheless, our task in this work is to explore the effect of prior (under small data) on uncertainty quantification and propagation such that a fully Bayesian framework - Bayesian multimodel inference instead of BIC or AIC is adopted herein. We also prefer an Monte Carlo(MC)-based non-asymptotic statistical estimator of the model evidence rather than Laplace-based approximation algorithm since the accuracy of model evidence is paramount critical and sensitive to the posterior model probabilities. MC-based integration method provides a accurate and robust estimate of the integral in Eq. (\ref{eq:evidence0}) by generating random samples from the parameter prior distribution as 

% where $N_k$ is the number of random samples. Once the model-form and its probabilities are specified, the model parameter estimation is another pivotal issue that will be discussed in the following section. 

%%%%%%%%%%%%%%%%%%%%%%%%%%%%%%%%%%%%%%%%%%%%%%%%%%%%%%%%%%%%%%%%%%%%%%%%%%%
\section{Bayesian model parameter estimation}
\label{sec:parametric}

The multimodel inference process discussed in the previous section identifies a set of candidate model forms and their associated probabilities. For each of these models $M_j\in\bm{\mathcal{M}}$ there are, of course, additional uncertainties associated with model parameters $\bm{\theta}_j$ These uncertainties are quantified using classical Bayesian inference applying Bayes' rule as
\begin{equation}
{ p }(\bm{\theta}_j|\bm{d},M_j)=\frac { p({ \bm{d}}|\bm{\theta}_j,M_j)p(\bm{\theta}_j|M_j) }{ p(\bm{d}|M_j) } \propto { p({ \bm{d}}|\bm{\theta}_j,M_j)p(\bm{\theta}_j|M_j) }, \quad j=1,\dots, m \label{eq: Bayesian_inference}
\end{equation}
where, again, $p({ \bm{d}}|\bm{\theta}_j,M_j)$ is the likelihood function and $p(\bm{\theta}_j|M_j)$ is the prior probability density. The evidence, $p(\bm{d}|M_j)$, (Eq.\ \eqref{eq:evidence0}) serves only as a normalizing constant in this case. Therefore, unlike in the model selection process, it does not need to be evaluated explicitly as the posterior ${ p }(\bm{\theta}_j|\bm{d},M_j)$ can be estimated from samples using MCMC.

The simplest and most commonly used MCMC algorithms are Metropolis-Hastings (MH) \cite{hastings1970} and Gibbs sampling \cite{geman1984}. In this work, we use an MH-based MCMC algorithm - the \emph{Affine-invariant ensemble sampler} \cite{goodman2010ensemble} proposed by Goodman and Weare \cite{goodman2010ensemble} and implemented in the emcee software package \cite{foreman2013emcee}. The main advantage of this algorithm is that it leverages an ensemble of chains to adapt the proposal density through an implicit affine transformation. This has the effect of greatly improving efficiency for anisotropic and degenerate densities (increasing the acceptance rate {\color{black}while maintaining sample quality}) and significantly reducing the correlation length of the Markov chains yielding independent samples more rapidly. An added benefit is that the method is largely ``self-tuning'' -- requiring only one or two tuning parameters compared with $\sim N^2$ tuning parameters for most MH-based algorithms. For brevity, the reader is referred to \cite{goodman2010ensemble,foreman2013emcee} for algorithm details.

Although the parameter estimation performed here is conventional (nothing new), an important distinction of the multimodel parameter estimation process used herein is that we retain, and propagate, the full joint parameter density. Other conventional methods typically select a single, maximum likelihood parameter value and retain only the corresponding distribution for uncertainty propagation. This has the effect of ignoring parametric uncertainty. In combination with the multimodel selection, the result of our UQ process is therefore a set of probability models $\bm{\mathcal{M}}$ (with associated probabilities $\hat{\pi}_j,\hspace{3pt}j=1,\dots,m$), each of which has an associated joint parameter pdf ${ p }(\bm{\theta}_j|\bm{d},M_j)$. A method for propagation of this complete uncertainty has been previously proposed by the authors \cite{zhang2018} and is reviewed in the following section.

\section{Efficient multimodel uncertainty propagation}
\label{sec:propagation}

Given a set of probability models $\bm{\mathcal{M}}=\{M_j\};\hspace{3pt}j=1,\dots,m$ with \textcolor{black}{posterior} model probabilities $\hat{\pi}_j$ and joint {\color{black}posterior} parameter densities ${ p }(\bm{\theta}_j|\bm{d},M_j)$, uncertainties {\color{black} associated with the random variable $X$} are propagated using a Monte Carlo approach that employs importance sampling reweighting to reduce a nested Monte Carlo analysis to a single Monte Carlo analysis. {\color{black}More specifically, consider that $X$ is now the input to some stochastic system $U=g(X)$. Given uncertainties in the form of the probability model of $X$ (represented by the posterior model probabilities $\hat{\pi}_j$) and uncertainty in the parameters of each candidate model (described by the joint posterior parameter densities ${ p }(\bm{\theta}_j|\bm{d},M_j)$), we aim to quantify uncertainties in the response quantity $U$.

Conventional approaches to solving this type of problem involving uncertain probability distributions require nested Monte Carlo simulations where first the probability model space is sampled. That is, $N_d$ probability models are sampled from the multimodel set according to their associated model probability masses $\hat{\pi}_j$. For each of the $N_d$ model-form samples, the parameter vector $\bm{\theta}_j$ is randomly sampled from the joint parameter pdf ${ p }(\bm{\theta}_j|\bm{d},M_j)$ to obtain a sample pdf (i.e.\ a realization of a specific model-form and it's associated parameters). The set of $N_d$ sample pdfs serves as a finite-dimensional approximation of the total uncertainty (both model-form and parametric) that can be propagated by Monte Carlo simulation. This is achieved by drawing $N_s$ samples from each of the $N_d$ distributions and evaluating the model $N_d\times N_s$ times. Clearly this is a highly inefficient process and is intractable for problems of even moderate computational expense. }

The method proposed in \cite{zhang2018} reduces this expense considerably to a single-loop Monte Carlo by employing importance sampling reweighting (also used in \cite{fetz2016} for propagation of random sets). First, an optimal sampling density is identified by minimizing the expected mean square differences between the sampling density $q(\bm{x})$ and the ensemble of $N_d$ probability models $p_i(\bm{x}|\bm{\theta}),i=1,...,N_d$. This corresponds to solving the following optimization problem under isoperimetric constraint $\hat{{I}}(q)$
\begin{equation}
\begin{aligned}
& \underset{q}{\text{minimize}}
& &\hat{{T}}(q)=E_{\theta}\left [ \int_{\Omega }{\hat{F}(\bm{x}, \bm{\theta}, q(\bm{x}))}d\bm{x} \right] \\
& \text{subject to}
& &\hat{{I}}(q) =  \int_{\Omega}{q(\bm{x})d\bm{x}}-1=0 
\end{aligned}
\label{eq:opt}
\end{equation}
where the action functional $\hat{F}$ is \textcolor{black}{the total square differences}:
\begin{equation}
{\hat{F}(\bm{x}, \bm{\theta}, q(\bm{x}))}={  \frac { 1 }{ 2 } \sum_{i=1}^{N_d}{ \left( { p_i(\bm{x} | \bm{\theta}_i) } - { q(\bm{x}) }  \right) ^{ 2 }}   } \label{eq:MSD_funcitonal} 
\end{equation}
\textcolor{black}{ and $E_{\theta}$ is the expectation with respect to the posterior probability of the model parameters $\bm{\theta}$.} $\hat{{I}}(q)$ ensures that $q(\bm{x})$ is a valid pdf. It is shown that the optimization problem in Eq.\ \eqref{eq:opt} has closed-form solution given by the mixture model \cite{zhang2018}
\begin{equation}
\hat{q}(\bm{x}) =\frac{1}{N_d}  \sum_{i=1}^{N_d}E_{\theta}\left[{p_i({\bm{x} | \bm{\theta}})}\right] \label{eq: opt_MSD2}
\end{equation}
It is straightforward to show that this solution generalizes as:
\begin{equation}
\hat{q}(\bm{x}) =  \sum_{i=1}^{N_d}\hat{\pi}_i E_{\theta} \left [ {  p_i(\bm{x}|\bm{\theta})} \right]  \label{eq: opt_MSD3}
\end{equation}
where $ \hat{\pi}_i$ is the posterior model probability for model $M_i$. 

Samples are drawn from the optimal sampling density $\hat{q}(\bm{x})$ and the \textcolor{black}{response of the system} $g(\bm{x})$ evaluated at each sample point. The statistical response of the \textcolor{black}{system} $g(\bm{x})$ is reweighted according to each of the $N_d$ sample pdfs using importance sampling as
\begin{equation}
{E}_{q}\left[g(\bm{X})\frac{p(\bm{X})}{q(\bm{X})}\right]  = \int_{\Omega}g(\bm{x})\frac{p(\bm{x})}{q(\bm{x})}q(\bm{x})d\bm{x} \approx \frac{1}{N} \sum_{i=1}^{N} g(\bm{x}_i) \frac{p(\bm{x}_i) }{q(\bm{x}_i)}\label{eq:E_ISs} 
\end{equation}
The result is simultaneous propagation of all $N_d$ probability models in the sampled set.

It is shown that this approach may be easily updated to accommodate new information from Bayesian inference as additional data are collected. This will typically reduce the uncertainty associated with the model form and parameters but will come at cost of a loss of optimality in the importance sampling density. If the change in optimal sampling density is considerable, the effect will be an increase in the statistical variance, or potential instability, of the importance sampling estimate. This can be addressed by resampling from the new optimal sampling density, but this is computationally prohibitive. In a parallel work \cite{Jiaxin_Shields_corrected}, a method is proposed to efficiently accommodate a measure change in Monte Carlo simulation that minimizes the impact on the sample set. In other words, it retains as many samples as possible from the original Monte Carlo set drawn from density $\hat{q}(\bm{x})$ and adds a minimal number of samples from a ``correction'' density such that the combined set follows the desired new density $\hat{q}^*(\bm{x})$ while keeping the sample size constant. The method is utilized herein to maintain efficiency as data are added but, because it is not essential to the objective of this work (investigation of the effect of priors), the details are not included. The interested reader is referred to \cite{Jiaxin_Shields_corrected}.

\section{Formulating model and parameter priors}

For a given set of models $\bm{\mathcal{M}}$, the effectiveness of the Bayesian method depends firmly on the specification of the prior model probability $p(M_j)$ and the parameter prior $p(\bm{\theta}_j | M_j)$.  Reasonable choices of prior distributions will have minor effects on posterior inference with well-identified parameters and large data size. However, when datasets are small and/or prior data is not entirely appropriate, specification of prior probabilities becomes very important. In this section, we briefly review approaches for formulating non-informative and data-driven, informative priors for multimodel inference.

% The most common and practical approach to prior specification in this context is to construct noninformative formulations using subjective and empirical Bayes considerations where needed. Theoretically, one often would fail to specify priors that allow the posterior to accumulate probability at the actual model that is determined by the data. However, lack of data  provides only indirect information about the parameters of interest, the prior knowledge becomes more important within the Bayesian framework. In order to reduce the high uncertainty resulting from small data associated with noninformative prior, we aim to investigate the information from past data in an explicitly model-based way to formulate a valid subjective prior in a Bayesian statistical approach. This section describes the prior specifications for prior model probability $p(M_j)$ and parameter prior $p(\bm{\theta}_j | M_j)$, as well as how to formulate the informative, data-driven priors. 

\subsection{Prior model probabilities}
A popular and simple choice for the prior model probability $p(M_j),j=1,...,m$, is the uniform prior 
\begin{equation}
\pi_j=p(M_j) = 1/m
\end{equation}
This prior is noninformative in the sense of favoring all models equally. Under this prior, the posterior model probability is equal to the ratio of the model evidence to the cumulative evidence, 
\begin{equation}
\hat{\pi}_j = p(M_j| \bm{d})  = \dfrac{p(\bm{d} | M_j)}{\sum_{k=1}^mp(\bm{d} | M_k)}\label{eq: model_prior}
\end{equation}
and, as mentioned, these asymptotically correspond to the BIC model weights. However, the apparent noninformativeness of Eq. (\ref{eq: model_prior}) can be deceptive since it is only uniform in probability and will typically not be uniform on the model characteristics. Hence, in setups where several models are very similar and only a few are different, Eq. \eqref{eq: model_prior} may bias the posterior model probability away from accurate models \cite{chipman2001practical}. 

Burnham and Anderson \cite{burnham2004multimodel} make a compelling case that model prior probabilities should depend on dataset size ($n$) and model complexity (i.e. number of model parameters, $K_j$). In other words, small datasets should have priors that favor less complex (lower-dimensional) models to avoid overfitting. This is a major motivation for the use of AIC as a model selection criterion given that it implies the use of savvy priors as discussed in Section \ref{sec:evidence_calc}. But, this effect is expected to be of minimal importance here as all of the considered probability models have comparable complexity.

Finally, model prior probabilities in real-world applications are often selected according to subjective preference, which may result from historical data, the modeler's experience, or solicited expert opinion. This is especially important as it will be shown that strong prior beliefs can greatly influence posterior model probabilities leading to very accurate (if the priors are correct) or inaccurate (if the priors are incorrect) assessments of uncertainty. 

In this work, we consider each of these respective prior model probabilities and aim to understand their influence on uncertainty quantification and propagation. 

% More importantly, the model prior in real application is highly possible selected associated with personal or subjective preference, which may result from previous information including data and experiences or expert opinions. An underlying and correct prediction on model can greatly reduce the uncertainty in system response. 

\subsection{Parameter prior probabilities}
Prior probabilities for model parameters also play an important role in multimodel uncertainty quantification and propagation. Here, we broadly distinguish between so-called noninformative and informative priors and elaborate how these various priors can be constructed under conditions of ignorance (no prior information is available), previously existing (often historical) data, and under subjective assumptions.

\subsubsection{Noninformative priors}
\label{Noninformative priors}
\textcolor{black}{One of the most common noninformative priors is the uniform prior that is flat, diffuse and often considered as ``vague''. It is worth noting that a diffuse or vague prior may not be uniform and sometimes a diffuse prior can be more informative than the uniform prior \cite{berger1989,gelman2006,tenorio2017}. The uniform prior can be expressed as}   
%(One of the most common noninformative priors is the uniform prior (also referred to as the diffuse, vague, or flat prior \cite{gelman2006, berger1989})  
\begin{equation}
p(\bm{\theta}_j | M_j) = \textup{Constant}, \ \ \ \ \bm{\theta}_j \in \Omega_{\theta_j} \ \ \ \ 1 \le j \le m 
\end{equation}
where the range of $\bm{\theta}_j$, $\Omega_{\theta_j}$ is a subset of the parameter space $\Theta_j$ ($\Omega_{\theta_j} \subset \Theta_j$).This indicates that there is no \emph{a priori} reason to favor any particular parameter value. Instead, we only know its range $\bm{\theta}_j \in \Omega_\theta$. Thus, the posterior distribution in Eq.\ (\ref{eq: Bayesian_inference}) is proportional to the likelihood, 
\begin{equation}
{ p }(\bm{\theta}_j|\bm{d},M_j) \propto { p({ \bm{d}}|\bm{\theta}_j,M_j)}, \ \ \ \ \bm{\theta}_j \in \Omega_{\theta_j} \ \ \ \ 1 \le j \le m  \label{eq:Bayesian_inference}
\end{equation}
If the range $\Omega_{\theta_j}$ is specified as the parameter space $\Omega_{\theta_j} = \Theta_j$, Bayesian inference assuming a flat prior may cause an improper prior if 
\begin{equation}
{ \int_{\Theta_j} {p(\bm{\theta}_j|M_j)d\bm{\theta}_j }  } = \infty, \ \ \ \ 1 \le j \le m \label{eq:evidence}
\end{equation}
In this case the normalizing constant sometimes does not exist. If an improper prior is employed, one needs to be sure that the posterior is proper. 

Another commonly used noninformative prior is the Jeffreys prior \cite{jeffreys1946}, which is defined to be proportional to the square root of the determinant of the Fisher information matrix
\begin{equation}
{ p }(\bm{\theta}_j|\bm{d},M_j)  \propto \left| J(\bm{\theta}_j) \right| ^{1/2},  \ \ \ \ 1 \le j \le m  
\end{equation}
The Fisher information is given as:
\begin{equation}
\  J(\bm{\theta}_j)  = -\int { E\left[ \frac { { \partial  }^{ 2 }\log p(\bm{x}|\bm{\theta}_j, M_j ) }{ \partial \bm{\theta}_j \partial { \bm{\theta}  }_{j}^{ T } }  \right]  } p(\bm{x}|\bm{\theta}_j, M_j )d \bm{x}, \ \ \ \ 1 \le j \le m 
\end{equation}
For certain models, the Jeffreys prior cannot be normalized and is therefore an improper prior in such cases. 

In this work, we employ proper uniform priors as representative noninformative priors. While this admittedly does not account for various nuances that may arise from assuming different noninformative priors, the intention here is to compare the effects of a suitably representative noninformative prior on multimodel uncertainty quantification and propagation against the effects of various informative priors.

% Many statisticians favor noninformative priors since they appear to be more objective, and most of the theoretical work on prior distribution has been on setting up rules for noninformative prior distribution that satisfy various invariance principles \cite{berger2006, gelman2006, berger1989}. However, it is unrealistic to expect that noninformative prior stands for total ignorance about the parameter of interest, instead, noninformative prior allow a more objective approach to inference by estimating the parameters from limited data rather than requiring them to be specified using subjective information. 

\subsubsection{Informative priors}
\label{sec:inform_prior}
At the other extreme, an informative prior is one that yields a posterior that is not dominated by the likelihood; instead an informative prior has an essential impact on the posterior distribution. This is especially true for inference from small datasets. The appropriate use of informative priors illustrates the power of the Bayesian approach: information gathered from previous studies, past experiences or expert opinions can be combined with new data in a natural way. We can therefore interpret an informative prior as the state of our subjective prior knowledge. However, in practice, prior specification of subjective knowledge might be biased as it is often difficult to specify precisely and \textcolor{black}{historical} data, experiments, and experiences may not be totally appropriate for the current problem. One objective of this study is to understand the influence of such imprecise and/or incorrect informative priors on multimodel uncertainty quantification and propagation.

% \subsection{Informative priors formation based on previous data}
In this work, we formulate data-driven informative priors by exploiting \textcolor{black}{historical} data, denoted $\hat{\bm{d}}$, as may be available in the literature. We specifically avoid formulating priors based on assumptions or intuition. In this sense, the \textcolor{black}{historical} data $\hat{\bm{d}}$ represents the existing state of knowledge as objectively as possible. Yet, as previously mentioned, these data may not be entirely appropriate for the problem at hand and therefore may or may not provide ``good'' priors.

The data-driven prior is quantified by applying Bayes' rule to the \textcolor{black}{historical} data, $\hat{\bm{d}}$. The posterior then becomes the prior for the analysis using the currently observed data, $\bm{d}$. This initial Bayesian inference starts with a suitable noninformative prior, termed the ``pre-prior''. Within this framework, the currently observed data $\bm{d}$ is effectively treated as an extension of the \textcolor{black}{historical} data $\hat{\bm{d}}$. If the \textcolor{black}{historical} dataset is relatively large, the resulting prior is referred to as strongly informative and dominates the pre-prior. If the \textcolor{black}{historical} dataset is small, the resulting prior is referred to as weakly informative and retains some influence of the noninformative pre-prior. 

The approach used in this work is summarized in three stages as follows: 

% Several methods for formulating an informative prior all seek to make use of previous data $\hat{\bm{d}}$ in determining a prior pdf to represent the prior knowledge \cite{guikema2007}.  A straightforward way to construct an informative data-driven prior $p^*(\bm{\theta}_j | M_j)$, is to identify the previous data $\hat{\bm{d}}$ as a separate data which is used for updating an appropriate prior. If no additional past information is assumed, it is reasonable to start with a noninformative prior $p(\bm{\theta}_j|M_j)$ as discussed in Section \ref{Noninformative priors}, update this prior with the previous data $\hat{\bm{d}}$, and use the posterior pdf $p(\bm{\theta}_j|\hat{\bm{d}},M_j)$ from this first update as the prior $p^*(\bm{\theta}_j | M_j) =p(\bm{\theta}_j|\hat{\bm{d}},M_j)$ that is to be updated with the observed data $\bm{d}$. The initial updated prior $p^*(\bm{\theta}_j | M_j)$ is named the ``pre-prior". Within this framework, the current observed data $\bm{d}$ is effectively treated as an extension of the previous data $\hat{\bm{d}}$, which is assumed to dominate the pre-prior. This approach can be characterized into four stages as follows: 
\begin{itemize}
\item {\it Stage 1: Noninformative pre-prior } - Noninformative pre-priors $\hat{p}(\bm{\theta}_j|M_j)$ can be developed in a number of different ways. When the likelihood function $\hat{p}(\hat{\bm{d}} |\bm{\theta}_j,M_j )$ is given, one can derive the noninformative prior based on Jeffrey's rule, or simple use a flat prior instead. 

\item {\it Stage 2: Pre-Bayesian inference} - A pre-Bayesian inference is employed herein to identify the the posterior distribution based on \textcolor{black}{historical} data $\hat{\bm{d}}$ combined with a given noninformative prior $\hat{p}(\bm{\theta}_j | M_j)$ and the specified model $M_j$
\begin{equation}
p^*(\bm{\theta}_j | M_j) =\hat{p}(\bm{\theta}_j|\hat{\bm{d}},M_j)=\frac { \hat{p}({\hat{\bm{d}}}|\bm{\theta}_j,M_j )\hat{p}(\bm{\theta}_j|M_j) }{ \int_{\Theta_j} {\hat{p}({ \hat{\bm{d}}}|\bm{\theta}_j,M_j)\hat{p}(\bm{\theta}_j|M_j)d\bm{\theta}_j }  }, \quad 1 \le j \le m \label{eq:pre_BI}
\end{equation}
The posterior distribution $\hat{p}(\bm{\theta}_j|\hat{\bm{d}},M_j)$ is taken as the prior distribution $p^*(\bm{\theta}_j | M_j)$ for the currently observed data $\bm{d}$.

\item {\it Stage 3: Nonparametric estimate from posterior samples} - Eq.\ \eqref{eq:pre_BI} is typically solved implicitly using MCMC. Therefore, the data-driven prior is not available in closed-form for Bayesian updating using the new data, $\bm{d}$. A nonparametric kernel density estimate is therefore used to approximate the unknown prior probability density function from the MCMC samples. 

% The posterior knowledge resulting from pre-Bayesian inference in Stage 2 is often identified as a form of posterior samples generated by MCMC algorithm.  A nonparametric estimate is necessary for approximating of unknown pdf utilizing only information from their samples. Kernel density estimation (KDE) is widely acknowledged as one of the most popular nonparametric techniques for this purpose \cite{elgammal2002,botev2010}. 

For multivariate density functions involving parameter vector  $\bm{\theta}_j$ with dimension $K_j$, the kernel density estimate $\tilde{f}(\bm{\theta}_j|M_j)$ has the form given a sample set $\bm{\theta}_j = \left\{\bm{\theta}_{j}^{1}, \bm{\theta}_{j}^{2},...,\bm{\theta}_{j}^{n} \right\}$ of size $n$ given model $M_j$ \cite{scott2015multivariate} as
\begin{equation}
\tilde { f } ({ \bm{\theta}  }_{ j }|M_j)=\frac { 1 }{ n } \sum _{ k=1 }^{ n } \prod _{ i=1 }^{ K_j }{ \left\{ \frac { 1 }{ w_{ i } } \phi\left (\frac { \bm{\theta} _{ j,i }-\bm{\theta} _{ j,i }^{k} }{ w_{ i } }  \right ) \right\}  } , \ \ \ \ 1 \le j \le m \label{eq:KDE}
\end{equation}
where $\bm{\theta} _{ j,i }^{k} $ is the $k^{th}$ sample in the $i^{th}$ dimension of $\bm{\theta}_j$ given model $M_j$, and $w_i$  the corresponding bandwidth. $\phi(\cdot)$ is a chosen Gaussian kernel given by 
\begin{equation}
\phi(x) = \frac{1}{\sqrt{2\pi}}e^{-x^2/2}
\end{equation}
The kernel bandwidth is then determined by minimizing the asymptotic mean integrated square error (AMISE) \cite{silverman1986density} such that, for the Gaussian kernel, the optimal bandwidth is
\begin{equation}
w_{ i }^{ opt }=\left[  \frac { 4 }{ K_j+2 } \right]^{ 1/(K_j+4) }n^{ -1/(K_j+4) }\sigma _{ i }
\end{equation}
where $\sigma_i$ is the standard deviation of the samples $ \left\{\bm{\theta}_{j,i}^{1}, \bm{\theta}_{j,i}^{2},...,\bm{\theta}_{j,i}^{n} \right\}$. The kernel density estimate $\tilde { f } ({ \bm{\theta}  }_{ j } |M_j)$ is then employed as the informative prior for Bayesian inference using the observed data $\bm{d}$.

\end{itemize}

\section{Application: plate buckling strength problem}
Uncertainty in the material and geometric properties of ship structural components can significantly impact the performance, reliability and safety of the structural system \cite{ClassNK}. In this work, we apply the proposed methodology to quantify and propagate the uncertainty in material properties for  buckling strength of a simply supported rectangular plate under uniaxial compression. An analytical formulation for the normalized buckling strength for a pristine plate was first proposed by Faulkner\cite{faulkner1973}
\begin{equation}
\psi =\frac { { \sigma  }_{ u } }{ { \sigma  }_{ 0 } } =\frac { 2 }{ \lambda  } -\frac { 1 }{ \lambda ^{ 2 } }  \label{eq:Faulkner } 
\end{equation}
where $\sigma_u$ is the ultimate stress at failure, $\sigma_0$ is the yield stress, and $\lambda$ is the slenderness of the plate with width \(b\), thickness \(t\), and elastic modulus \(E\) given by 
\begin{equation}
\lambda =\frac { b }{ t } \sqrt { \frac { \sigma _{ 0 } }{ E }  } \label{eq:slenderness}
\end{equation}
Eq. (\ref{eq:slenderness}) was further modified by Carlsen \cite{carlsen1977} to study the effect of residual stresses and non-dimensional initial deflections \(\delta_{0}\)  associated with welding
\begin{equation}
\psi =\left( \frac { 2.1 }{ \lambda  } -\frac { 0.9 }{ \lambda ^{ 2 } }  \right) \left( 1 -\frac { 0.75\delta_{0} }{ \lambda }  \right)\left( 1 -\frac { 2\eta t }{b}  \right) \label{eq:buckling_strenth} 
\end{equation}
where $\eta t$ is the width of the zone of tension residual stress.

The design buckling strength is based on nominal values for the six variables in Eq. (\ref{eq:buckling_strenth}) provided in Table \ref{tab:defination_variables}. However, the actual values of these variables often differ from the design values due to uncertainties in the material properties and ``as built" geometry yielding uncertainty in the buckling strength. We are therefore interested in investigating the effect of the six uncertain variables shown in Table \ref{tab:defination_variables} on the buckling strength of simply supported mild steel plates. Emphasis is placed on assessing the influence of uncertainty in the yield strength $\sigma_0$ since it is the most sensitive variable identified by Global sensitivity analysis (see Table \ref{tab:defination_variables}) and for clarity of demonstration. 

\begin{table}[!ht] \footnotesize
\centering
	\caption{ Statistical properties of plate material, geometry and imperfection variables from Hess et al. \cite{hess2002} and Guedes Soares \cite{soares1988}}
	\label{tab:defination_variables}
	\begin{tabular}{cccccc}
		\hline
		Variables & Physical Meaning   & Nominal Value & Mean        & COV    & Global Sensitivity \\ \hline
		\(b\)         & width              &36            & 0.992*36    & 0.028  & 0.017              \\
		\(t\)         & thickness          & 0.75           & 1.05*0.75    & 0.044  & 0.045              \\
		\(\sigma_0\)    & yield strength           & 34            & 1.023*34      & 0.116 & 0.482              \\
		\(E\)         & Young's modulus      & 29000         & 0.987*29000 & 0.076  & 0.194              \\
		\(\delta_0\)    & initial deflection & 0.35          & 1.0*0.35    & 0.05   & 0.043              \\
		\(\eta\)      & residual stress    & 5.25          & 1.0*5.25    & 0.07   & 0.233              \\ \hline
	\end{tabular}
\end{table}

\subsection{Description of \textcolor{black}{historical} data}

The work of Hess et al. \cite{hess2002} presented a review of uncertainties in material and geometric properties for mild steel plates for ship building applications. They conducted statistical analysis of data compiled from tests/measurements sponsored by the Ship Structure Committee (SSC) \cite{mansour1984,atua1996} as part of an effort to establish a database of marine steel properties and tests/measurements performed by the Naval Surface Warfare Center, Carderock Division (NSWCCD). These past sources of yield strength data are very important because they provide a valuable source of prior information. However, it remains difficult to represent the uncertainties since the measured data are scarce. Hence, quantification of uncertainties and variations is necessary to determine the probabilistic characteristics of these random variables. In this work, we make use of the \textcolor{black}{historical} experimental data to predict the probabilistic characteristics of yield strength of mild steel. The source of the material property data are a series of historical reports from the SSC-352 \cite{kufman1990}, SSC-142 \cite{gabriel1962} and SSC-145 \cite{boulger1962}, which include material from four classes of structural steels summarized as follows

\begin{itemize}
\item {\it ABS-A} - plates with thickness not exceeding 1/2 inch and all shapes
\item {\it ABS-B} - plates with thickness over 1/2 inch but not exceeding 1 inch
\item {\it ABS-C} - plates with thickness  over 1 inch 
\item {\it ASTM-A7} - Historical conventional structural steel alloy replaced by ASTM-A36
\end{itemize}

The three ABS steels are typical ship-building and marine steels and vary somewhat in chemical composition but possess nominally the same design properties (most notably $\sigma_0 = 34$ ksi) while the ASTM-A7 is a historical carbon steel having design yield strength in the range $\sigma_0 = [30,33]$ ksi. The statistical analysis of these data are reproduced from Hess et al. \cite{hess2002} in Table  \ref{tab: data_summary}. These data are useful for our purposes since they are representative of the type of historical data (these tests data back to 1948) that may be available for assigning prior distributions in Bayesian inference but are not truly representative of what may be expected from modern materials. Thus, the statistical analysis of the four materials provided by \cite{hess2002} give us different priors from which to initiate our investigation. 

\begin{table}[!ht] \footnotesize
\centering
\caption{Statistical information and comments of informative knowledge from \textcolor{black}{historical} data}
\label{tab: data_summary}
\begin{tabular}{cccccccc}
\hline
Steel type & Min  & Max  & Mean   & COV   & Distribution & \# of tests & Comments                                                                              \\ \hline
ABS-A      & 31.9 & 39.6 & 36.091 & 0.059 & Lognormal    & 33          & \begin{tabular}[c]{@{}c@{}}Weakly informative \\ but incorrect\end{tabular}  \\
ABS-B      & 27.6 & 46.8 & 34.782 & 0.116 & Lognormal    & 79          & Informative and correct                                                                     \\
ABS-C      & 30.9 & 41.5 & 33.831 & 0.081 & Lognormal    & 13          & \begin{tabular}[c]{@{}c@{}}Weakly informative \\ but incorrect\end{tabular}                                                                \\
ASTM-A7       & 28.6 & 49.4 & 38.197 & 0.108 & Normal       & 58          & Informative but incorrect \\ \hline
\end{tabular}
\end{table}

The application of interest here is a ship structural plate with thickness $t = 0.75$ inch. It is therefore of ABS-B material class and we assume that the ``true'' model for the ABS-B material is that given in Table 2. Note that, in reality, this is not in fact the true model for ABS-B material but for our purposes it provides a baseline from which we have an {\it informative} and {\it correct} prior. The ABS-A and ABS-C materials are similar to the ``true" ABS-B material and their datasets smaller so are considered to provide weakly informative but technically incorrect priors. Finally, the ASTM-A7 material is considerably different. Given that it is a comparatively large dataset, we consider it to give an informative but incorrect prior. Note that, under practical conditions of limited data, an analyst may consider any one of these data sets to be ``close enough" so as to define a prior for UQ (justifiably or not). Our objective is to study the influence of using these different priors in the context of multimodel Bayesian UQ.

% and for the sake of illustration the yield strength is assumed to vary according to the Lognormal distribution for ABS-B in Table \ref{tab: data_summary}. This     

\begin{figure}[!ht]	
	\centering
	\subfigure[]{\includegraphics[height=1.6in]{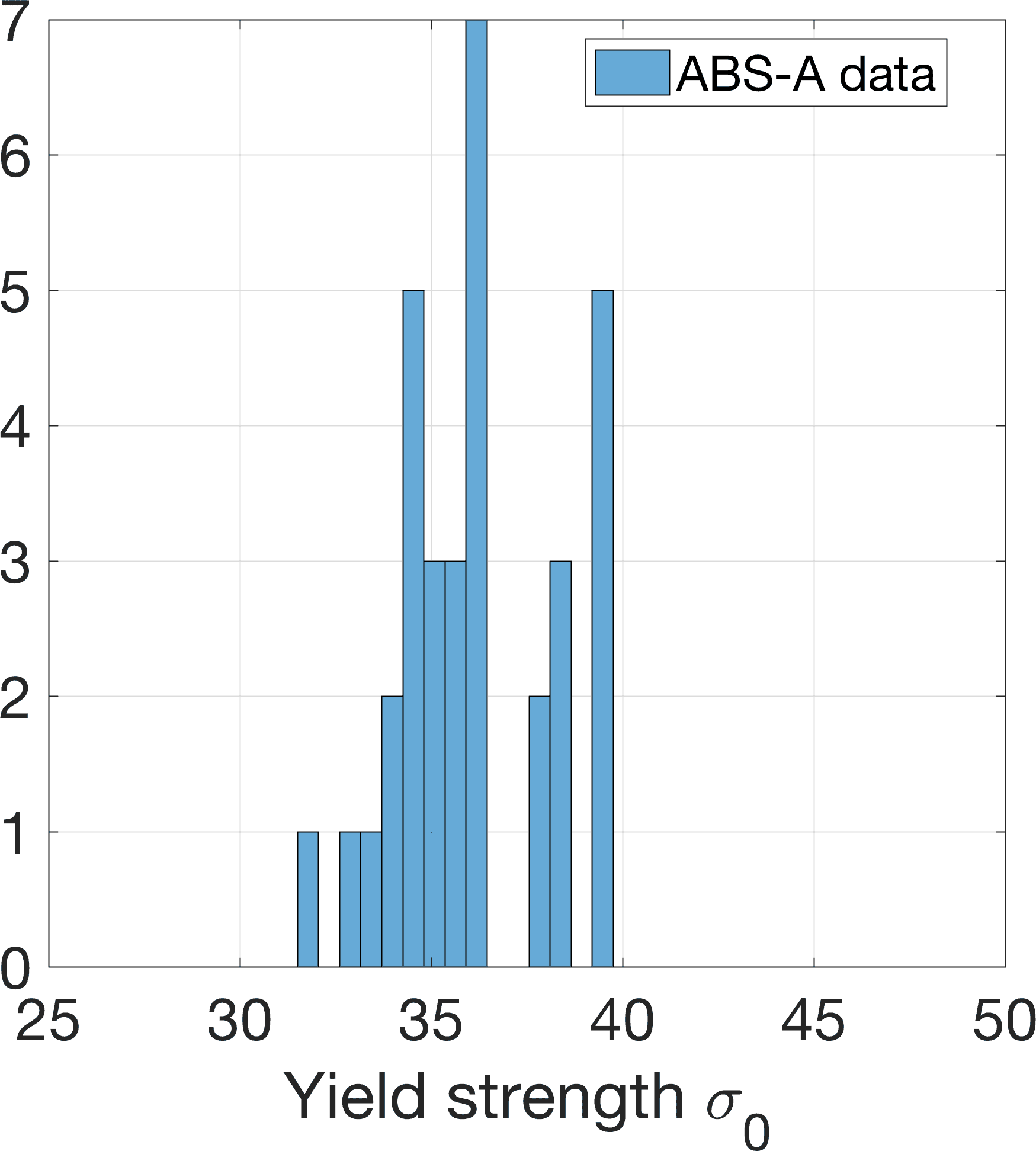}}
	\subfigure[]{\includegraphics[height=1.6in]{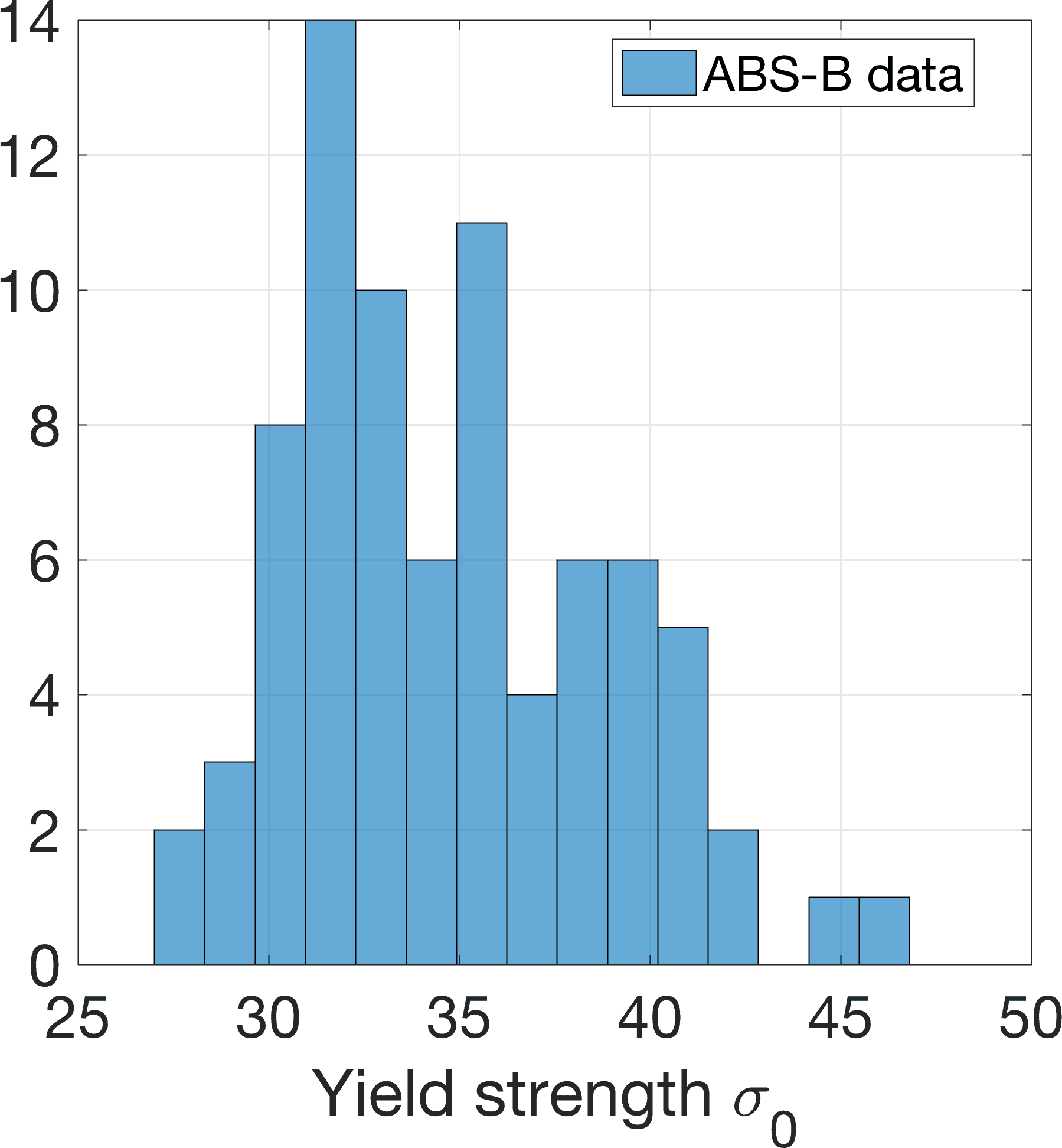}} 
	\subfigure[]{\includegraphics[height=1.6in]{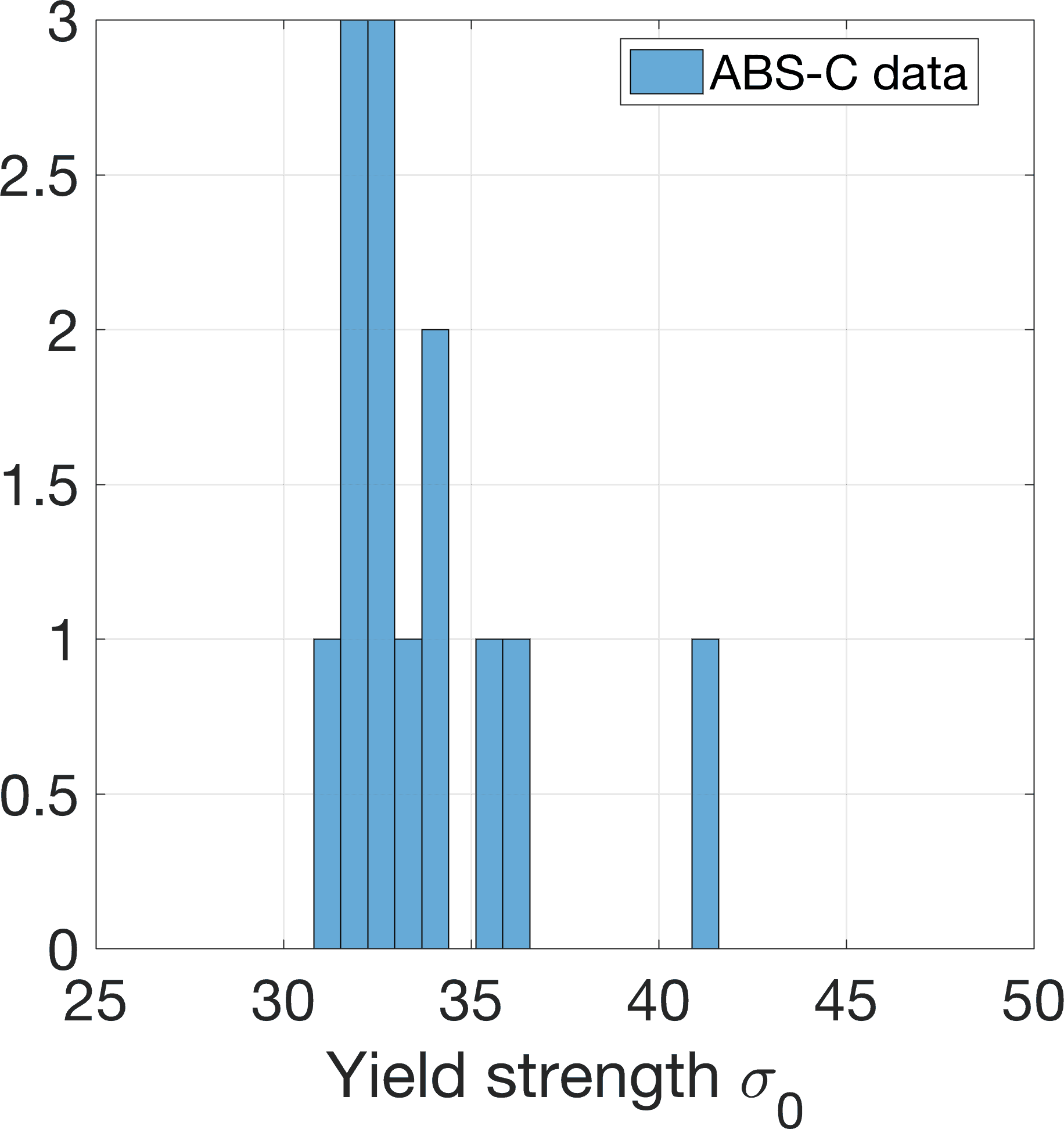}}
	\subfigure[]{\includegraphics[height=1.6in]{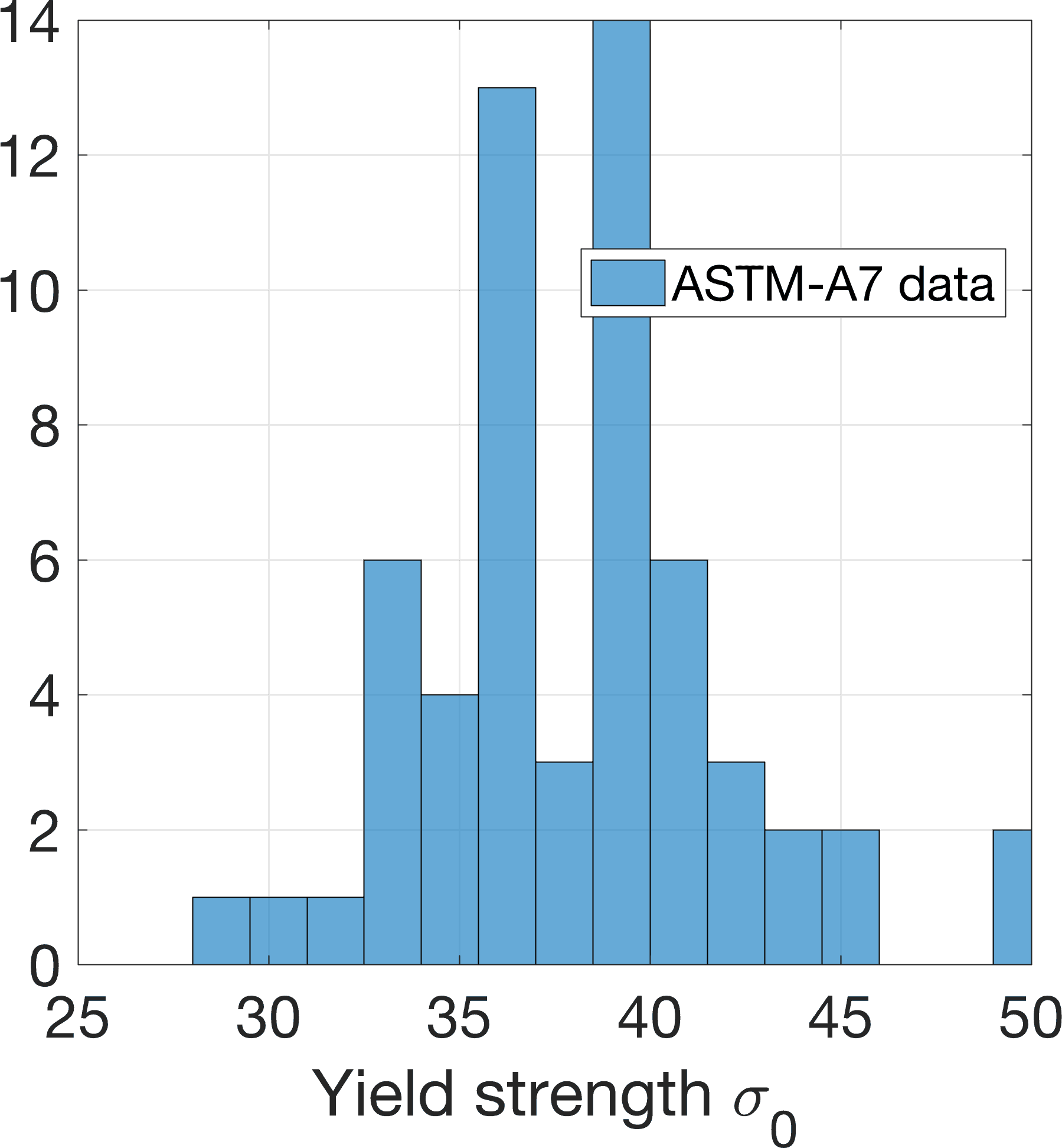}}
	\caption[]{Histograms of material data for (a) ABS-A, (b) ABS-B, (c) ABS-C and (d) ASTM-A7}  \label{fig:ABS_hist}
\end{figure}

\begin{figure}[!ht]	
	\centering
	{\includegraphics[height=1.8in]{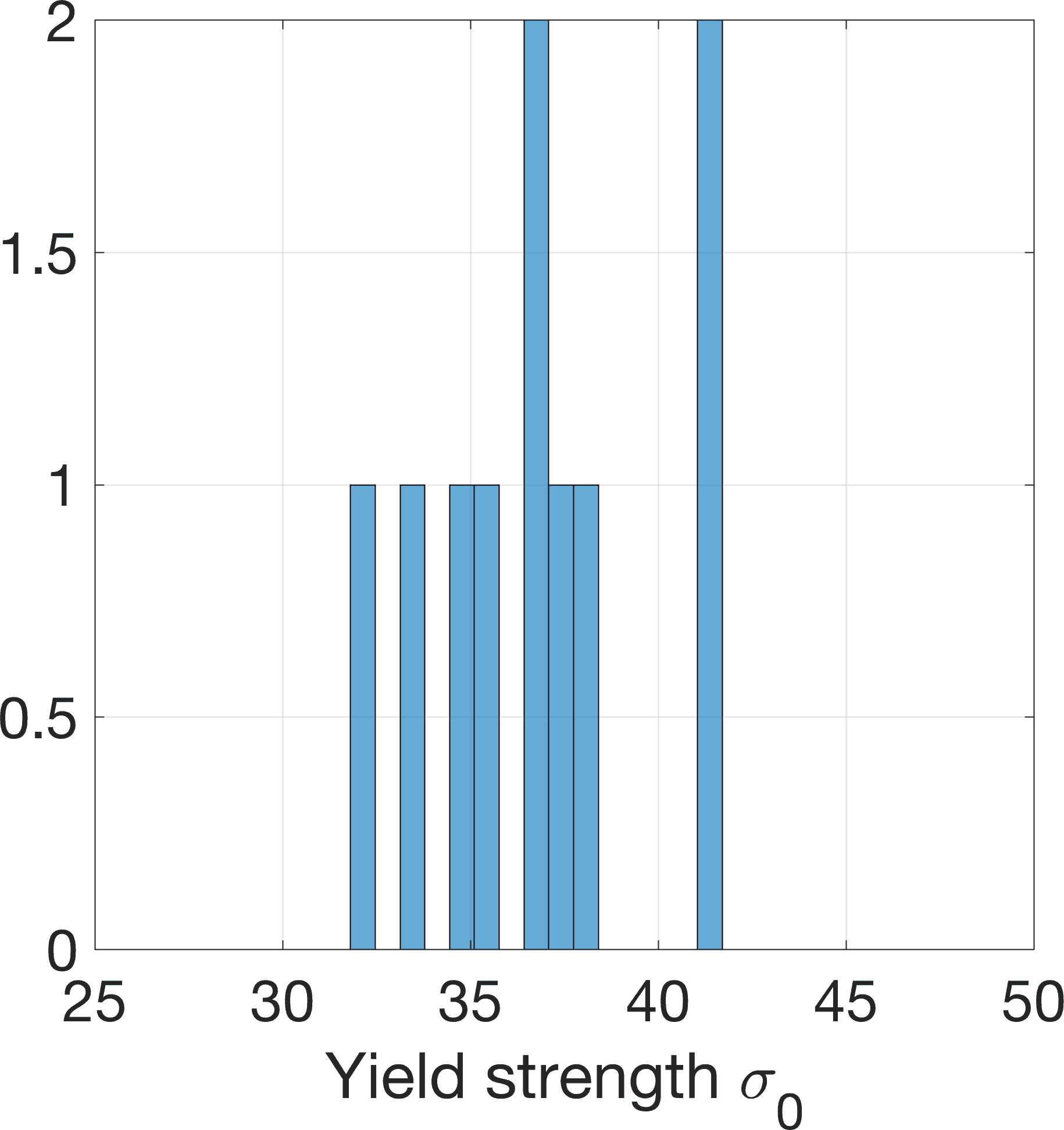}}
	\caption[]{Ten randomly sampled yield strength data that serve as the initial dataset}  \label{fig:sigma_10data_hist}
\end{figure}

Figure \ref{fig:ABS_hist} shows histograms of the material data for each of the four classes: ABS-A, ABS-B, ABS-C and ASTM-A7.  
The ABS-B material data collected from the technical report SSC-142 \cite{gabriel1962} has mean $\mu=34.782$, coefficient of variation 0.116, and is assumed to follow a lognormal distribution. Again, we assume this to be the ``true'' model and, for our investigation, all ``data'' are synthetically generated from $\sigma_0 \sim Lognormal(\mu_{{\sigma}_0}=34.782, \sigma_{{\sigma}_0}=0.116*34.782)$. The initial 10 yield strength values are shown in Figure \ref{fig:sigma_10data_hist}. 
Given these 10 values and the prior data, we contend that a single probability model form cannot be precisely identified. Therefore, we select seven candidate probability models including Gamma, Inverse Gaussian, Logistic, Loglogistic, Lognormal, Normal and Weibull. 
For each of these models, prior parameter densities are derived from each dataset in Figure \ref{fig:ABS_hist} as described in Section \ref{sec:inform_prior} and Bayesian inference performed in the following.

\subsection{Influence of data-driven priors on uncertainty quantification}

\begin{figure}[!ht]	
	\centering
	{\includegraphics[height=1.4in]{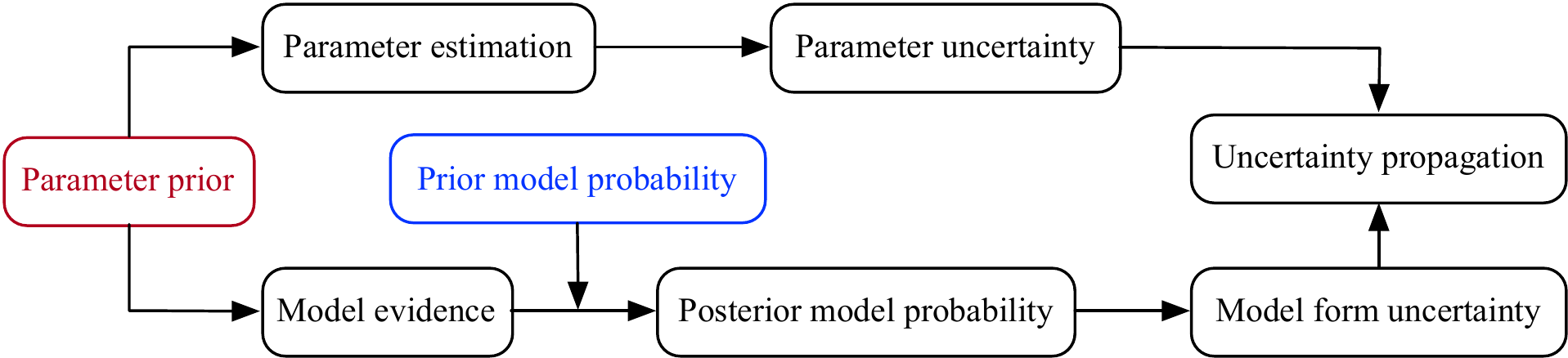}}
	\caption[]{Influence of parameter prior and model prior probability on uncertainty quantification and propagation}  \label{fig: effect of prior on UQ}
\end{figure}

In multimodel Bayesian UQ, there are two stages of inference related to model-form uncertainty (Section \ref{sec:multimodel}) and model parameter uncertainty (Section \ref{sec:parametric}). There is an interesting interplay between these two stages of inference as suggested by Eqs.\ \eqref{eq: evidence}-\eqref{eq:evidence0} and the flowchart in Figure \ref{fig: effect of prior on UQ}.
In the first stage, multiple candidate models are considered (i.e.\ the seven model listed above) and some assumptions are made regarding their prior model probabilities - perhaps informed by expert opinion. As data are collected, the model probabilities are updated using Bayes' rule (Eqs.\ \eqref{eq: evidence}-\eqref{eq:evidence0}). But, these updated probabilities are influenced by the selection of the parameter prior in Eq.\ \eqref{eq:evidence0} and, as we will see, this can play an important role in model selection. In the second stage, the model parameter distributions are inferred from the data for each model form. These are obviously strongly dependent on the prior parameter densities. These inference processes combine to provide the posterior information used to quantify uncertainty in the parameter of interest (here $\sigma_0$). The forthcoming Sections \ref{sec:model-form_effect}
and \ref{sec:parameter_effect} aim to answer the question: What influence do prior assumptions (in model-form and model parameters) have on the accuracy and convergence of posterior probabilities?

\subsubsection{Effect of priors on model-form uncertainty}
\label{sec:model-form_effect}

In Bayesian model selection, it is common to assume equal prior probability (i.e.\ $\pi_j=P(M_j)=1/m=1/7$ \cite{chipman2001practical}. In certain instances, subjective non-equal probabilities may be assigned. In fact, for our problem the existing literature suggests a ``preferred'' distribution for $\sigma_0$ (Hess et al.\ \cite{hess2002} suggest a lognormal distribution). With this information, we assign a prior model probability $\pi_{LN}=0.9$ and assume equal weight ($\pi_j=\frac{1-0.9}{6},\hspace{3pt}j\ne LN$) for the other models. Because there is a strong belief in the correct prior model, we refer to this as the ``strong correct'' prior. This strong correct prior will be compared against the uniform prior of equal probabilities as well as a ``strong incorrect'' prior wherein there is strong belief in the incorrect log-logistic model such that it has prior probability $\pi_{LL}=0.9$ and all other prior probabilities are equal. The three model prior cases we consider are summarized in Table \ref{tab:model_prior}.

\begin{table}[!ht] \footnotesize
\centering
\caption{Prior model probabilities.}
\label{tab:model_prior}
\begin{tabular}{cccc}
\hline
 & Uniform  & ``Strong Correct'' & ``Strong Incorrect'' \\ \hline
Gamma        		& 1/7	& 0.0167	& 0.0167                     \\
Inverse Gaussian    & 1/7	& 0.0167	& 0.0167                     \\
Logistic            & 1/7	& 0.0167	& 0.0167                     \\
Log-logistic        & 1/7	& 0.0167	& 0.9                     \\
Lognormal           & 1/7	& 0.9		& 0.0167                     \\ 
Normal           	& 1/7	& 0.0167	& 0.0167                     \\
Weibull           	& 1/7	& 0.0167	& 0.0167                     \\
\hline
\end{tabular}
\end{table}

As data are collected, posterior model probabilities are updated according to Eqs.\ \eqref{eq: evidence}-\eqref{eq:evidence0}. These probabilities depend on the parameter prior assumption and, as a result, they will differ based on the \textcolor{black}{historical} data we use to derive the prior. In the small data case, it can be quite difficult to make any meaningful conclusions regarding model probabilities as evidenced by the data in Table \ref{tab: model_weights_10data}, which gives the posterior model probabilities from 10 yield stress data for each of the parameter priors given equal prior model probabilities. Note that in these cases, the posterior is simply equal to the model evidence. 
\begin{table}[!ht]  \footnotesize
\centering
\caption{Posterior model probabilities given initial 10 data and different parameter priors.}
\label{tab: model_weights_10data}
\begin{tabular}{ccccccc}
\hline
Distribution     & AIC   & Noninformative & ABS-A & ABS-B & ABS-C & ASTM-A7 \\ \hline
Gamma            & 0.168 & 0.167          & 0.159 & 0.157 & 0.170 & 0.166   \\
Inverse Gaussian & 0.172 & 0.184          & 0.142 & 0.150 & 0.132 & 0.191   \\
Logistic         & 0.119 & 0.115          & 0.161 & 0.118 & 0.064 & 0.136   \\
Loglogistic      & 0.128 & 0.125          & 0.182 & 0.096 & 0.063 & 0.163   \\
Lognormal        & 0.167 & 0.162          & 0.184 & 0.140 & 0.182 & 0.176   \\
Normal           & 0.154 & 0.149          & 0.147 & 0.178 & 0.189 & 0.130   \\
Weibull          & 0.091 & 0.098          & 0.024 & 0.160 & 0.201 & 0.037   \\ \hline
\end{tabular}
\end{table}
This is a classic small data case where a precise ``best'' model is impossible to identify. Moreover, as suggested by the definition of model evidence, these posterior model probabilities are strongly dependent on the parameter prior with considerable differences across different priors.  

\begin{figure}[!ht]	
	\centering
	\subfigure[]{\includegraphics[height=2in]{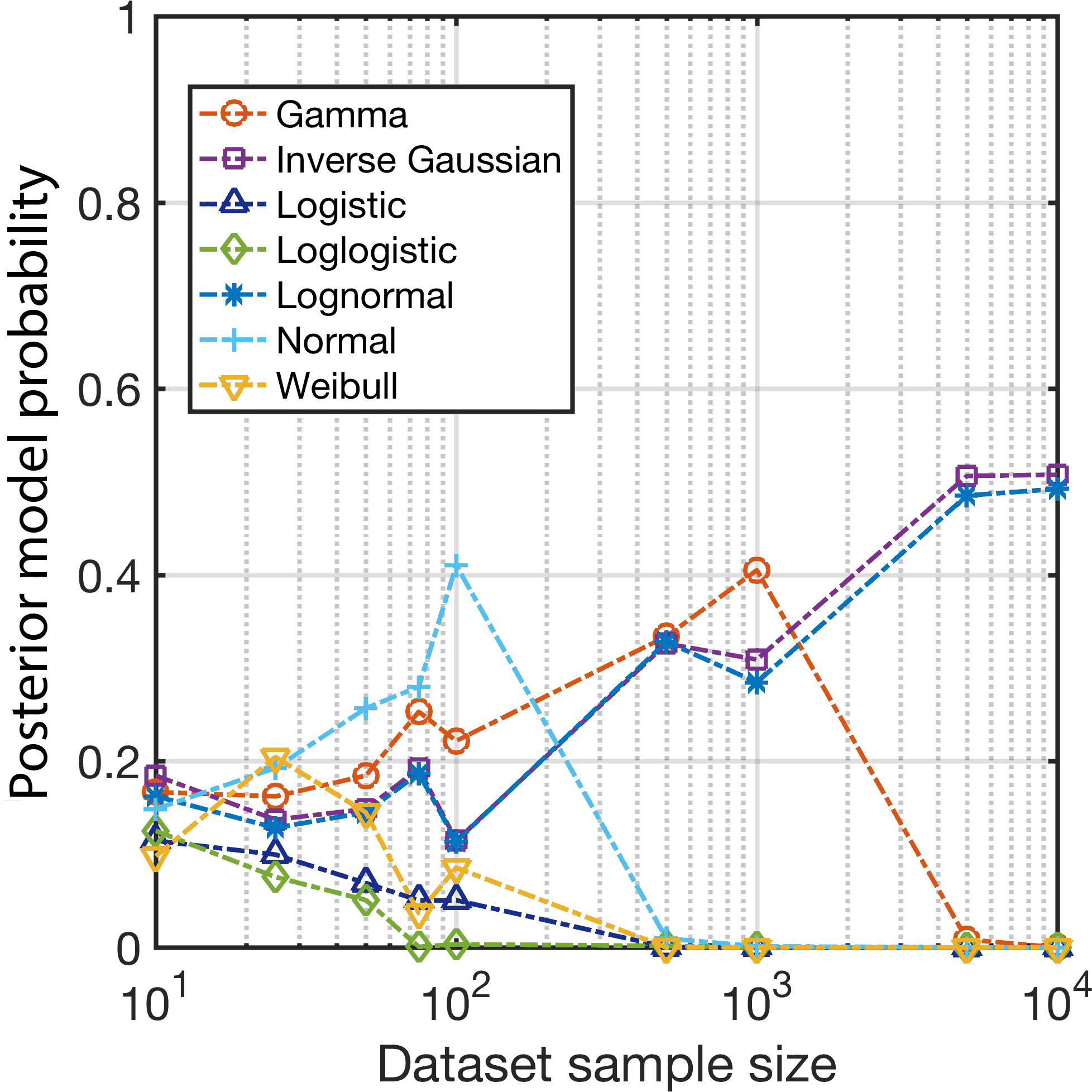}}
    \subfigure[]{\includegraphics[height=2in]{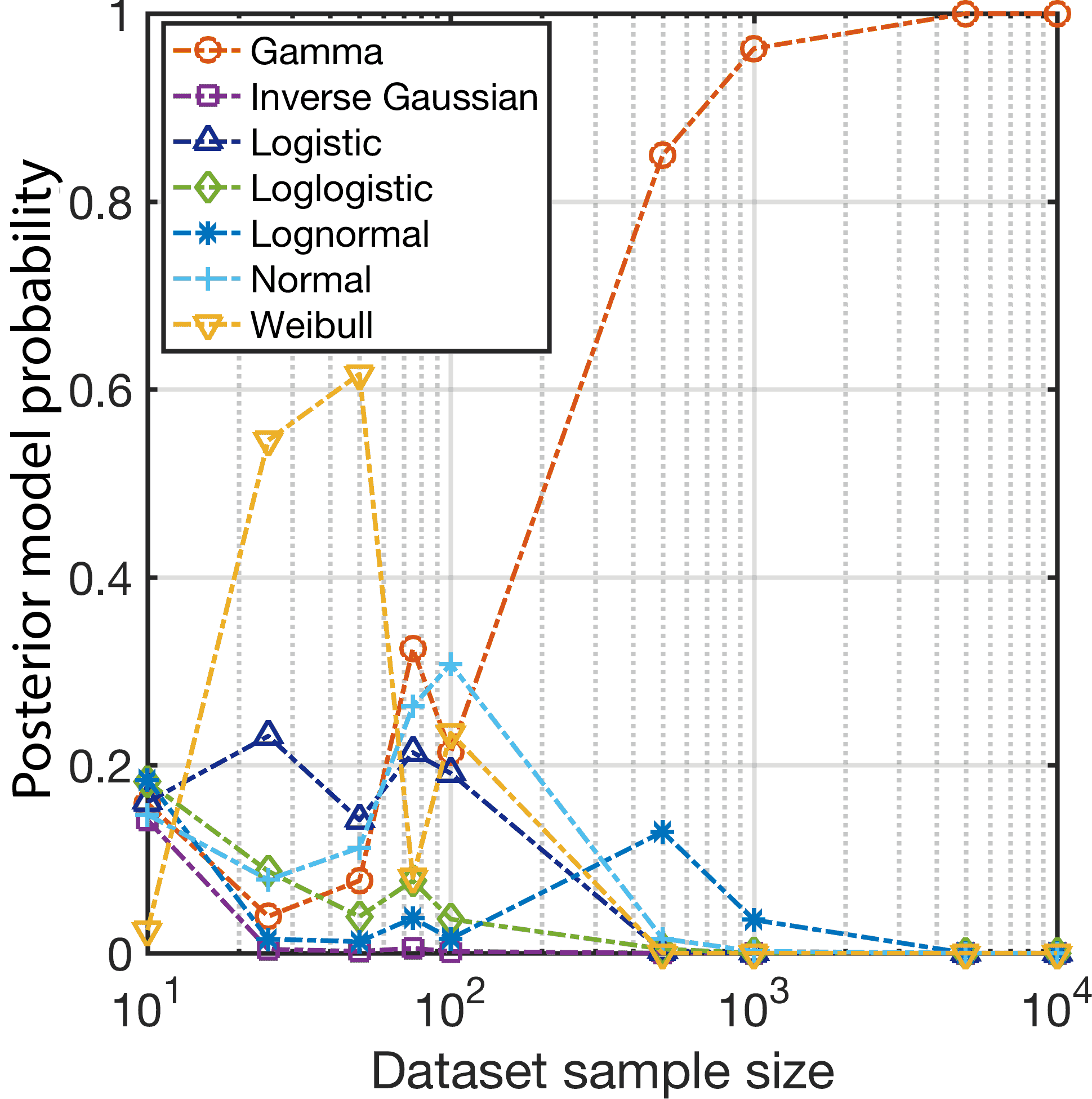}}
	\subfigure[]{\includegraphics[height=2in]{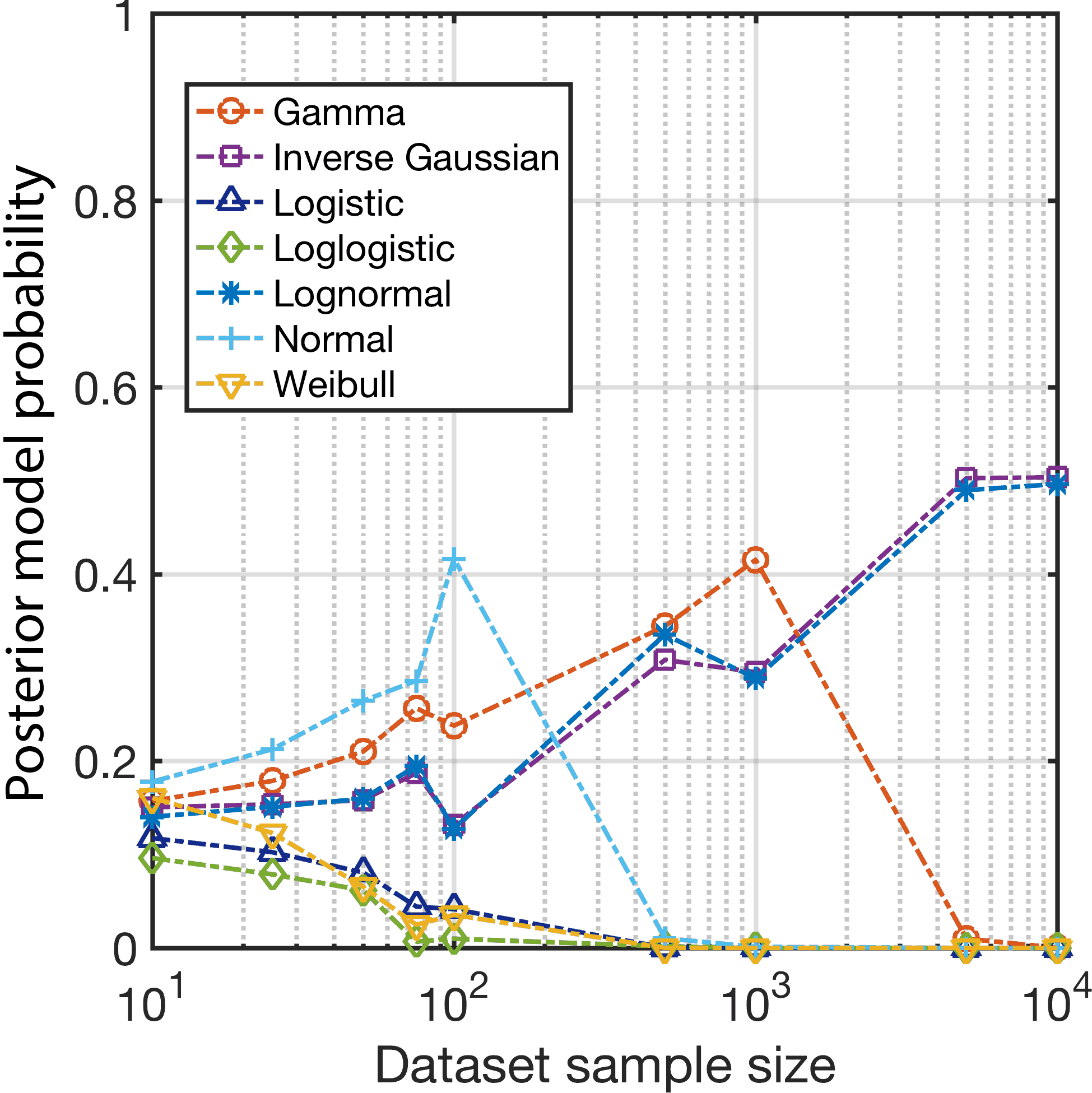}} 
	\subfigure[]{\includegraphics[height=2in]{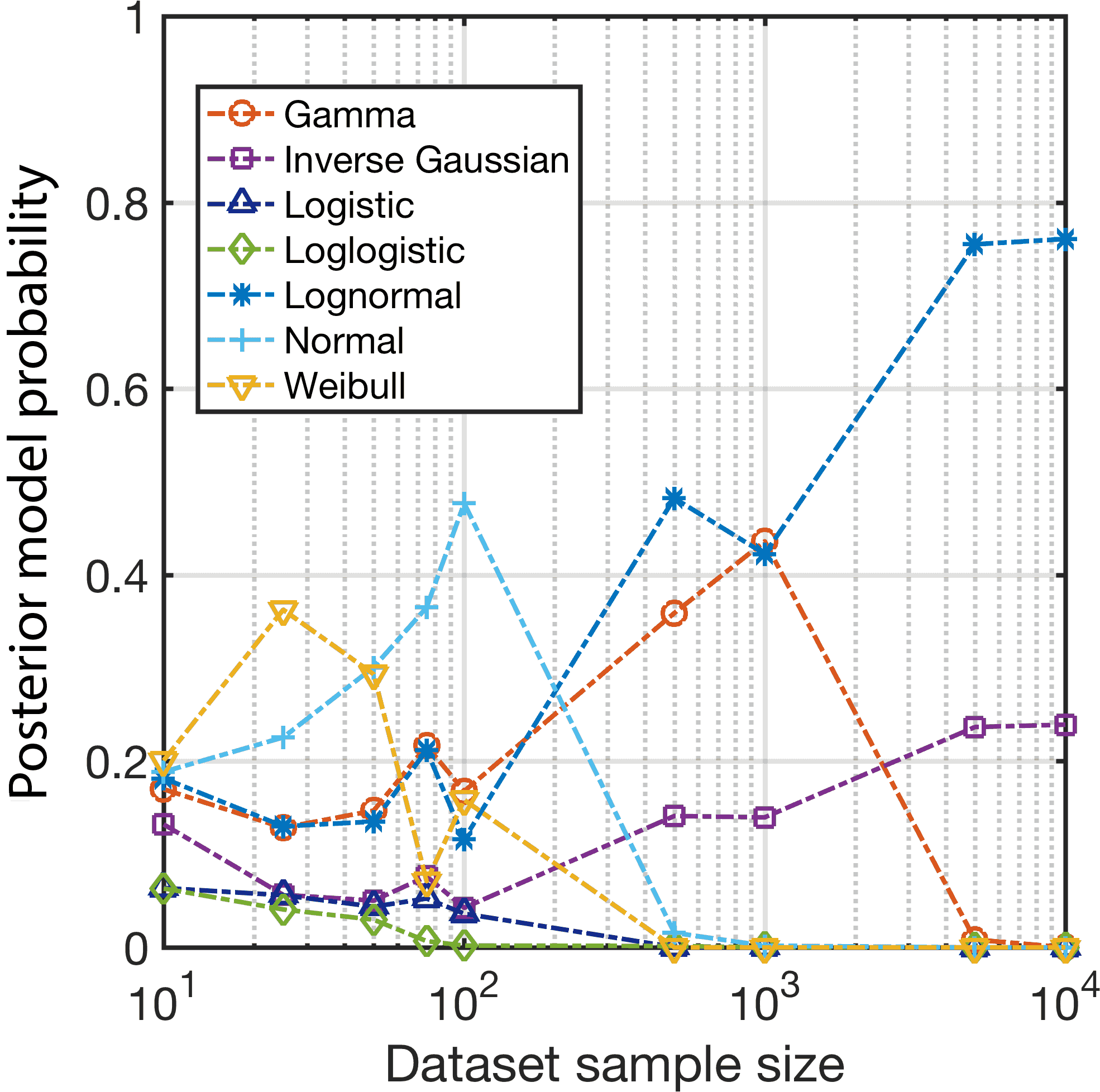}}
	\subfigure[]{\includegraphics[height=2in]{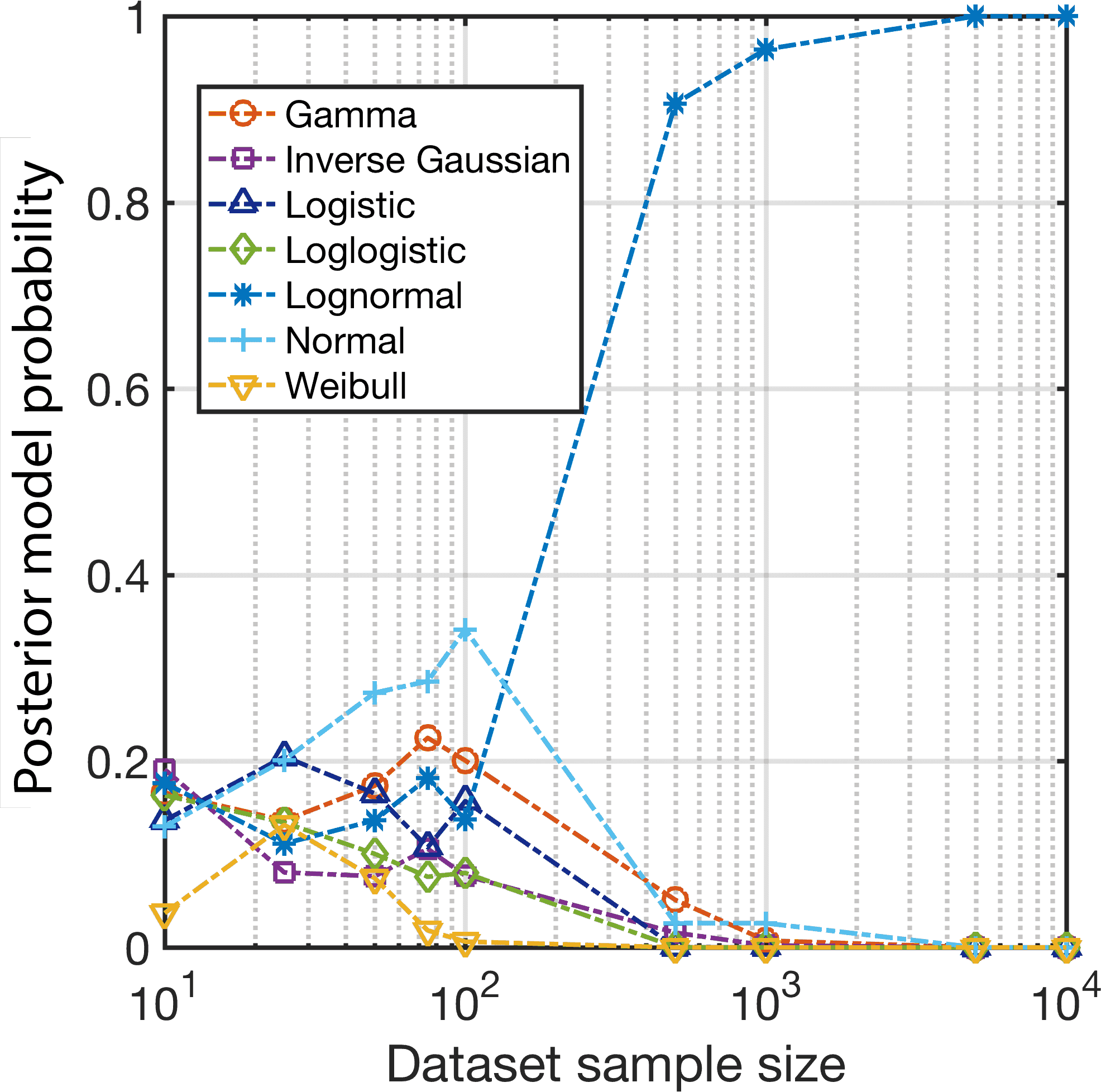}}
	\caption[]{Posterior model probabilities given equal prior model probabilities as a function of dataset size for different parameter priors: (a) Noninformative prior (b) ABS-A prior (c) ABS-B prior (d) ABS-C prior (e) ASTM-A7 prior} \label{fig:model_evidence1}
\end{figure}

\begin{figure}[!ht]
\centering
\includegraphics[height=2in]{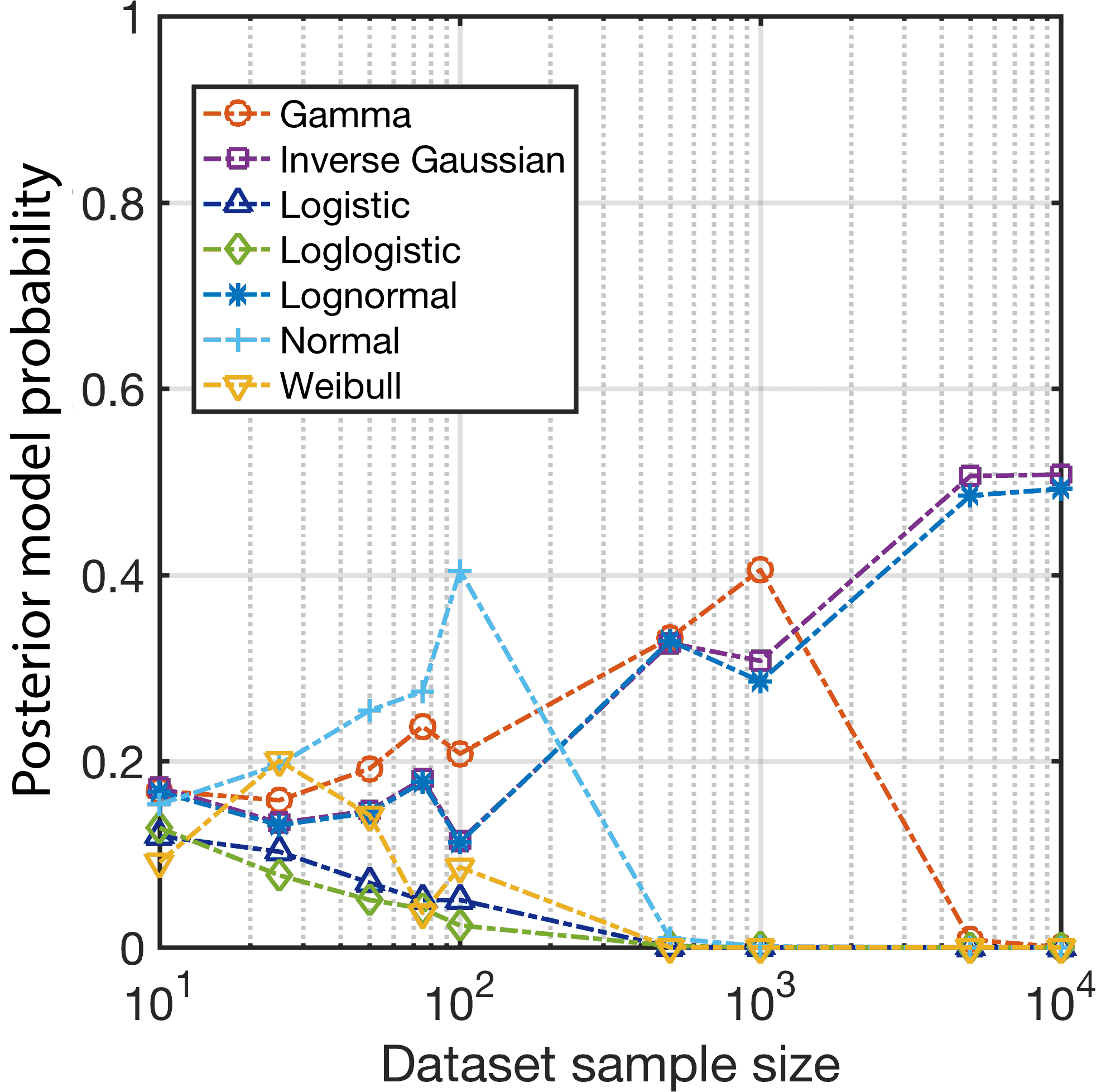}
\caption{Posterior model probabilities from AIC model selection.}
\label{fig:model_prob_AIC}
\end{figure}

We are also interested in the convergence of the model-form uncertainty as a function of the amount of data collected. The previous discussion in Table \ref{tab: model_weights_10data} highlighted how very small datasets lead to large model-form uncertainties -- with further uncertainty introduced by the selection of the parameter prior. But how much data is necessary to reduce this uncertainty and how does the performance change given different parameter priors? Figure \ref{fig:model_evidence1} shows the posterior model probabilities as a function of dataset size for different parameter priors (given equal prior model probabilities). 
For comparison, Figure \ref{fig:model_prob_AIC} shows the posterior model probabilities using AIC model selection (i.e. using savvy prior probabilities).
Notice that the AIC, noninformative, and ABS-B priors show nearly identical trends with added data. Although they fail to identify a unique model (they essentially identify the lognormal and the inverse Gaussian with equal probability), we will see that they are in fact among the ``best'' priors in terms of convergence toward the true probability model. Of particular interest is the fact that the ABS-A parameter prior converges toward the incorrect Gamma model and effectively discounts the lognormal model entirely.

\begin{figure}[!ht]	
	\centering
    \subfigure[]{\includegraphics[height=2in]{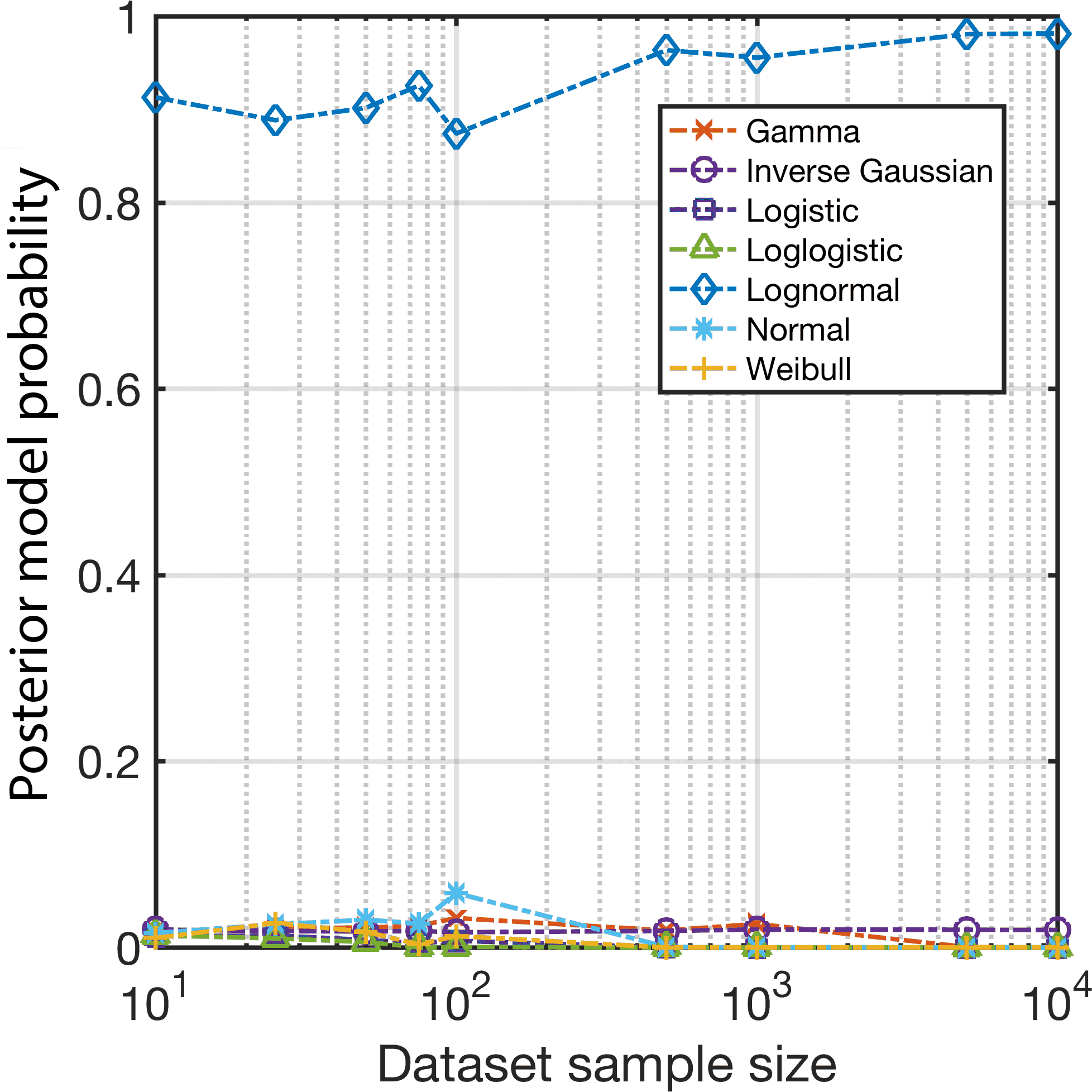}}
	\subfigure[]{\includegraphics[height=2in]{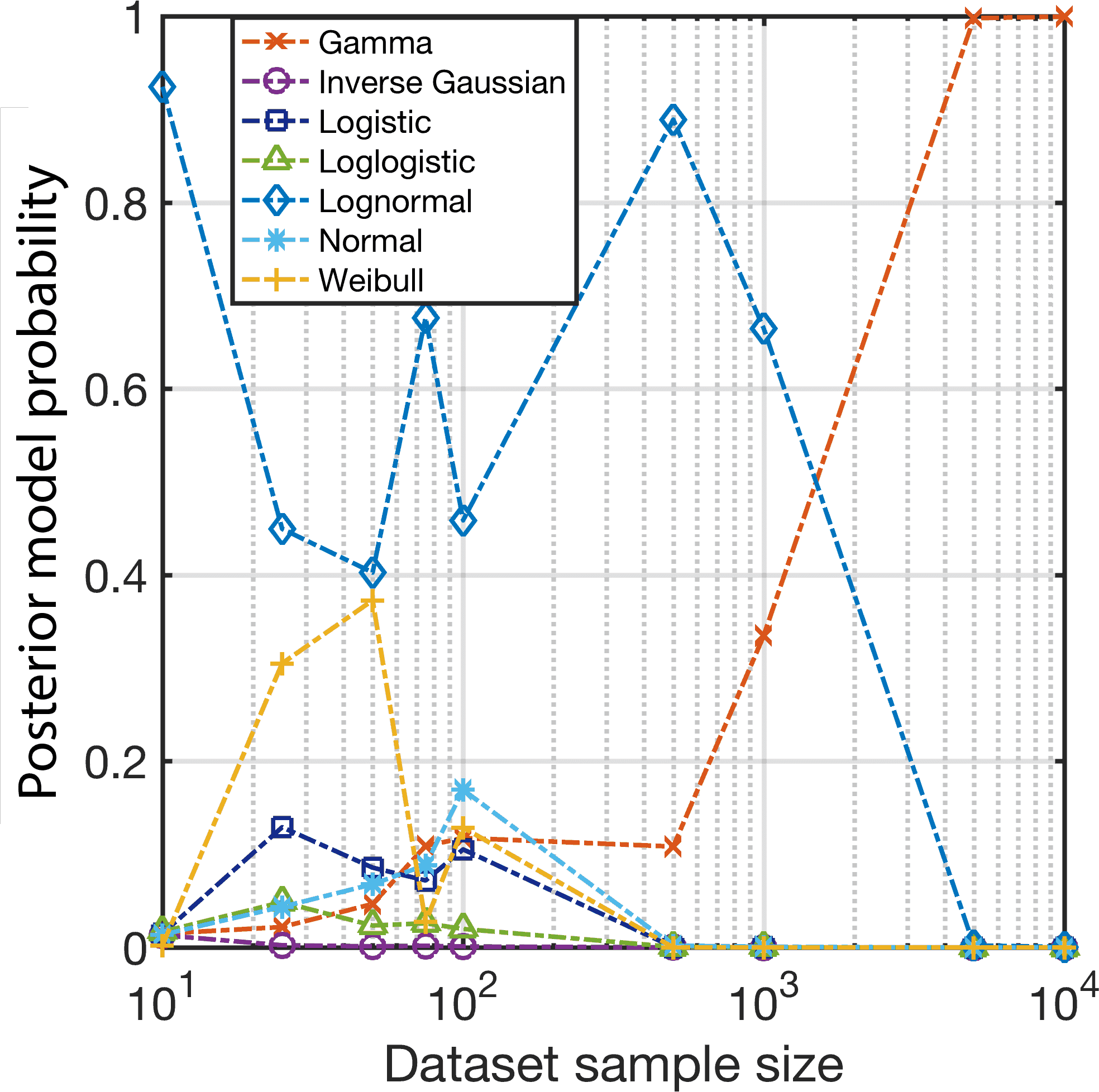}}
	\subfigure[]{\includegraphics[height=2in]{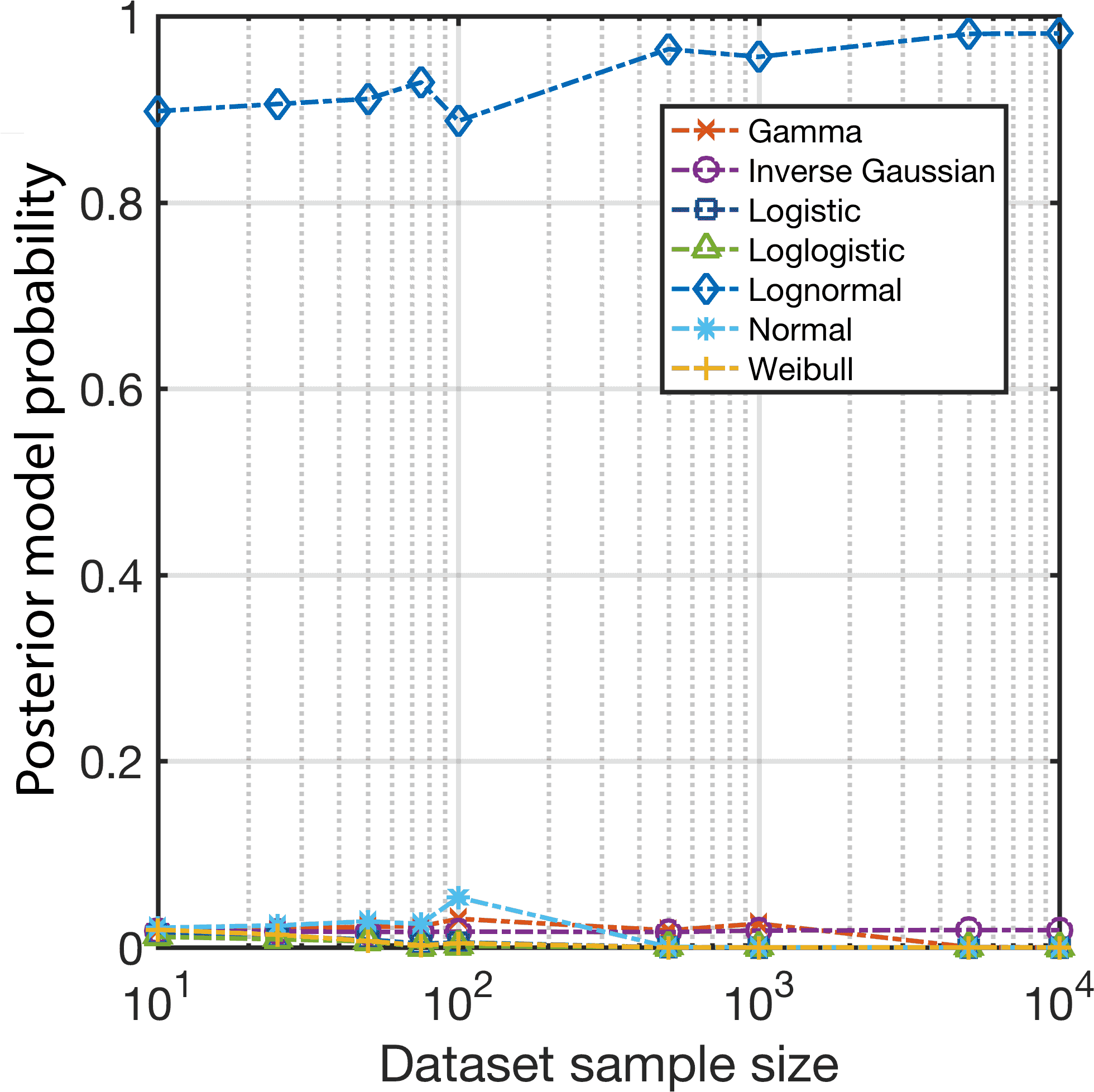}}
	\subfigure[]{\includegraphics[height=2in]{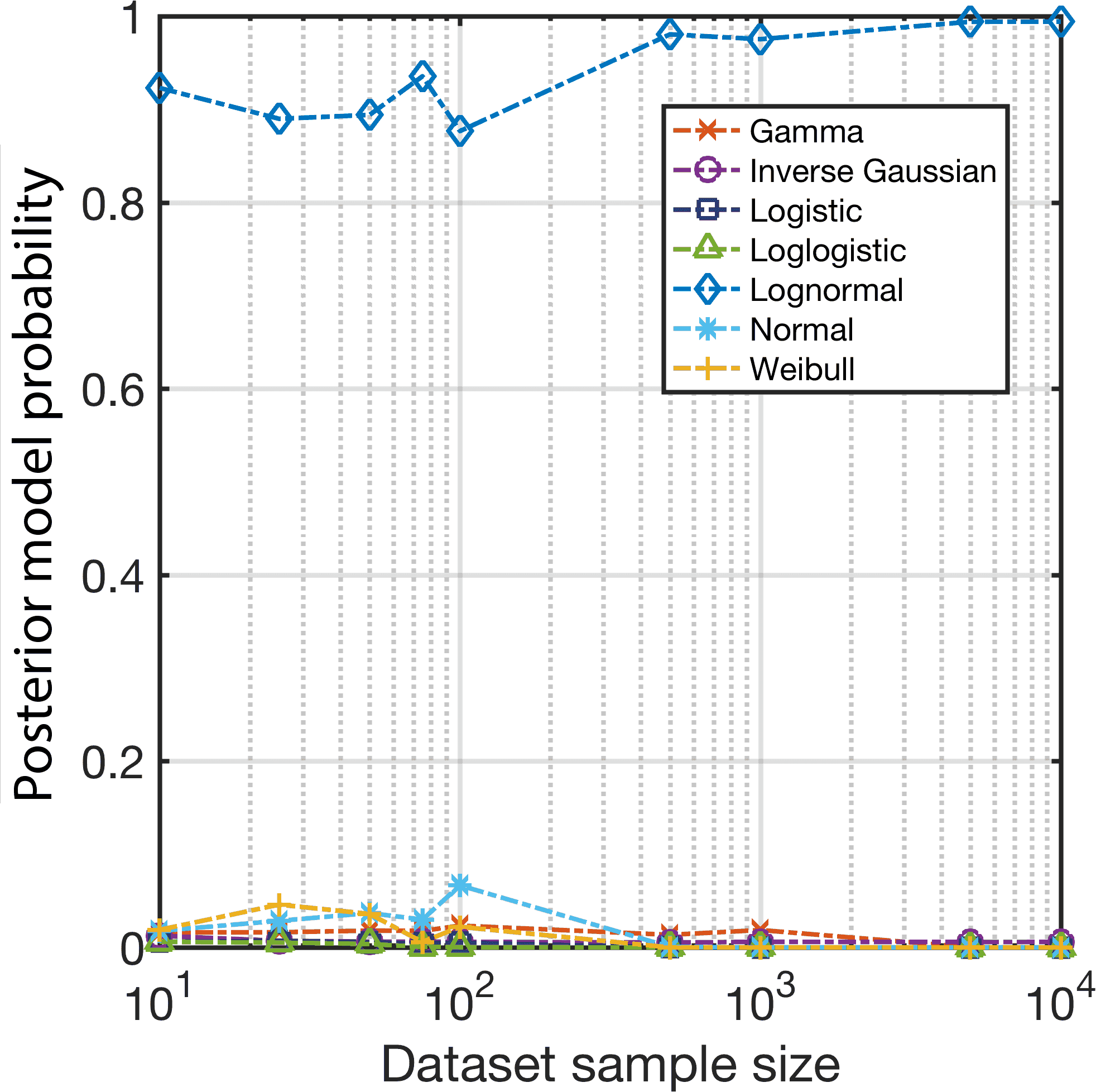}}
	\subfigure[]{\includegraphics[height=2in]{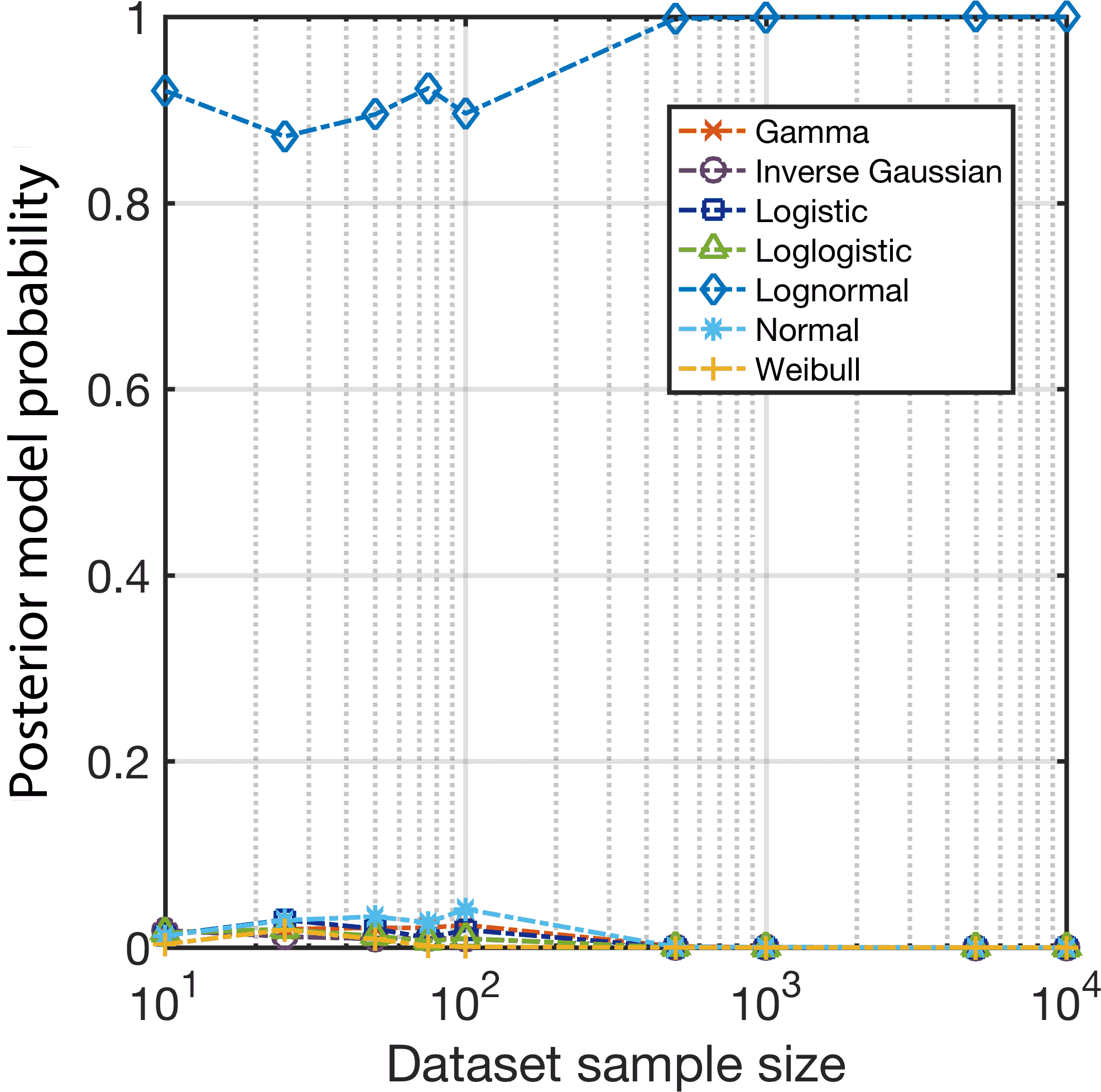}}
	\caption[]{Posterior model probabilities given ``strong correct'' prior model probabilities as a function of dataset size for different parameter priors: (a) Noninformative prior (b) ABS-A prior (c) ABS-B prior (d) ABS-C prior (e) ASTM-A7 prior.}  \label{fig:model_evidence3}
\end{figure}

\begin{figure}[!ht]	
	\centering
    \subfigure[]{\includegraphics[height=2in]{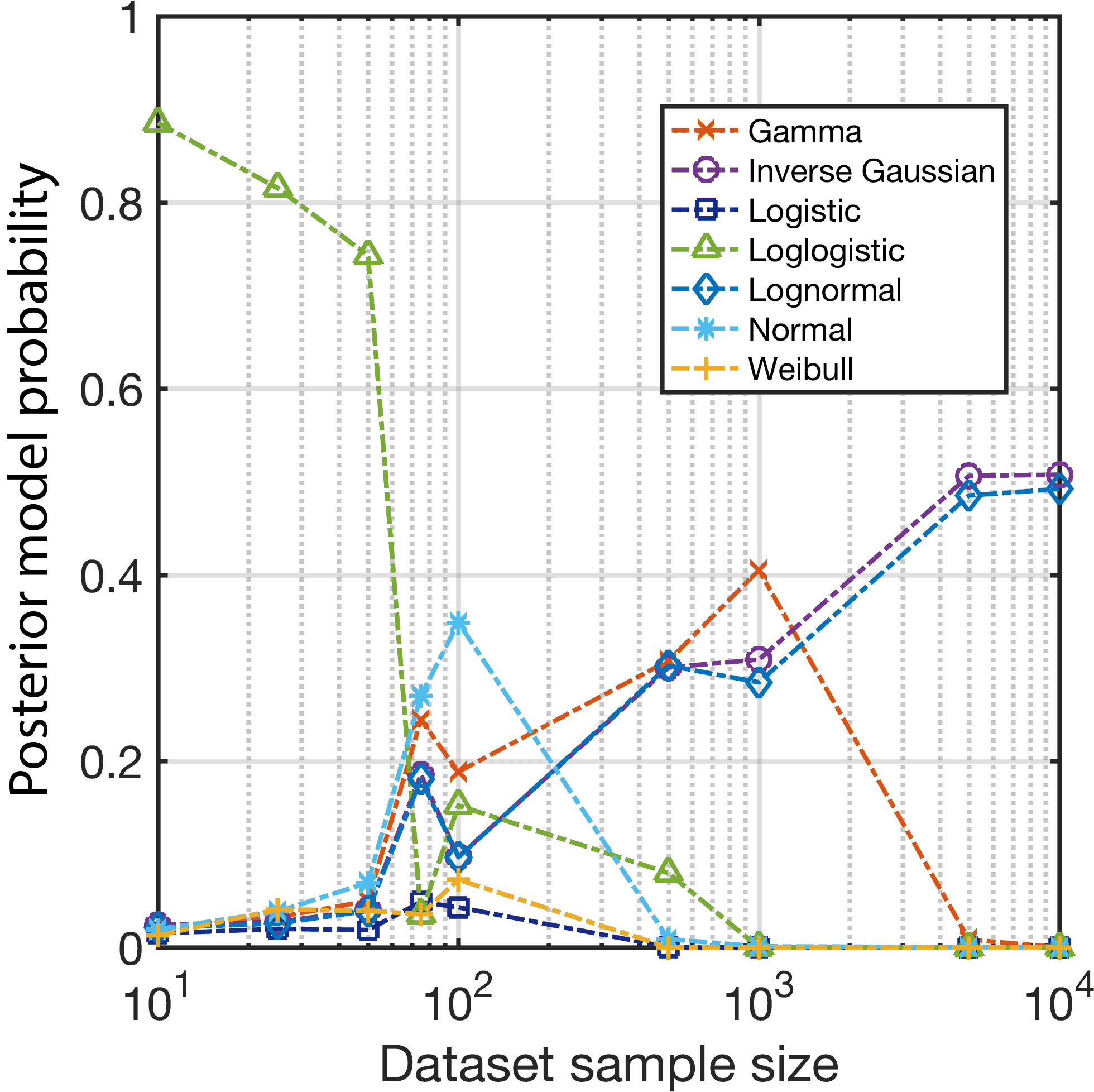}}
	\subfigure[]{\includegraphics[height=2in]{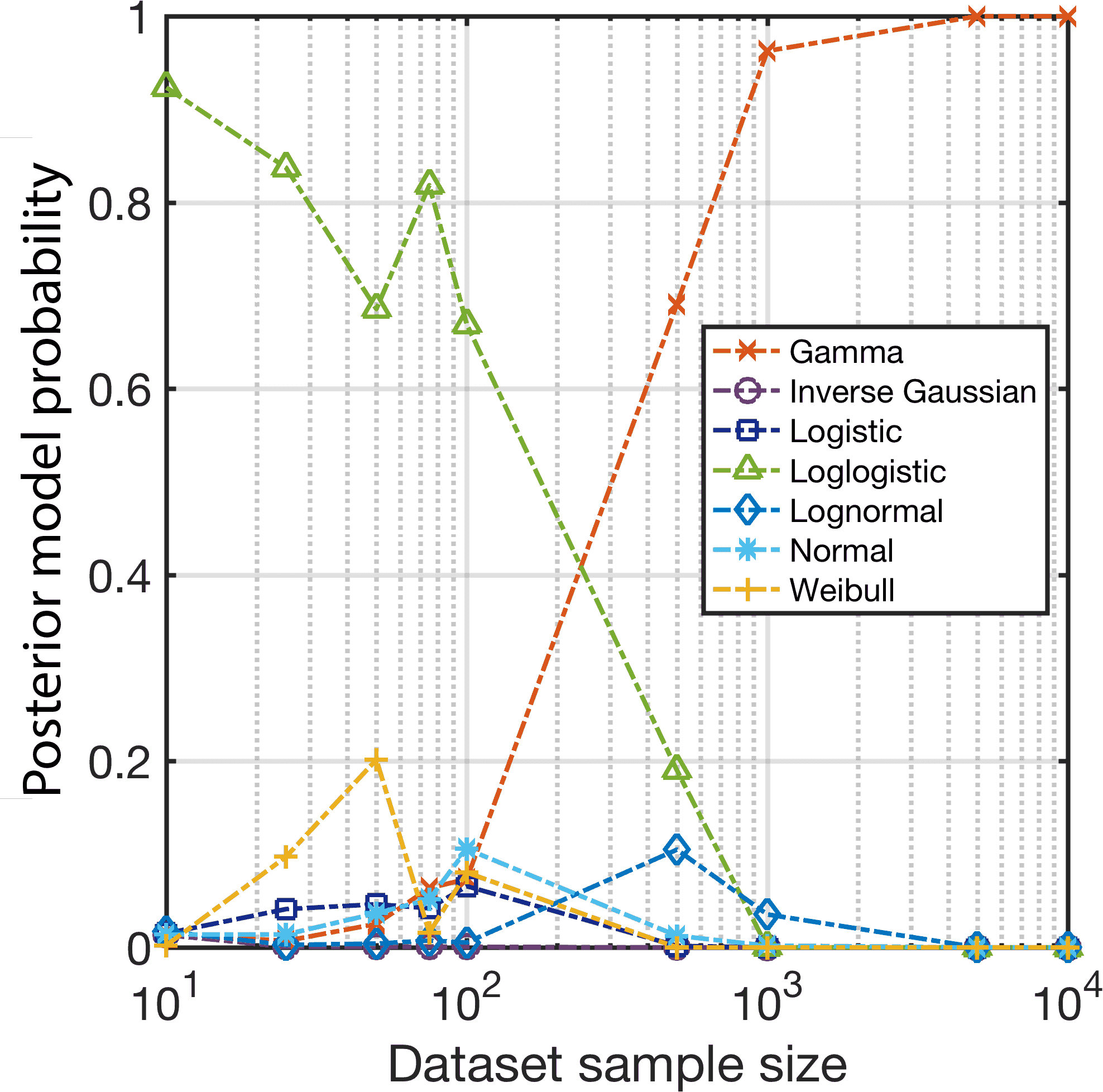}}
	\subfigure[]{\includegraphics[height=2in]{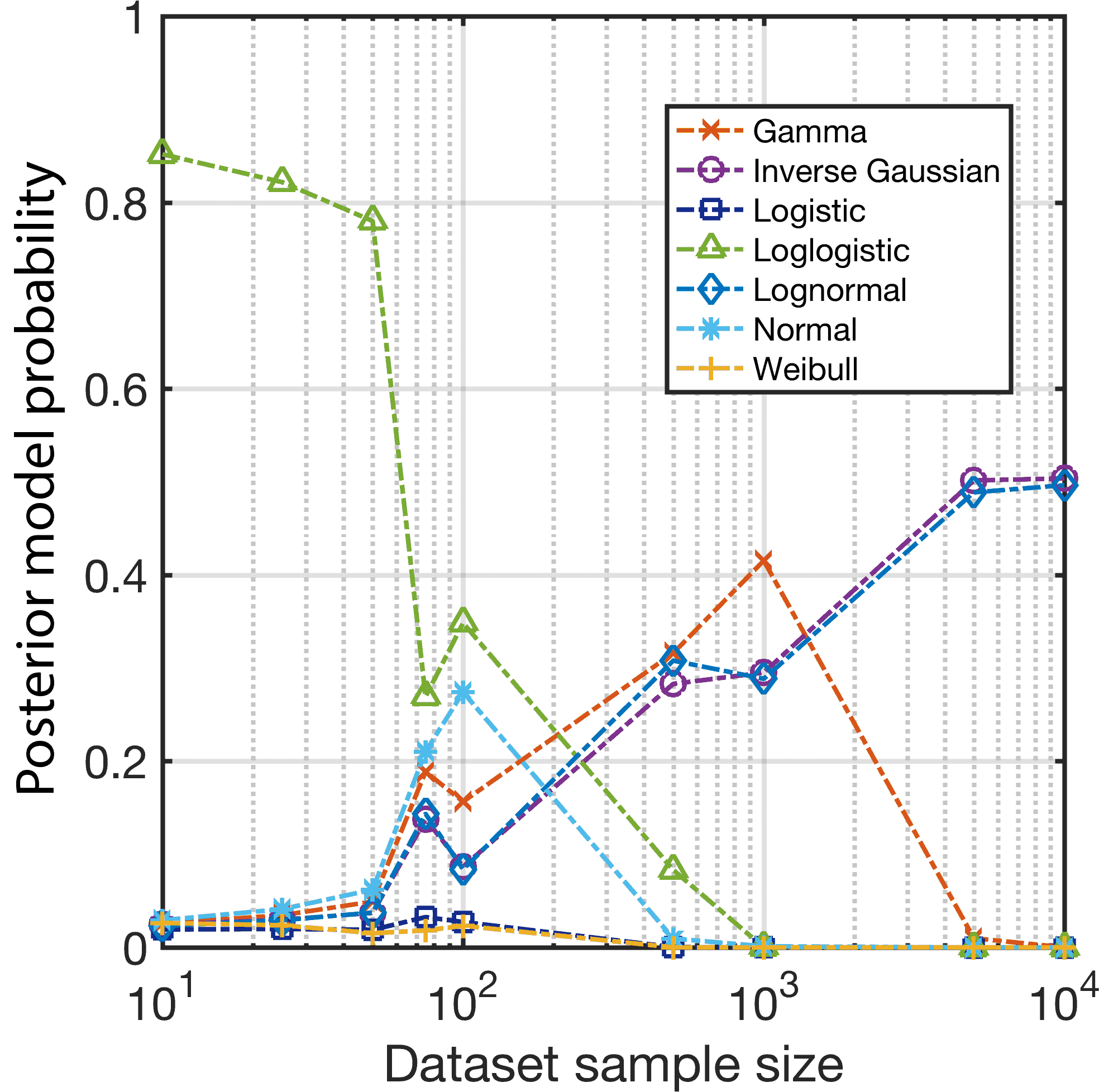}}
	\subfigure[]{\includegraphics[height=2in]{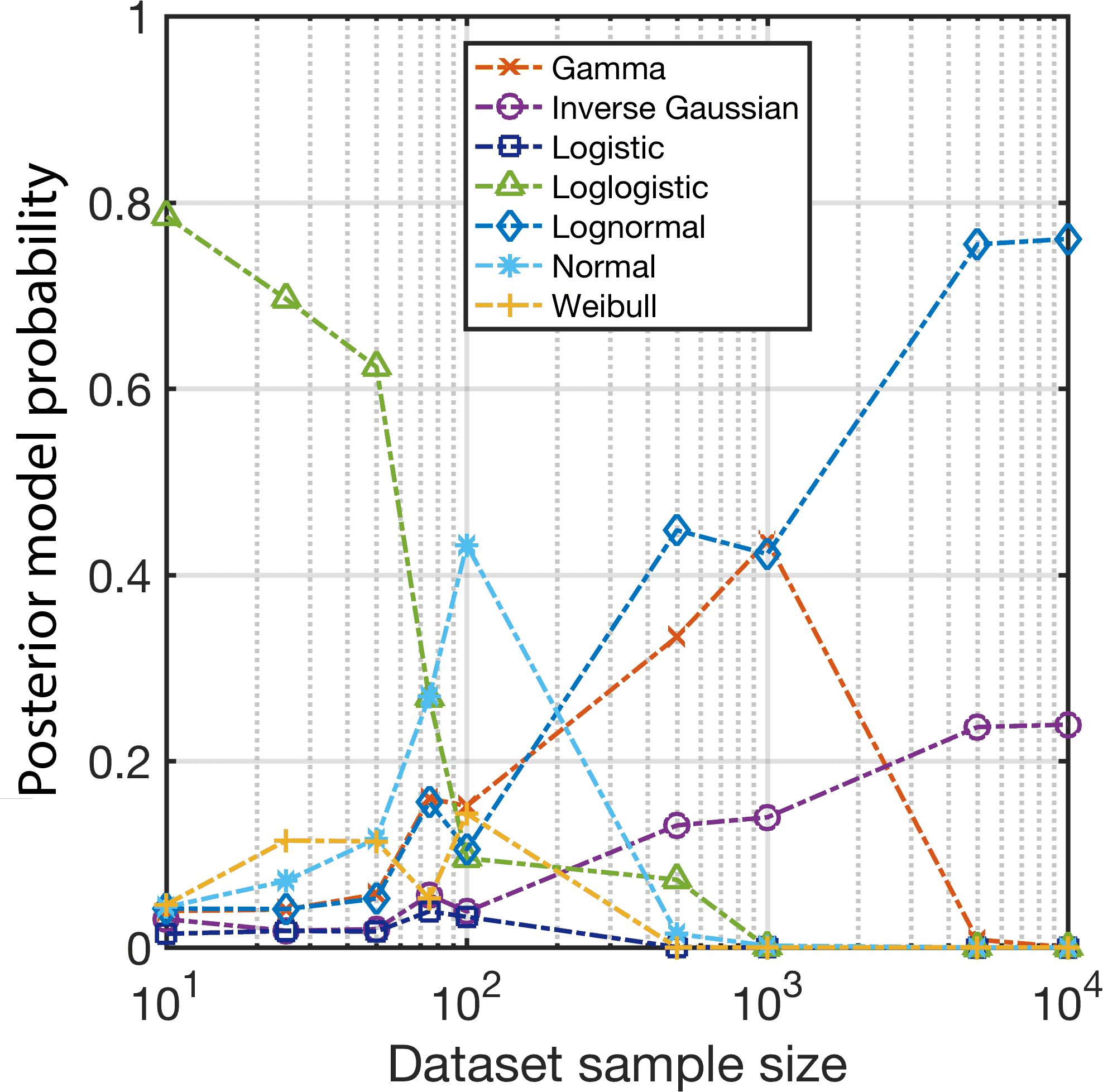}}
	\subfigure[]{\includegraphics[height=2in]{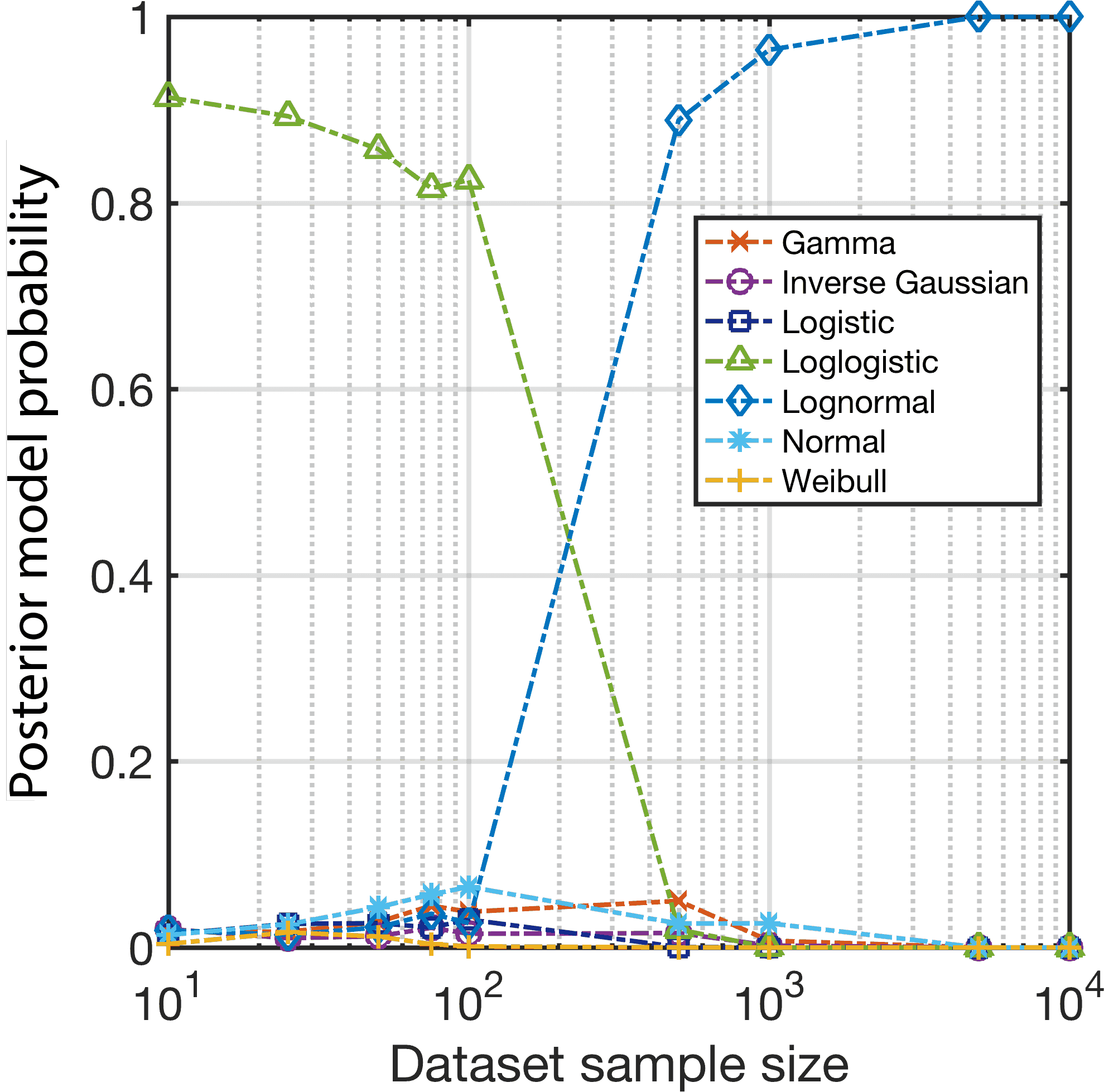}}
	\caption[]{Posterior model probabilities given ``strong incorrect'' prior model probabilities as a function of dataset size for different parameter priors: (a) Noninformative prior (b) ABS-A prior (c) ABS-B prior (d) ABS-C prior (e) ASTM-A7 prior.} \label{fig:model_evidence4}
\end{figure}

Use of informative model prior probabilities can change this convergence behavior considerably. Figures \ref{fig:model_evidence3} and \ref{fig:model_evidence4} show the posterior model probabilities with dataset size for each of the seven models given the strong correct prior (Figure \ref{fig:model_evidence3}) and the strong incorrect prior (Figure \ref{fig:model_evidence4}) for each of the considered parameter priors. 

For the strong correct model prior, four of the five cases show convergence toward the true lognormal model from which the data are drawn as the dataset grows large. Even with the strong incorrect model prior probabilities, the multimodel inference eventually suppresses the incorrect log-logistic model and identifies the correct lognormal form in these cases -- indicating a degree of robustness for these parameter priors. Note also the ABS-B with strong incorrect prior yield essentially equally probable lognormal and inverse Gaussian models because, under this prior the distributions are nearly identical in shape and the inference cannot discern between them. 

In both cases, the ABS-A parameter prior causes the inference to converge to the wrong Gamma model form even when 10,000 yield stress values are collected. The reason for this will be explored later but this points to the important conclusion that if the parameter prior is not wisely chosen, it may not be possible to infer even the correct model form for the data. This can have significant practical implications for uncertainty quantification and propagation.

\subsubsection{Effect of parameter prior on parameter uncertainty}
\label{sec:parameter_effect}

For each model form, the selection of the parameter prior will significantly impact the convergence of the posterior. Here, we focus on the case of the lognormal distribution as a representative case to illustrate this effect. Similar results for the other probability models were observed. 

Data are generated according to the ``true'' lognormal distribution and Bayesian inference conducted to infer the parameters of the lognormal model using each of the five considered parameter priors. Table \ref{tab: lognormal_small} shows the joint parameter pdf for ``small'' datasets ($\le$ 100 data) along with the true parameters (indicated by a $\star$).
\begin{table}[!ht] \footnotesize
\centering
\caption{Posterior parameter joint probability densities for the lognormal distribution with different priors considering small dataset size ($\le 100$ data).}
\label{tab: lognormal_small}
\begin{tabular}{crrrrr}
\hline 
Data & Noninformative & ABS-A (33) & ABS-B (79) & ABS-C (13) & ASTM-A7 (58)\\ \hline
Prior 
&\parbox[c][1.1in]{1in}{\includegraphics[height=1in]{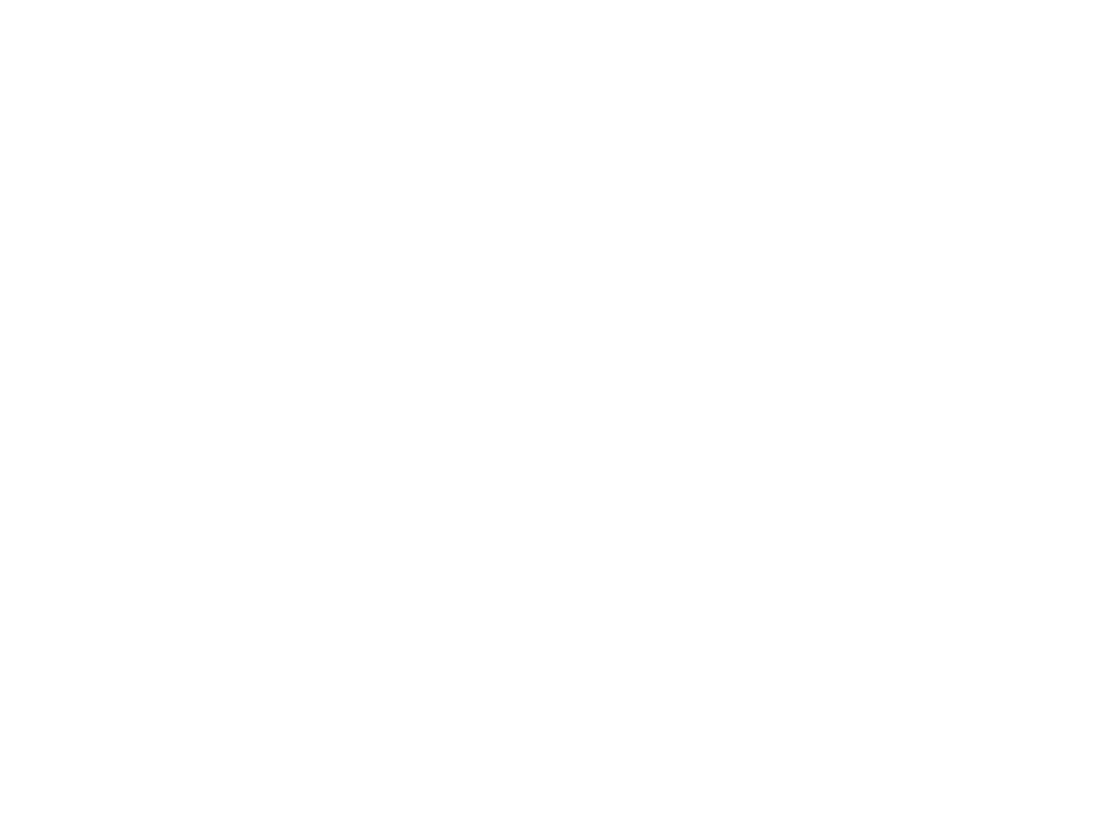}}
&\parbox[c][1.1in]{1in}{\includegraphics[height=1in]{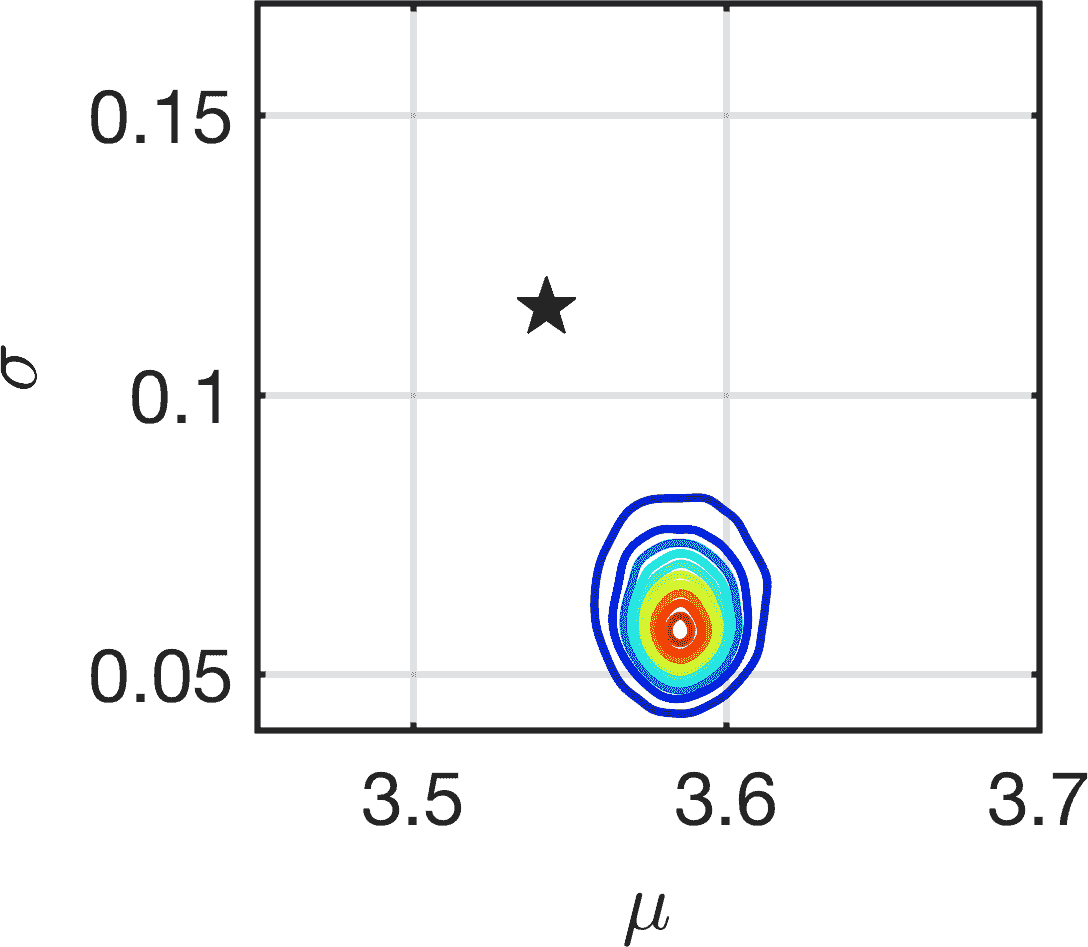}} 
&\parbox[c][1.1in]{1in}{\includegraphics[height=1in]{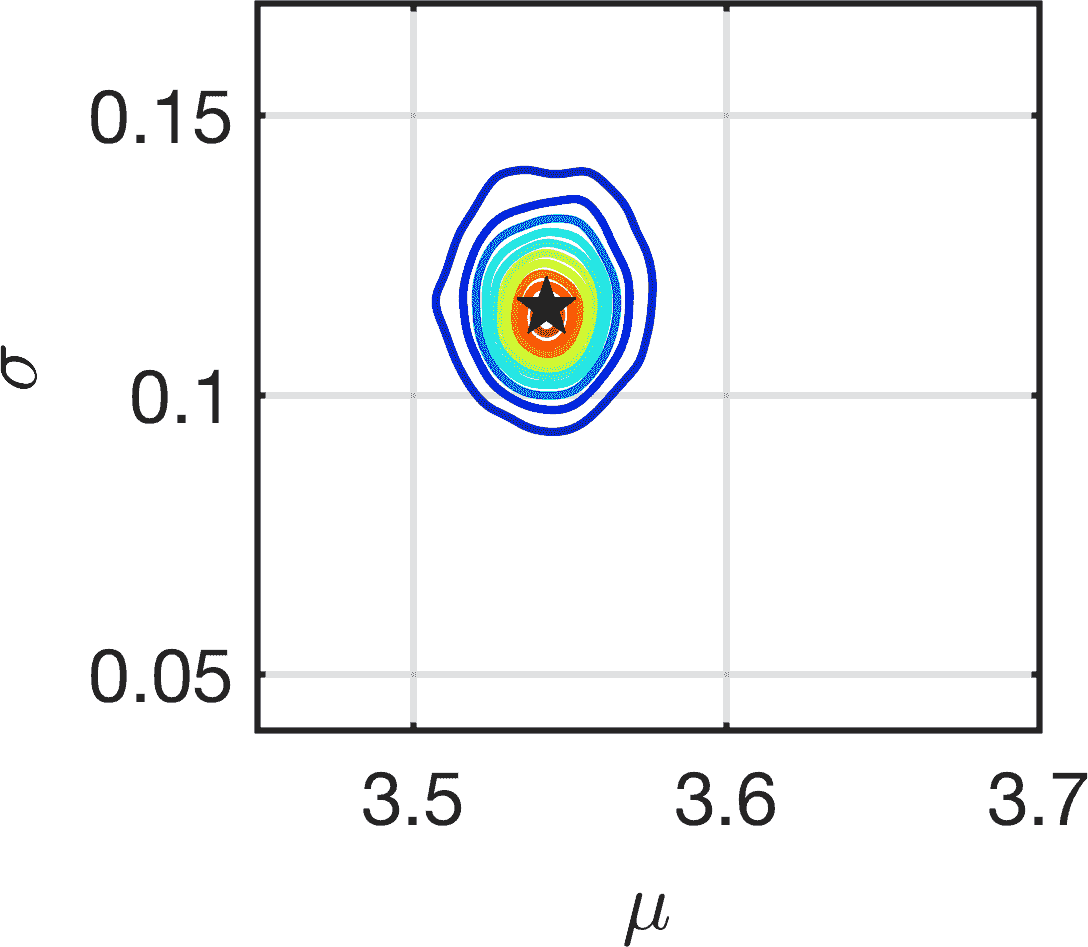}}  
&\parbox[c][1.1in]{1in}{\includegraphics[height=1in]{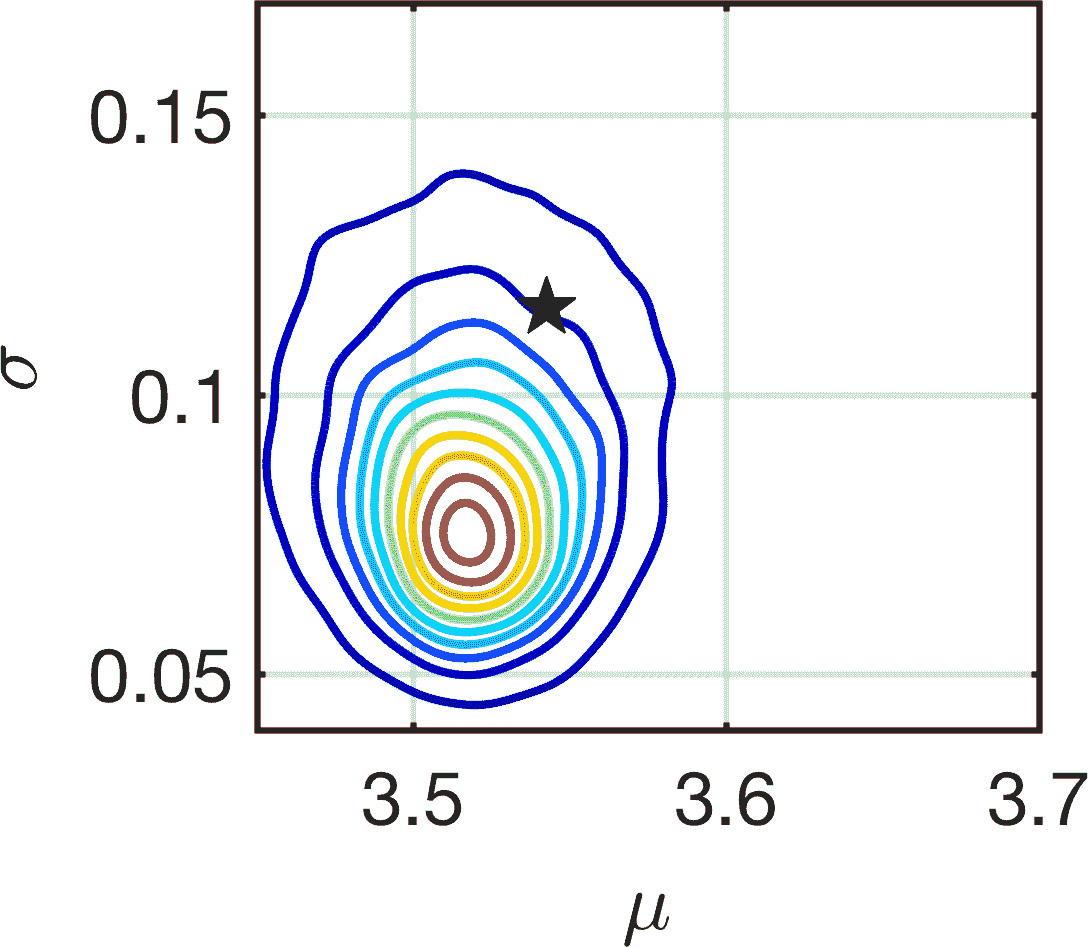}}  
&\parbox[c][1.1in]{1in}{\includegraphics[height=1in]{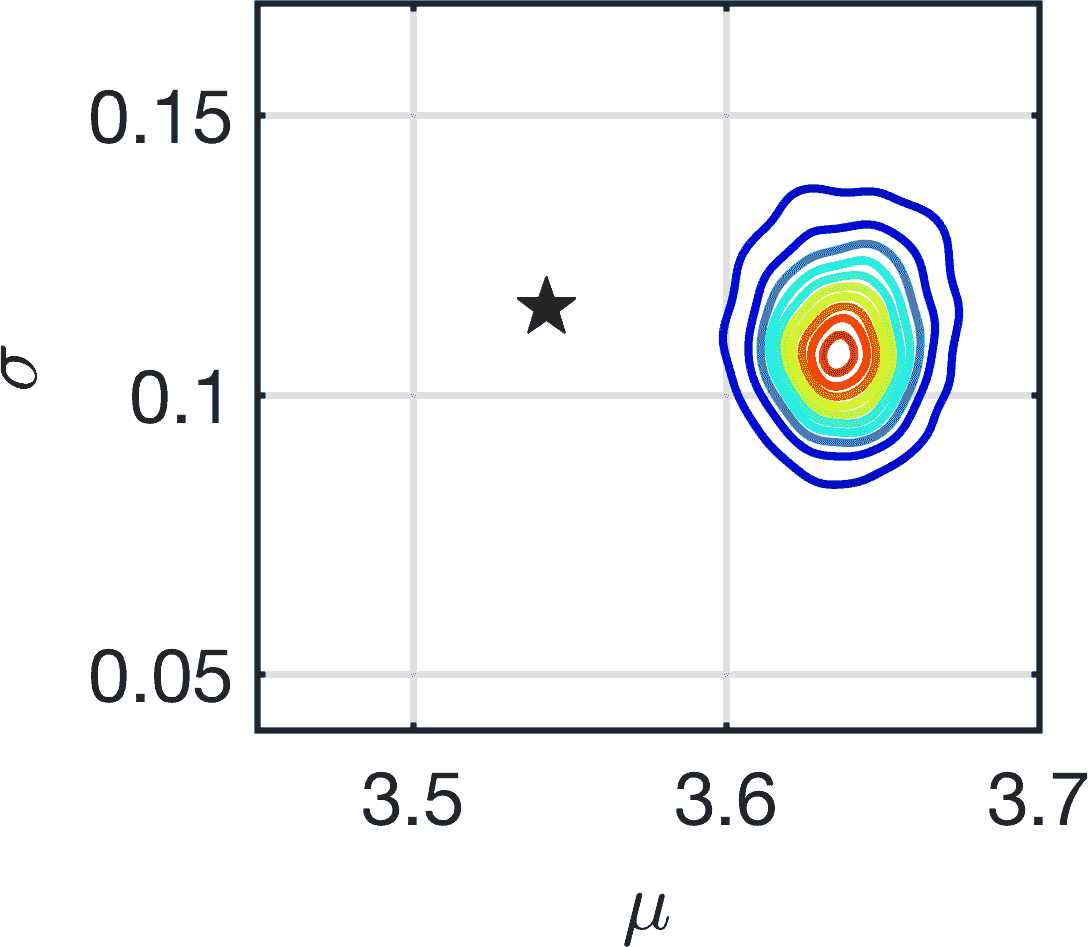}} \\
10
&\parbox[c][1.1in]{1in}{\includegraphics[height=1in]{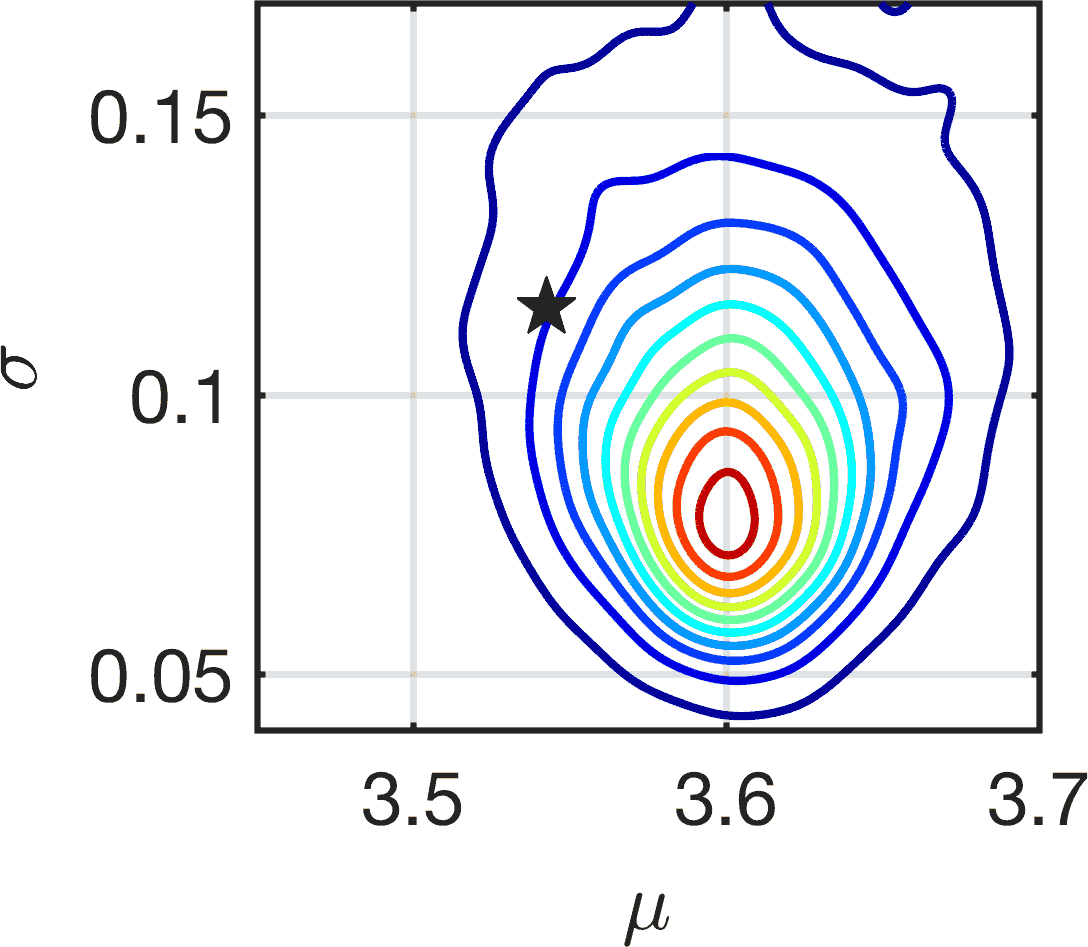}}
&\parbox[c][1.1in]{1in}{\includegraphics[height=1in]{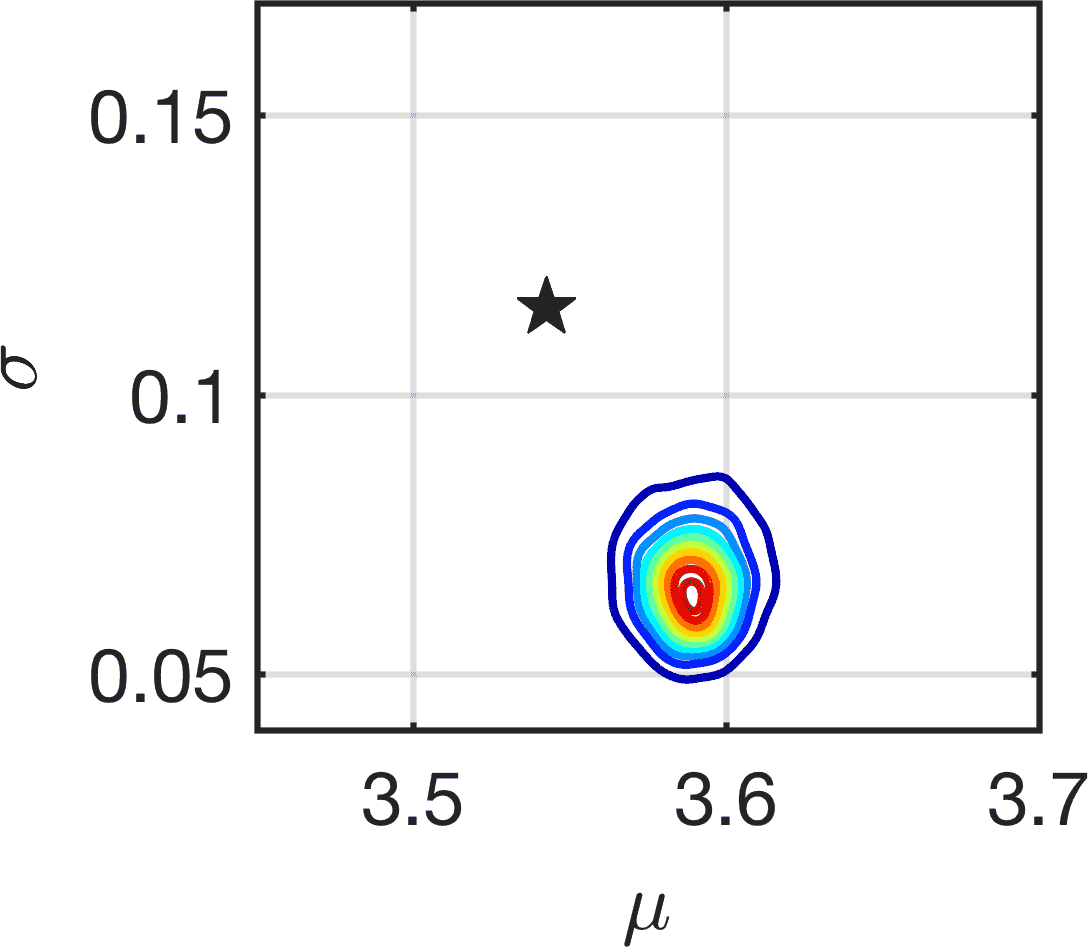}} 
&\parbox[c][1.1in]{1in}{\includegraphics[height=1in]{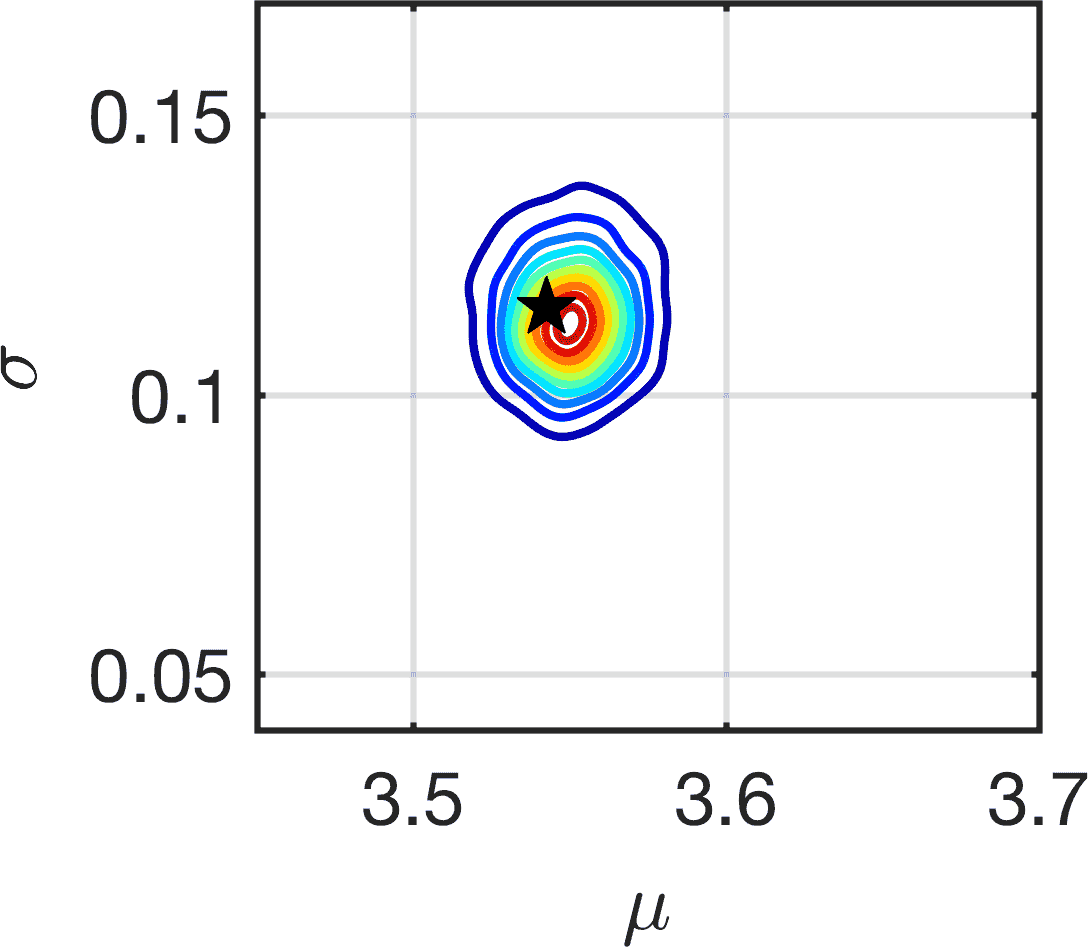}}  
&\parbox[c][1.1in]{1in}{\includegraphics[height=1in]{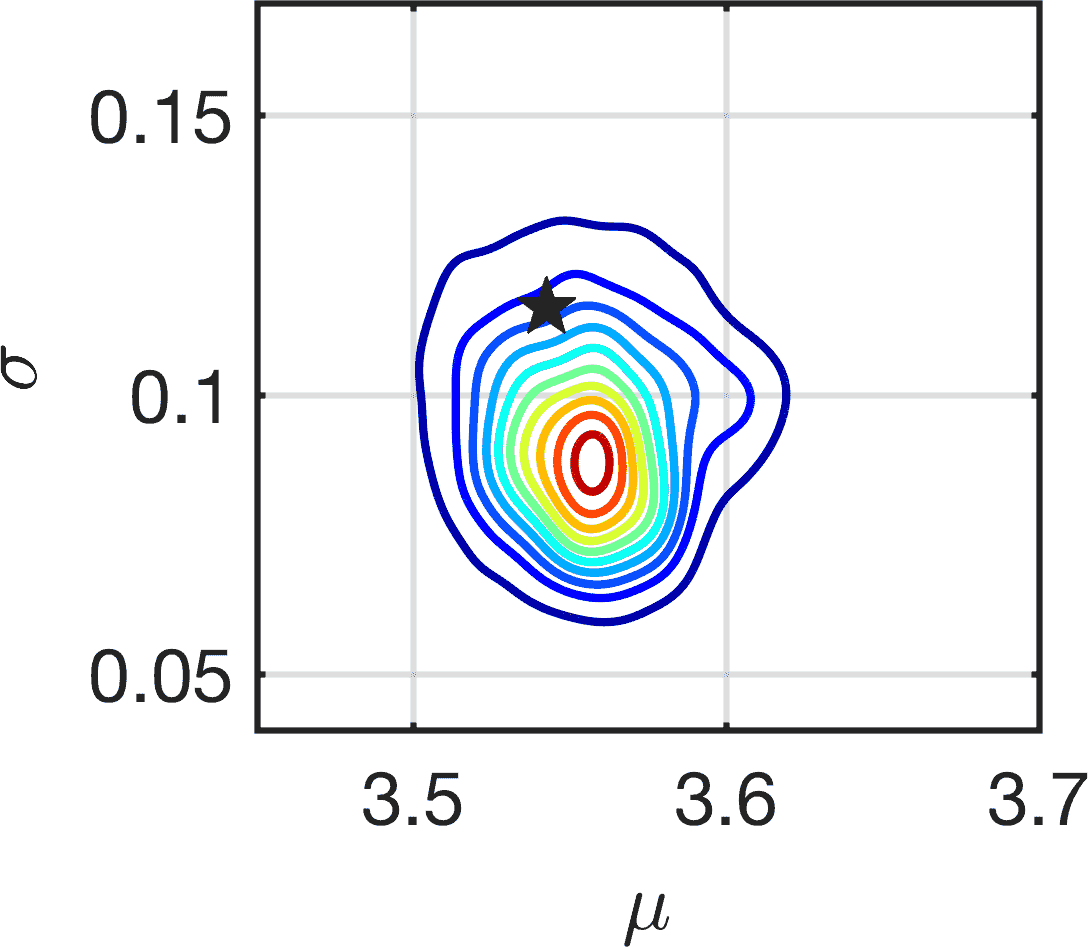}}  
&\parbox[c][1.1in]{1in}{\includegraphics[height=1in]{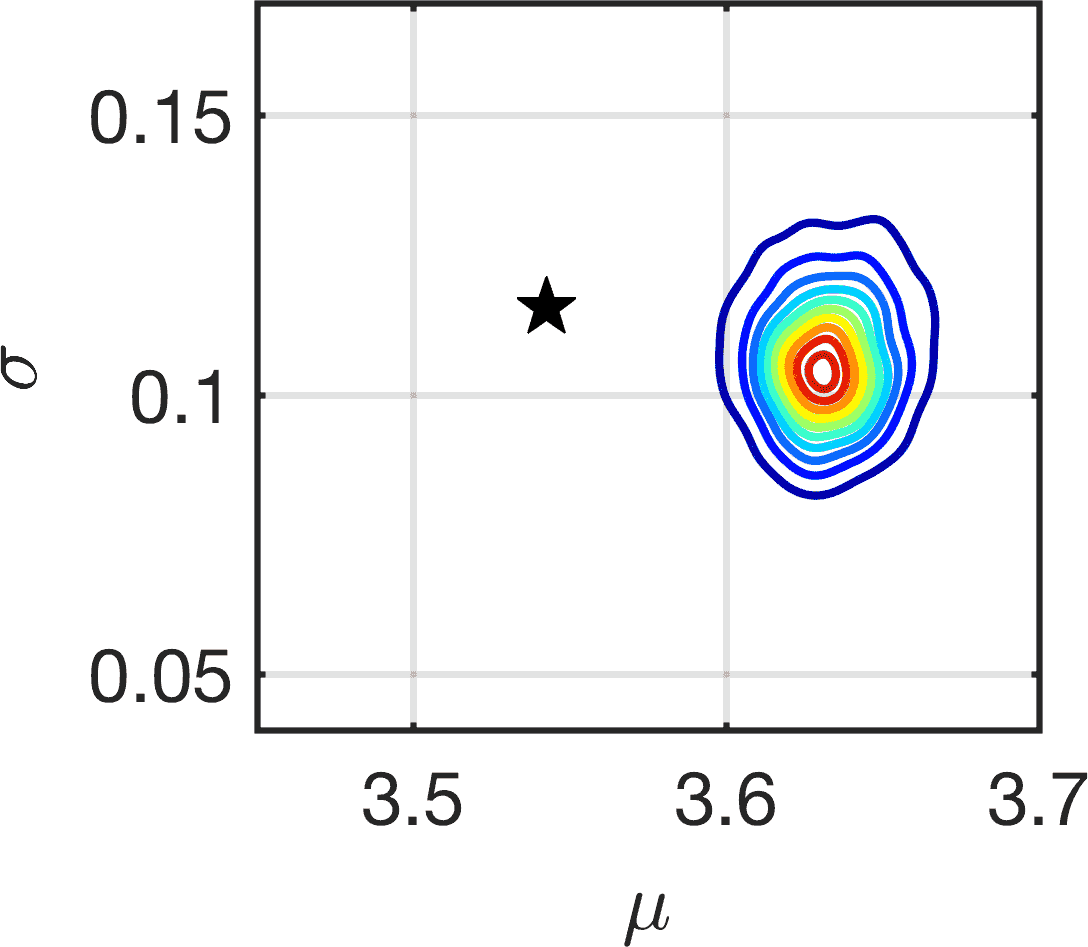}} \\
25
&\parbox[c][1.1in]{1in}{\includegraphics[height=1in]{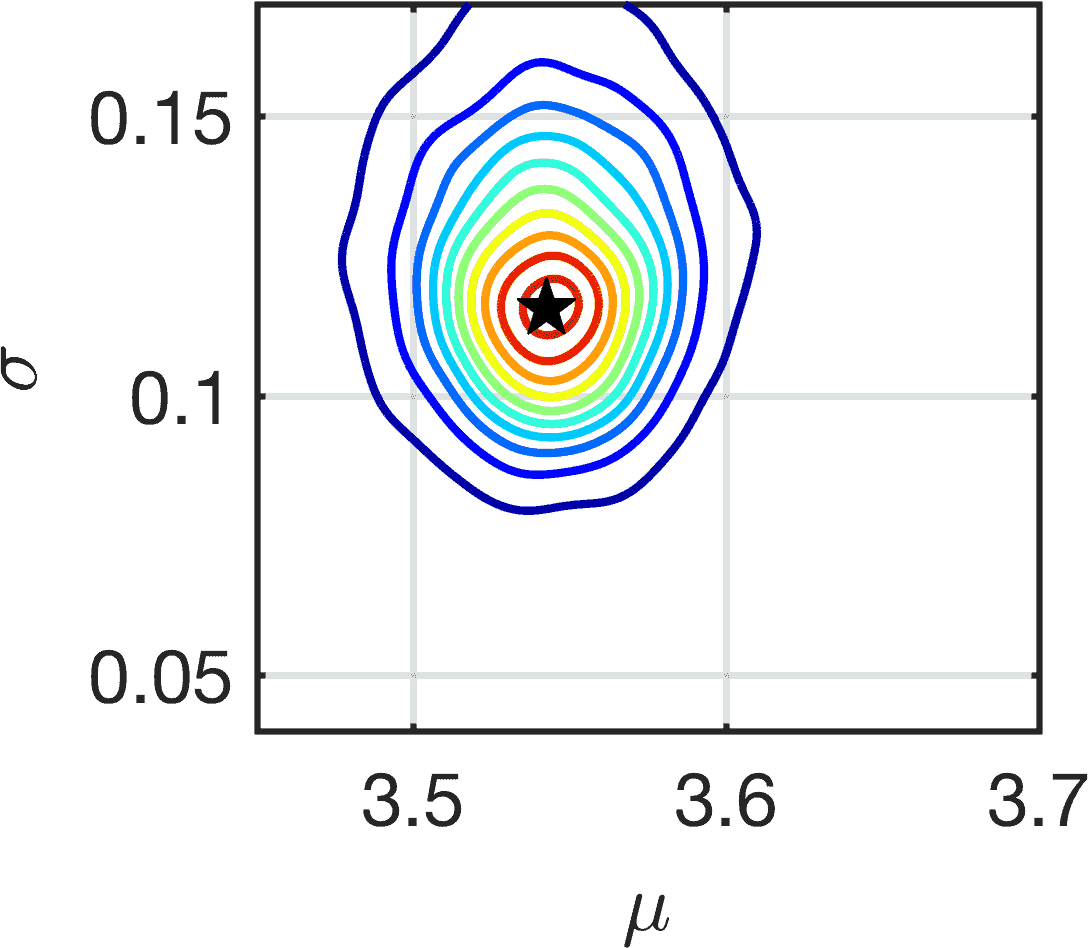}}
&\parbox[c][1.1in]{1in}{\includegraphics[height=1in]{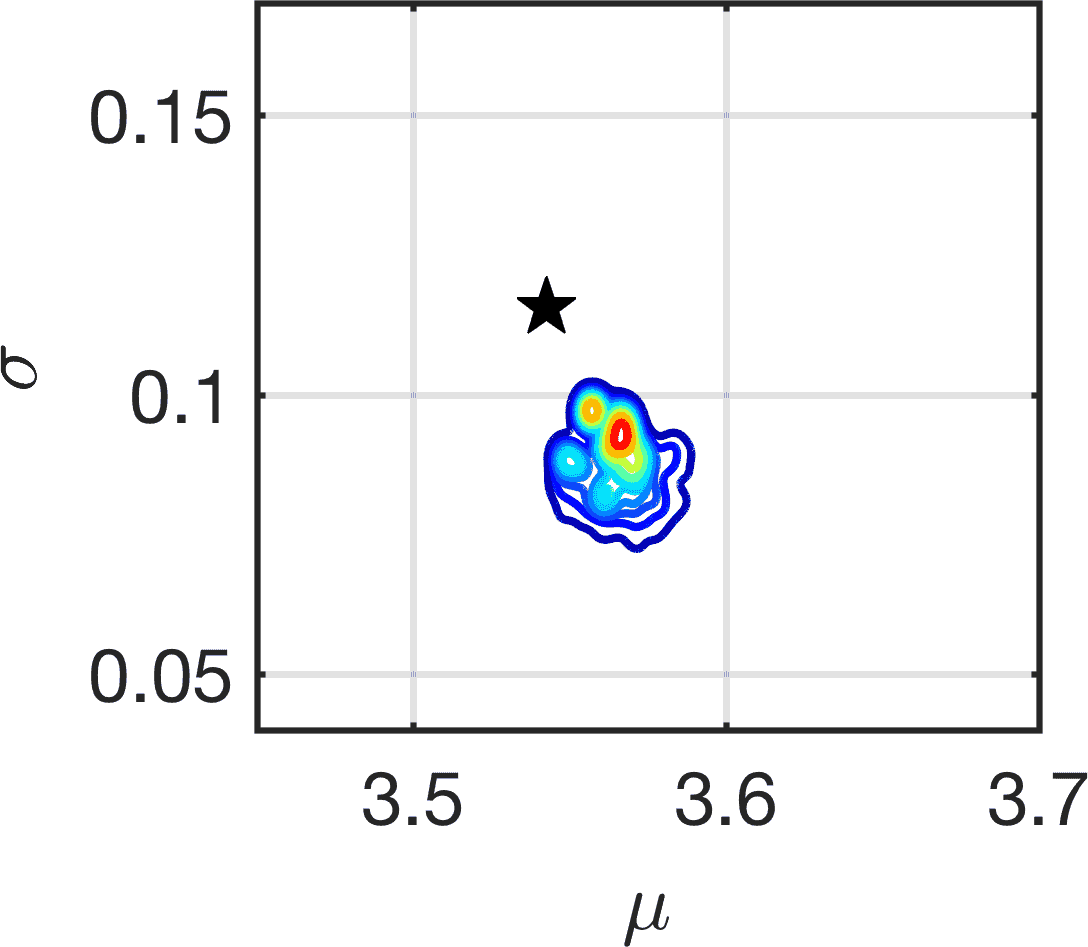}} 
&\parbox[c][1.1in]{1in}{\includegraphics[height=1in]{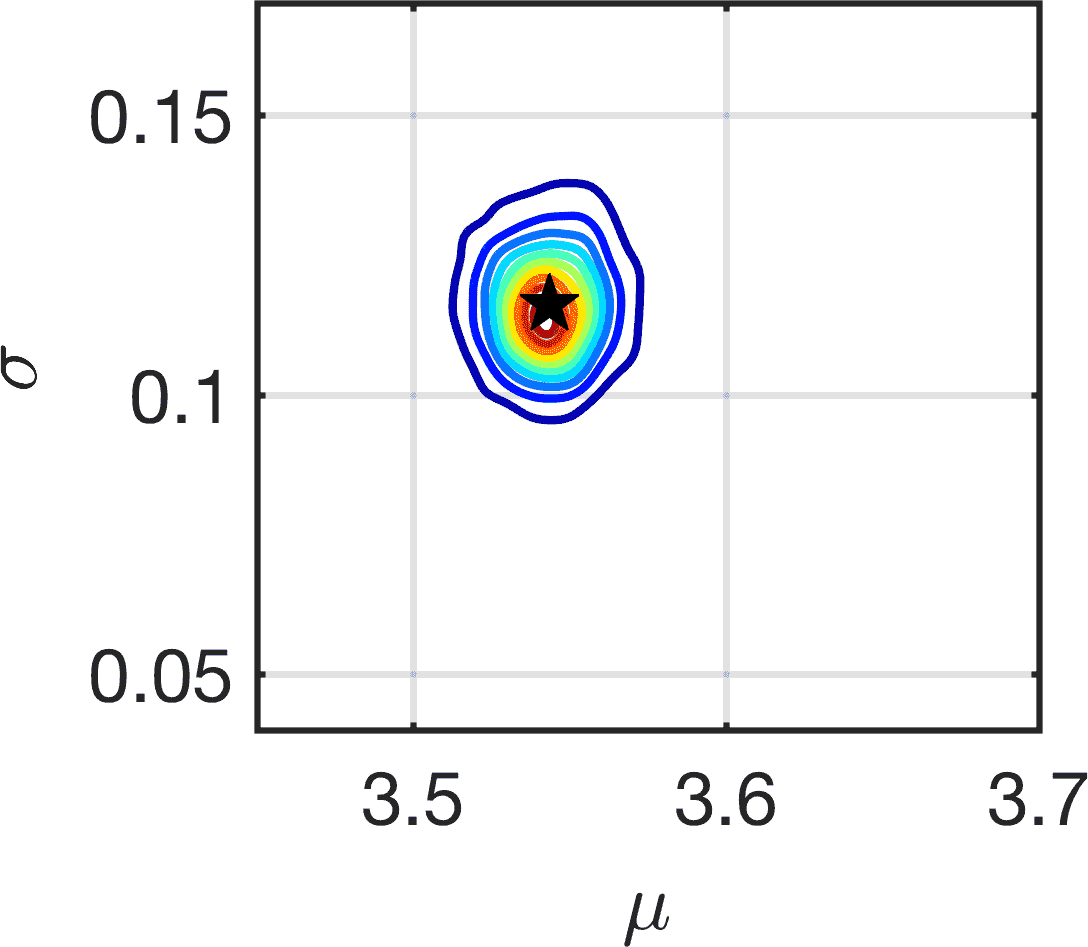}}  
&\parbox[c][1.1in]{1in}{\includegraphics[height=1in]{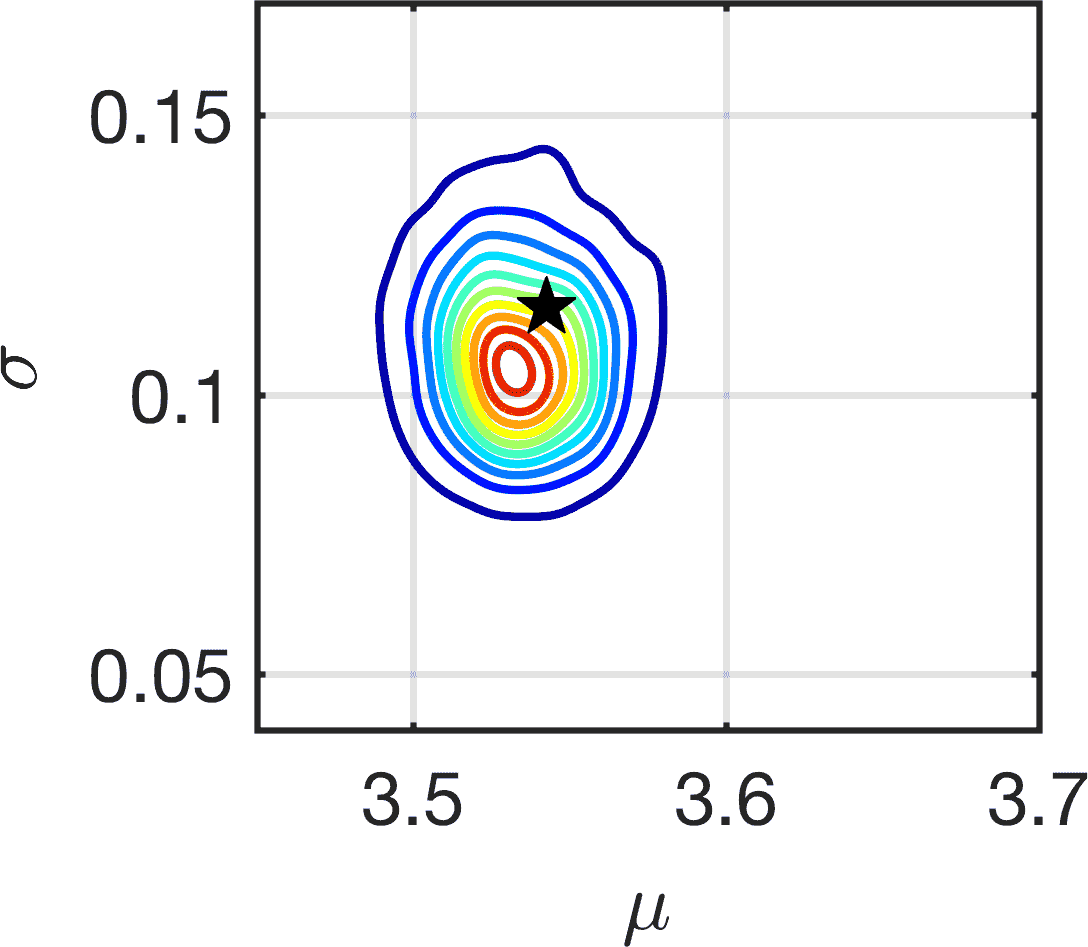}}  
&\parbox[c][1.1in]{1in}{\includegraphics[height=1in]{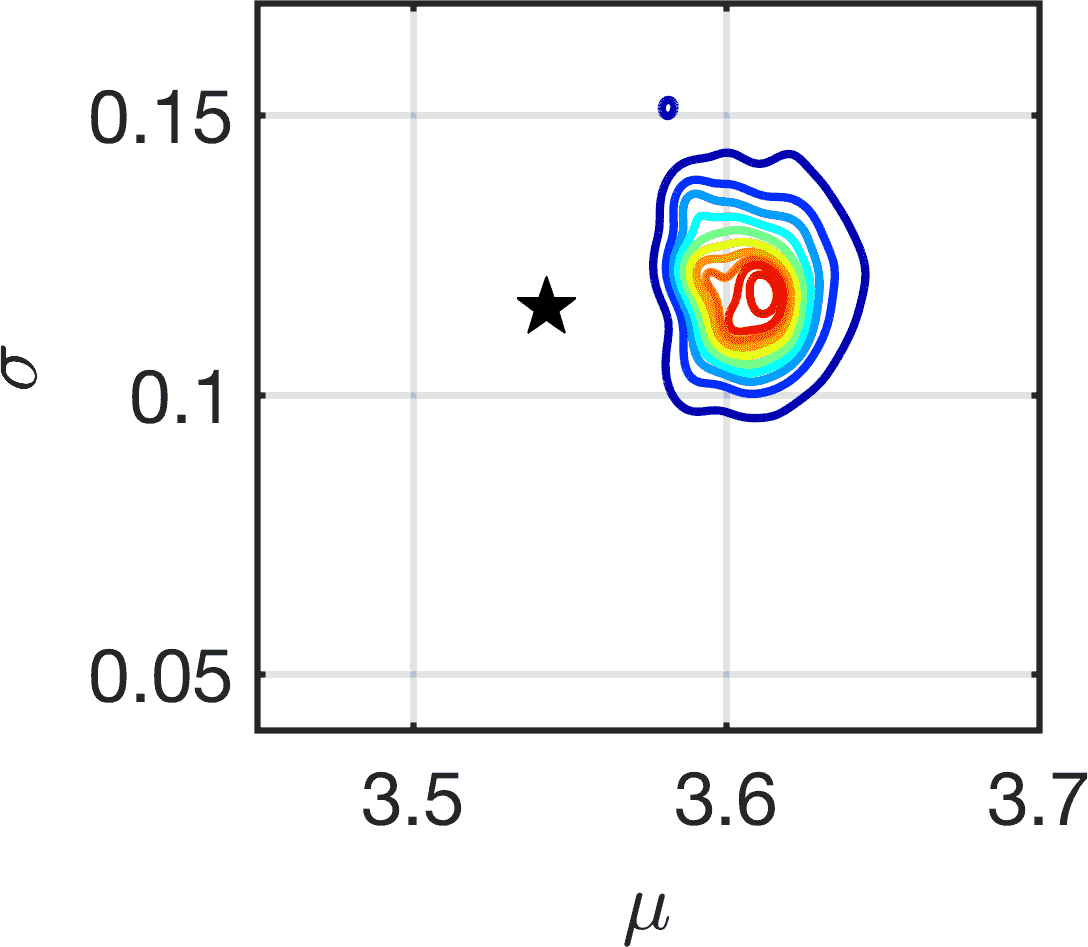}} \\
50
&\parbox[c][1.1in]{1in}{\includegraphics[height=1in]{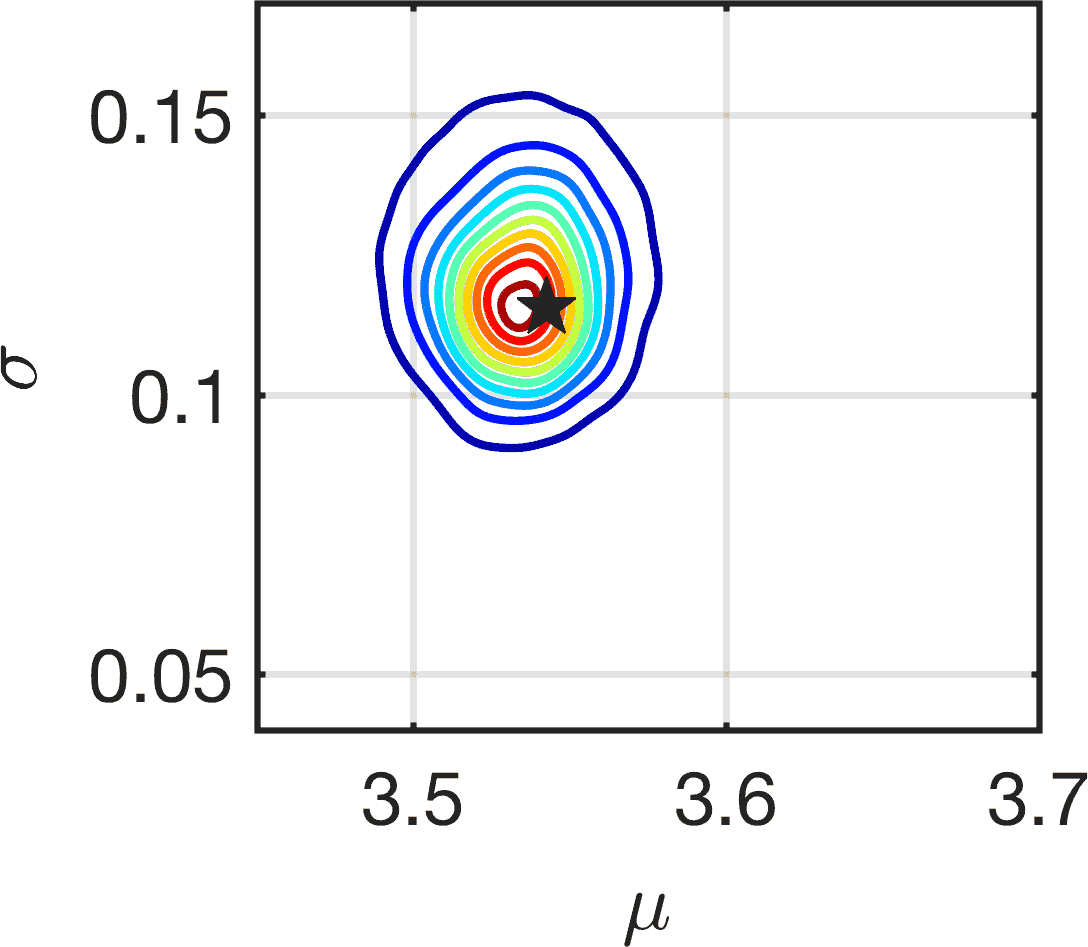}}
&\parbox[c][1.1in]{1in}{\includegraphics[height=1in]{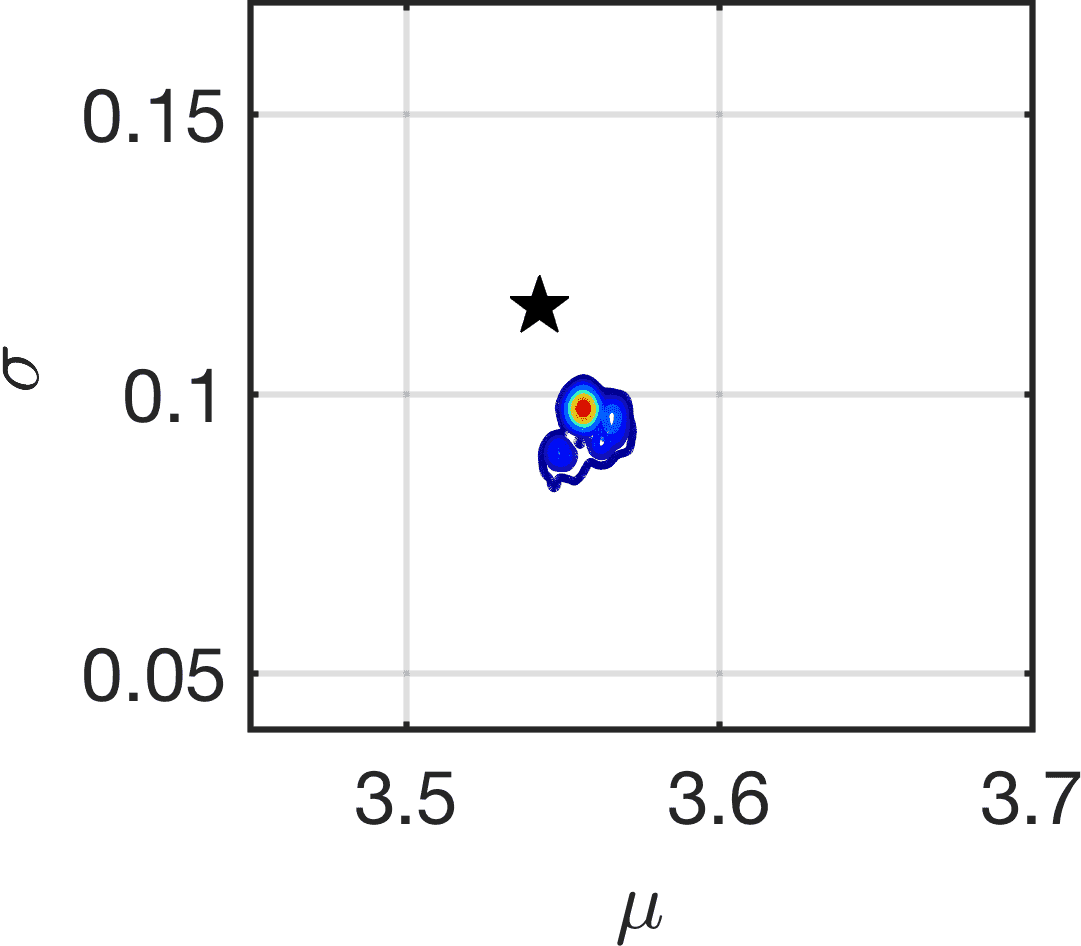}} 
&\parbox[c][1.1in]{1in}{\includegraphics[height=1in]{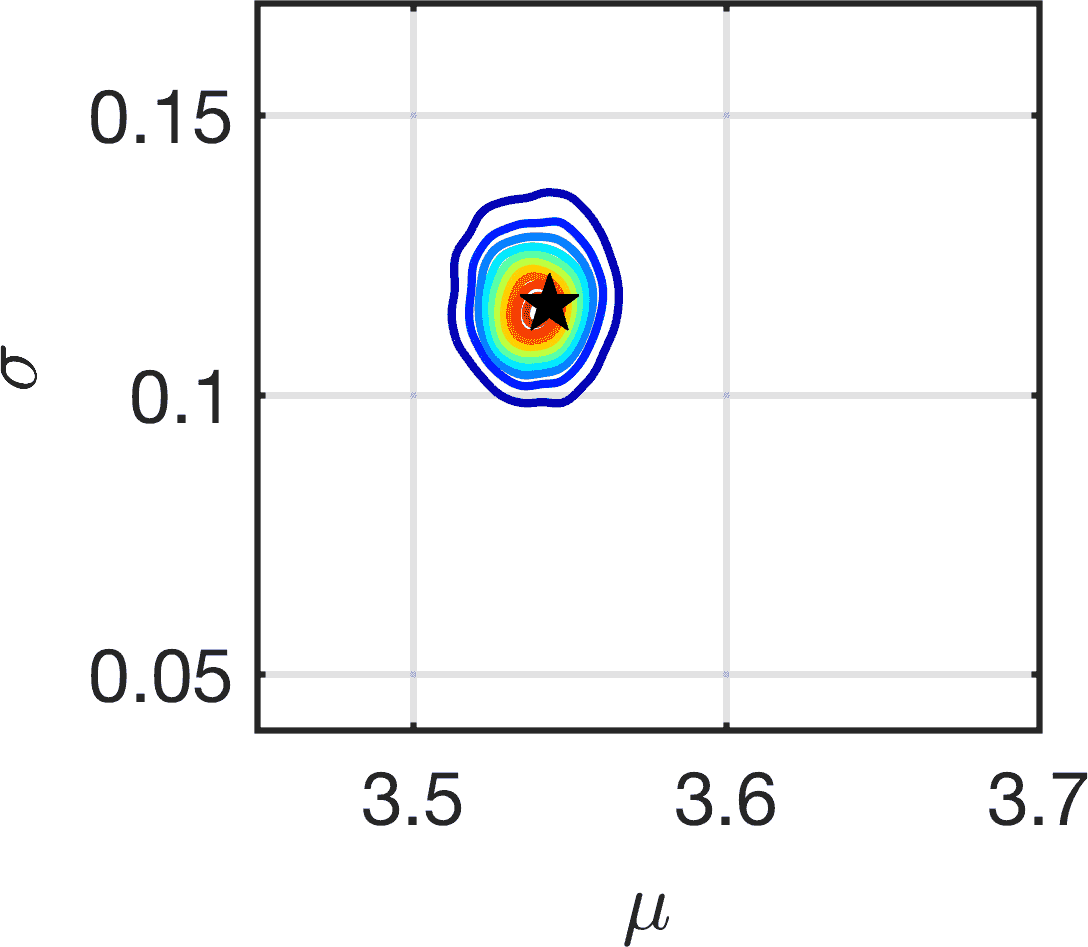}}  
&\parbox[c][1.1in]{1in}{\includegraphics[height=1in]{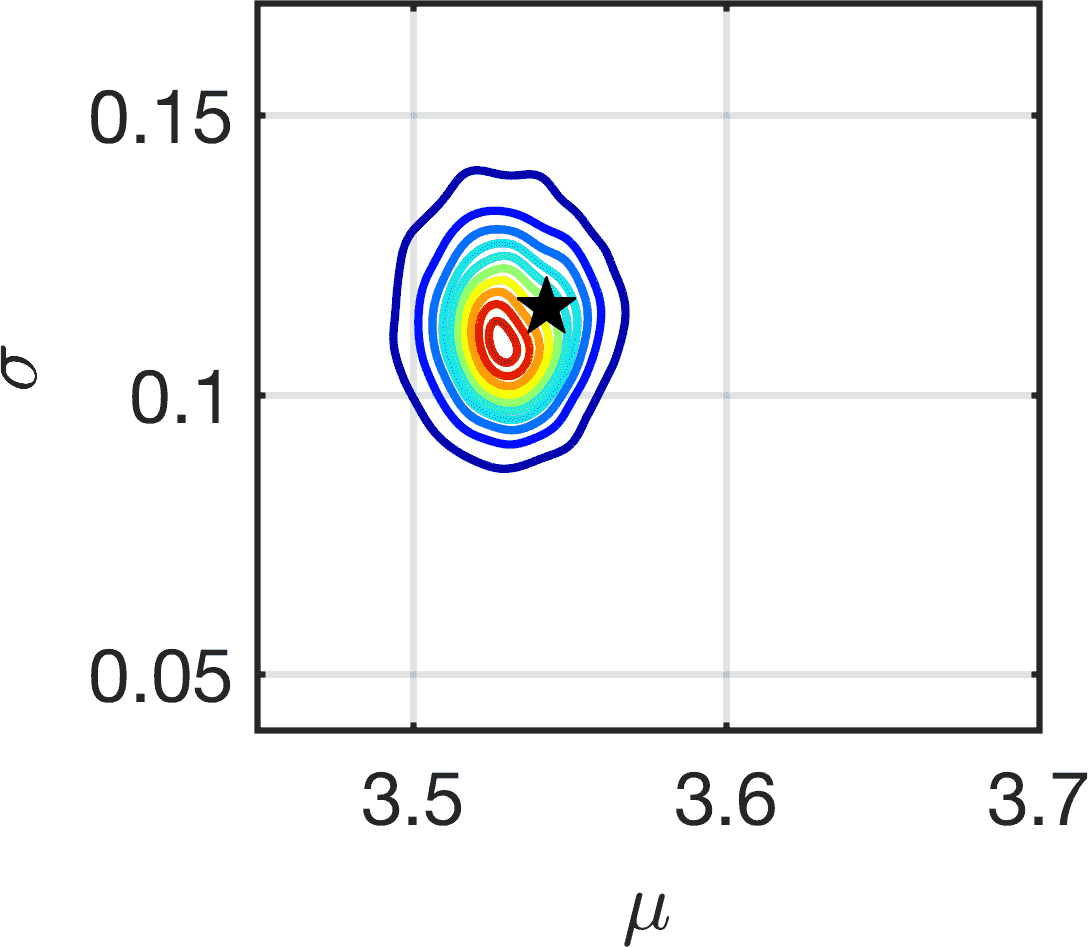}}  
&\parbox[c][1.1in]{1in}{\includegraphics[height=1in]{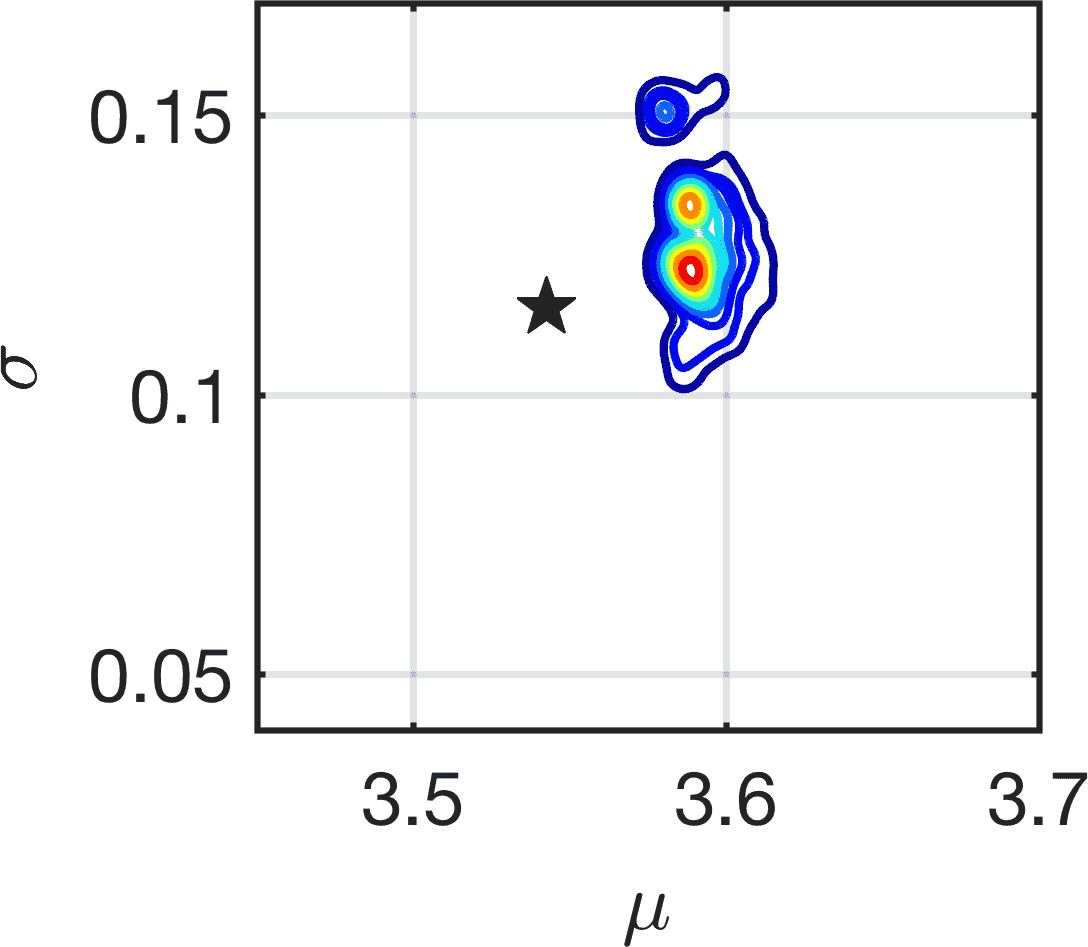}} \\
100
&\parbox[c][1.1in]{1in}{\includegraphics[height=1in]{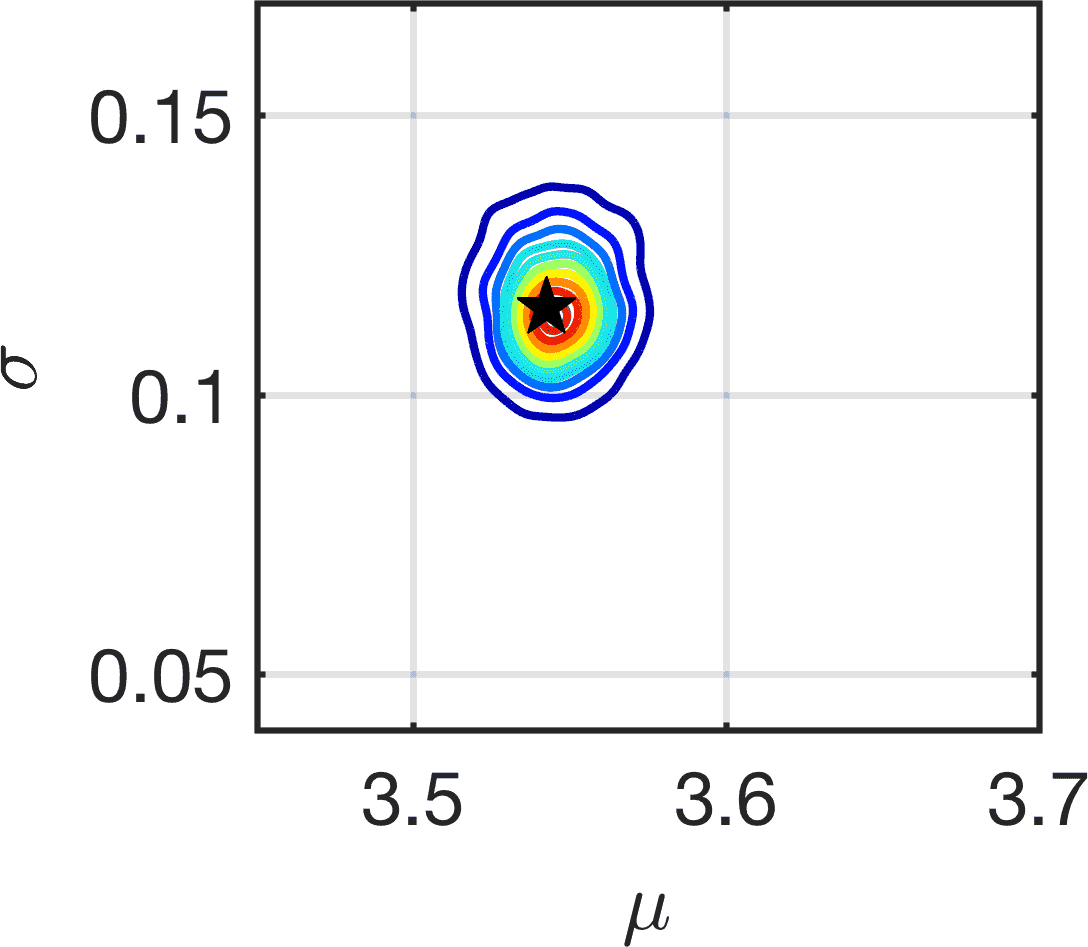}}
&\parbox[c][1.1in]{1in}{\includegraphics[height=1in]{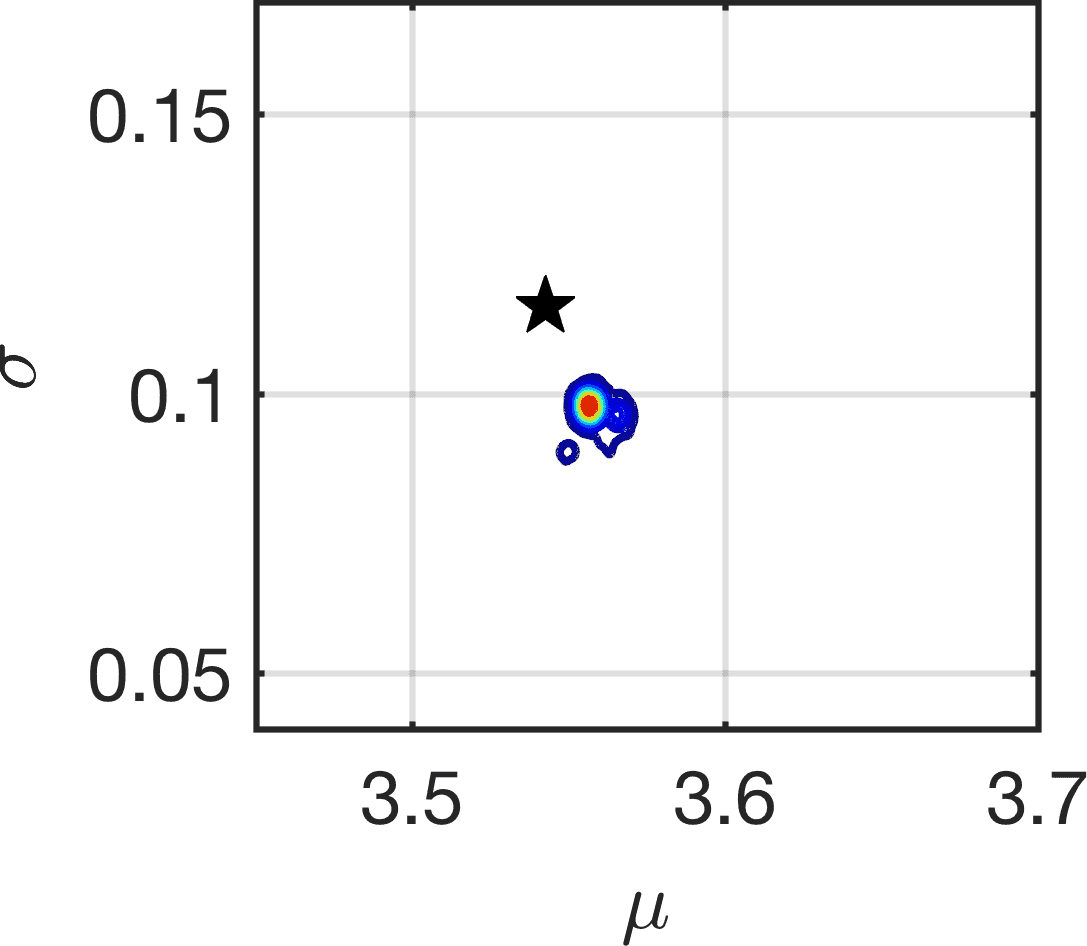}} 
&\parbox[c][1.1in]{1in}{\includegraphics[height=1in]{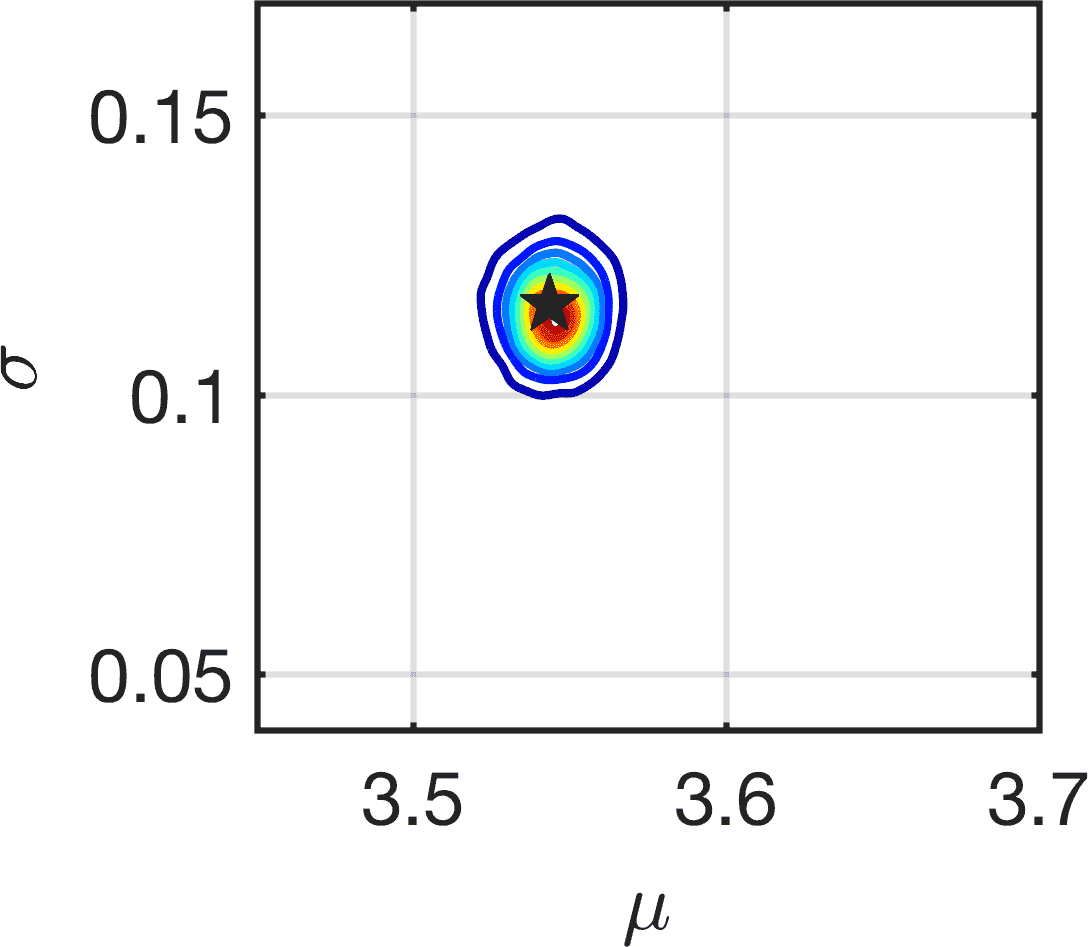}}  
&\parbox[c][1.1in]{1in}{\includegraphics[height=1in]{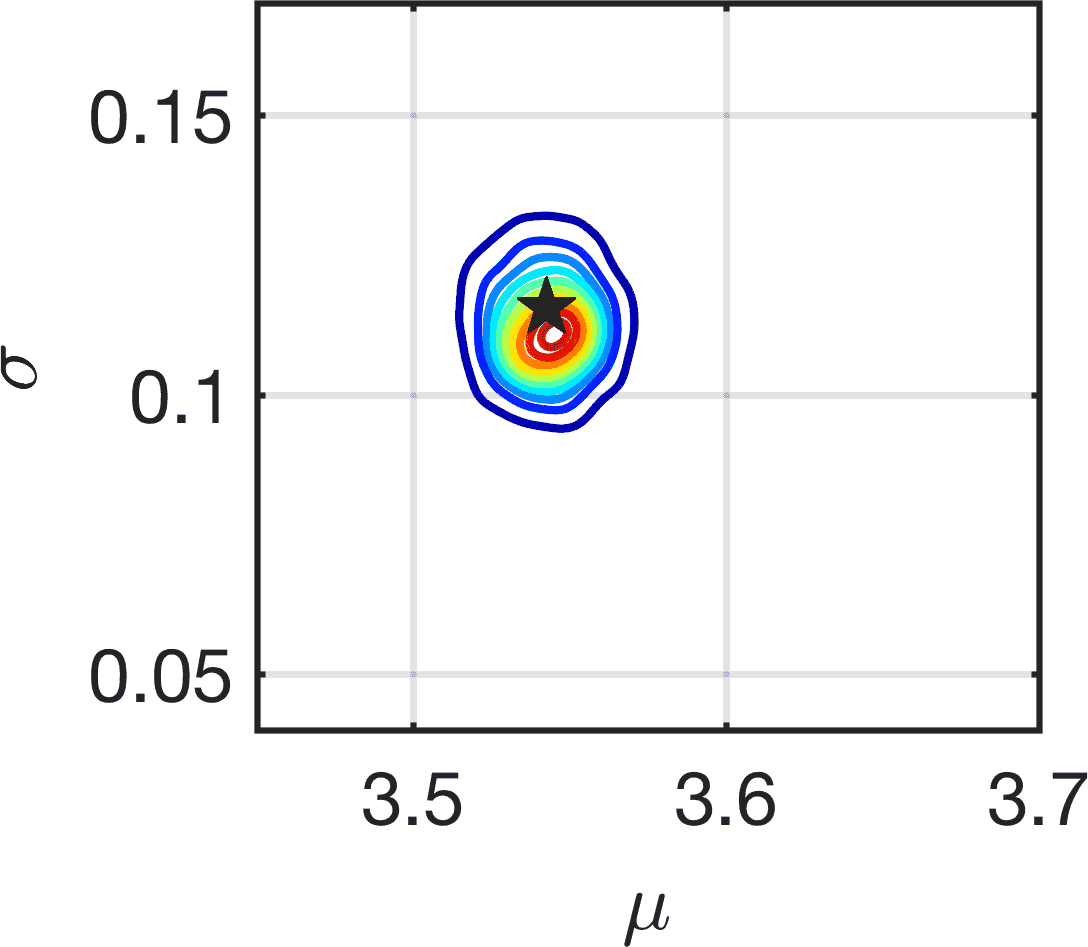}}  
&\parbox[c][1.1in]{1in}{\includegraphics[height=1in]{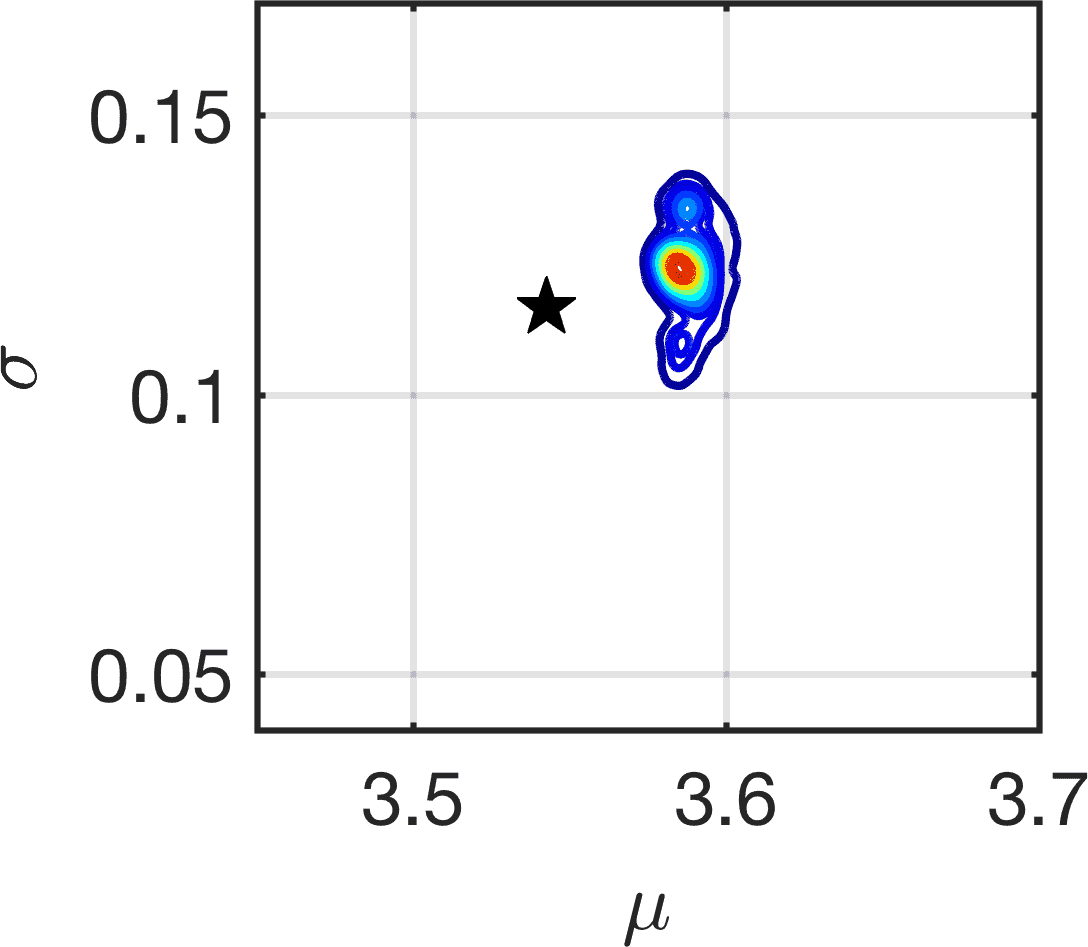}} \\
 \hline
\end{tabular}
\end{table}
Notice that the ABS-A and ASTM-A7 priors yield the most rapid convergence but neither their priors nor their posteriors include the true parameter values. Indeed, these models are unable to infer the correct distribution despite the prior belief that they may serve as reasonable priors. The noninformative and ABS-C priors provide relatively similar levels of convergence and include the true model. This is because they are sufficiently weak in the amount of incorrect information they provide. Lastly, as expected the correct ABS-B prior exhibits the best convergence to the true model.

Table \ref{tab: lognormal_large} provides similar plots for ``large'' datasets ($\ge$ 500 data). 
\begin{table}[!ht] \footnotesize
\centering
\caption{Posterior parameter joint probability densities for the lognormal distribution with different priors considering large dataset size ($\ge 500$ data).}
\label{tab: lognormal_large}
\begin{tabular}{crrrrr}
\hline 
Data & Noninformative & ABS-A (33) & ABS-B (79) & ABS-C (13) & ASTM-A7 (58)\\ \hline
500
&\parbox[c][1.1in]{1in}{\includegraphics[height=1in]{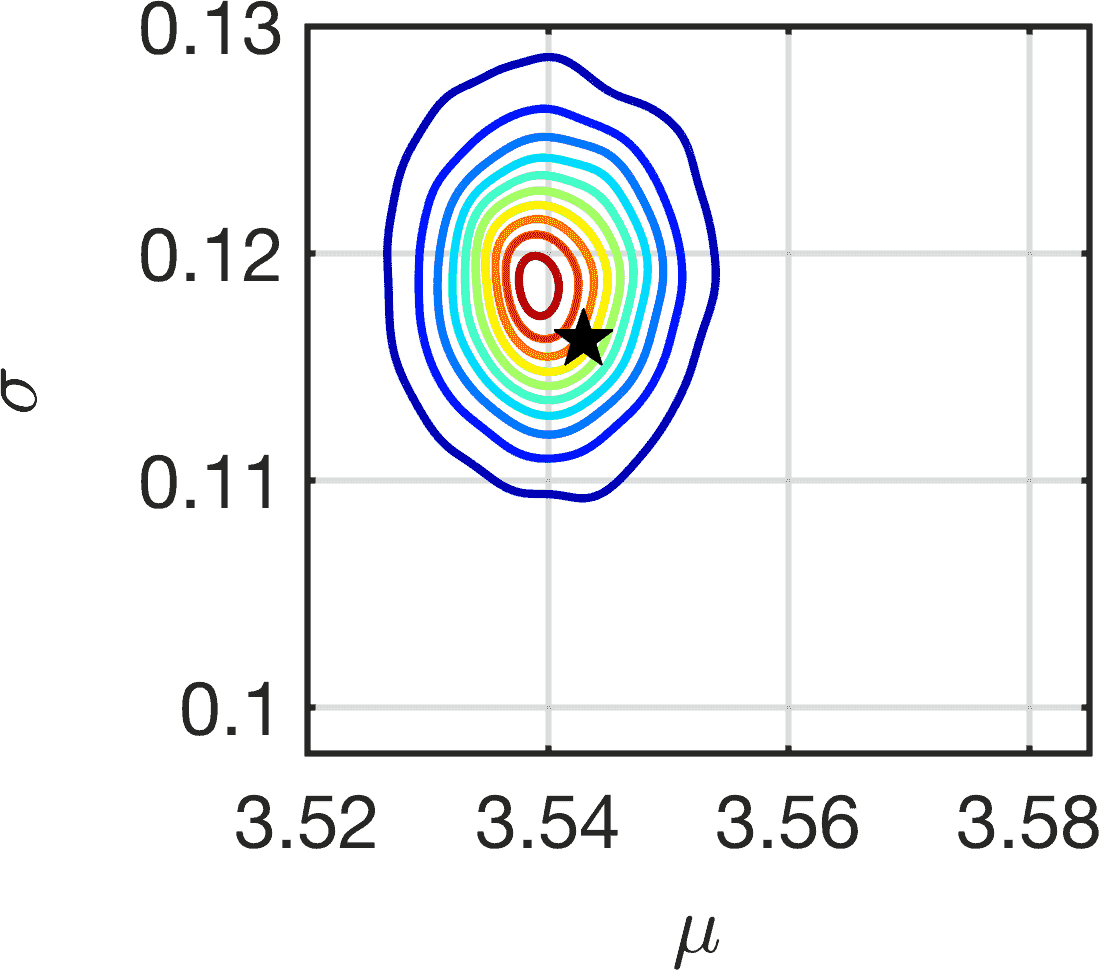}}
&\parbox[c][1.1in]{1in}{\includegraphics[height=1in]{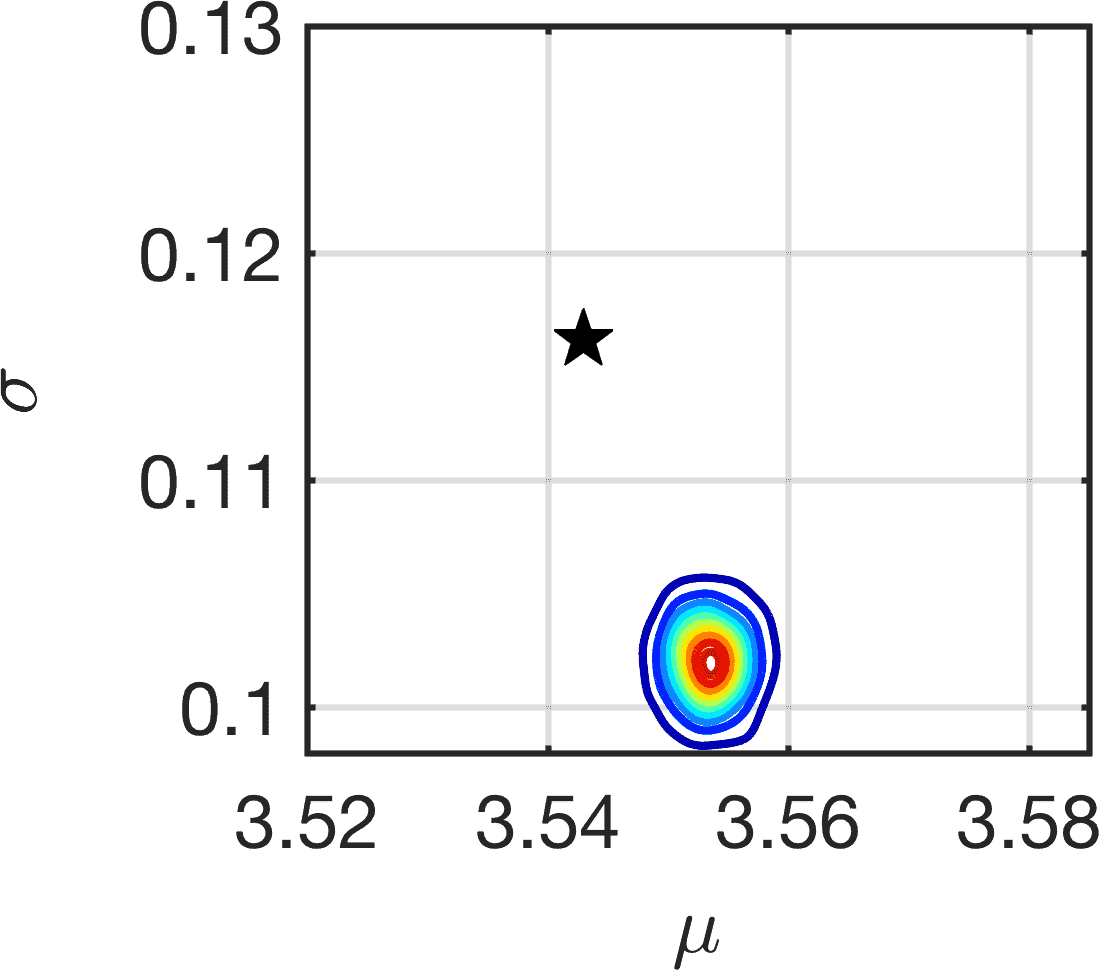}} 
&\parbox[c][1.1in]{1in}{\includegraphics[height=1in]{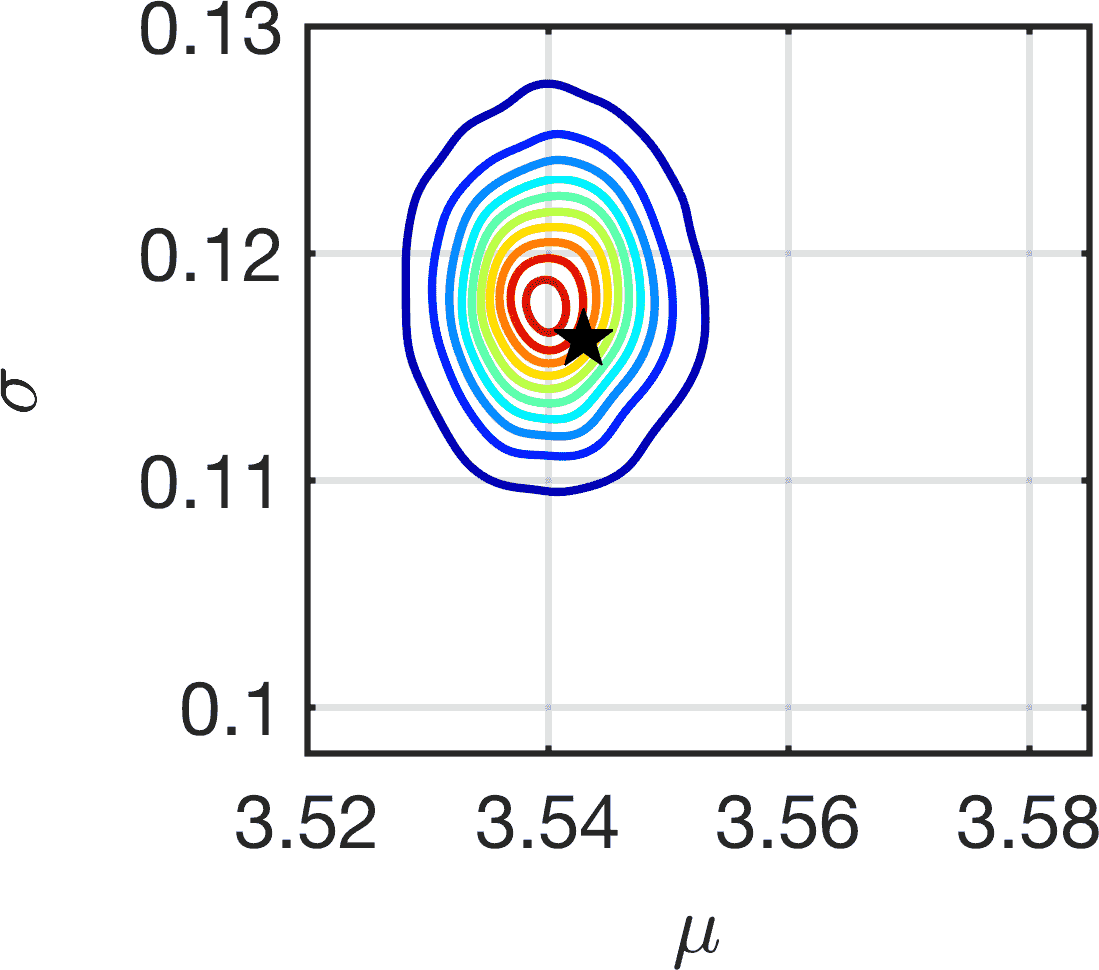}}  
&\parbox[c][1.1in]{1in}{\includegraphics[height=1in]{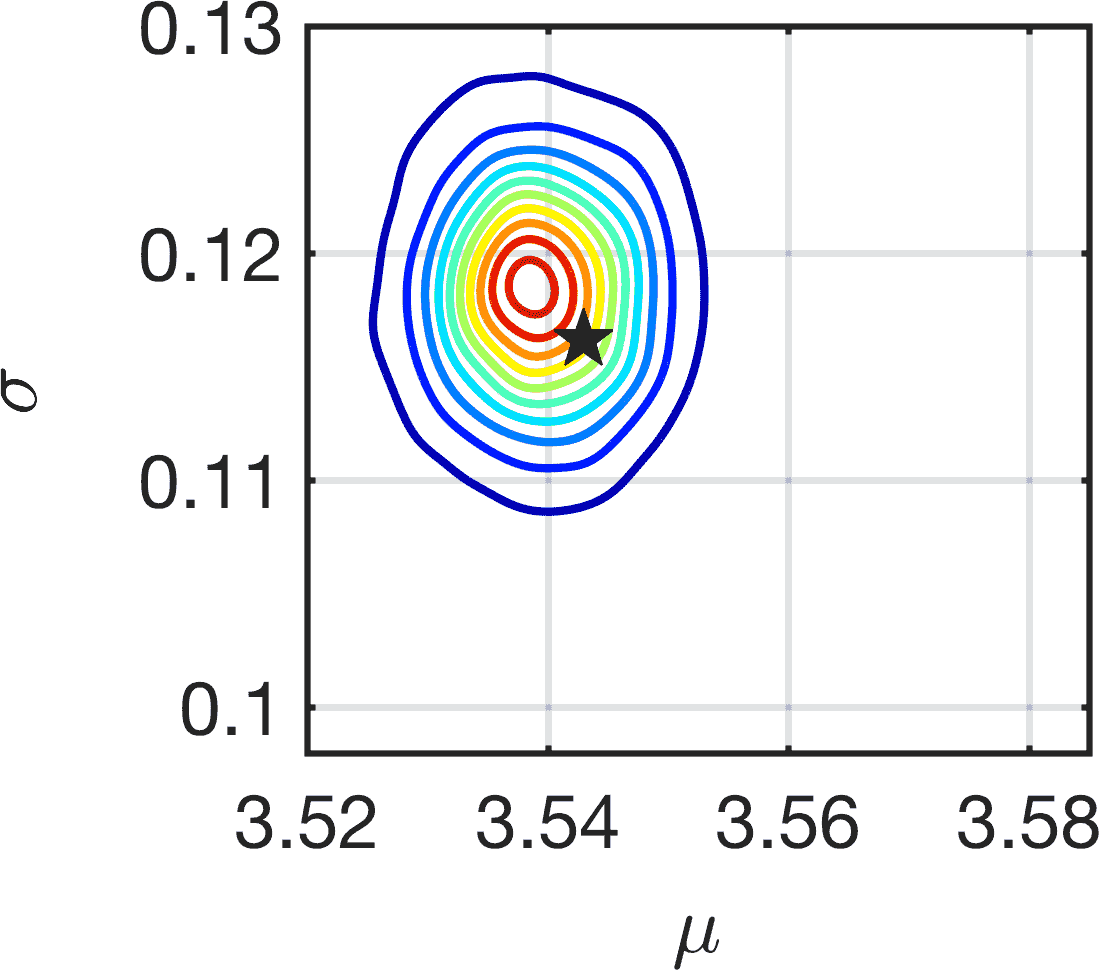}}  
&\parbox[c][1.1in]{1in}{\includegraphics[height=1in]{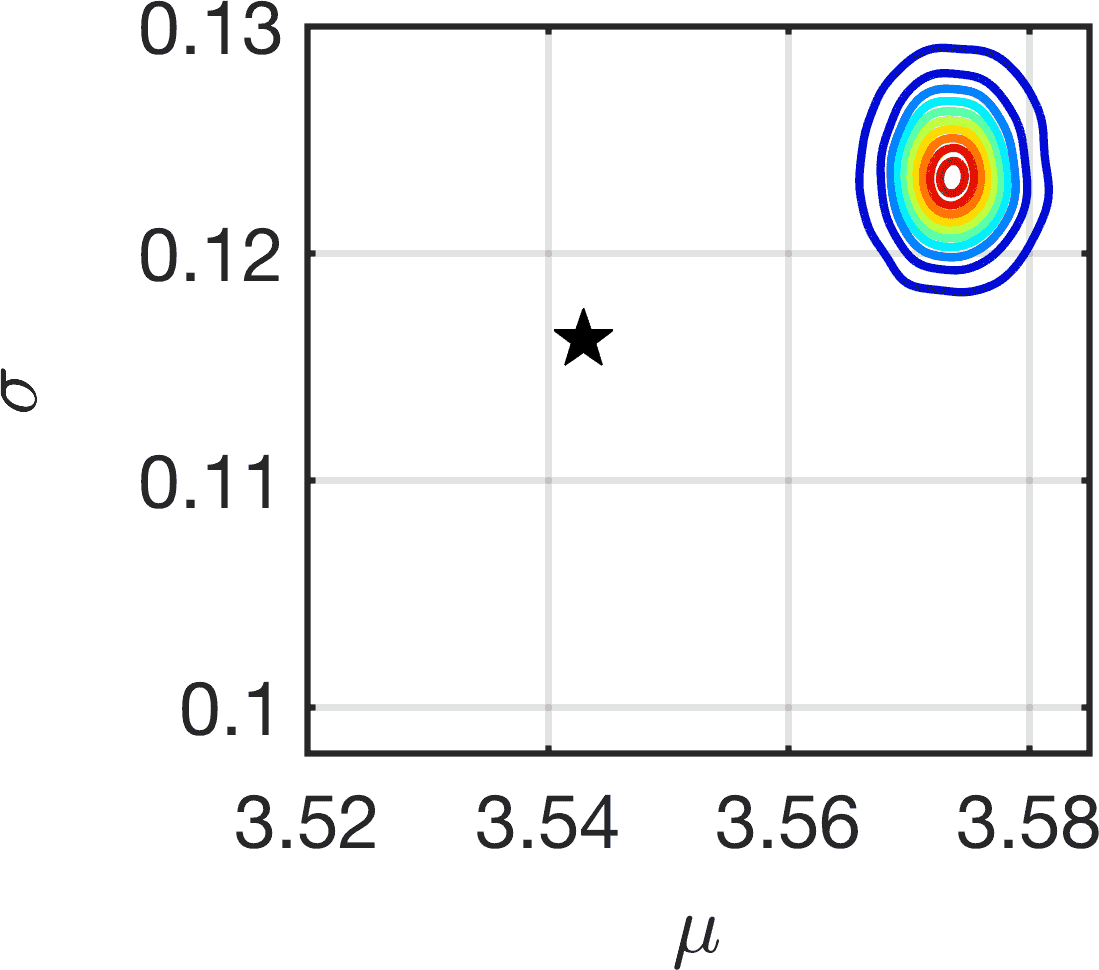}} \\
1000
&\parbox[c][1.1in]{1in}{\includegraphics[height=1in]{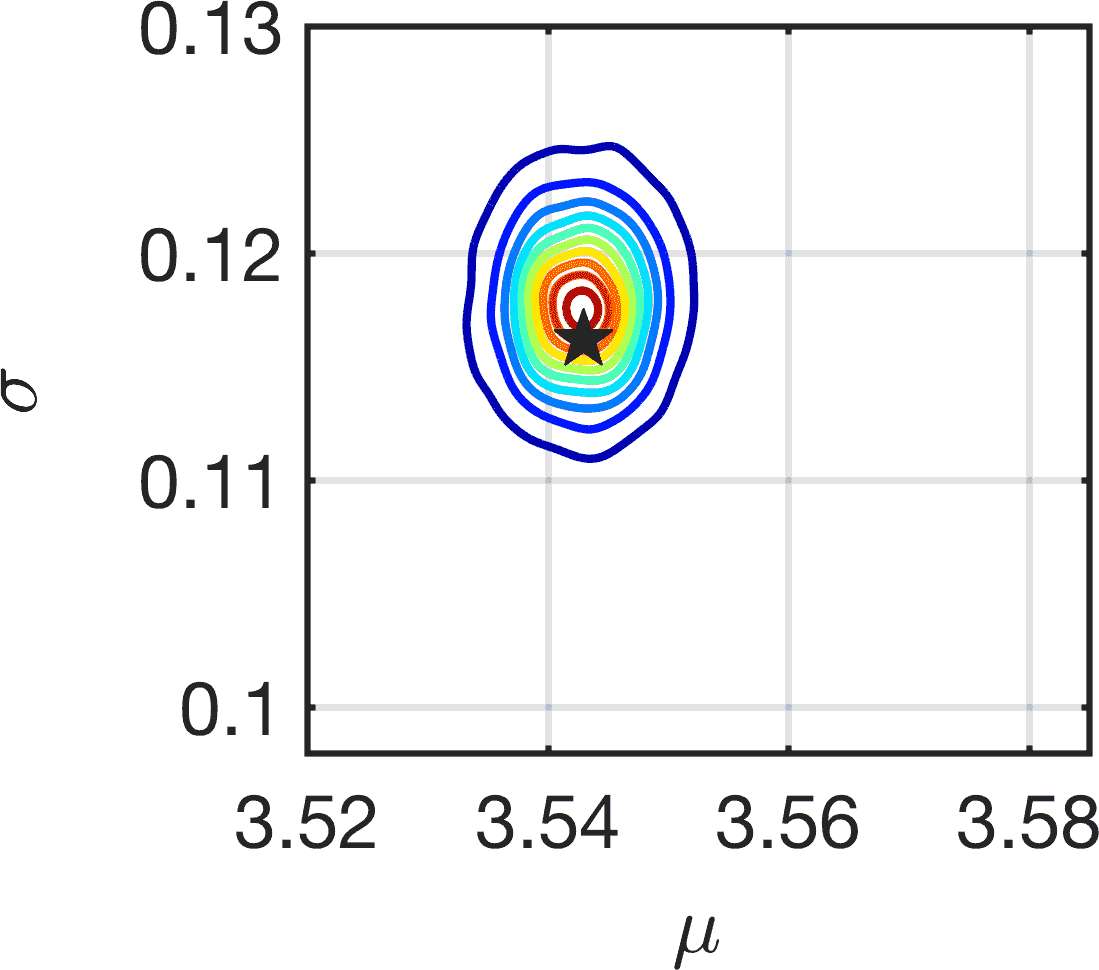}}
&\parbox[c][1.1in]{1in}{\includegraphics[height=1in]{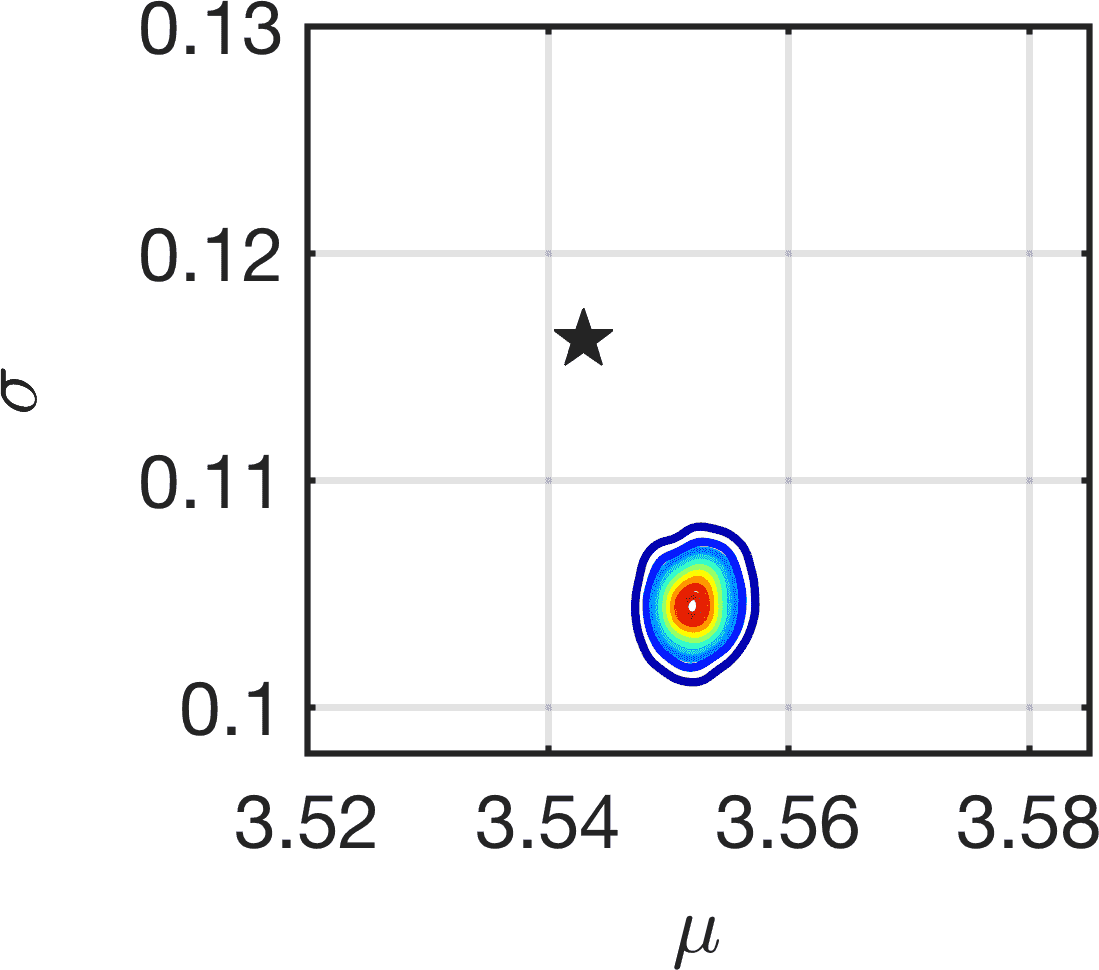}} 
&\parbox[c][1.1in]{1in}{\includegraphics[height=1in]{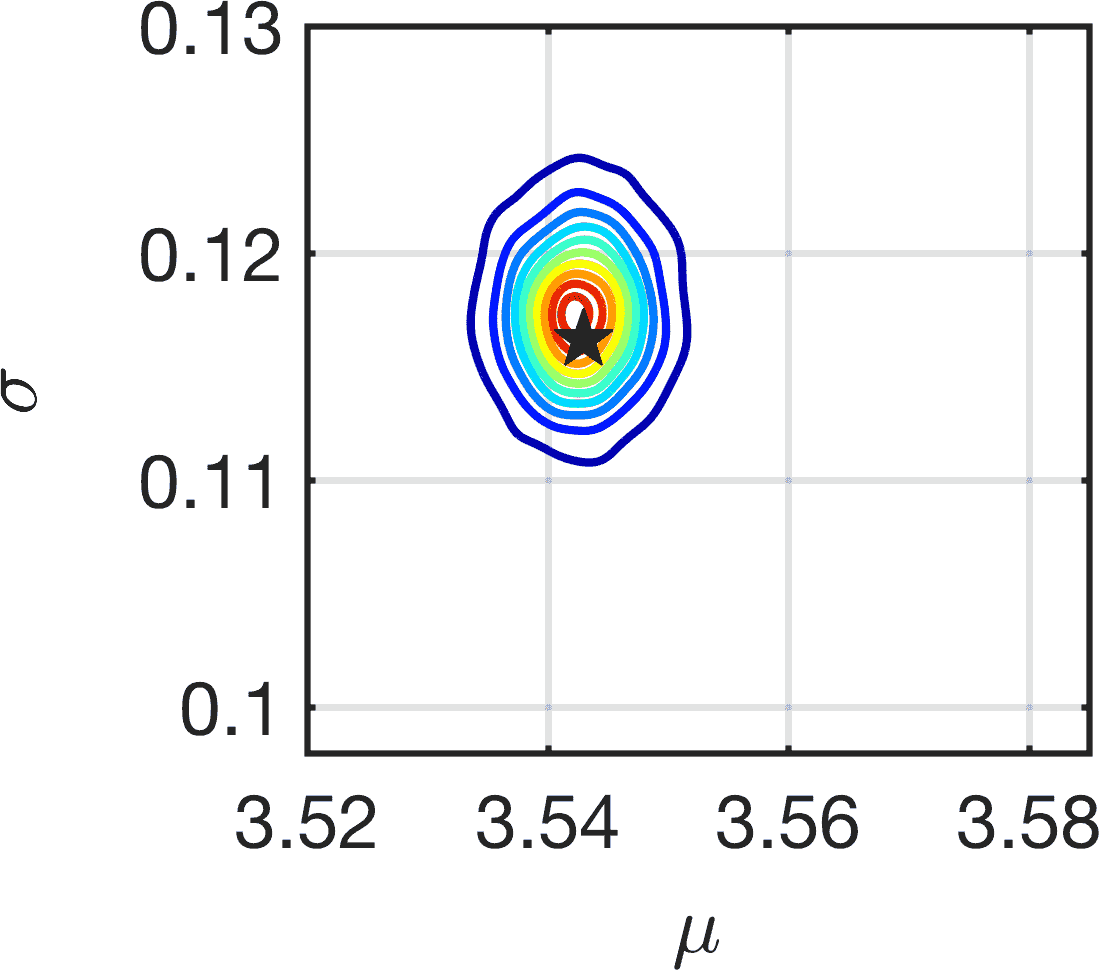}}  
&\parbox[c][1.1in]{1in}{\includegraphics[height=1in]{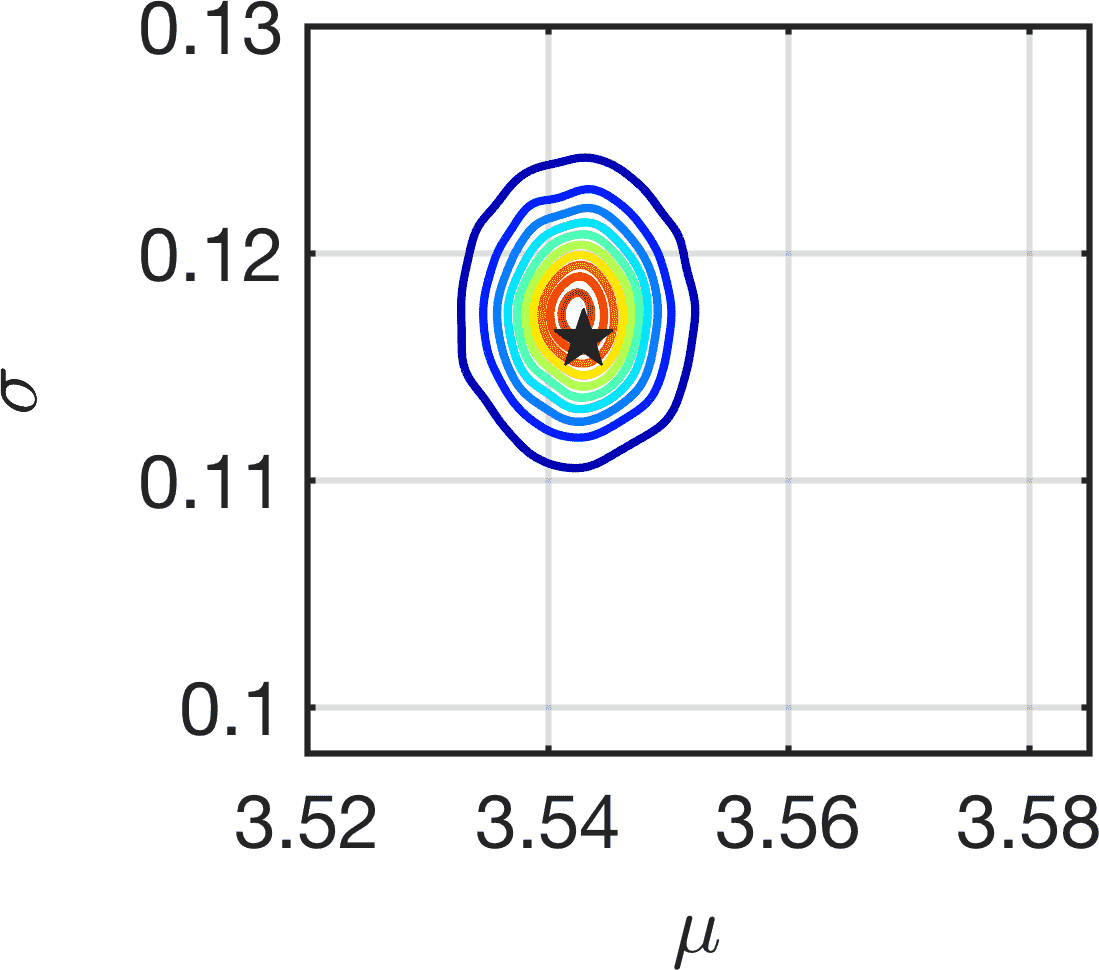}}  
&\parbox[c][1.1in]{1in}{\includegraphics[height=1in]{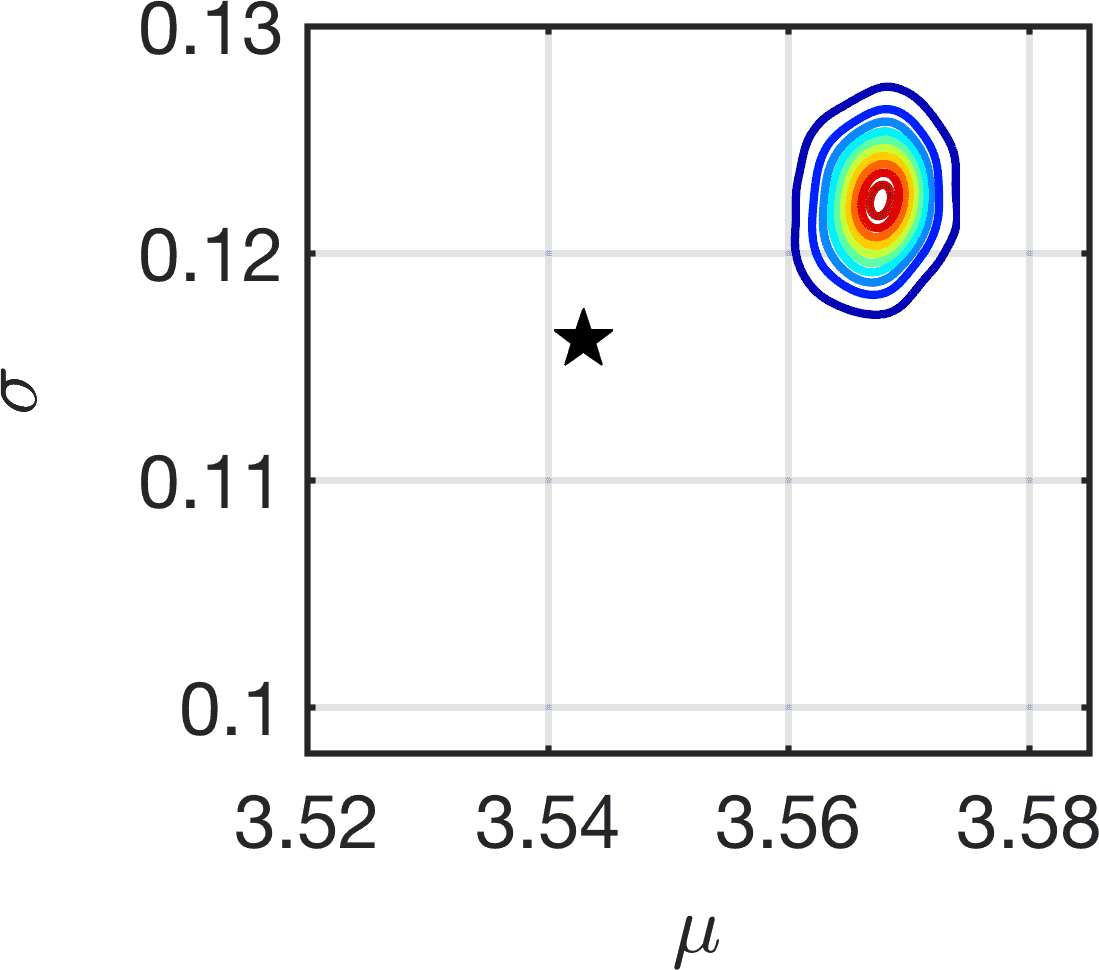}} \\
5000
&\parbox[c][1.1in]{1in}{\includegraphics[height=1in]{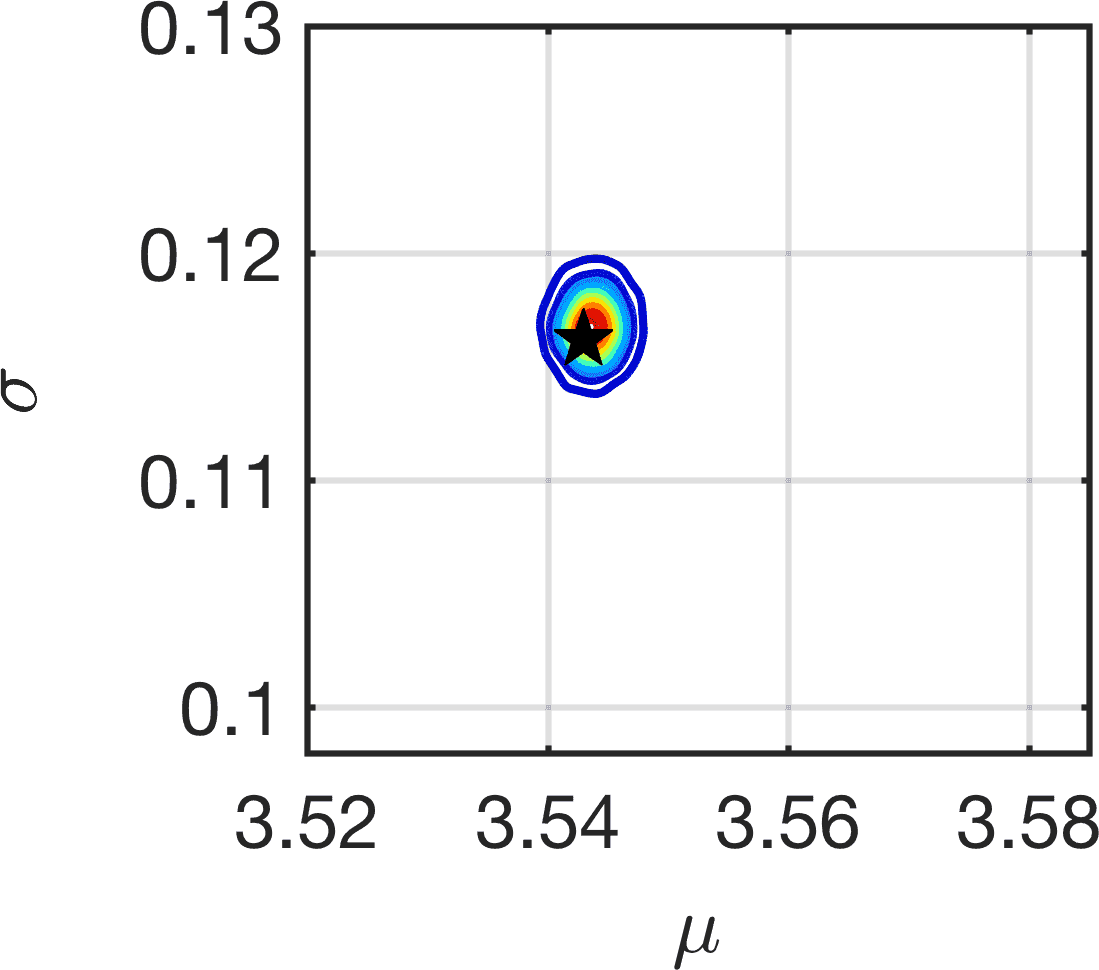}}
&\parbox[c][1.1in]{1in}{\includegraphics[height=1in]{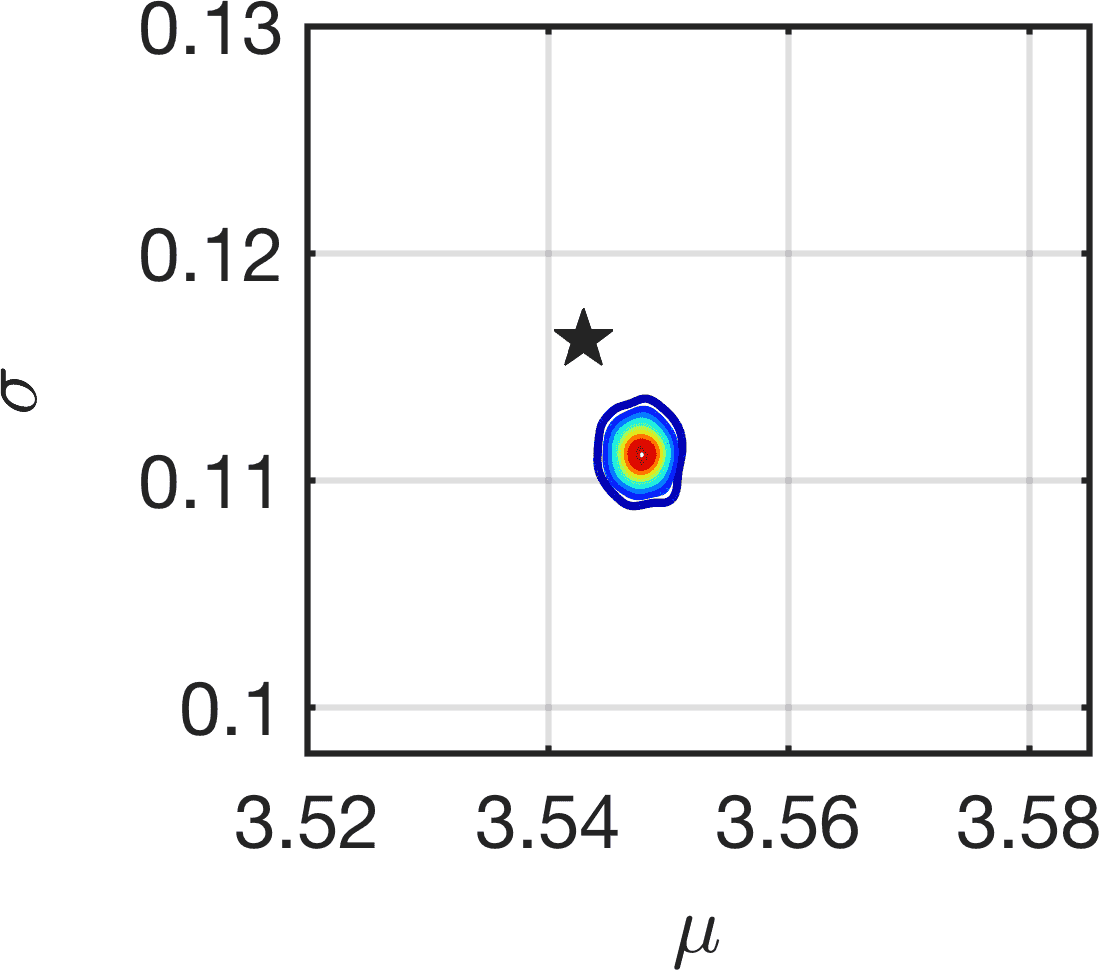}} 
&\parbox[c][1.1in]{1in}{\includegraphics[height=1in]{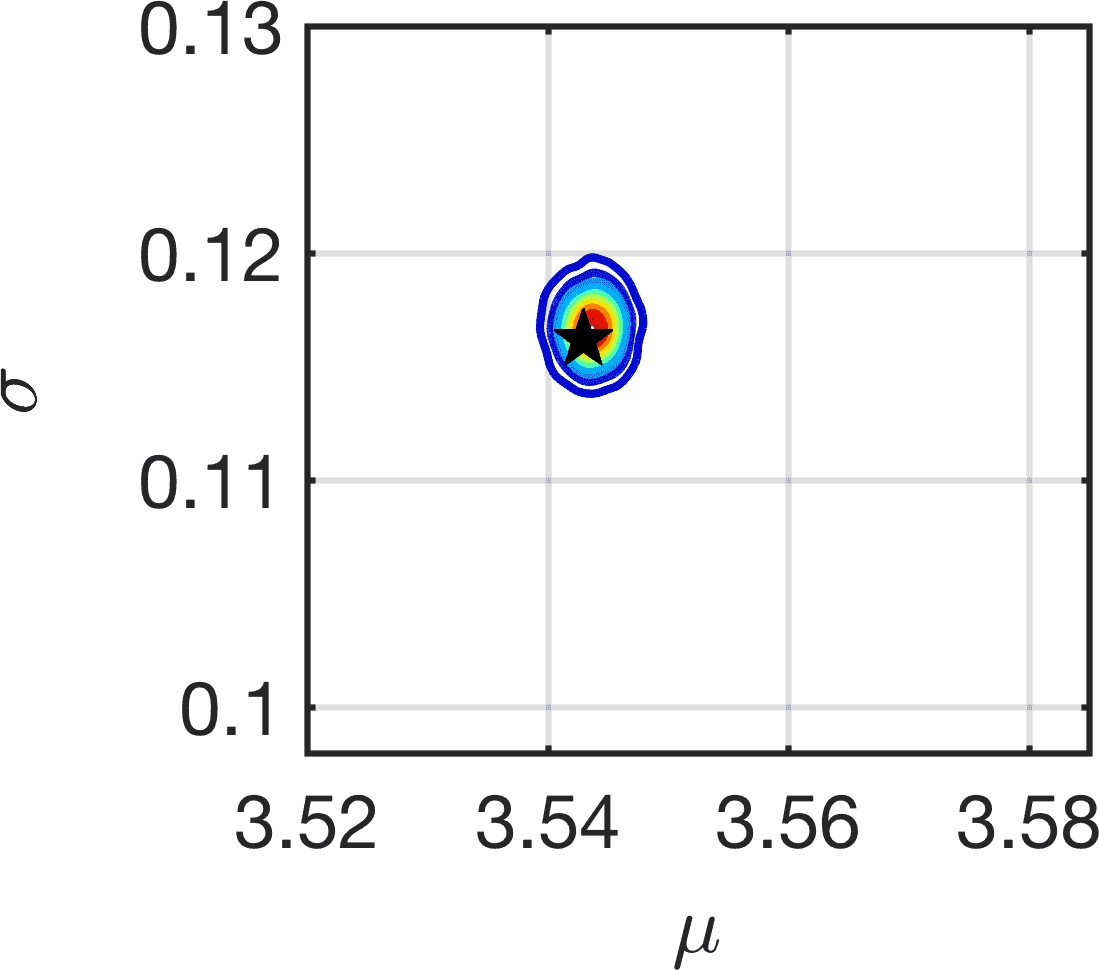}}  
&\parbox[c][1.1in]{1in}{\includegraphics[height=1in]{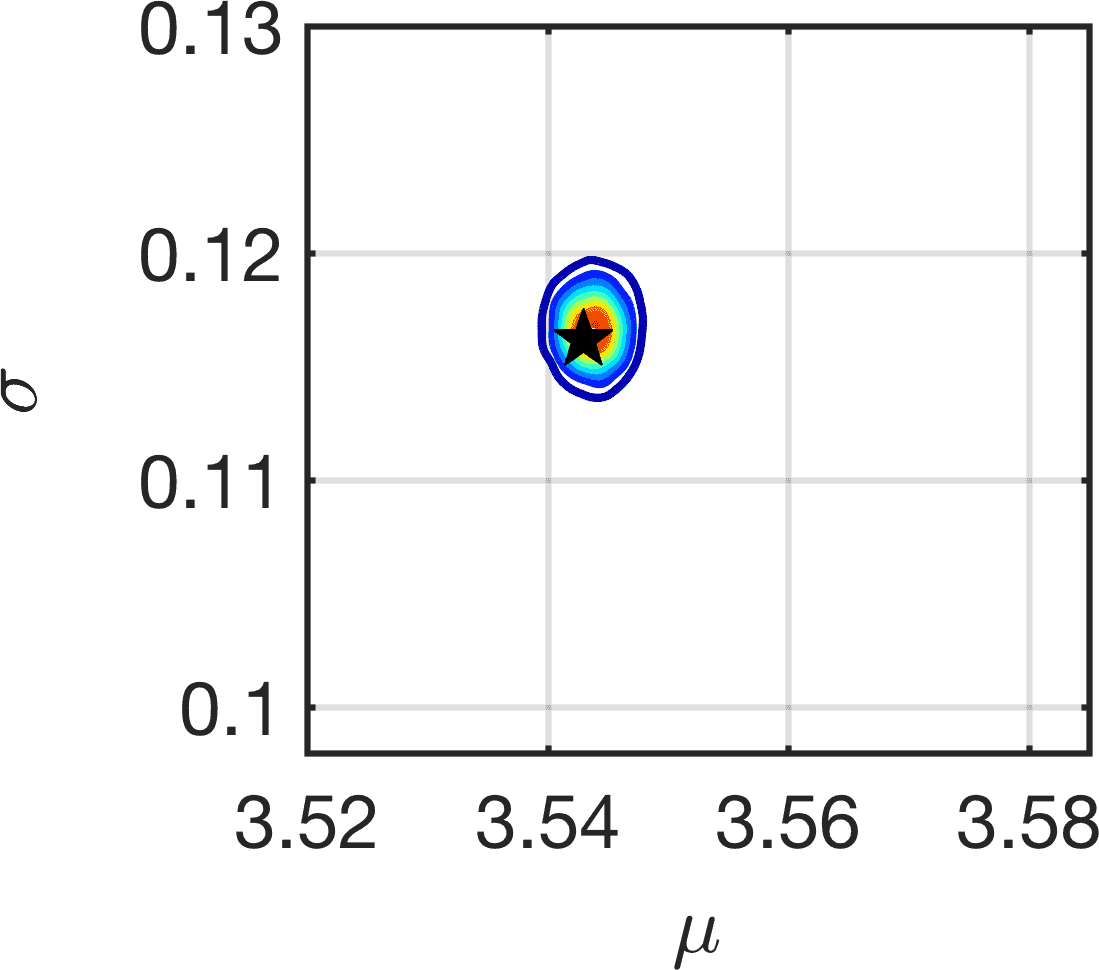}}  
&\parbox[c][1.1in]{1in}{\includegraphics[height=1in]{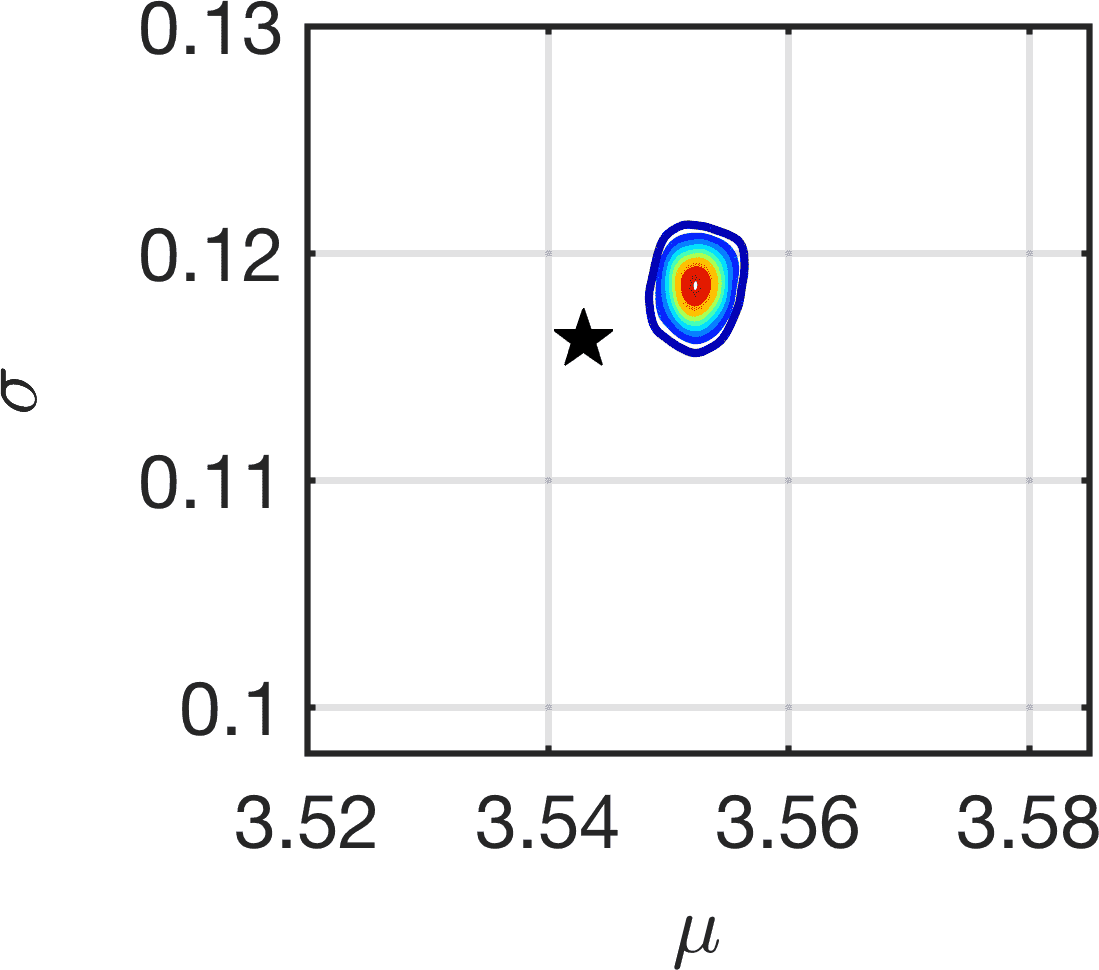}} \\
10000
&\parbox[c][1.1in]{1in}{\includegraphics[height=1in]{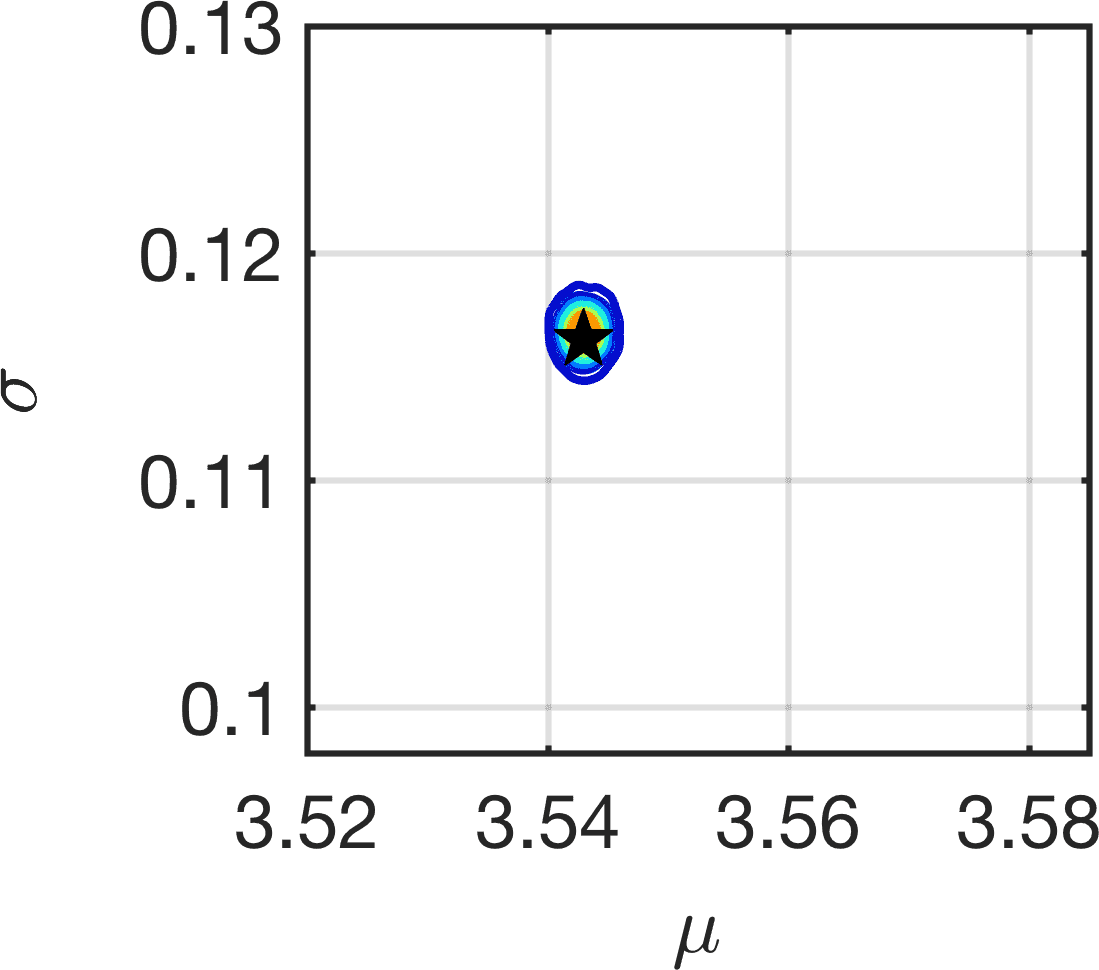}}
&\parbox[c][1.1in]{1in}{\includegraphics[height=1in]{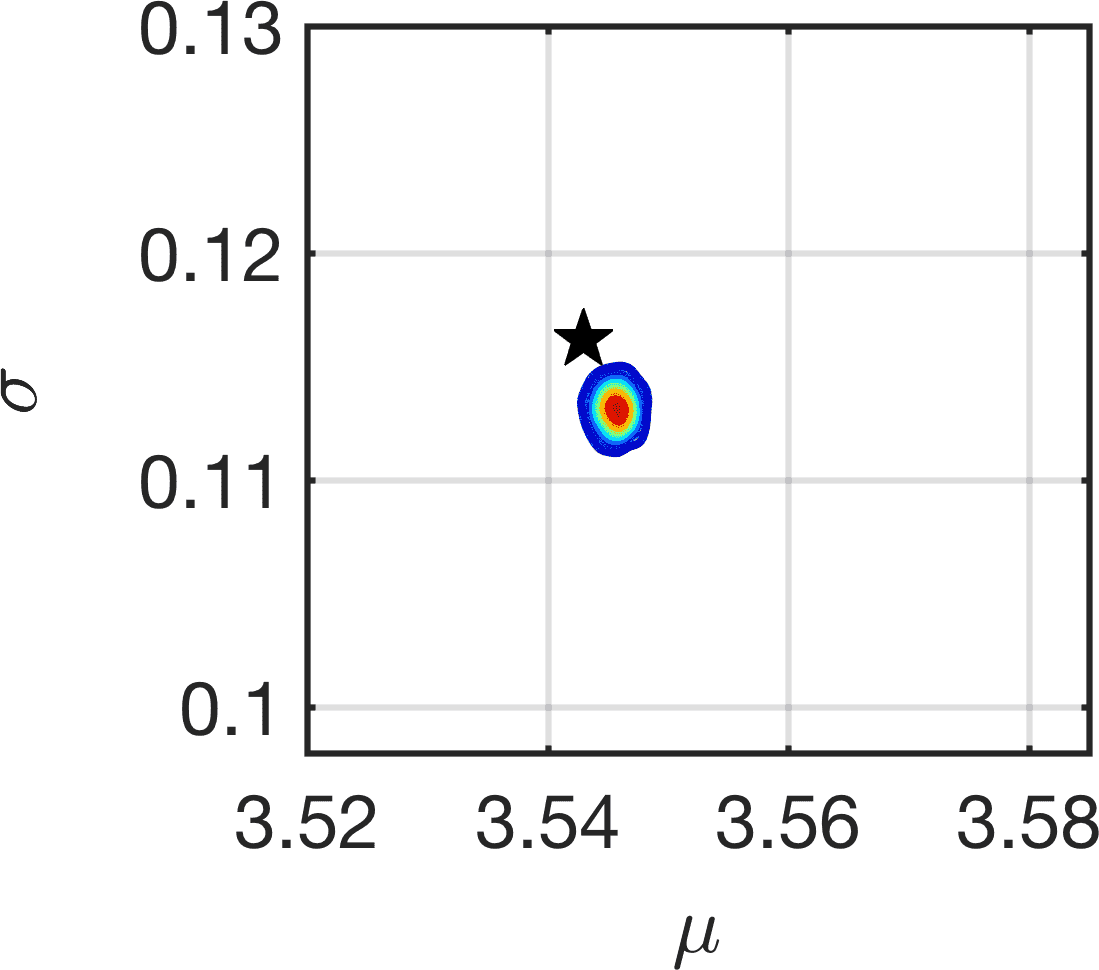}} 
&\parbox[c][1.1in]{1in}{\includegraphics[height=1in]{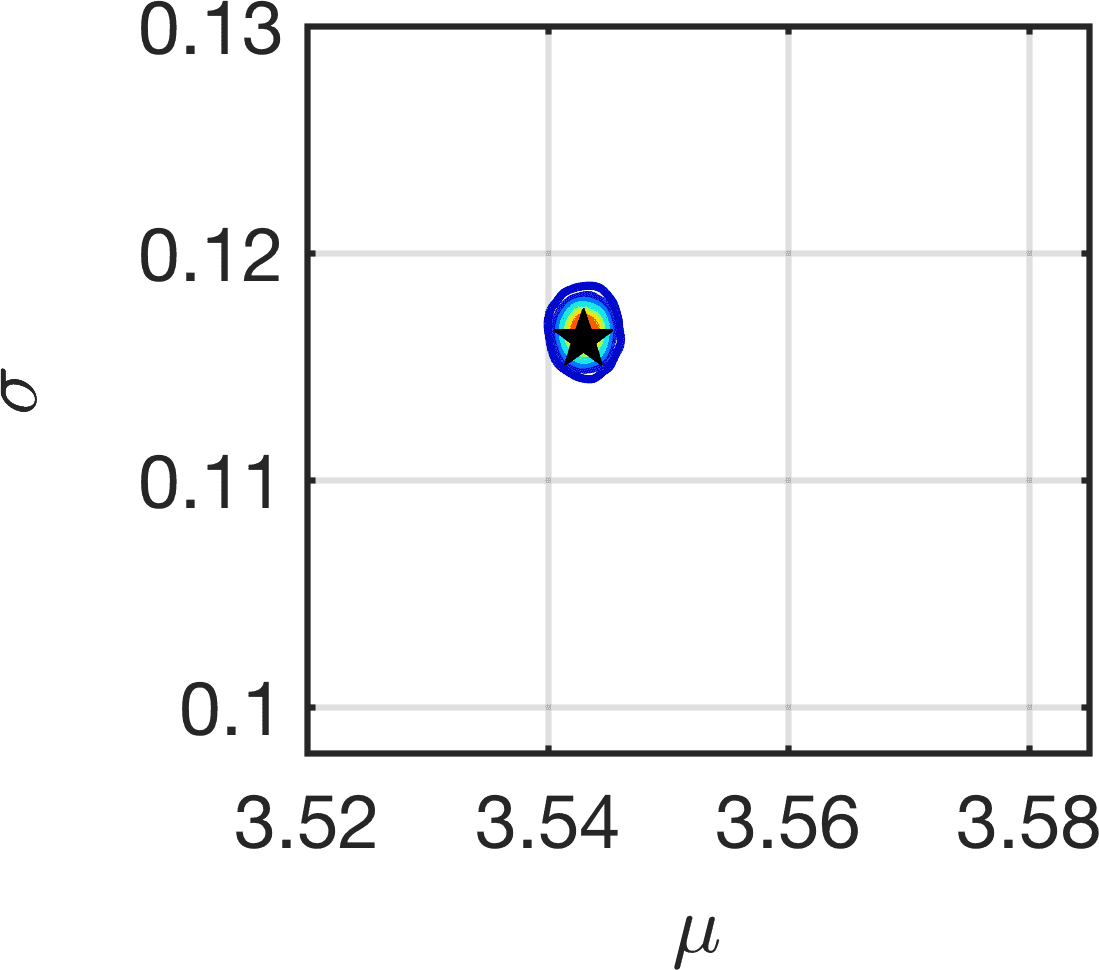}}  
&\parbox[c][1.1in]{1in}{\includegraphics[height=1in]{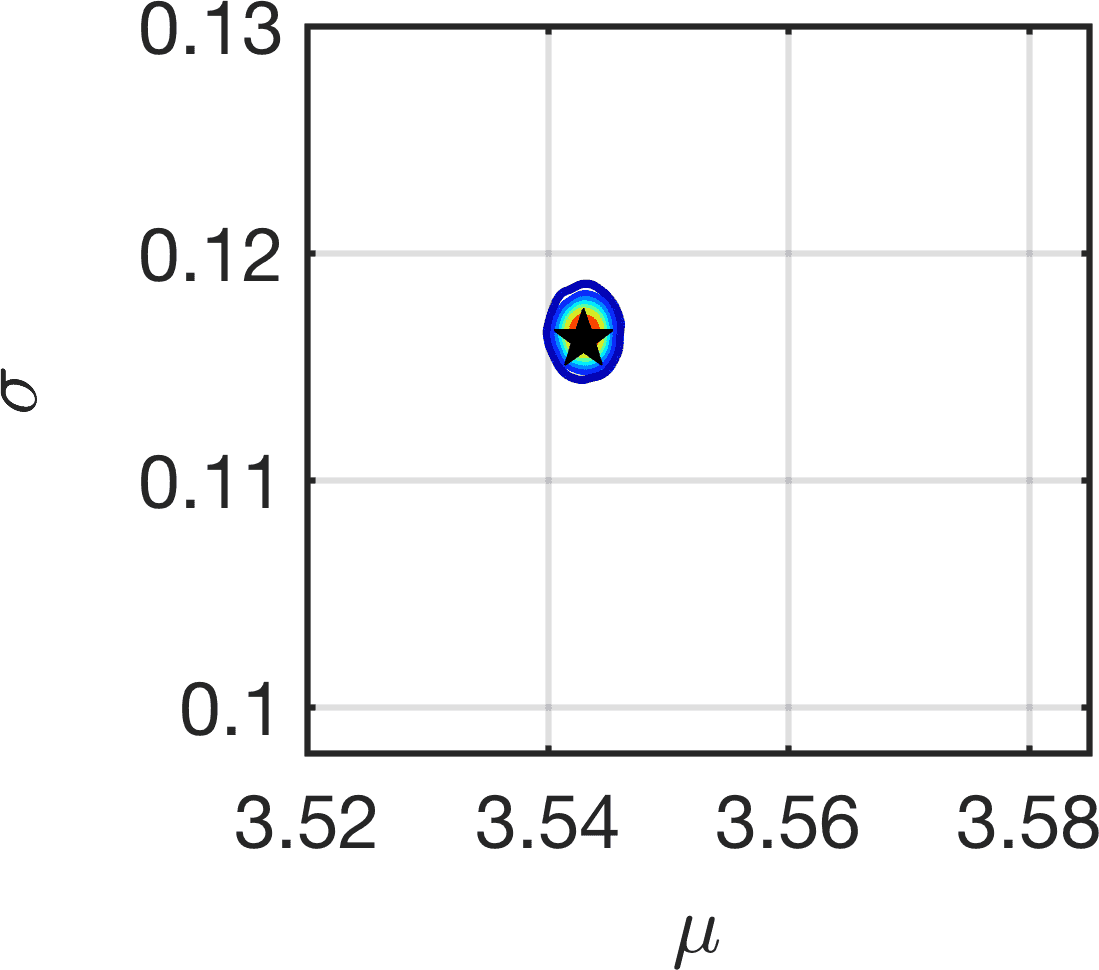}}  
&\parbox[c][1.1in]{1in}{\includegraphics[height=1in]{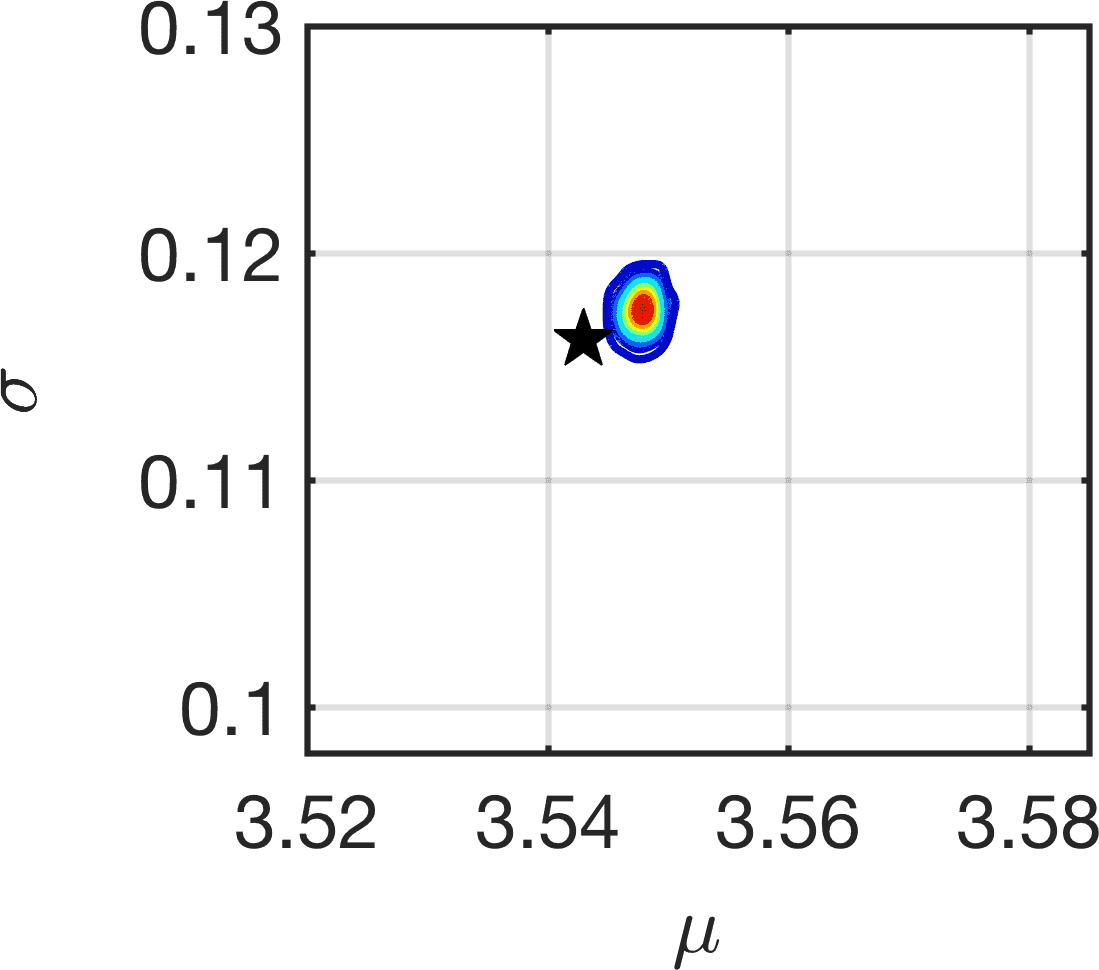}} \\
 \hline
\end{tabular}
\end{table}
We see that the models with ABS-A and ASTM-A7 priors continue to narrow and move slowly toward the correct parameters. However, even after 10,000 data are collected they do not come to include the correct parameters in their joint density with any significant probability. The noninformative, ABS-B, and ABS-C meanwhile continue to converge correctly at similar rates (there is very little improvement from using the ABS-B prior for large datasets). 

An alternative way to look at this is to populate a set of possible distributions by Monte Carlo sampling from the joint parameter densities. This is useful both for illustration purposes and because, as shown in the following section, we do this in order to propagate the total uncertainty. Table \ref{tab:lognormal_distributions} shows how these distributions change with dataset size for the noninformative, ABS-B, and ABS-A parameter priors. 
\begin{table}[!ht] \footnotesize
\centering
\caption{Monte Carlo sets of lognormal distributions drawn from the posterior parameter densities given noninformative, ABS-A, and ABS-B prior parameter densities.}
\label{tab:lognormal_distributions}
\begin{tabular}{cccc}
\hline 
Data & Noninformative & ABS-A (33) & ABS-B (79) \\ \hline
10
&\parbox[c][1.1in]{1in}{\includegraphics[height=1in]{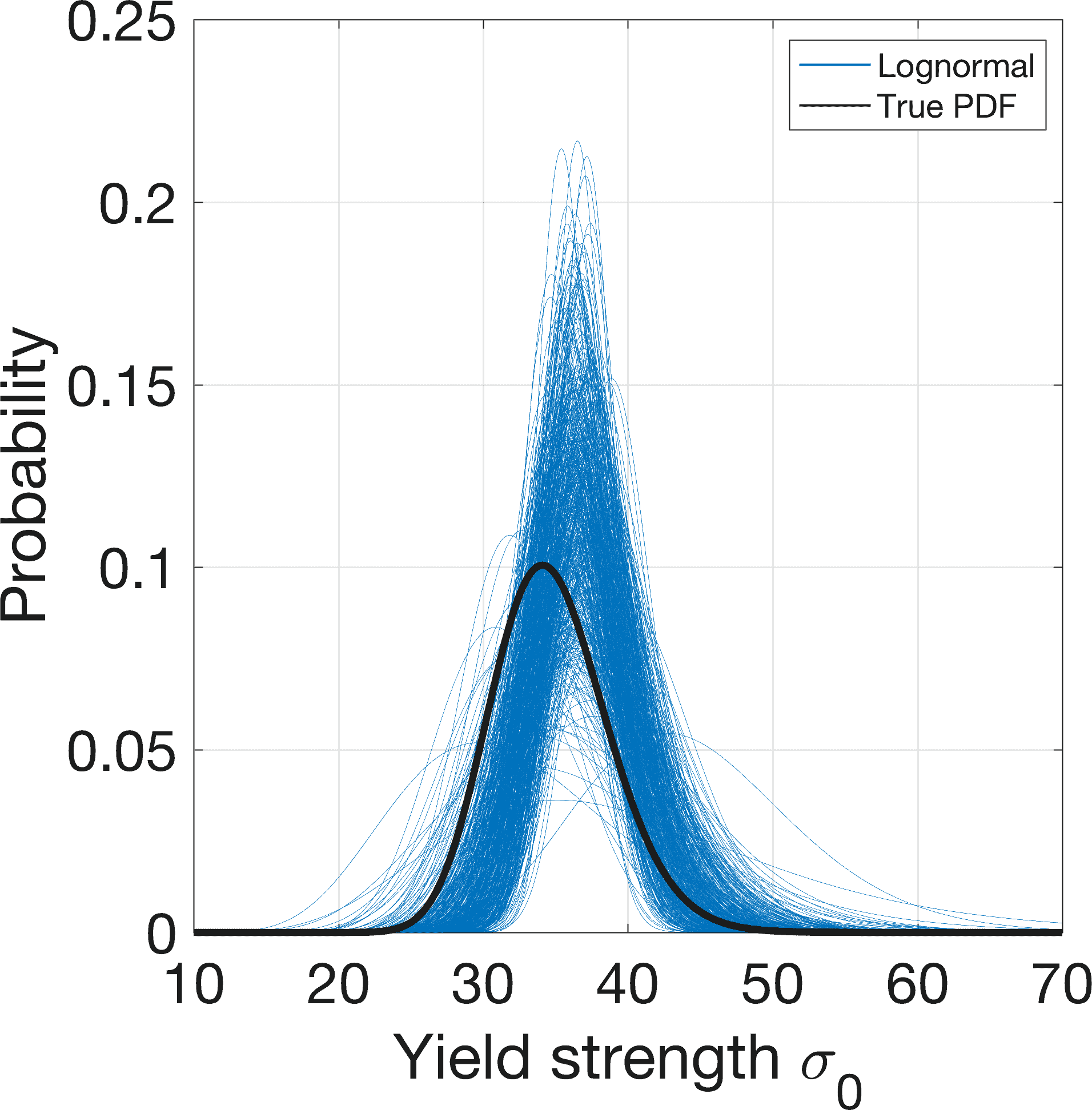}}
&\parbox[c][1.1in]{1in}{\includegraphics[height=1in]{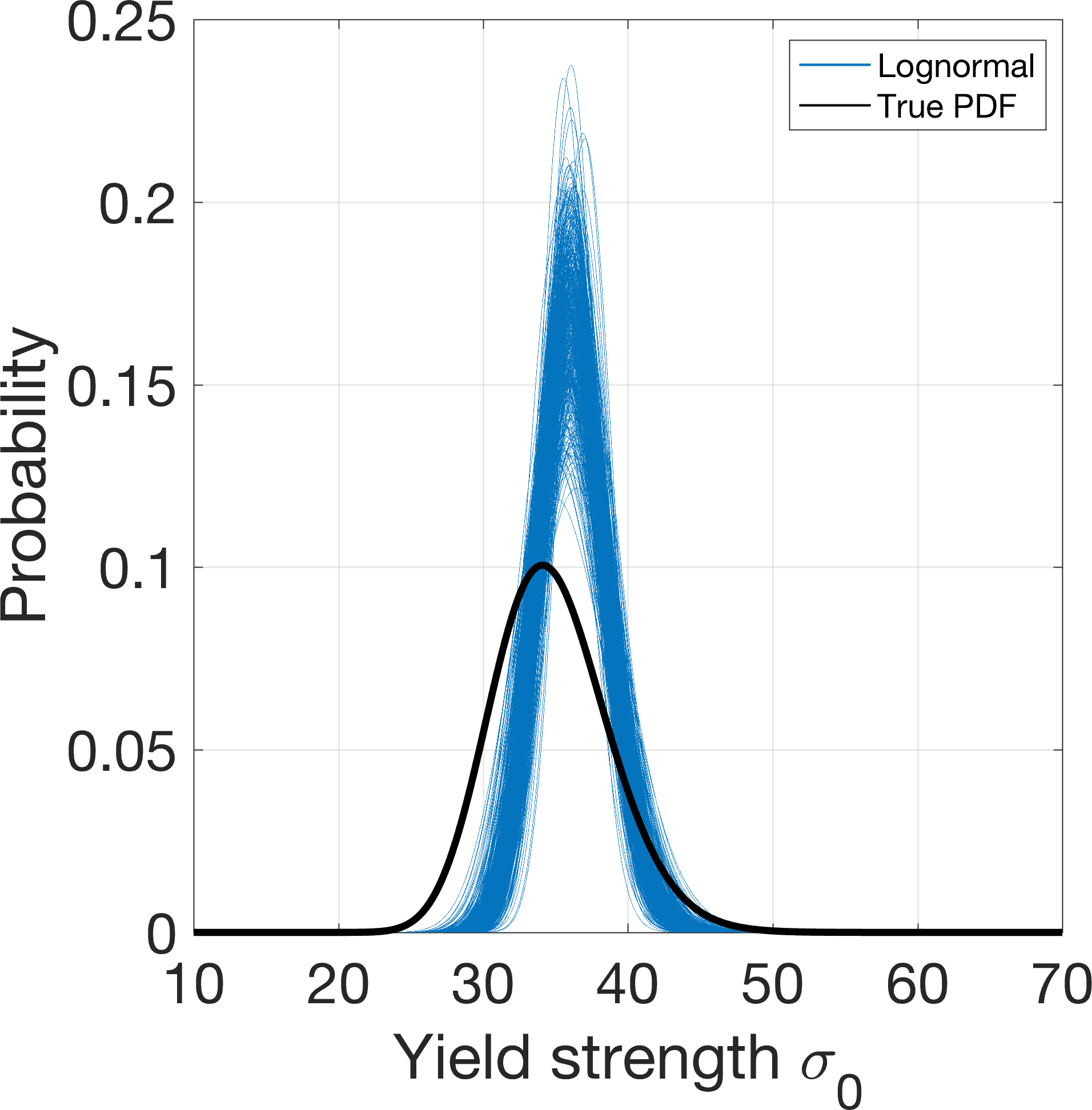}} 
&\parbox[c][1.1in]{1in}{\includegraphics[height=1in]{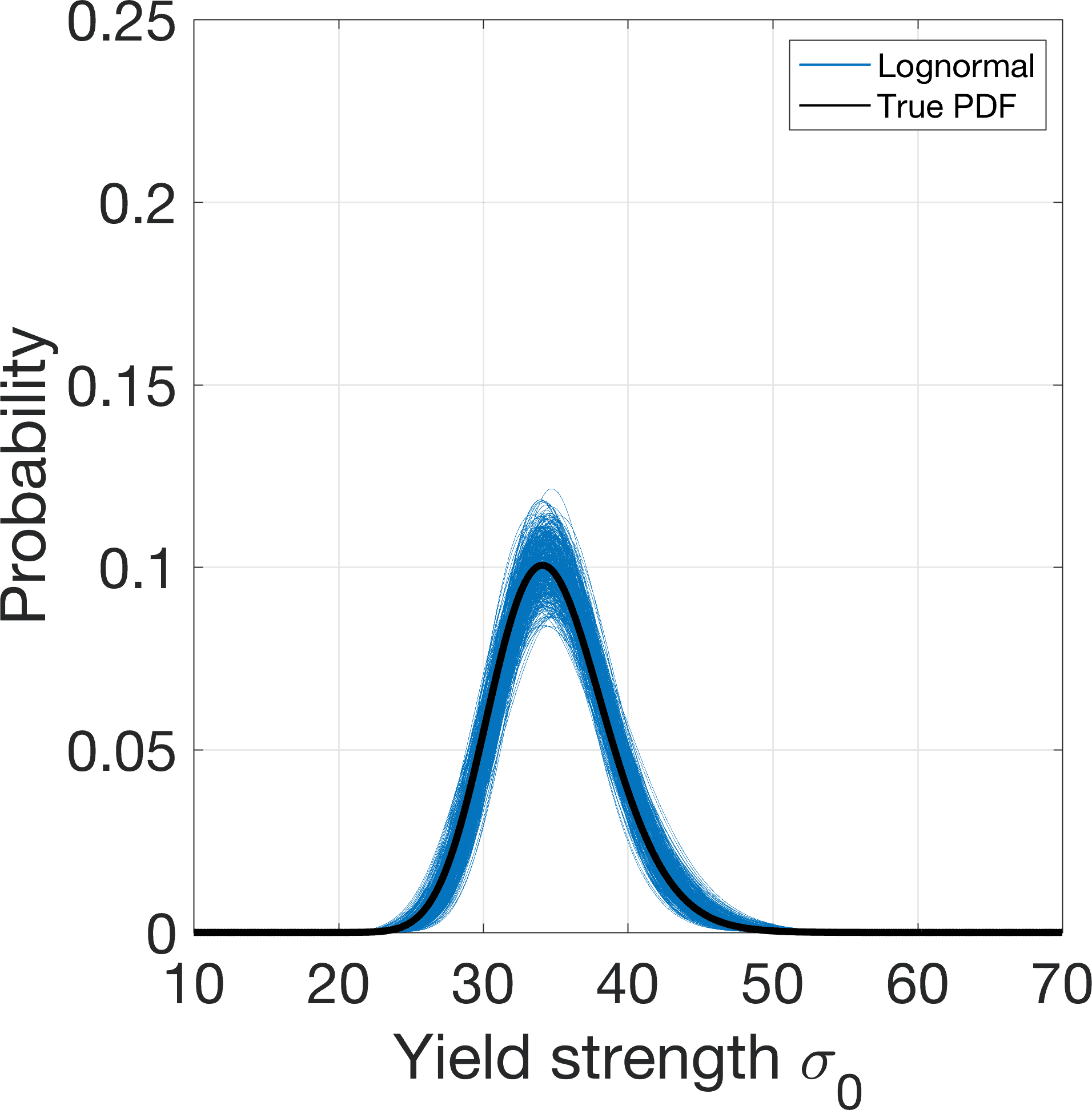}}   \\
25
&\parbox[c][1.1in]{1in}{\includegraphics[height=1in]{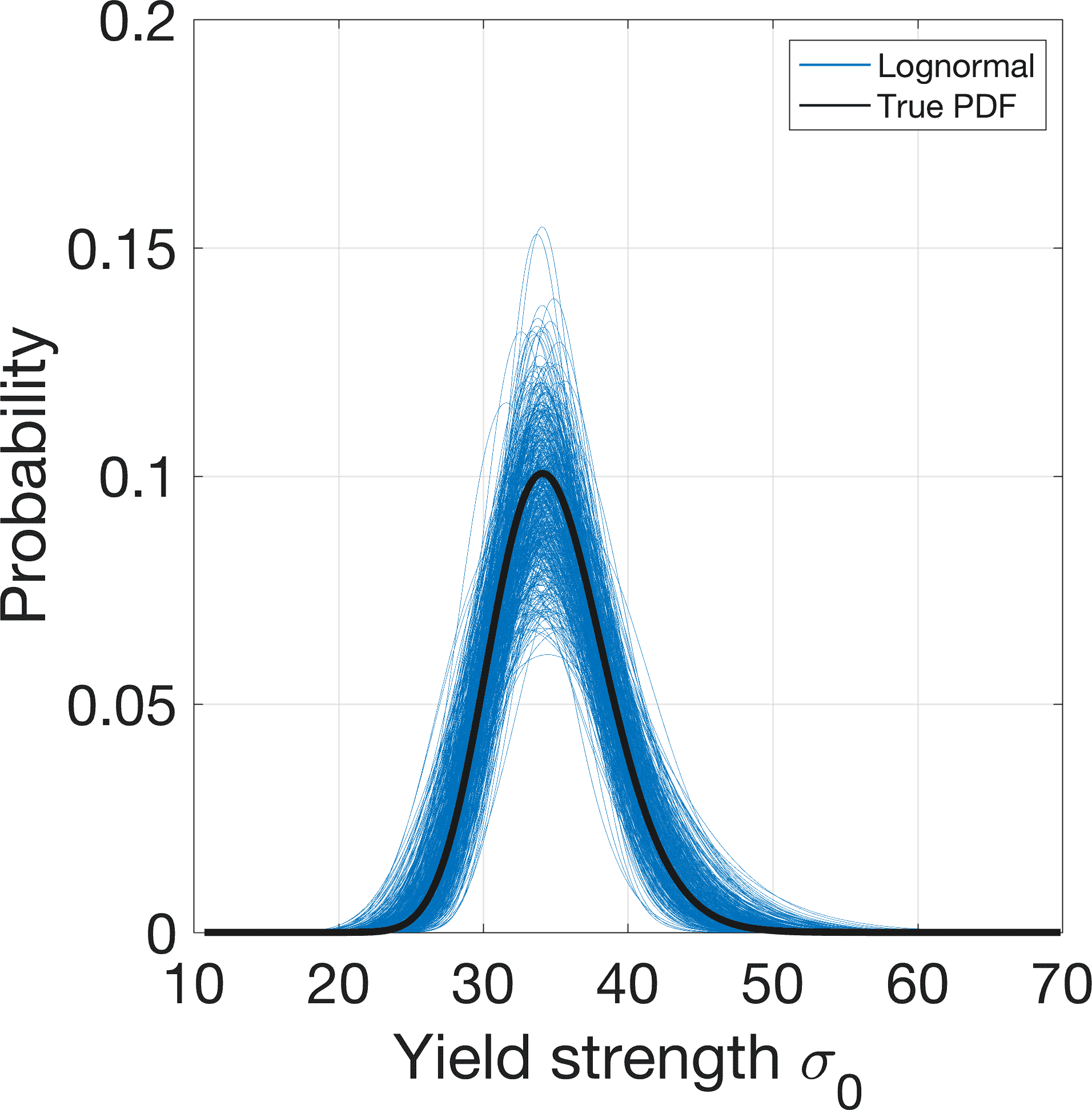}}
&\parbox[c][1.1in]{1in}{\includegraphics[height=1in]{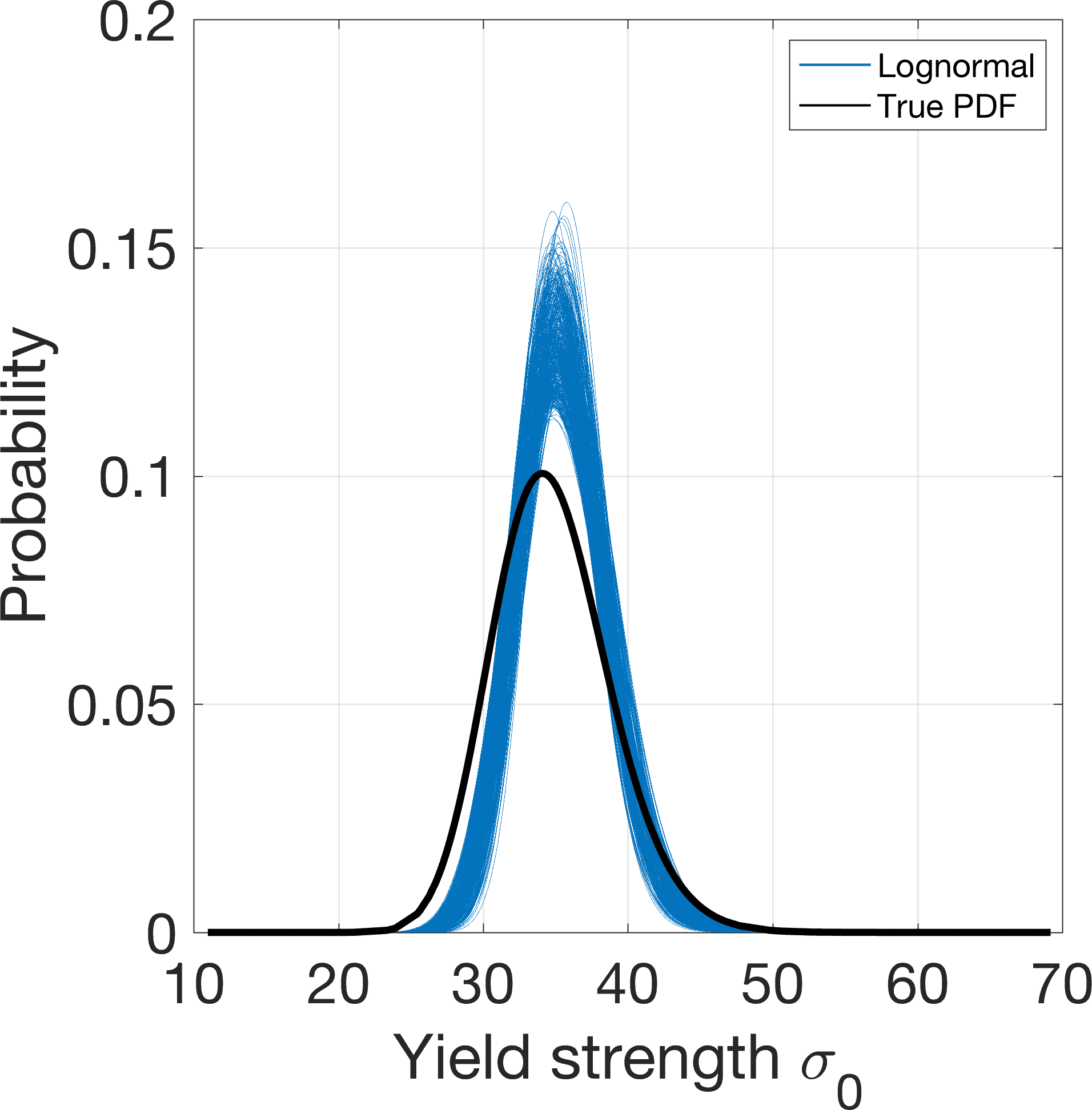}} 
&\parbox[c][1.1in]{1in}{\includegraphics[height=1in]{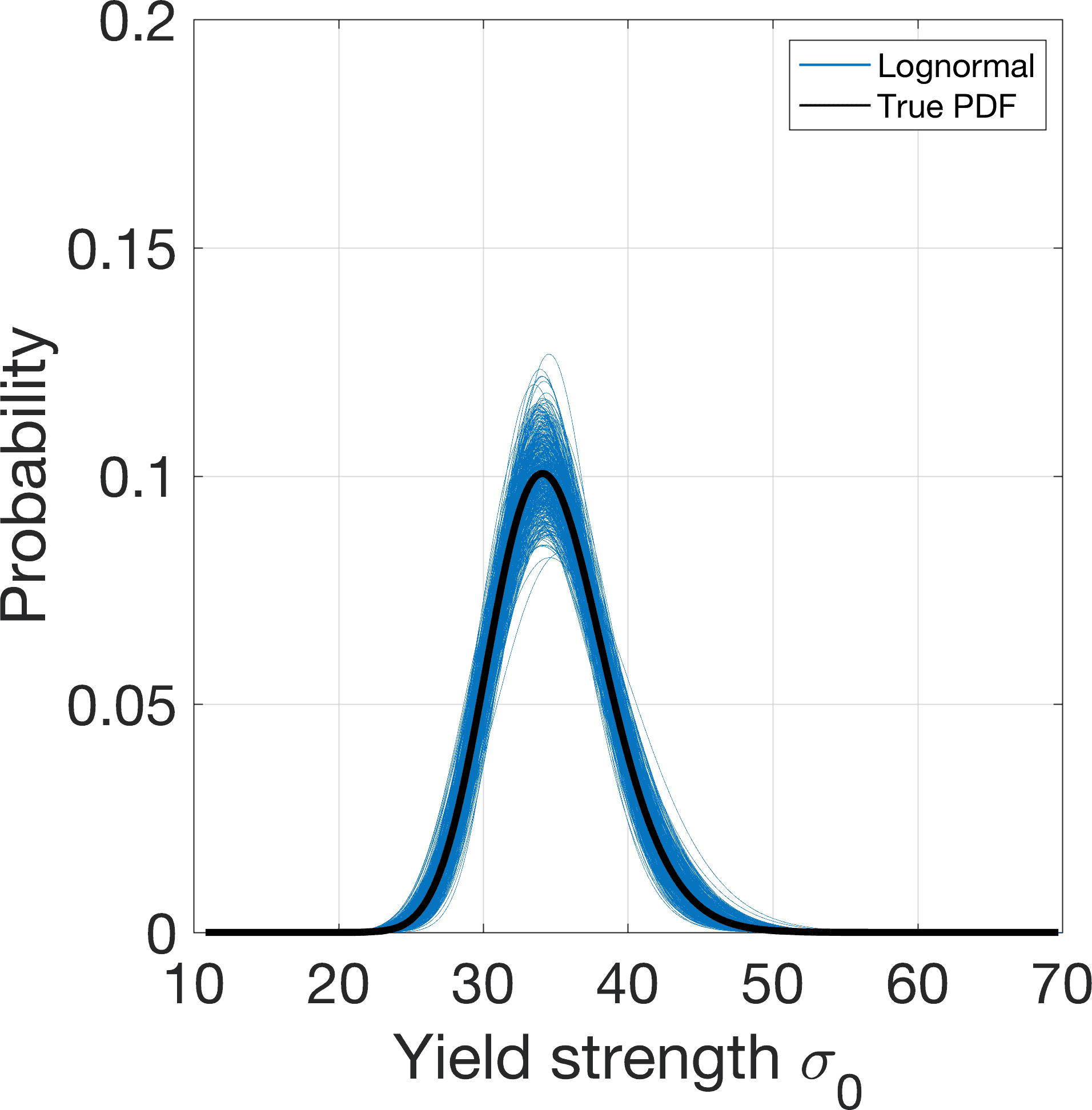}}   \\
50
&\parbox[c][1.1in]{1in}{\includegraphics[height=1in]{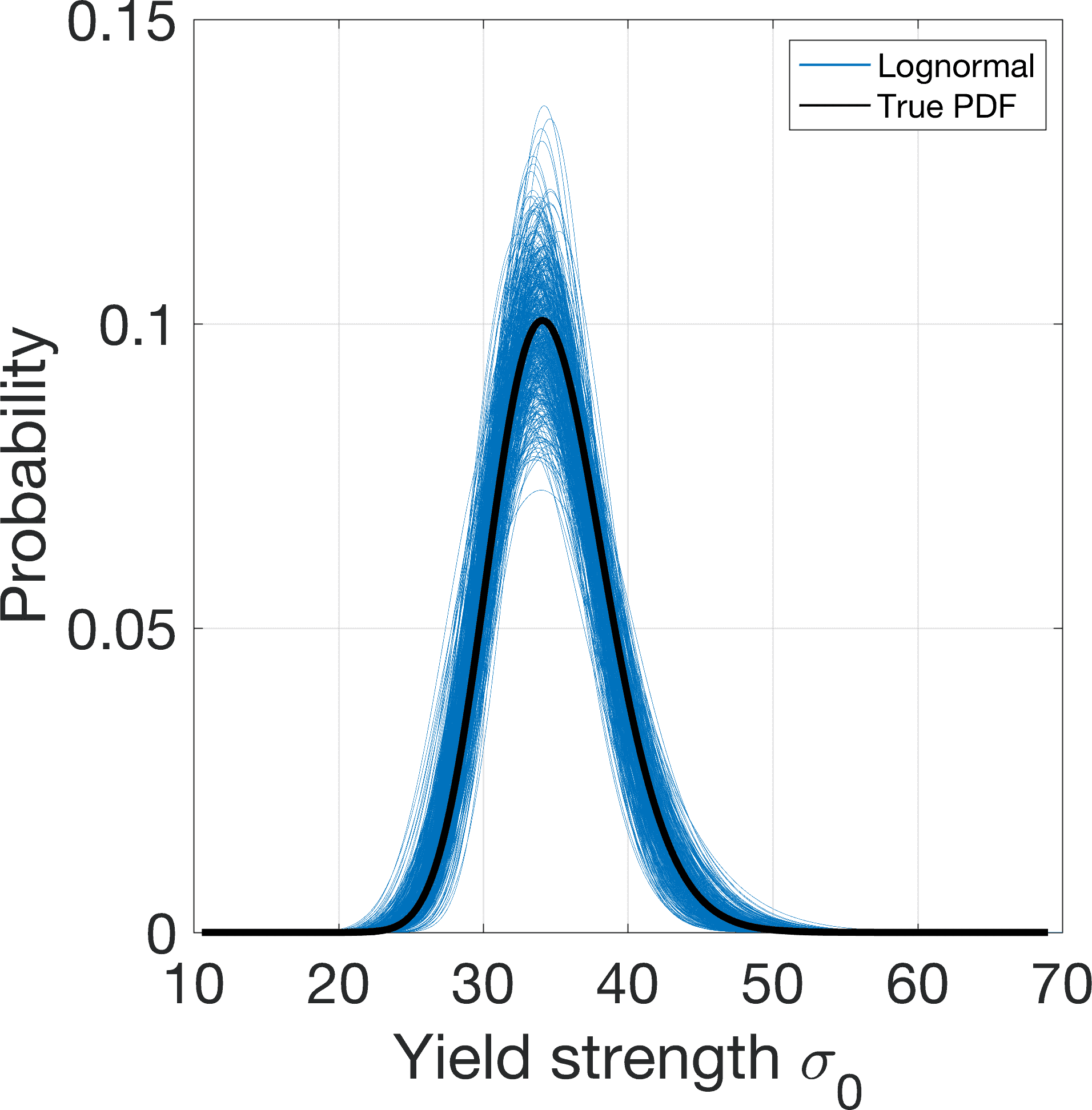}}
&\parbox[c][1.1in]{1in}{\includegraphics[height=1in]{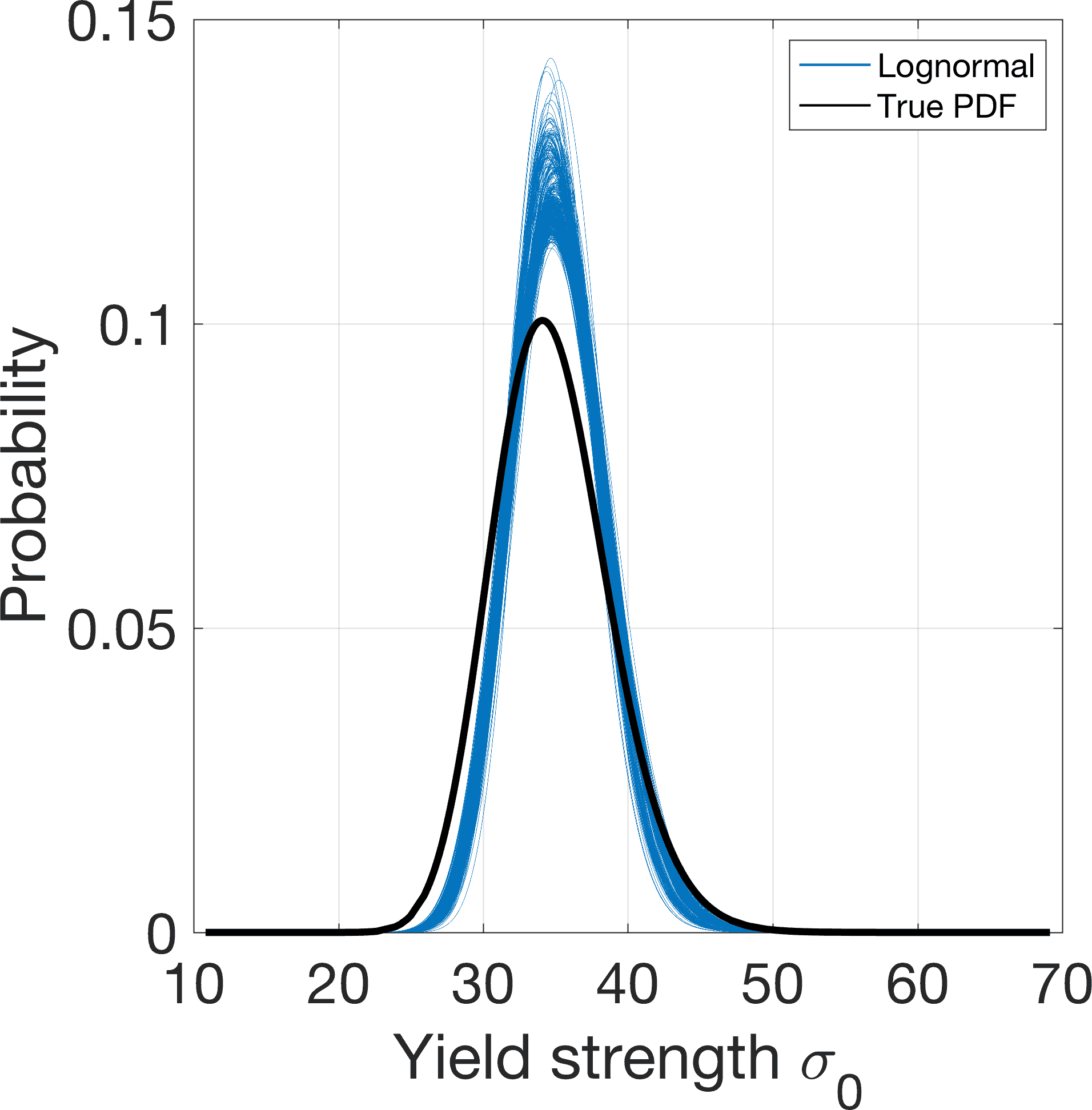}} 
&\parbox[c][1.1in]{1in}{\includegraphics[height=1in]{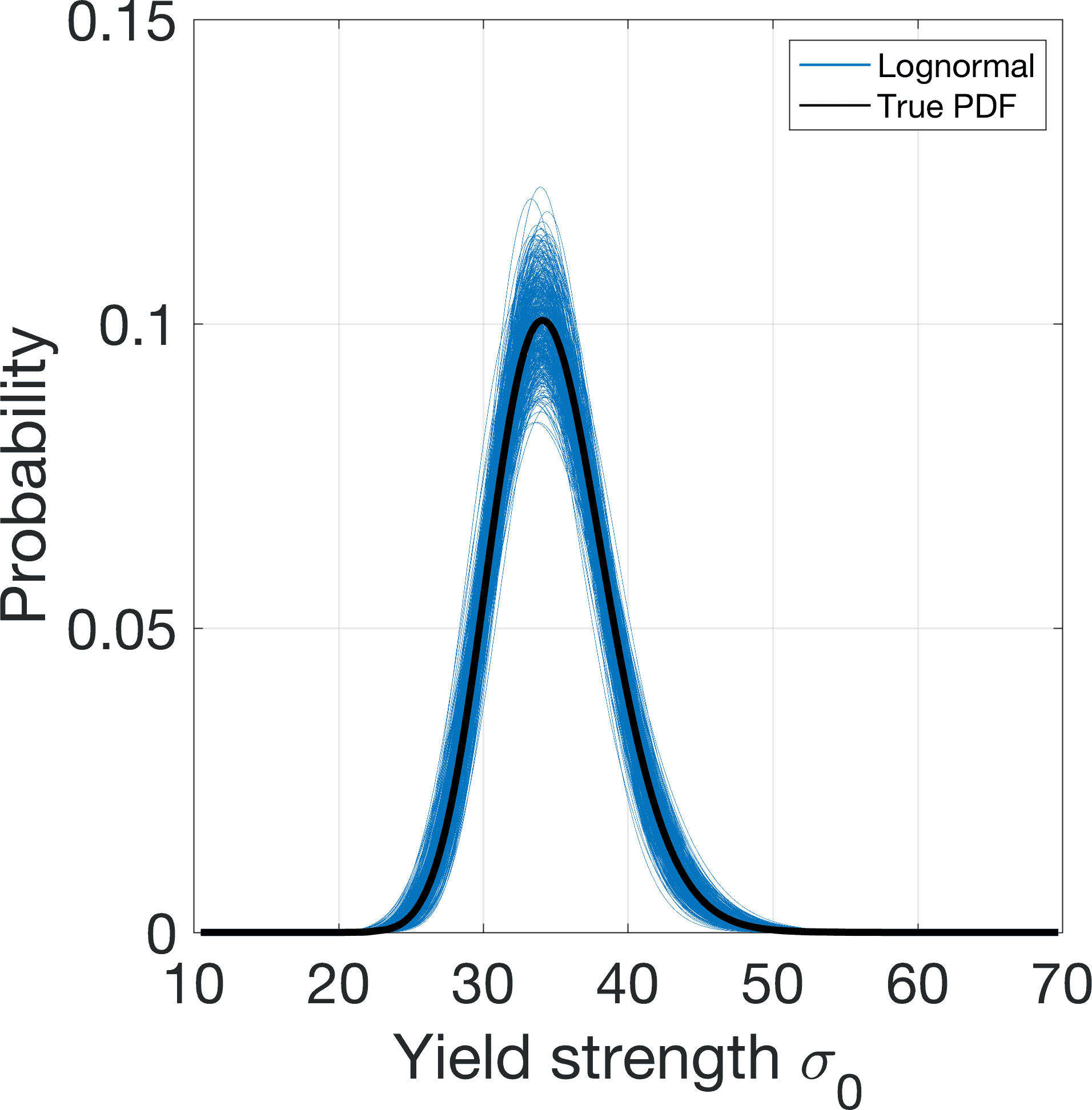}}   \\
100
&\parbox[c][1.1in]{1in}{\includegraphics[height=1in]{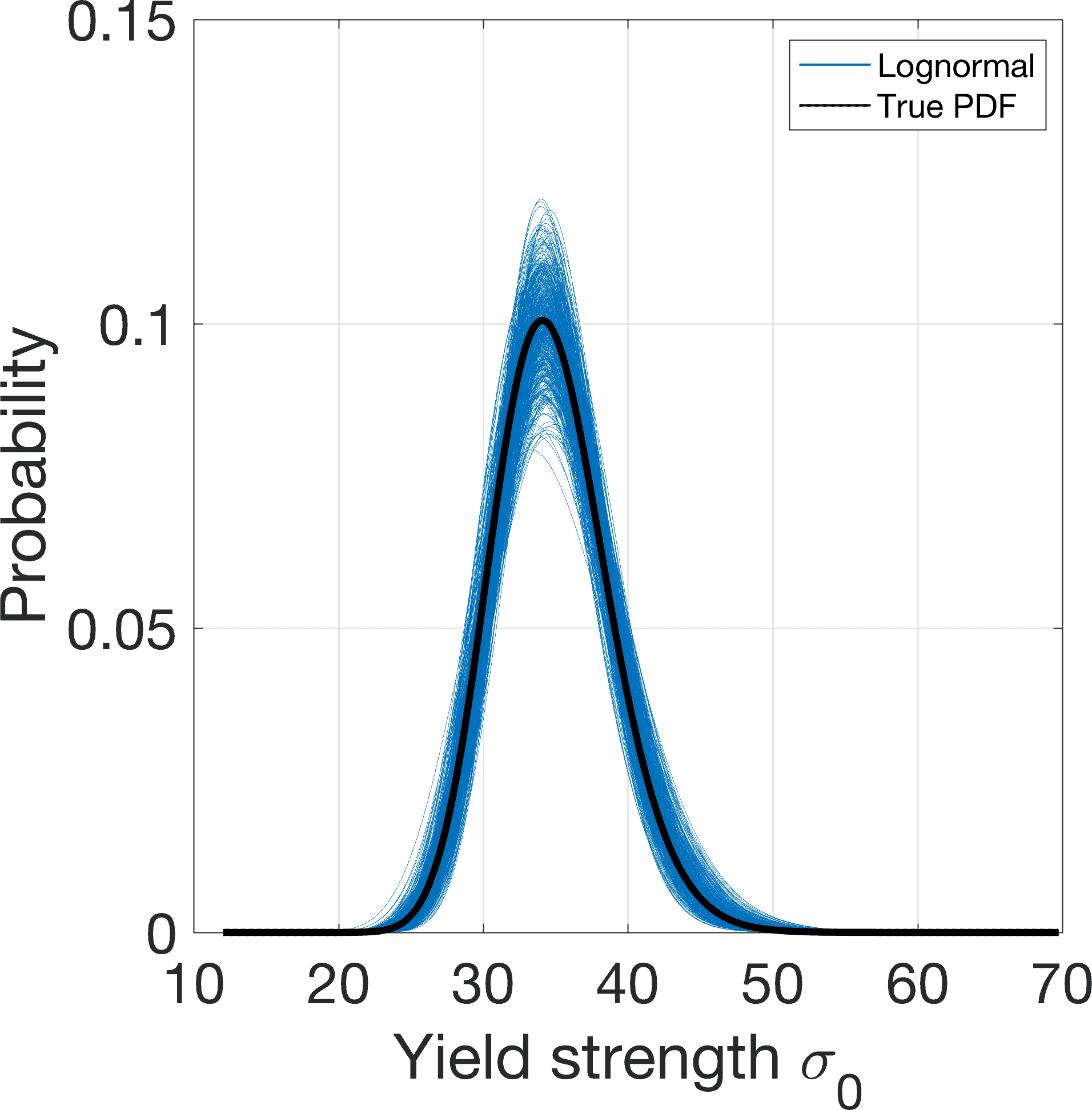}}
&\parbox[c][1.1in]{1in}{\includegraphics[height=1in]{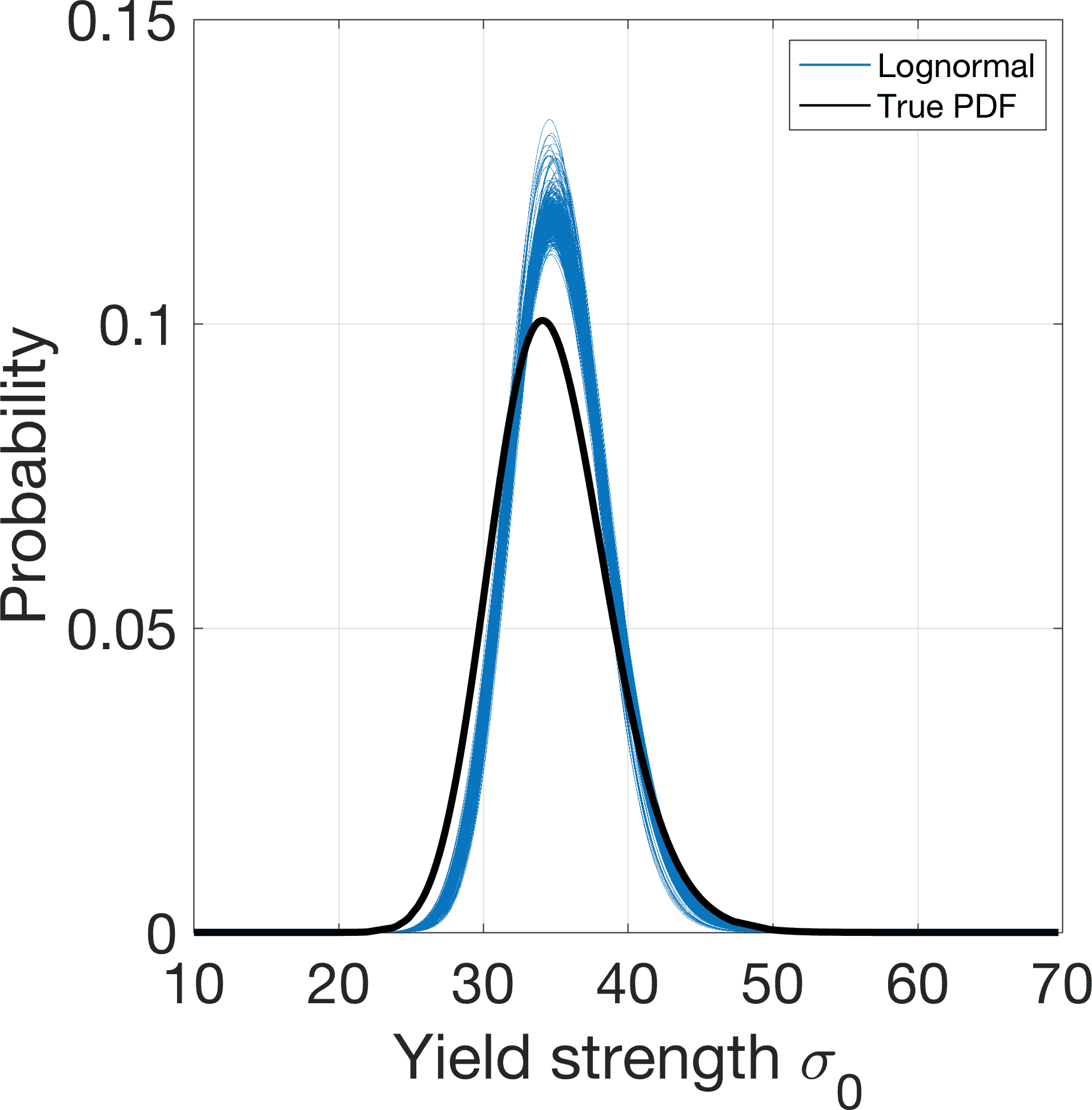}} 
&\parbox[c][1.1in]{1in}{\includegraphics[height=1in]{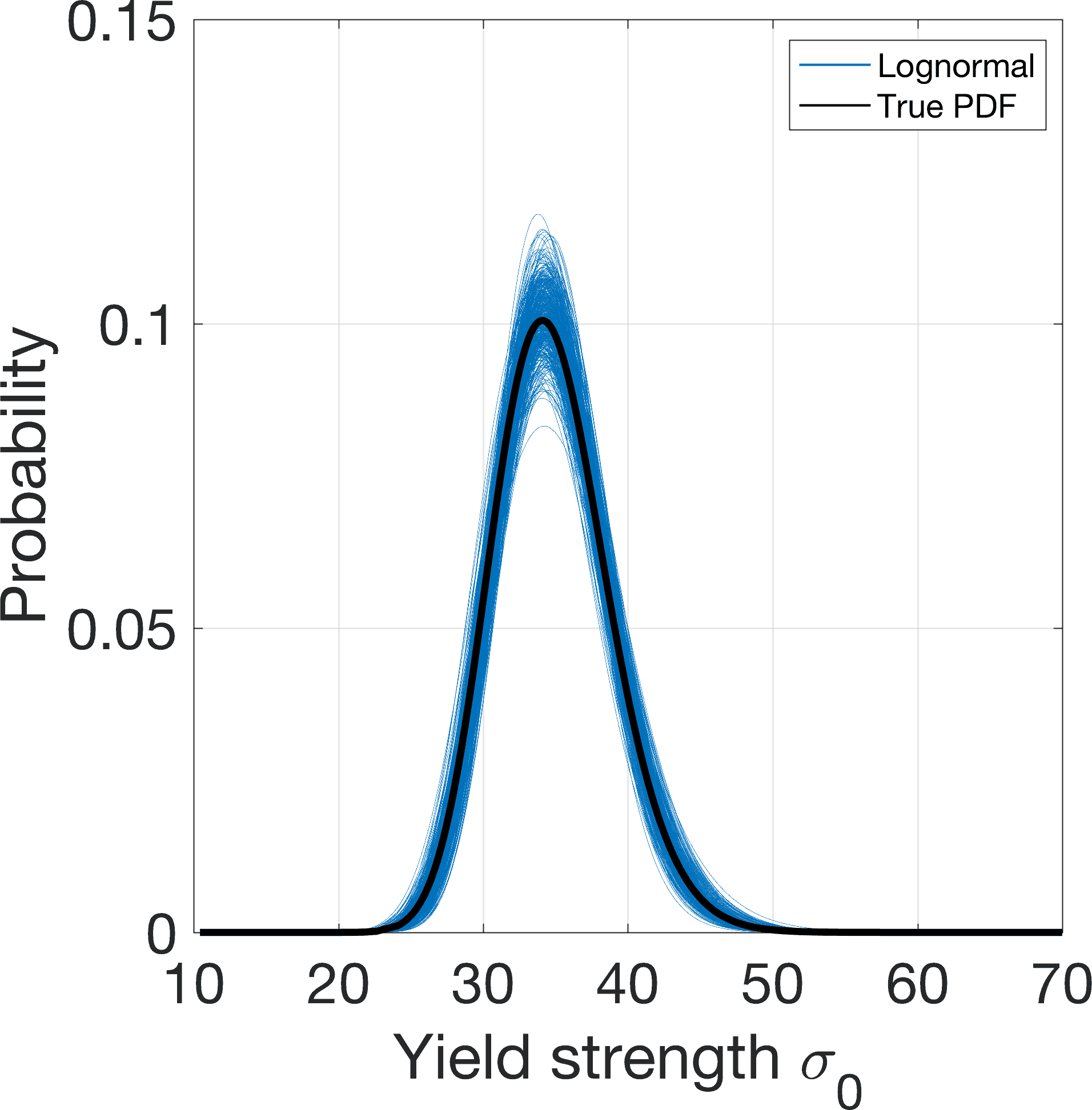}}   \\
 \hline
\end{tabular}
\end{table}
Notice that the band of distributions for the noninformative and ABS-B priors include the true distribution (bold) while the ABS-A case does not. Moreover, the band of distributions from the ABS-A prior is significantly narrower than those from both the noninformative and ABS-B priors, which implies confidence in the distribution. In other words, this prior gives false confidence in the wrong set of models. This may have major implications on uncertainty propagation, which is taken up in Section \ref{sec:effect_propagation}.

% Through the above analysis and discussion, we simply summarized as follows: 1) correct informative prior, like ABS-B, narrows the estimator, particularly in the small dataset stage, and always converges to the ``true" estimator; 2) incorrect but weak informative prior, like ABS-C, presents a limited improvement on estimation due to little informative data, but converges to the ``true" estimator at last; 3) incorrect but strong informative prior, like ABS-A and ASTM-A7, associated with significant statistical deviation, performs the narrowest but incorrect estimation which can be improved by collection of more data.  

\subsubsection{Effect of priors on total uncertainty}
\label{sec:total_uncertainty}

\begin{figure}[!ht]	
	\centering
	\subfigure[]{\includegraphics[height=2in]{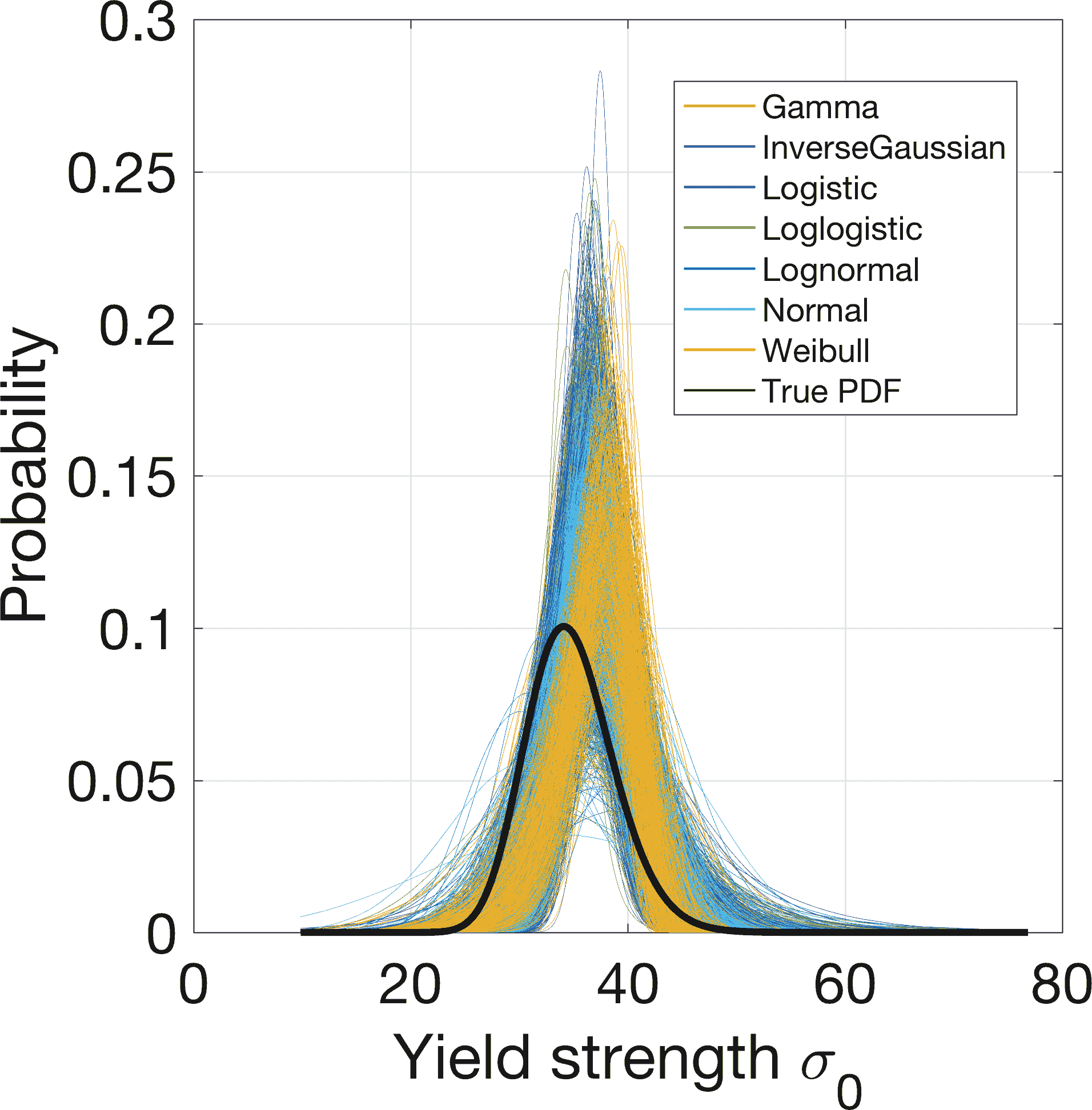}}
	\subfigure[]{\includegraphics[height=2in]{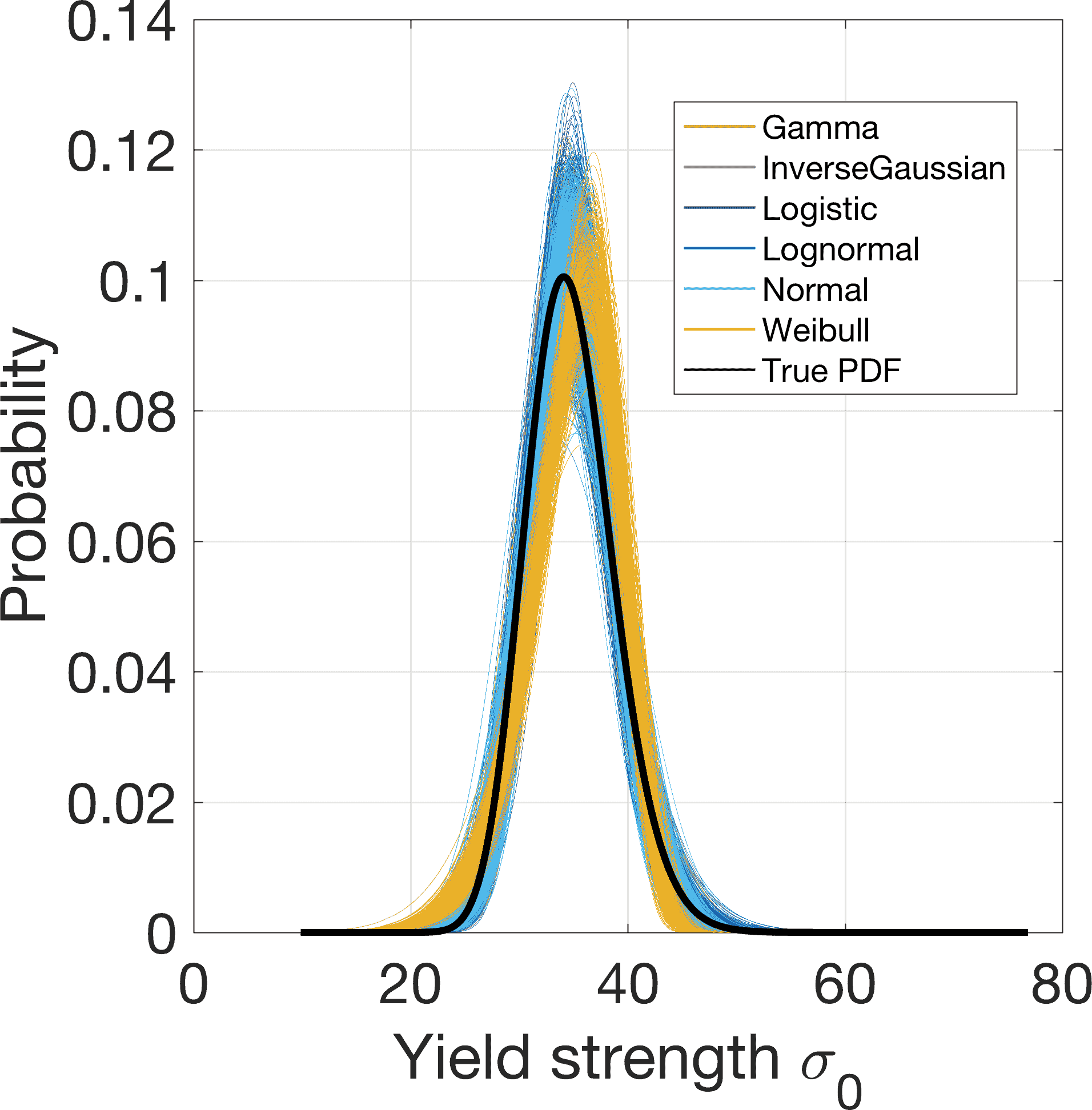}}
    \subfigure[]{\includegraphics[height=2in]{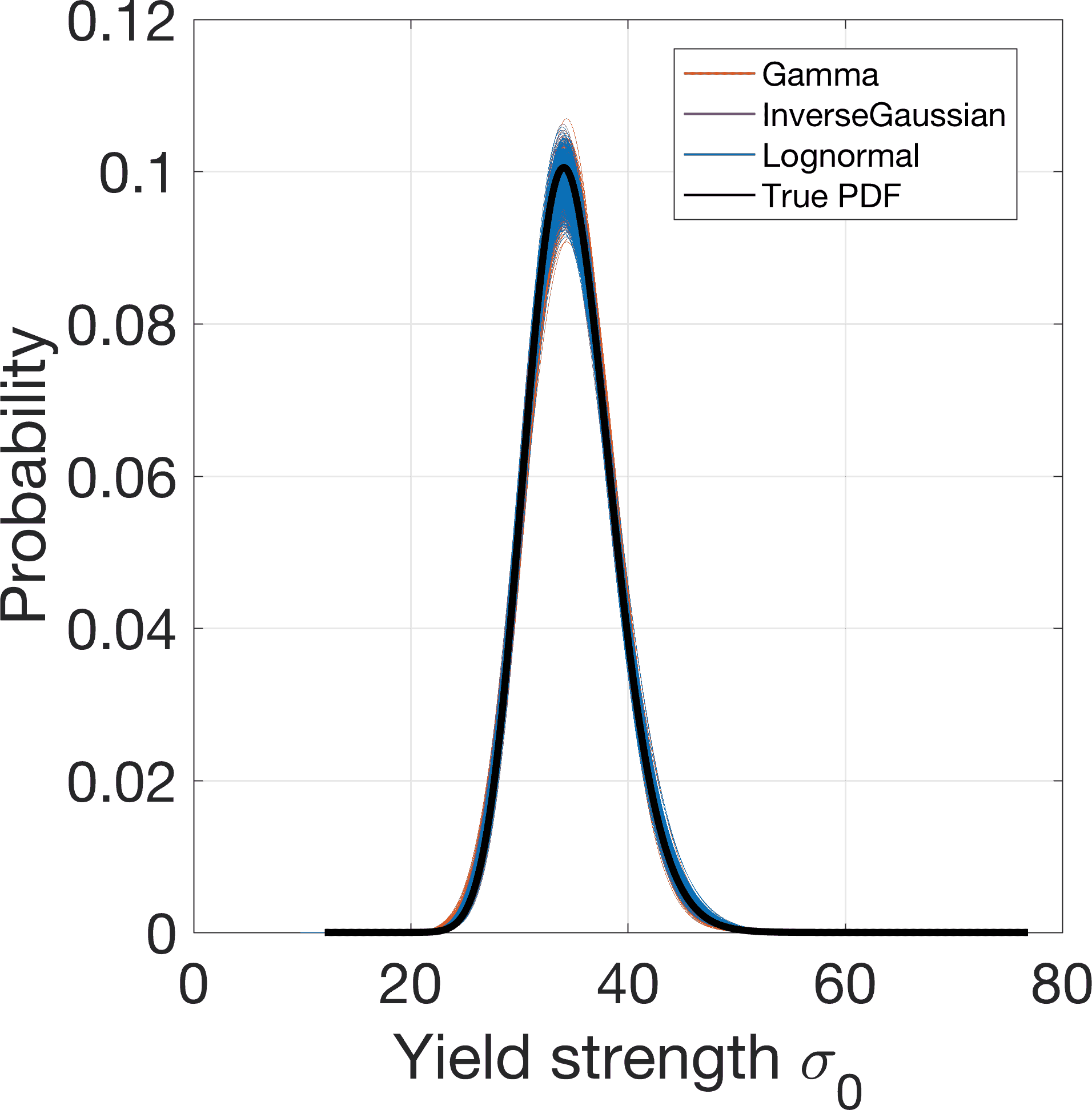}}
	\caption[]{ 5000 distributions given equal model prior probabilities with noninformative parameter priors for (a) 10data, (b) 100 data and (c) 1000 data } \label{fig:opt_vague_total_uq}
\end{figure}

The total uncertainty in the yield strength is represented by Monte Carlo sampling from the candidate distributions as described in Section \ref{sec:propagation}. For each sample (pdf), a probability model is randomly selected according to the posterior model  probabilities. The parameters of this model are then randomly sampled from its posterior joint pdf. The result is a Monte Carlo set of distributions as shown in Figure \ref{fig:opt_vague_total_uq}. This particular example shows a set of 5000 distributions given equal prior model probabilities with noninformative parameter priors for cases having datasets of size 10, 100, and 1000 data. 

\begin{figure}[!ht]	
	\centering
	\subfigure[]{\includegraphics[height=2in]{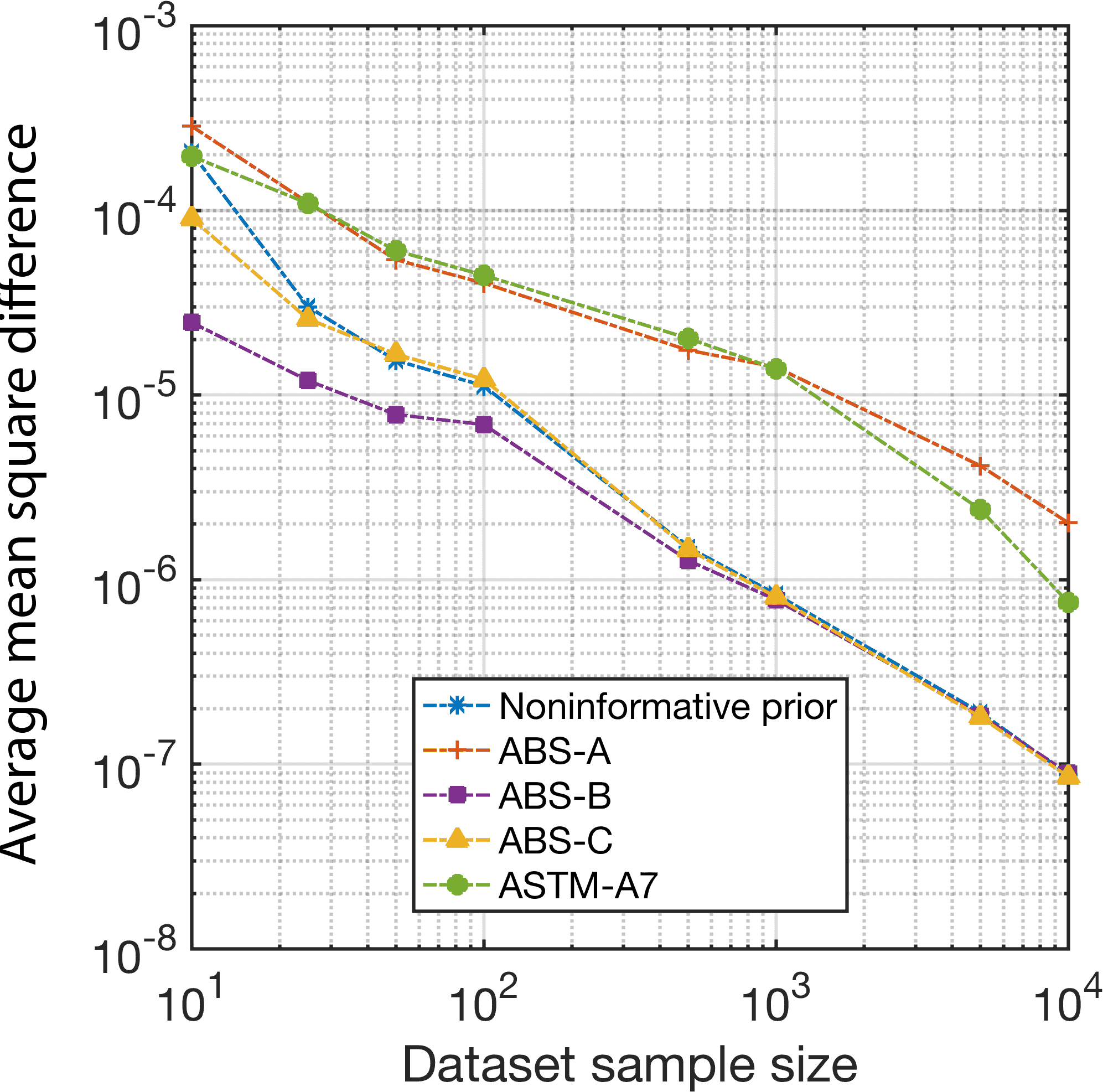}}
	\subfigure[]{\includegraphics[height=2in]{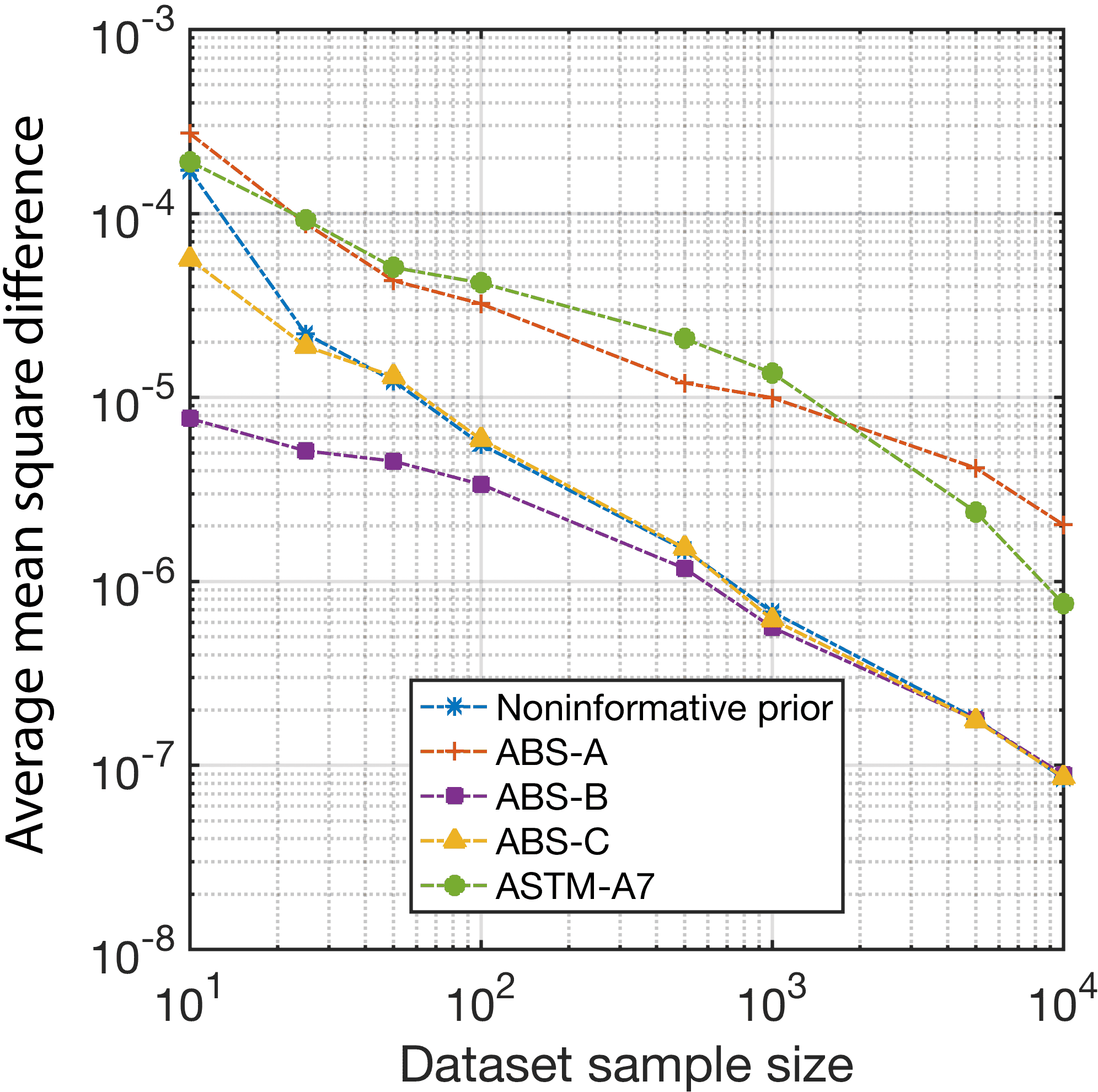}}
    \subfigure[]{\includegraphics[height=2in]{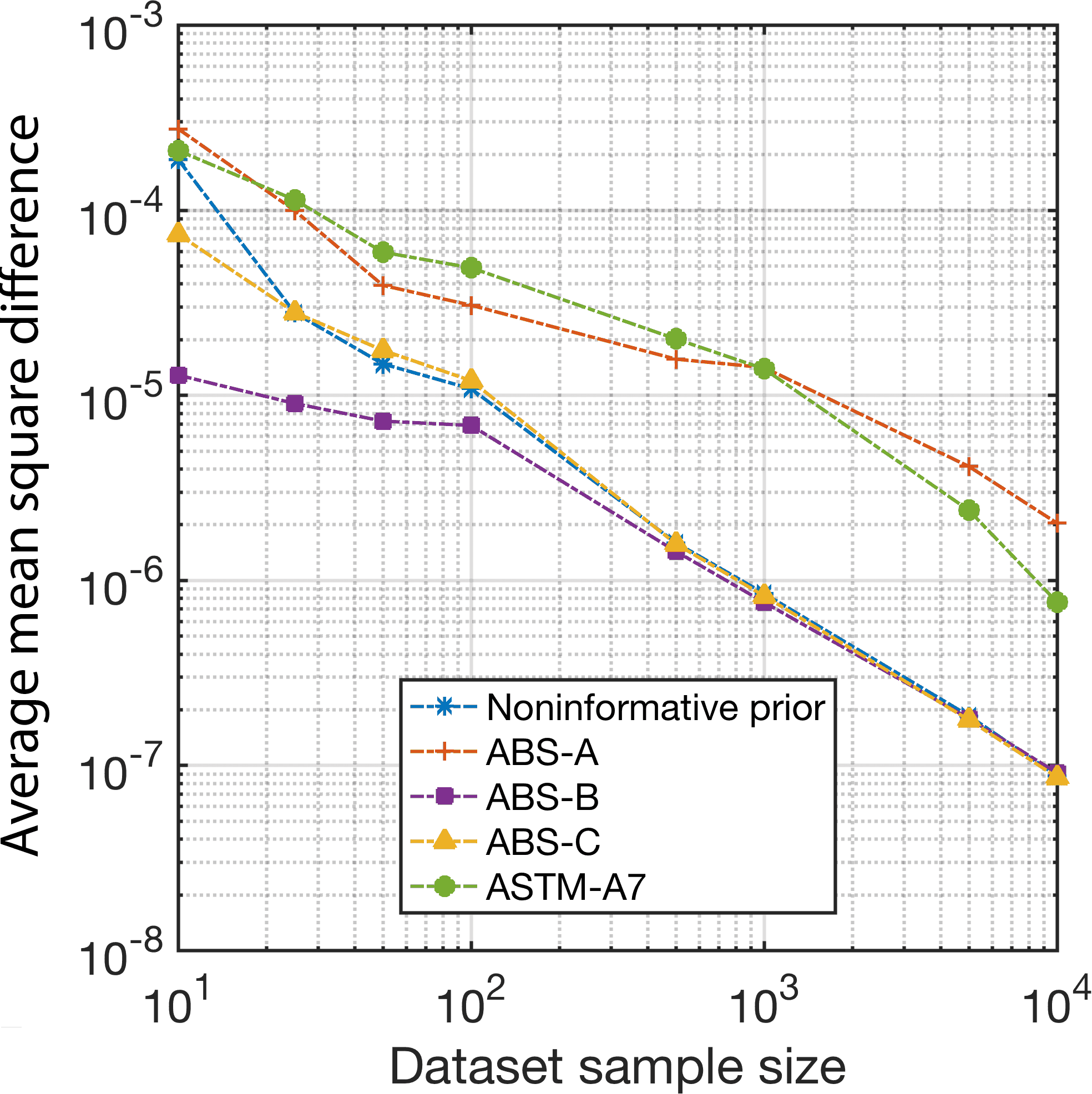}}
	\caption[]{ Convergence of average mean square distance for (a) equal model prior, (b) strong correct model prior and (c) strong incorrect model prior } \label{fig:convergence_distance}
\end{figure}

To measure the degree of uncertainty in a given model set, we compute the average mean square distance between the 5000 models in the set and the true lognormal density given by:
\begin{equation}
\delta = \dfrac{1}{2}\dfrac{1}{5000}\sum_{i=1}^{5000}(p_i(\bm{x}|\boldsymbol{\theta})-p(\bm{x}))^2
\label{eq:distance}
\end{equation}
where $p_i(\mathbf{x}|\boldsymbol{\theta})$ are the distributions in the set and $p(\mathbf{x})$ is the lognormal pdf of the true model. This distance is plotted as a function of dataset size in Figure \ref{fig:convergence_distance} for each parameter prior and for the three cases of model prior probabilities.
When the dataset size is small, there is a major benefit to using the correct ABS-B prior. That is, the set of distributions is comparatively close to the true distribution. All other priors, meanwhile, start by poorly representing the true distribution (relatively large average mean square distance to the true distribution) but the noninformative and ABS-C priors come to be almost as good as the ABS-B prior after $\sim100$ data are collected. The ABS-A and ASTM-A7 priors, on the other hand, cannot achieve the same level of accuracy as the other priors - even for very large datasets. The result is that the set of distributions generated from the ABS-A and ASTM-A7 priors can have residual errors that effectively results in identification/propagation of incorrect probability models (as evidenced again by the lognormal distributions from the ABS-A prior in Table \ref{tab:lognormal_distributions} which do not include the true model). 

\subsection{Influence of data-driven priors on uncertainty propagation}
\label{sec:effect_propagation}

The results of the multimodel inference process are used to identify a set of probability models (e.g.\ Figure \ref{fig:opt_vague_total_uq}) that can be propagated through a physics-based model using the method in Section \ref{sec:propagation} \cite{zhang2018}. Given the sensitivity of the posterior probabilities to the selection of the prior, this begs the question: What impact do prior assumptions have on response quantities from the model? If the prior yields rapid convergence, it stands to reason that we should expect rapid convergence (i.e. small uncertainty) in response quantities. But, how much of an improvement can be gained through good prior selection and how poor are the results if a bad prior is selected? The results of the previous section seem to imply that a poor prior can yield not just large uncertainties, but incorrect probabilistic response. We explore these issues in the context of our plate buckling problem in this section.

Total uncertainty is propagated using the IS reweighting method proposed in \cite{zhang2018} and reviewed in Section \ref{sec:propagation}. For illustration, consider quantification based on the ABS-B parameter prior with equal prior model probabilities. Table \ref{tab:uq_lognormal} shows the results of propagation for uncertainties quantified from different size datasets. 
\begin{table}[!ht] \footnotesize
\centering
\caption{Optimal sampling density (OSD), CDFs, mean and probability of failure for ABS-B prior associated with equal model prior probability as a function of dataset size from 10, 25, 50, 500, to 5000}
\label{tab:uq_lognormal}
\begin{tabular}{ccccc}
\hline
Data & OSD & CDFs & Mean & Probability of failure \\ \hline
10  
& \parbox[c][1.35in]{1.3in}{\includegraphics[height = 1.3in]{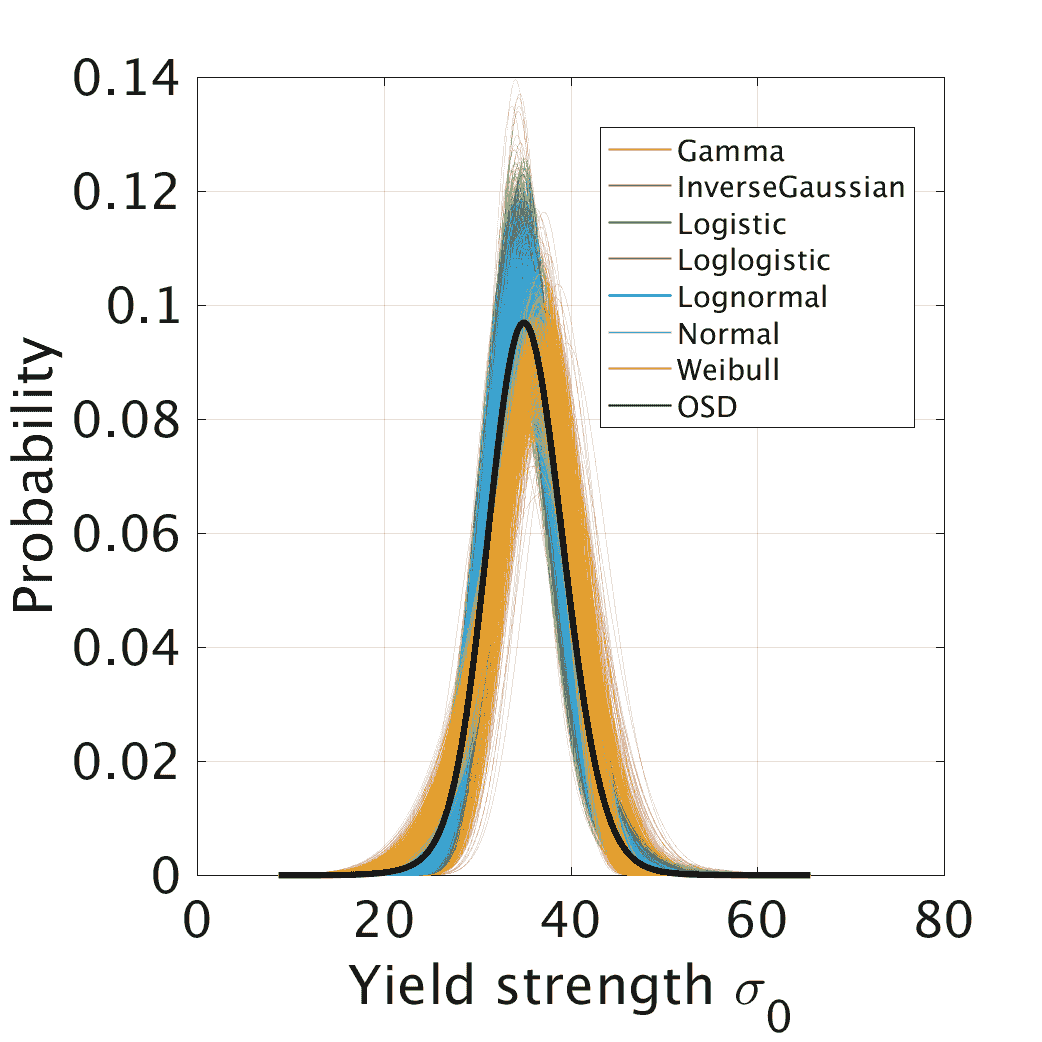}} 
&\parbox[c][1.35in]{1.3in}{\includegraphics[height=1.3in]{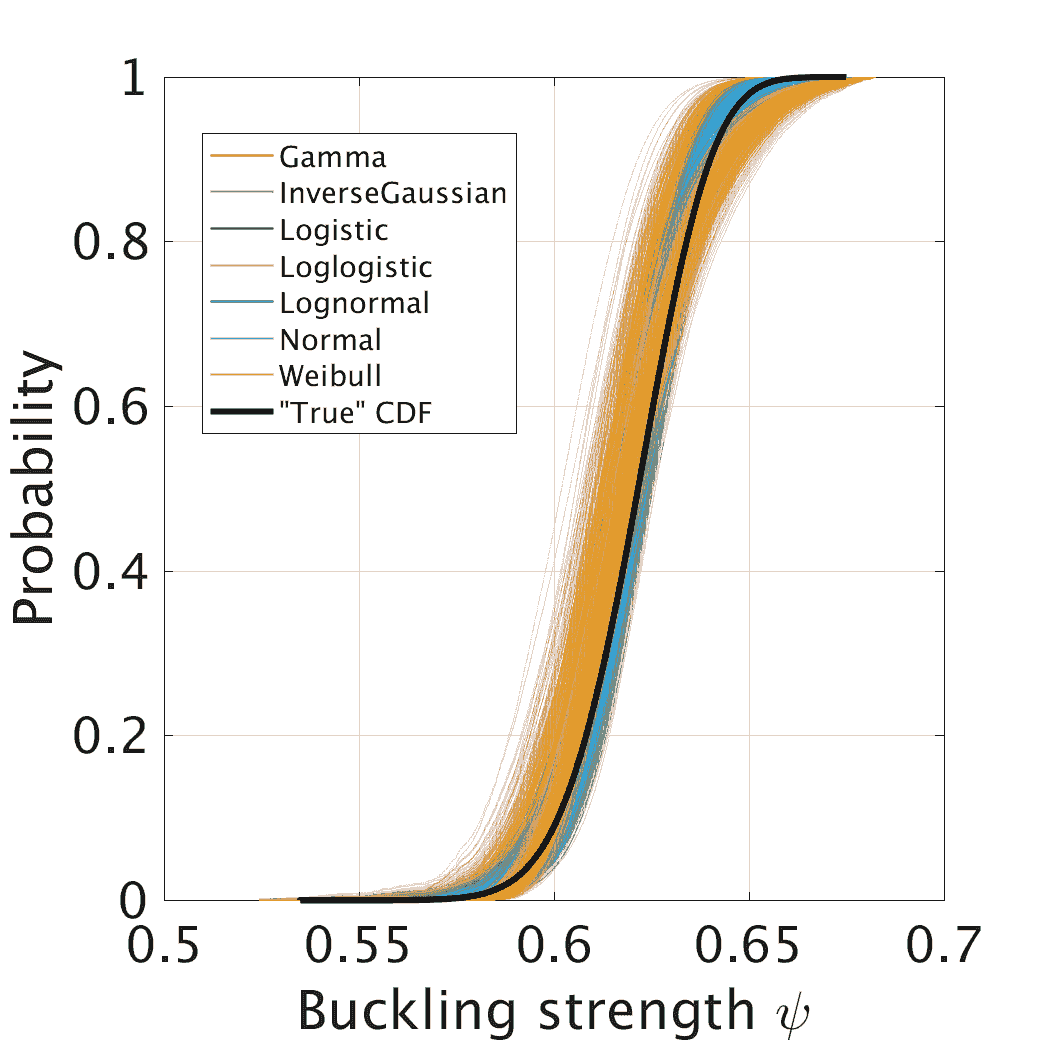}} 
&\parbox[c][1.35in]{1.3in}{\includegraphics[height=1.2in]{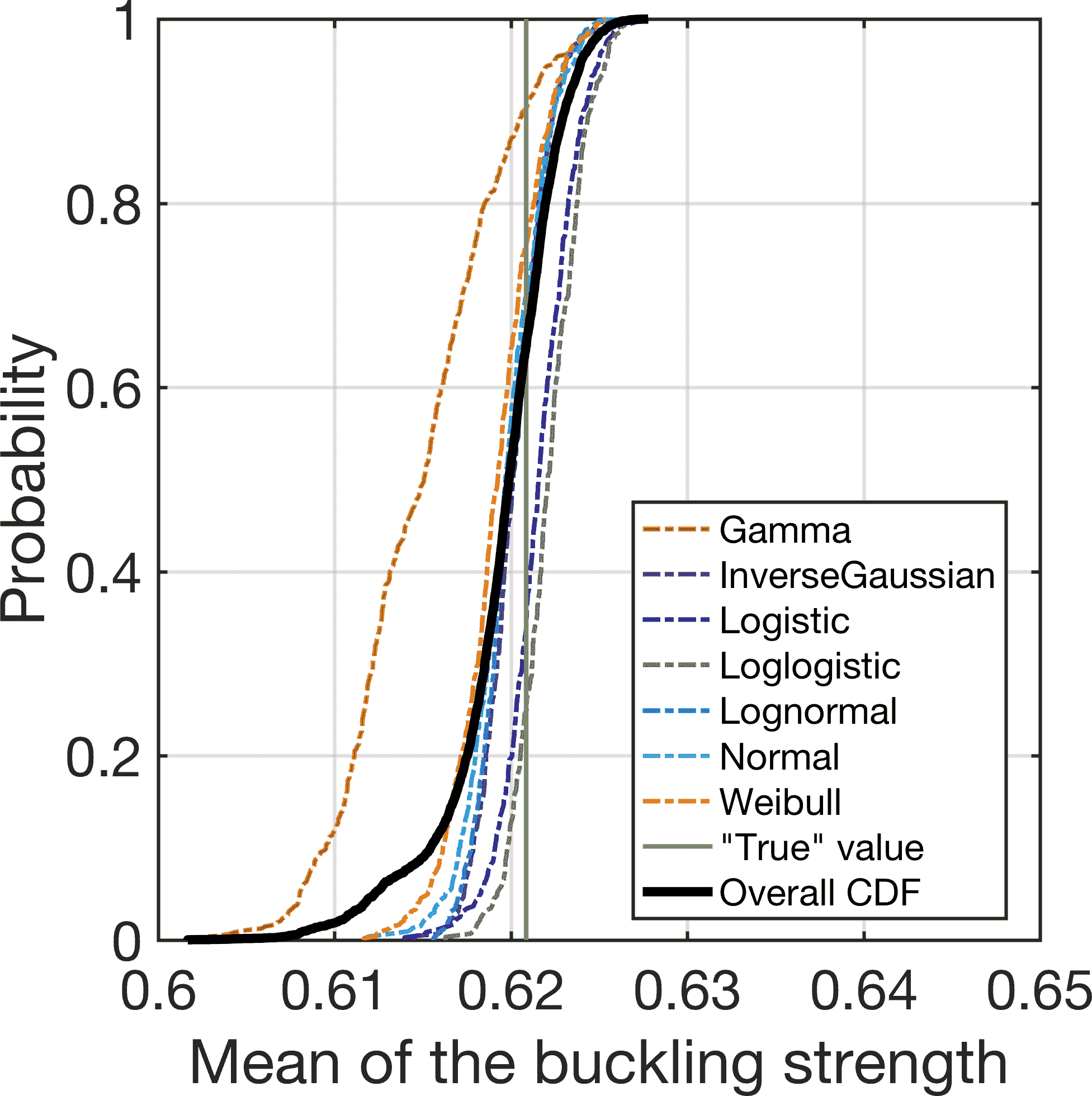}} 
&\parbox[c][1.35in]{1.3in}{\includegraphics[height=1.2in]{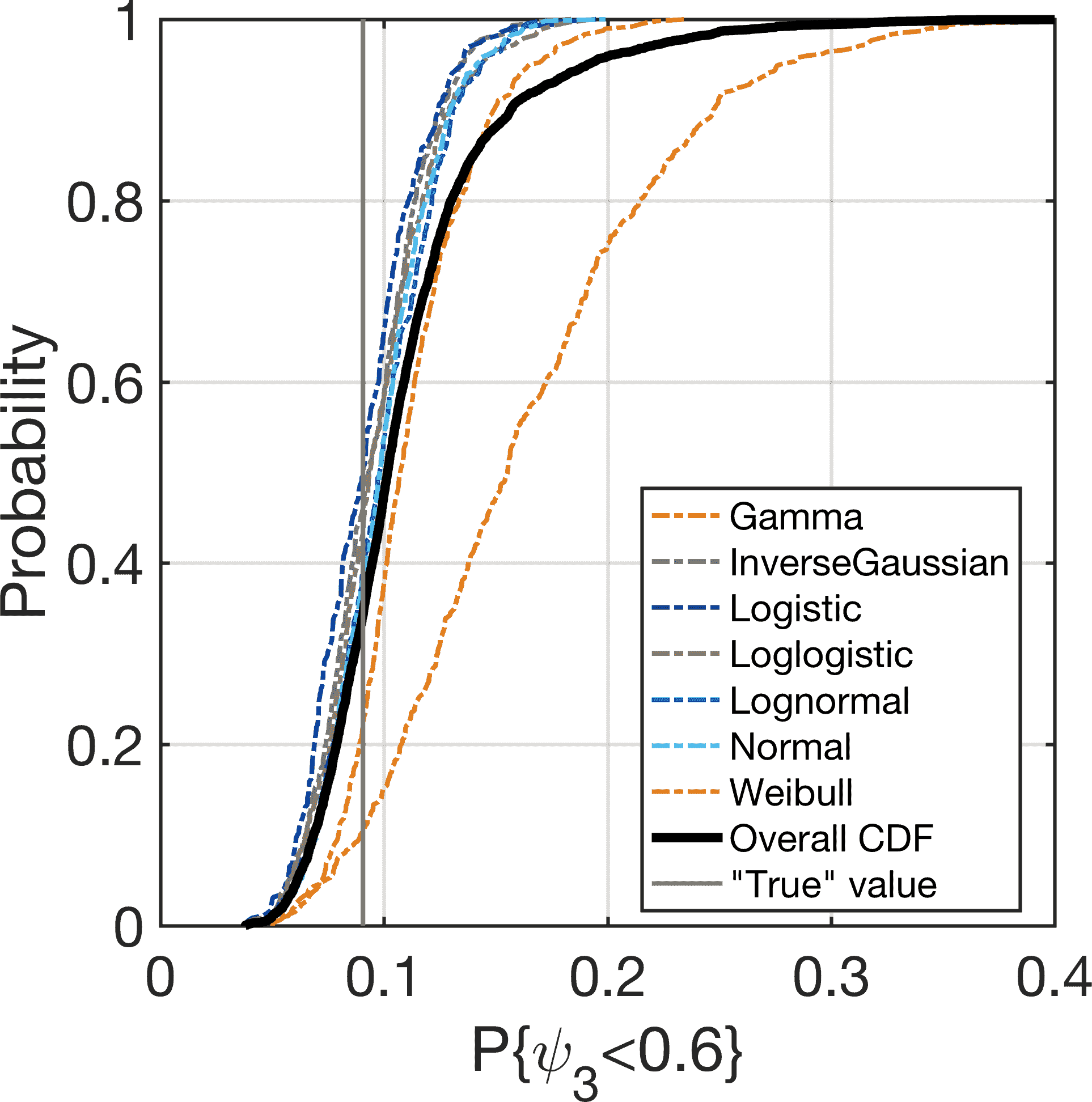}} \\
25
&\parbox[c][1.35in]{1.3in}{\includegraphics[height=1.3in]{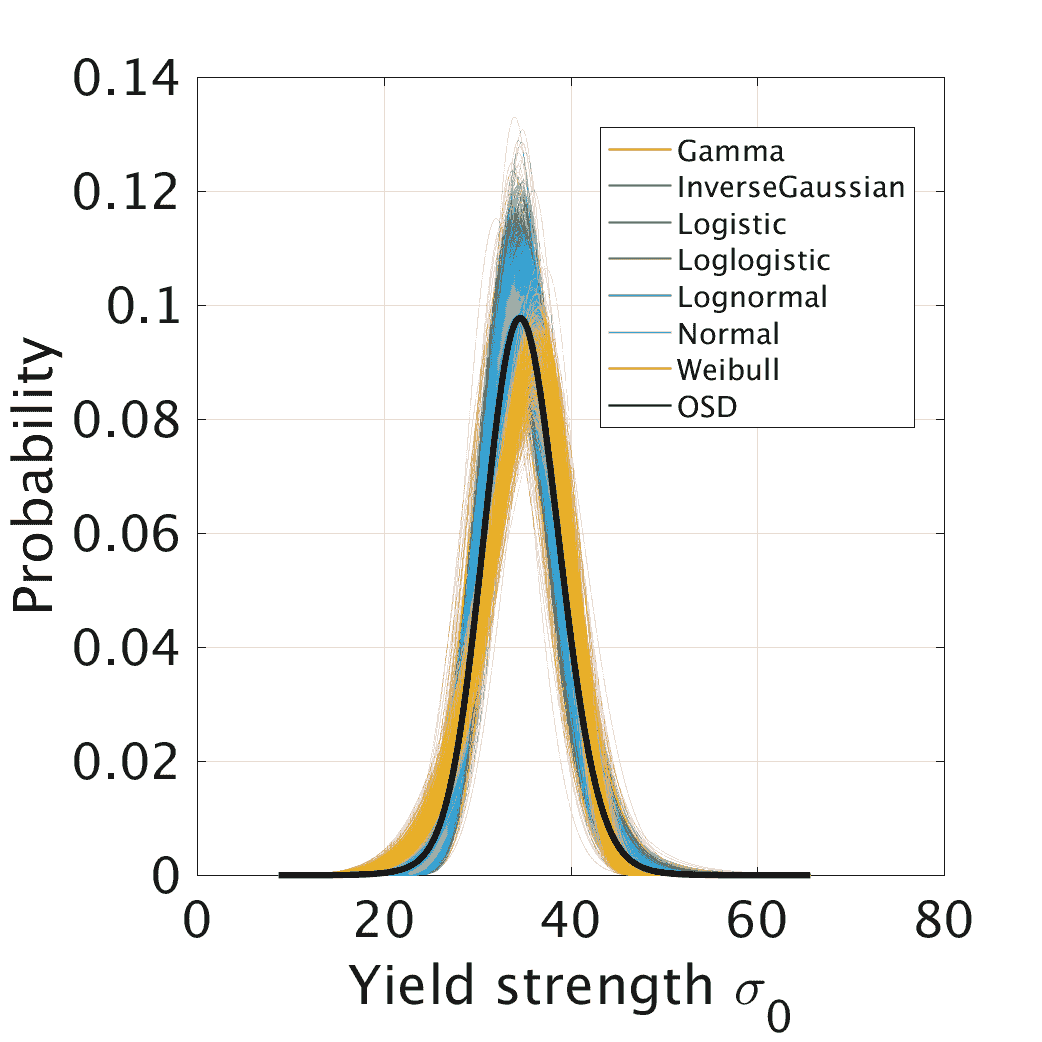}} 
&\parbox[c][1.35in]{1.3in}{\includegraphics[height=1.3in]{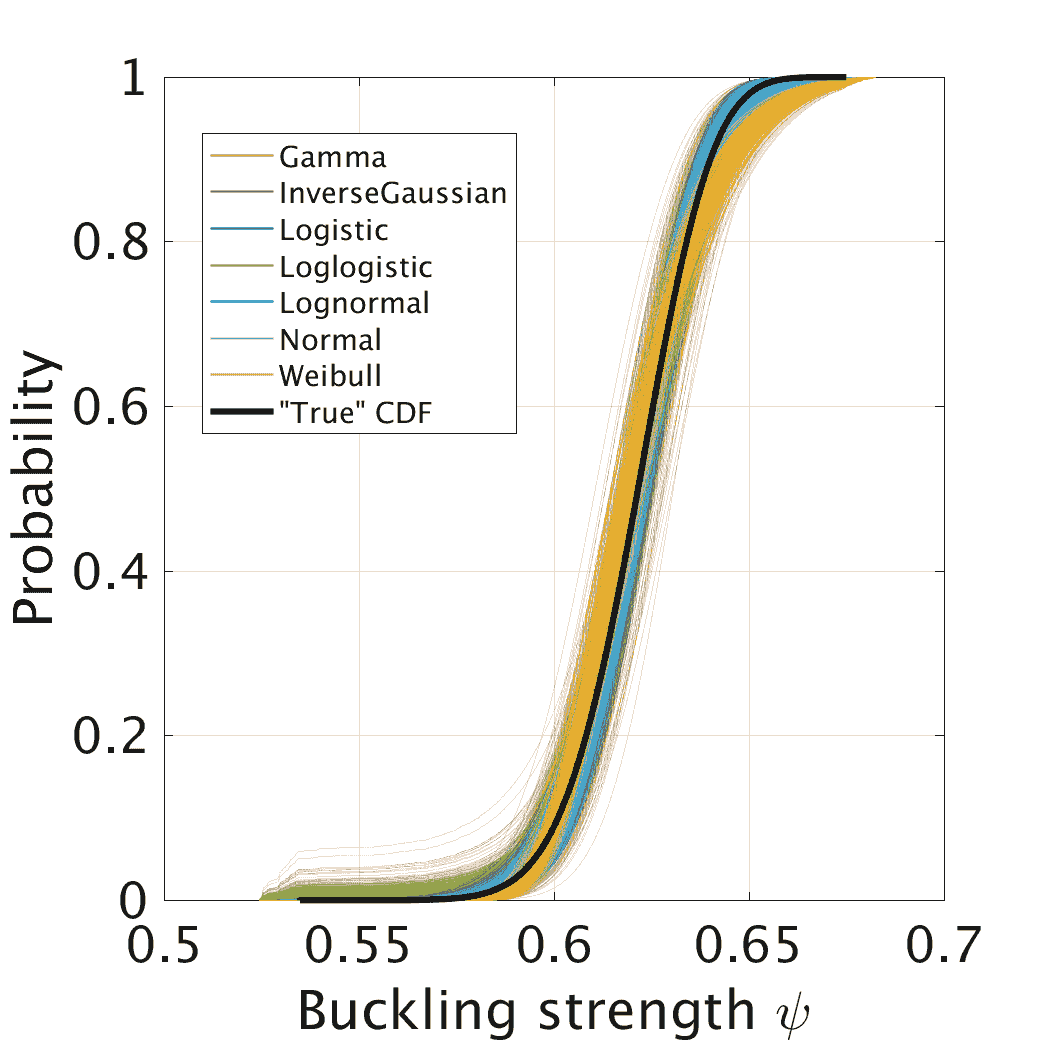}} 
&\parbox[c][1.35in]{1.3in}{\includegraphics[height=1.2in]{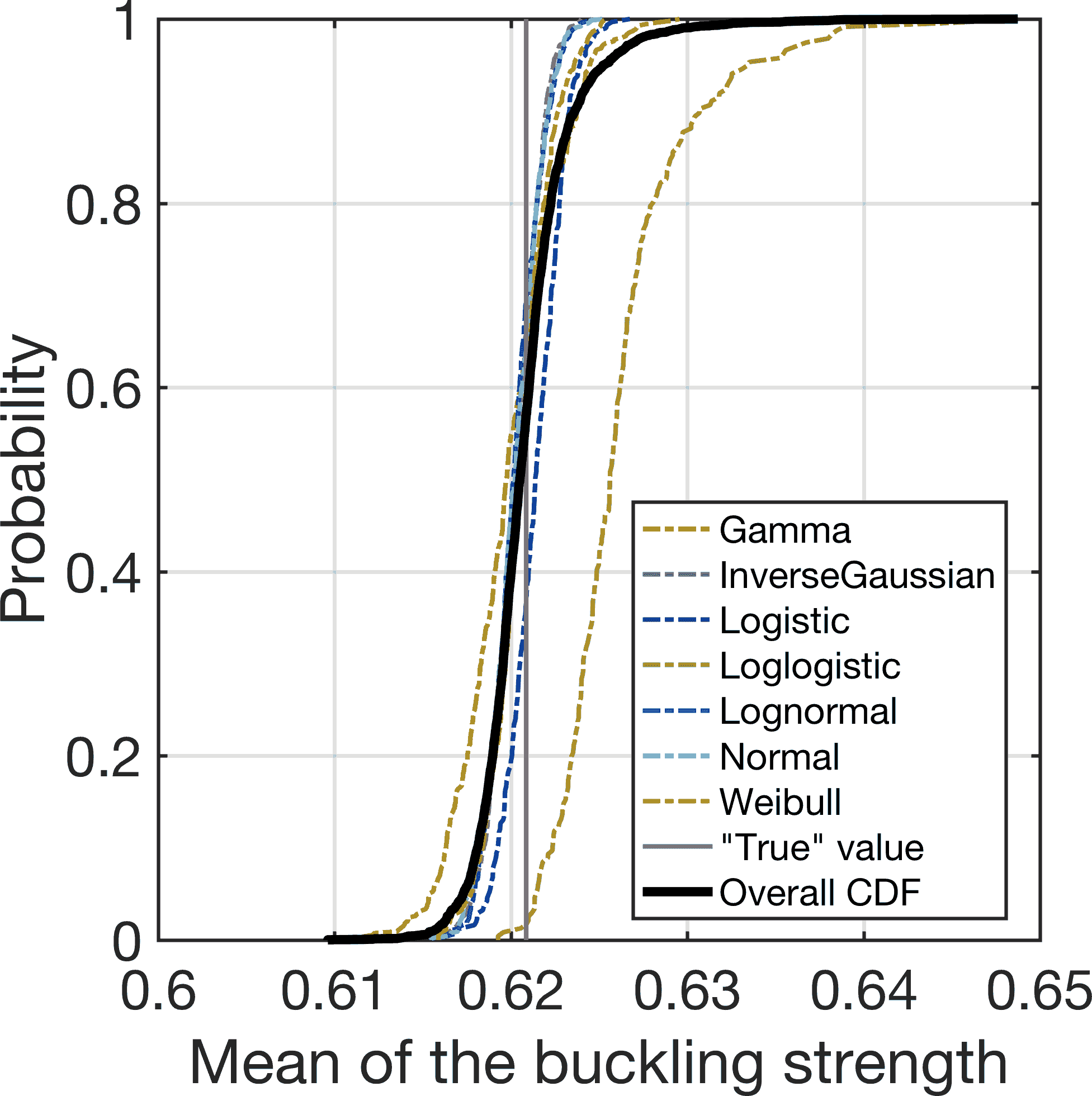}} 
&\parbox[c][1.35in]{1.3in}{\includegraphics[height=1.2in]{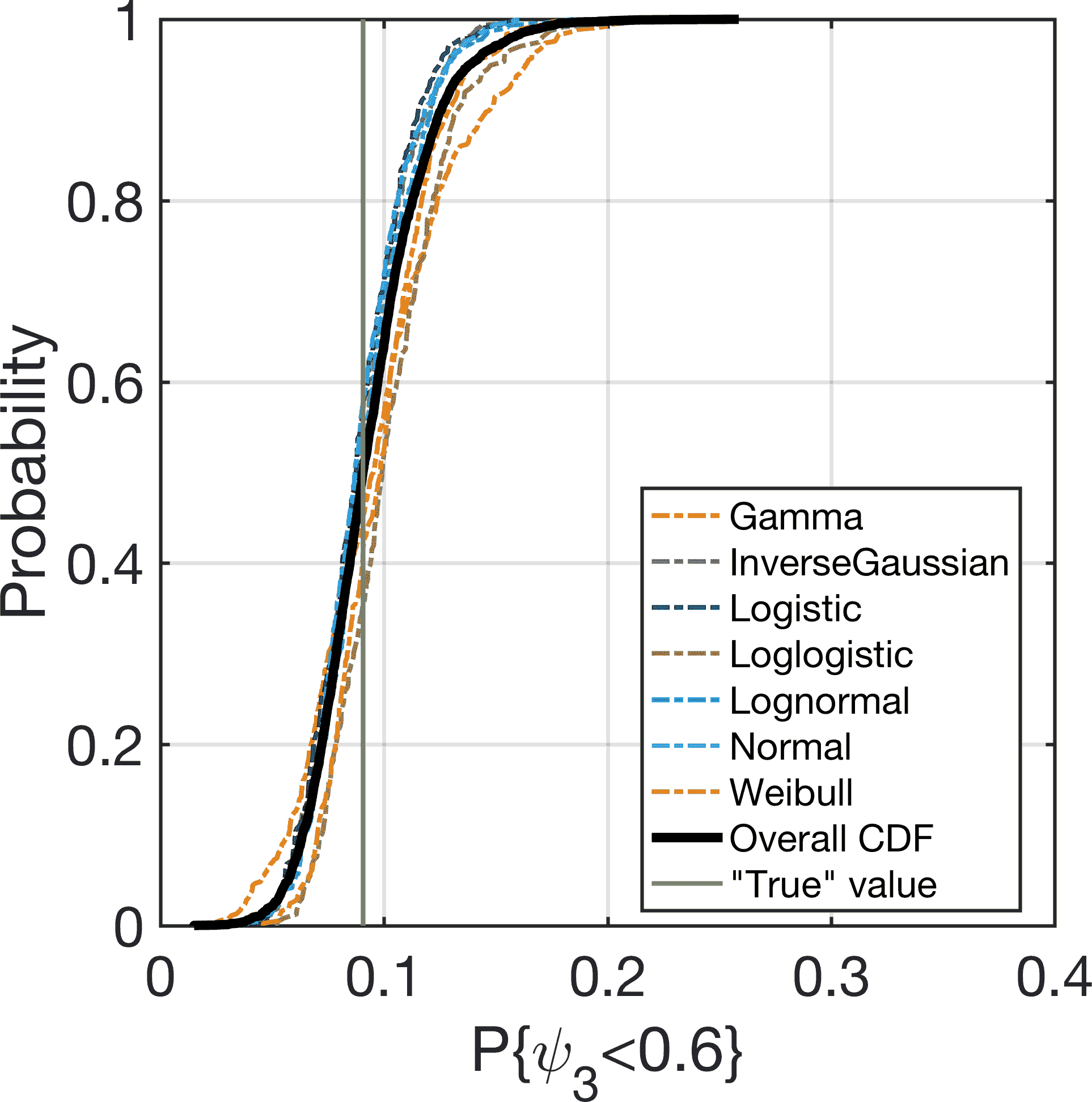}} \\
50
&\parbox[c][1.35in]{1.3in}{\includegraphics[height=1.3in]{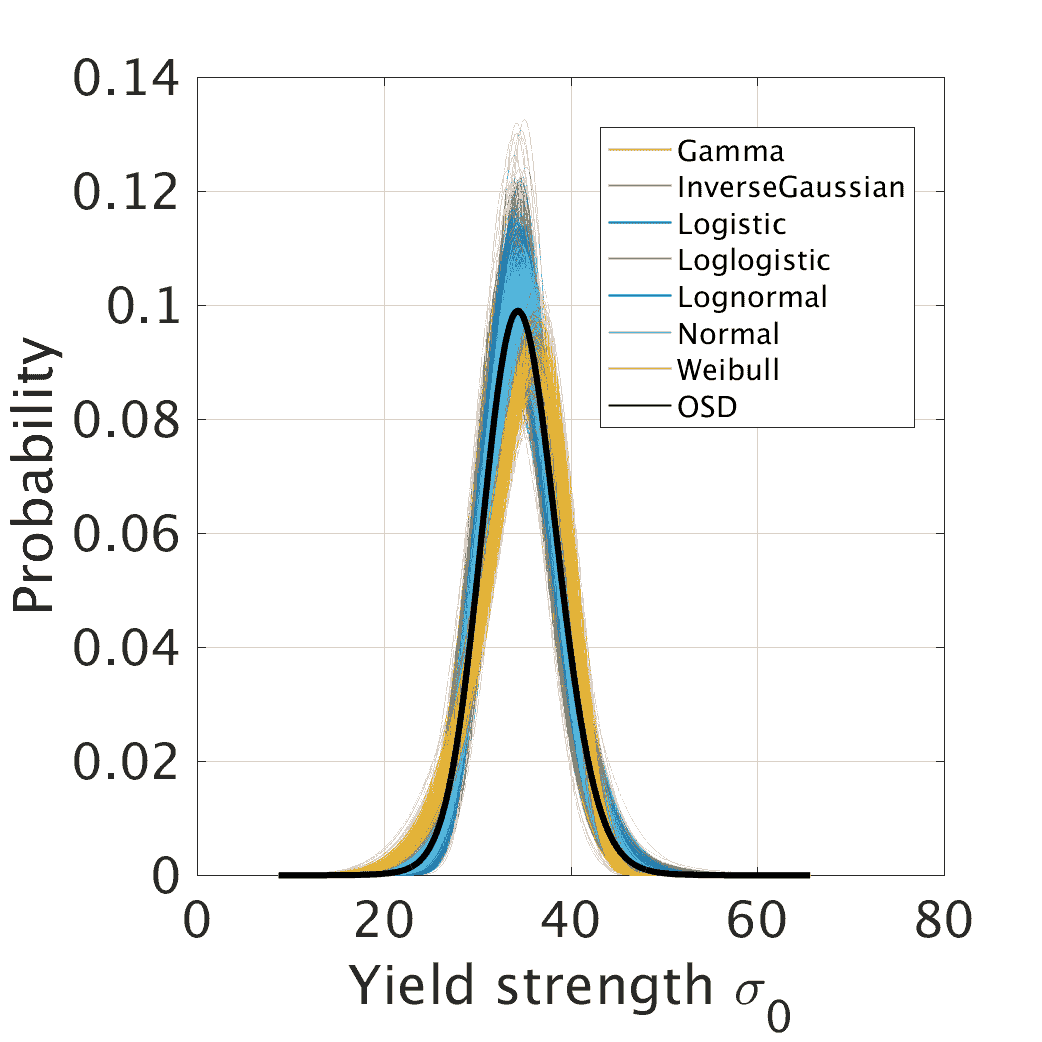}} 
&\parbox[c][1.35in]{1.3in}{\includegraphics[height=1.3in]{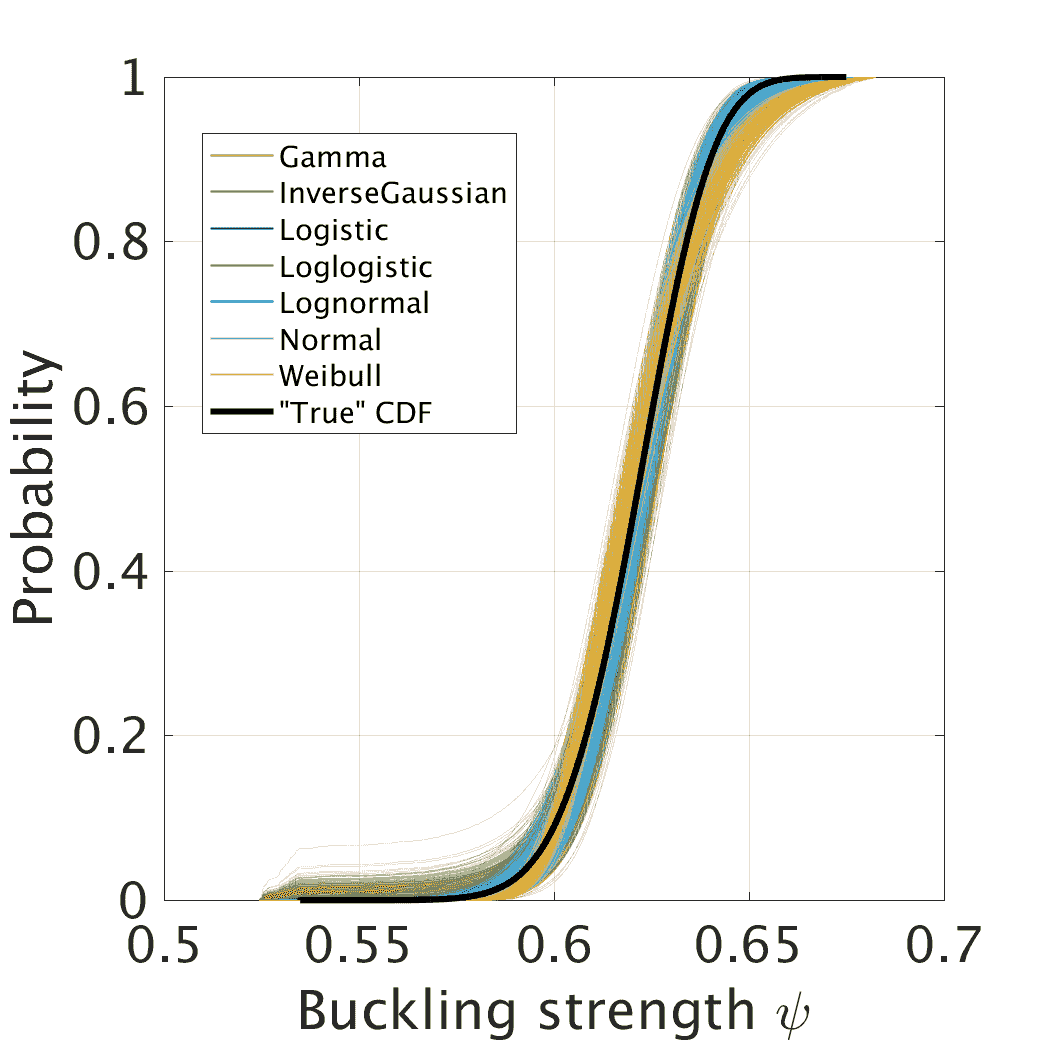}} 
&\parbox[c][1.35in]{1.3in}{\includegraphics[height=1.2in]{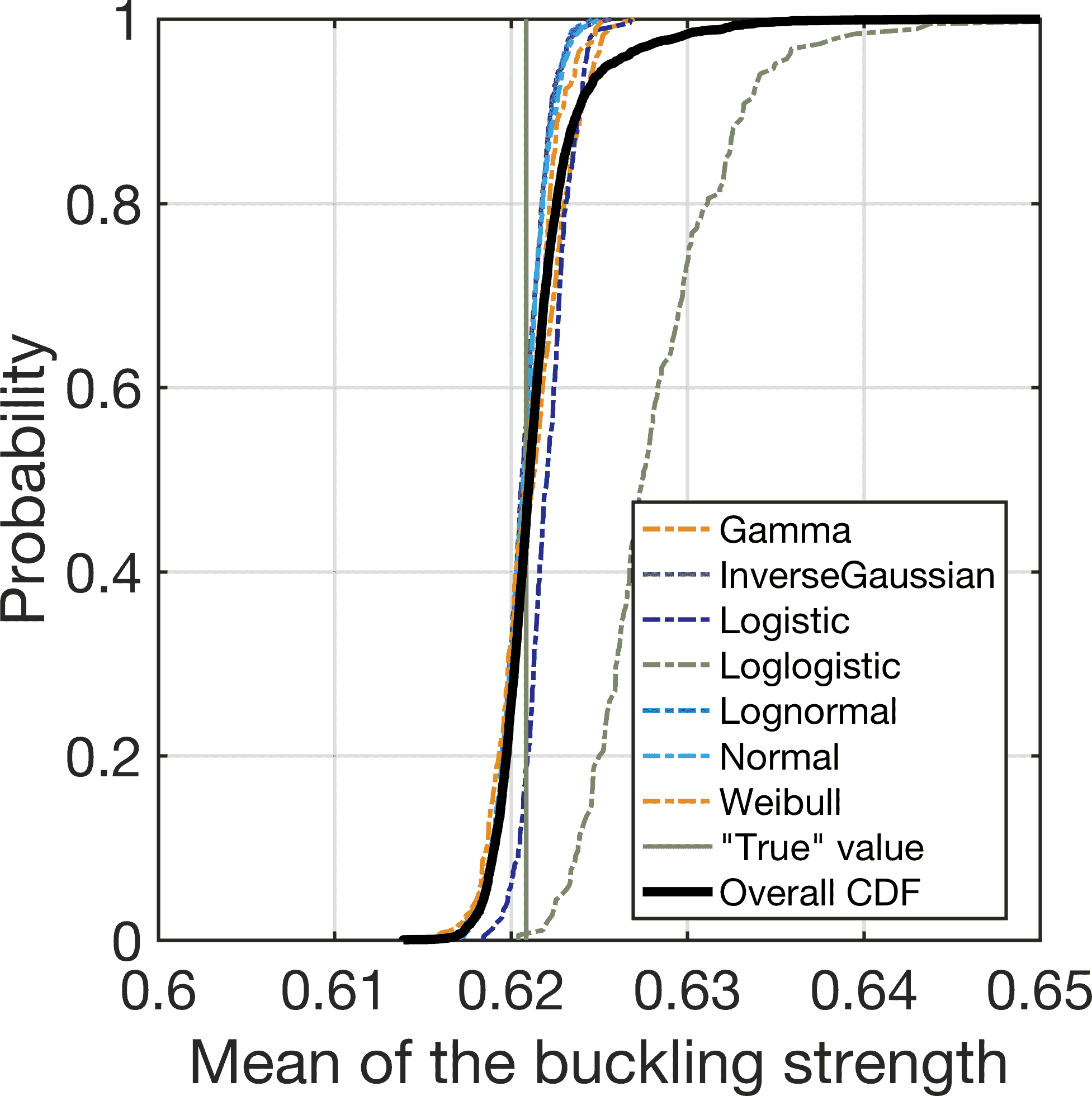}} 
&\parbox[c][1.35in]{1.3in}{\includegraphics[height=1.2in]{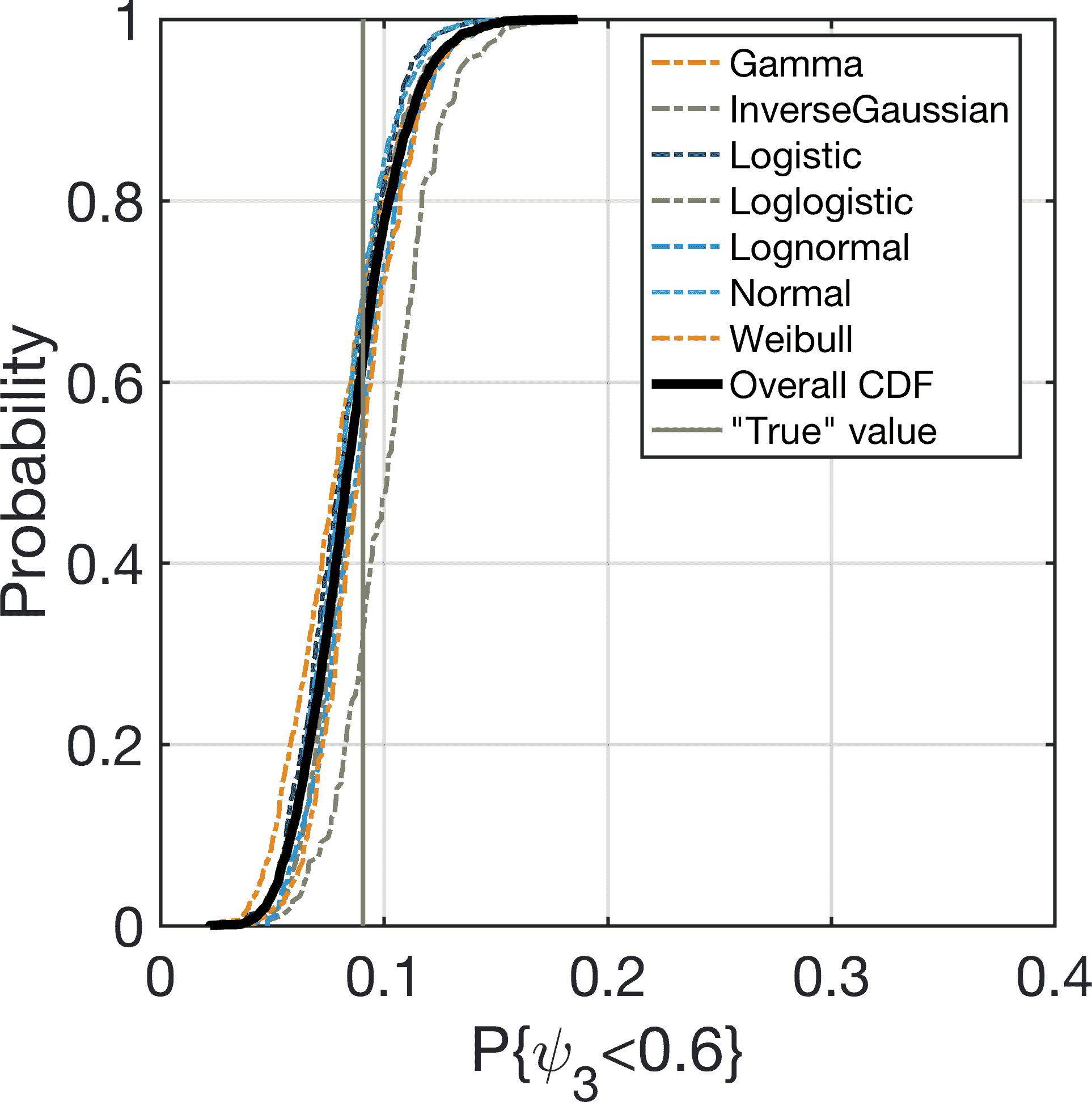}} \\
500
&\parbox[c][1.35in]{1.3in}{\includegraphics[height=1.3in]{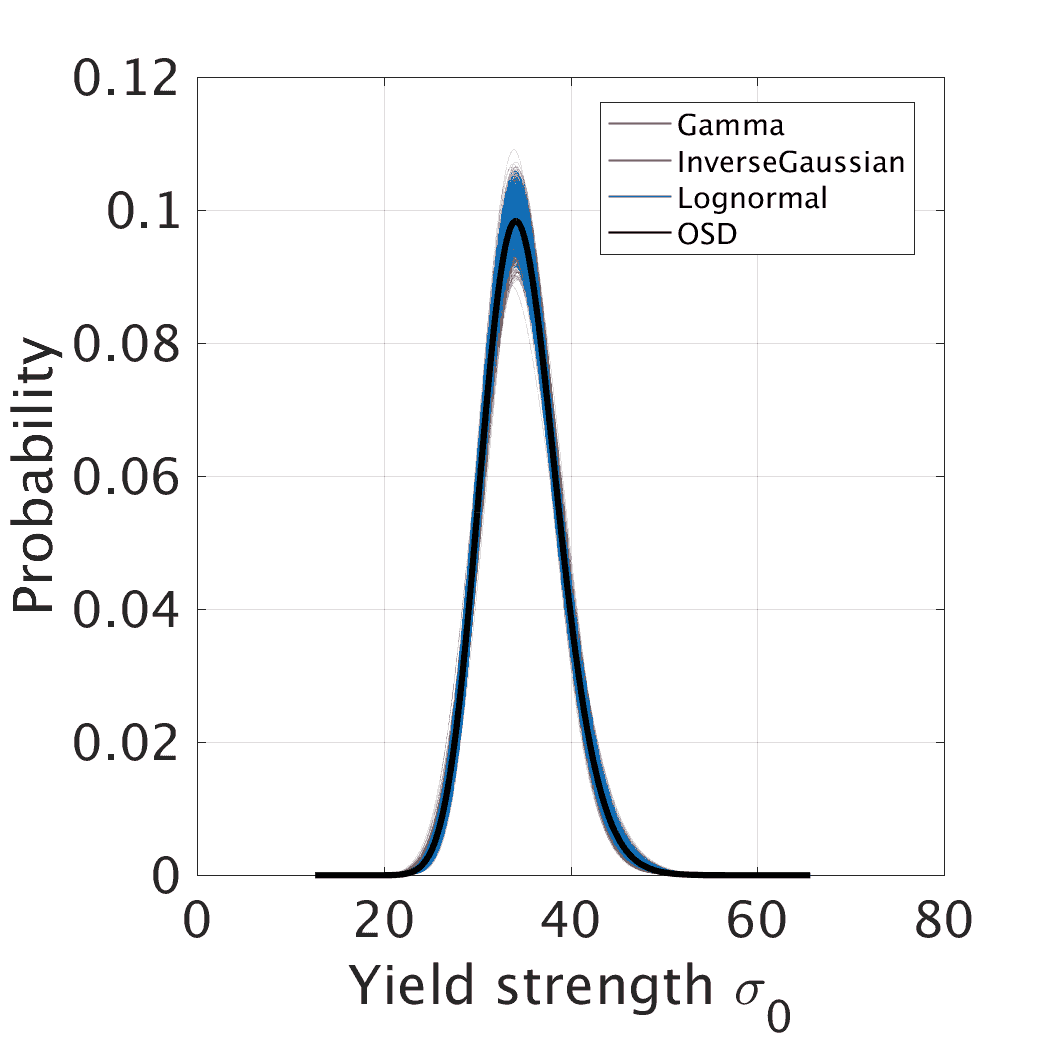}} 
&\parbox[c][1.35in]{1.3in}{\includegraphics[height=1.3in]{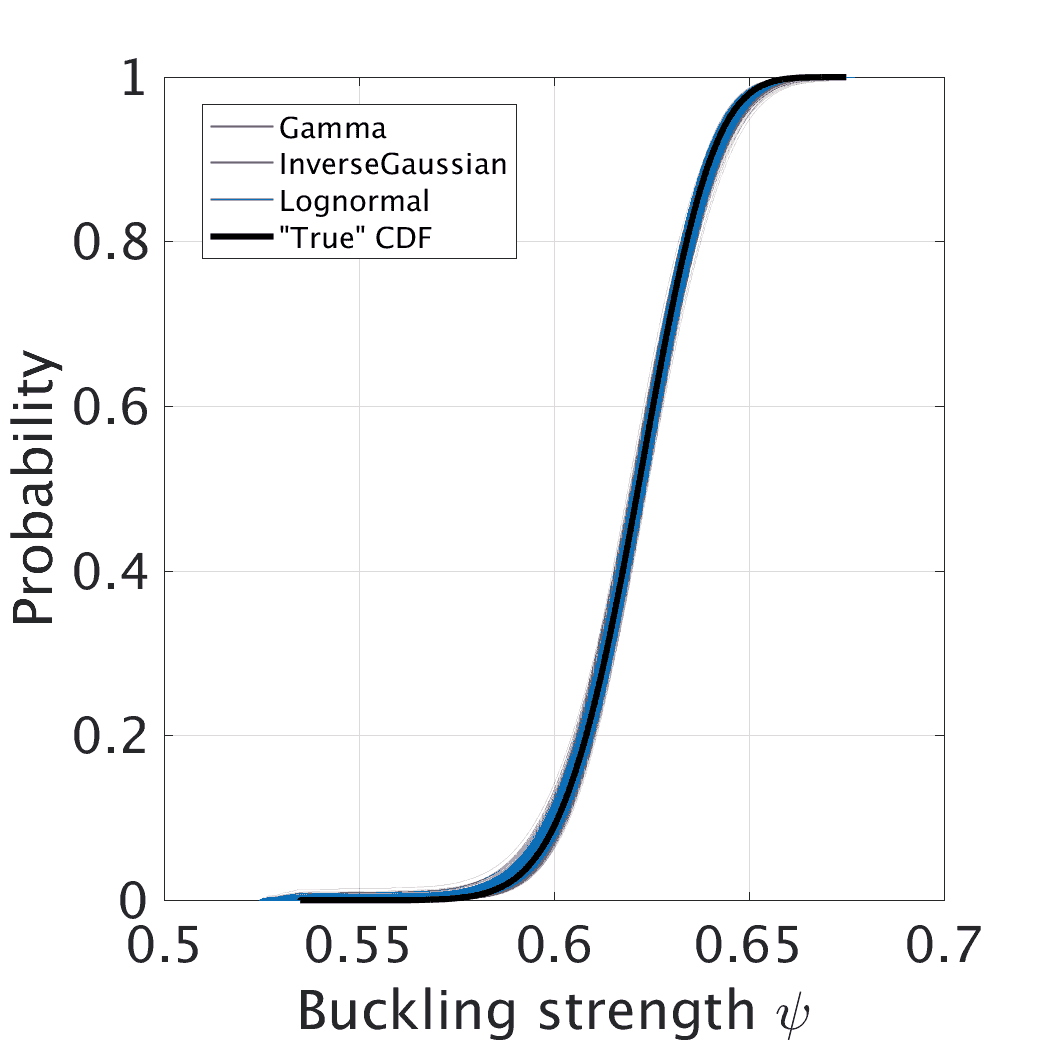}} 
&\parbox[c][1.35in]{1.3in}{\includegraphics[height=1.2in]{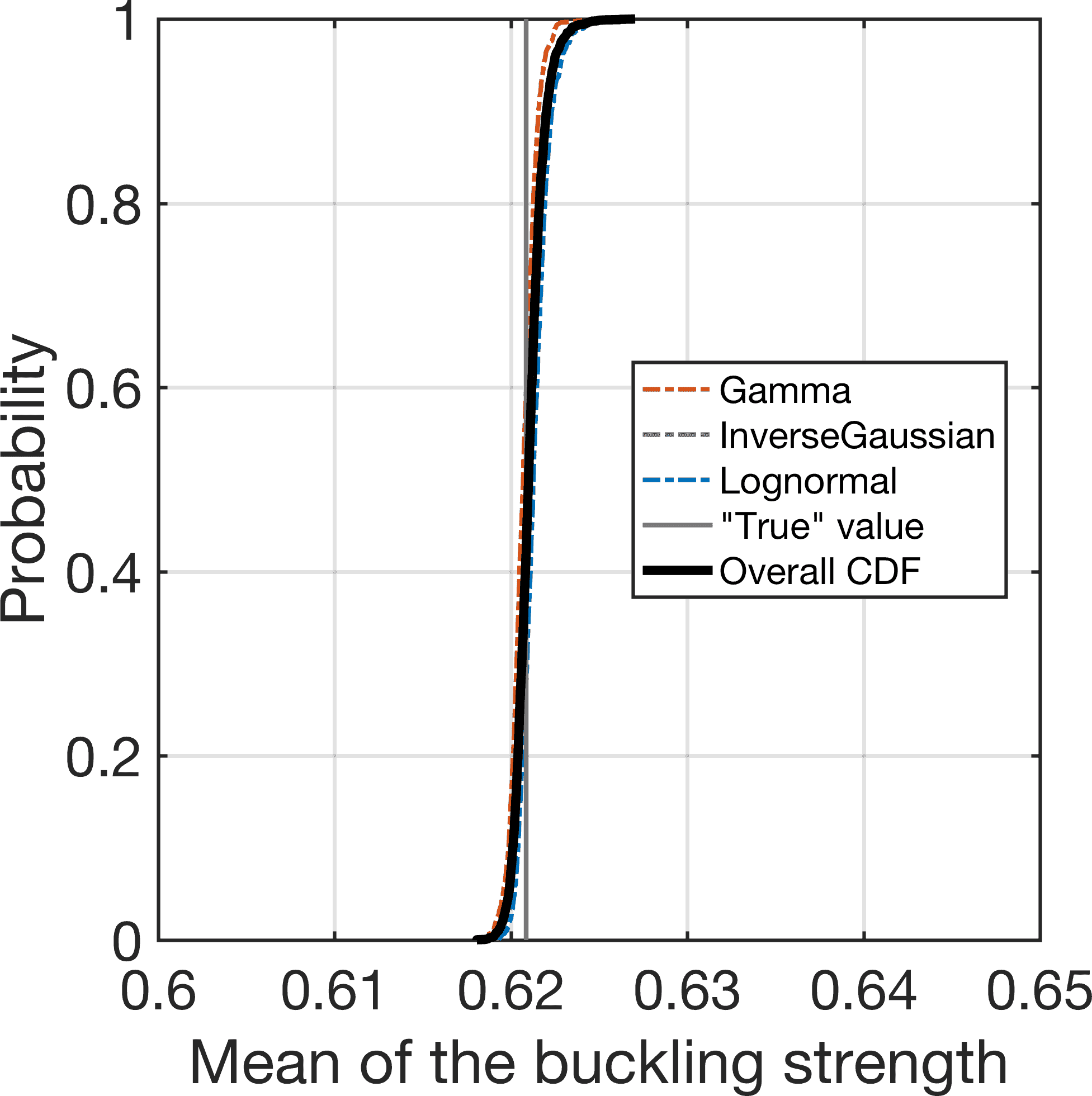}} 
&\parbox[c][1.35in]{1.3in}{\includegraphics[height=1.2in]{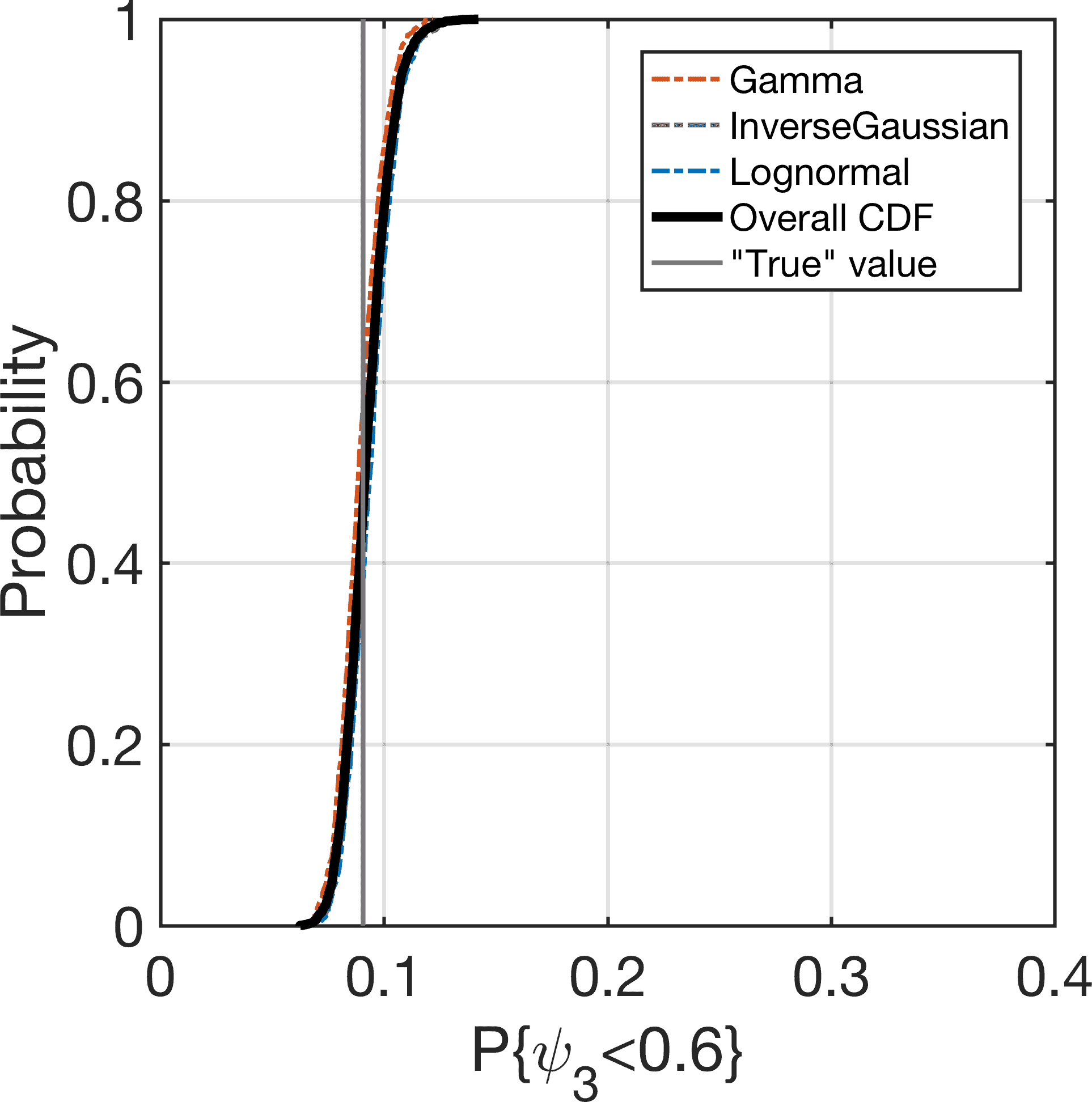}} \\
5000
&\parbox[c][1.35in]{1.3in}{\includegraphics[height=1.3in]{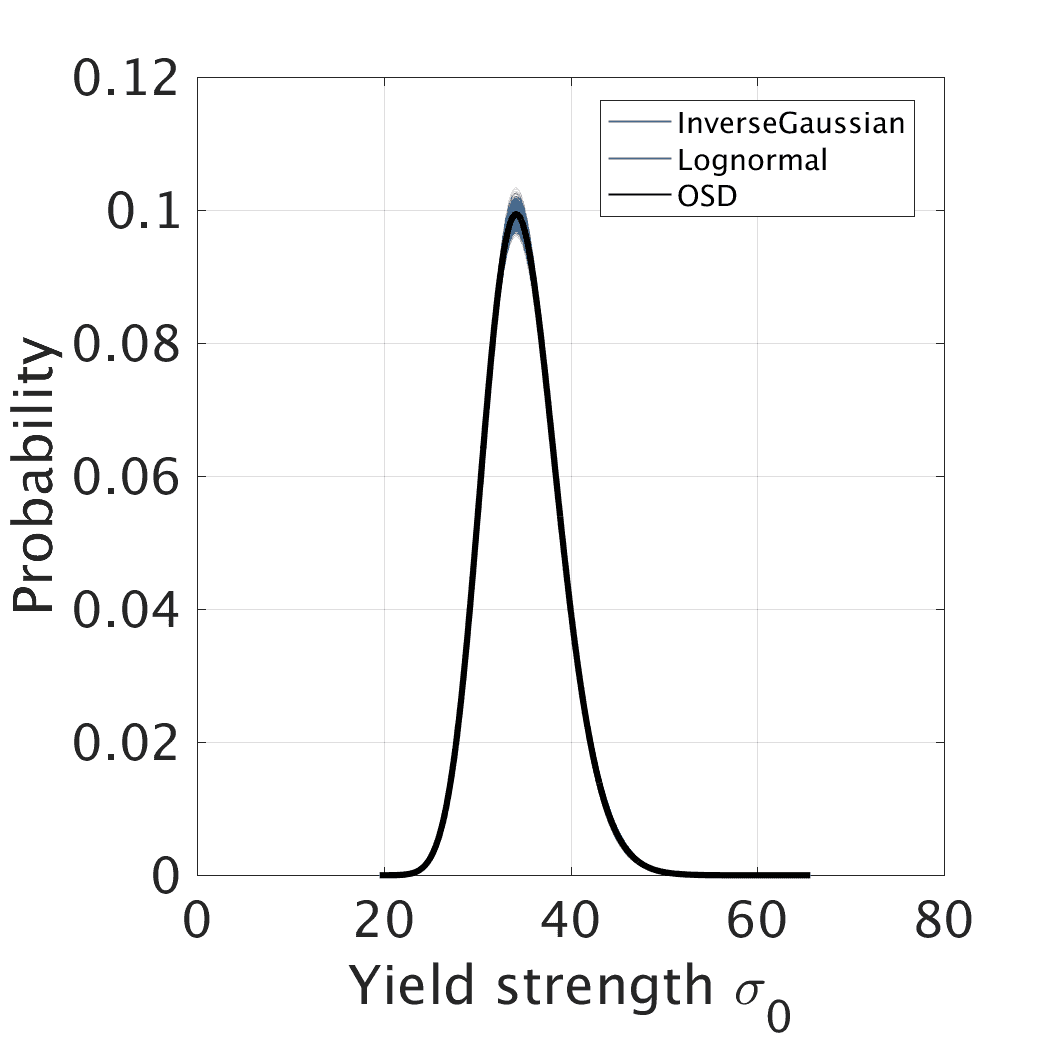}} 
&\parbox[c][1.35in]{1.3in}{\includegraphics[height=1.3in]{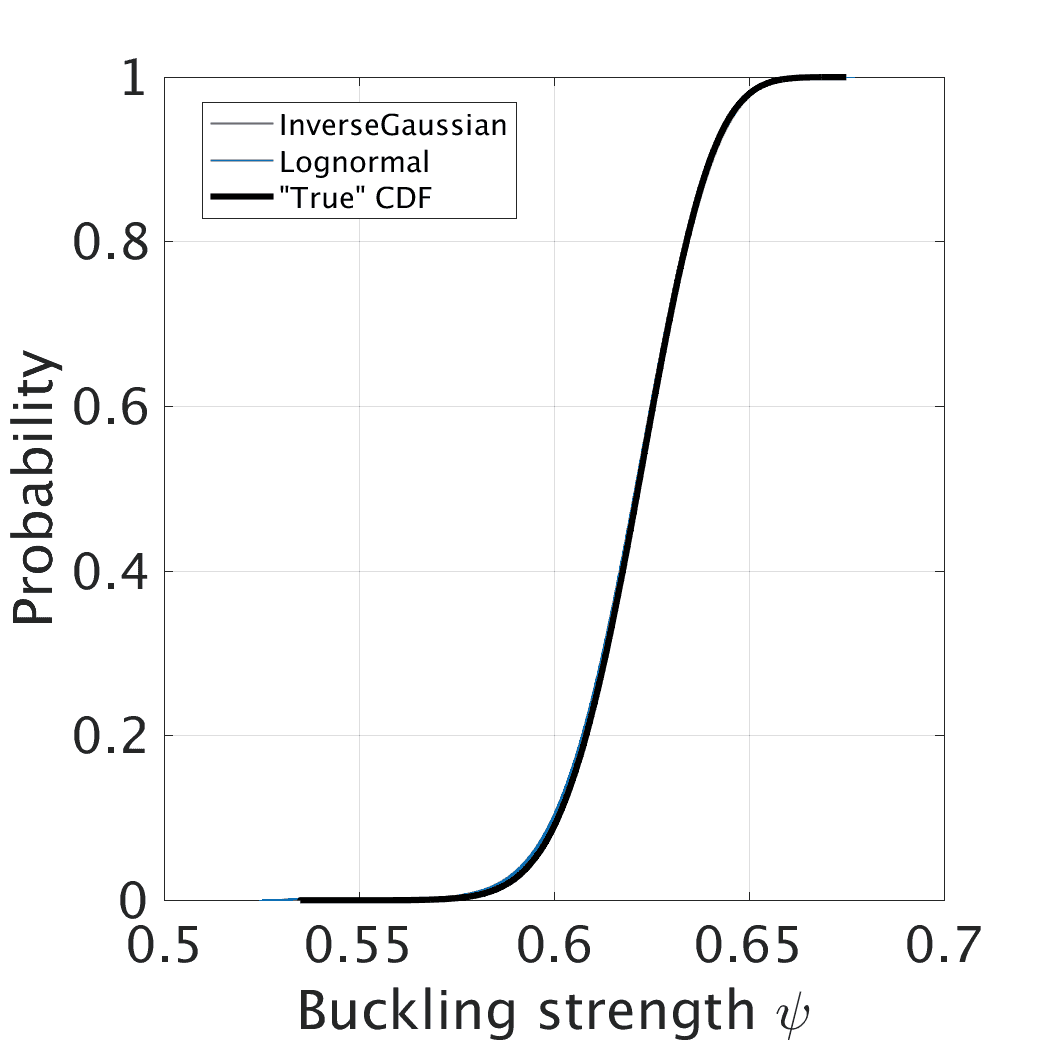}} 
&\parbox[c][1.35in]{1.3in}{\includegraphics[height=1.2in]{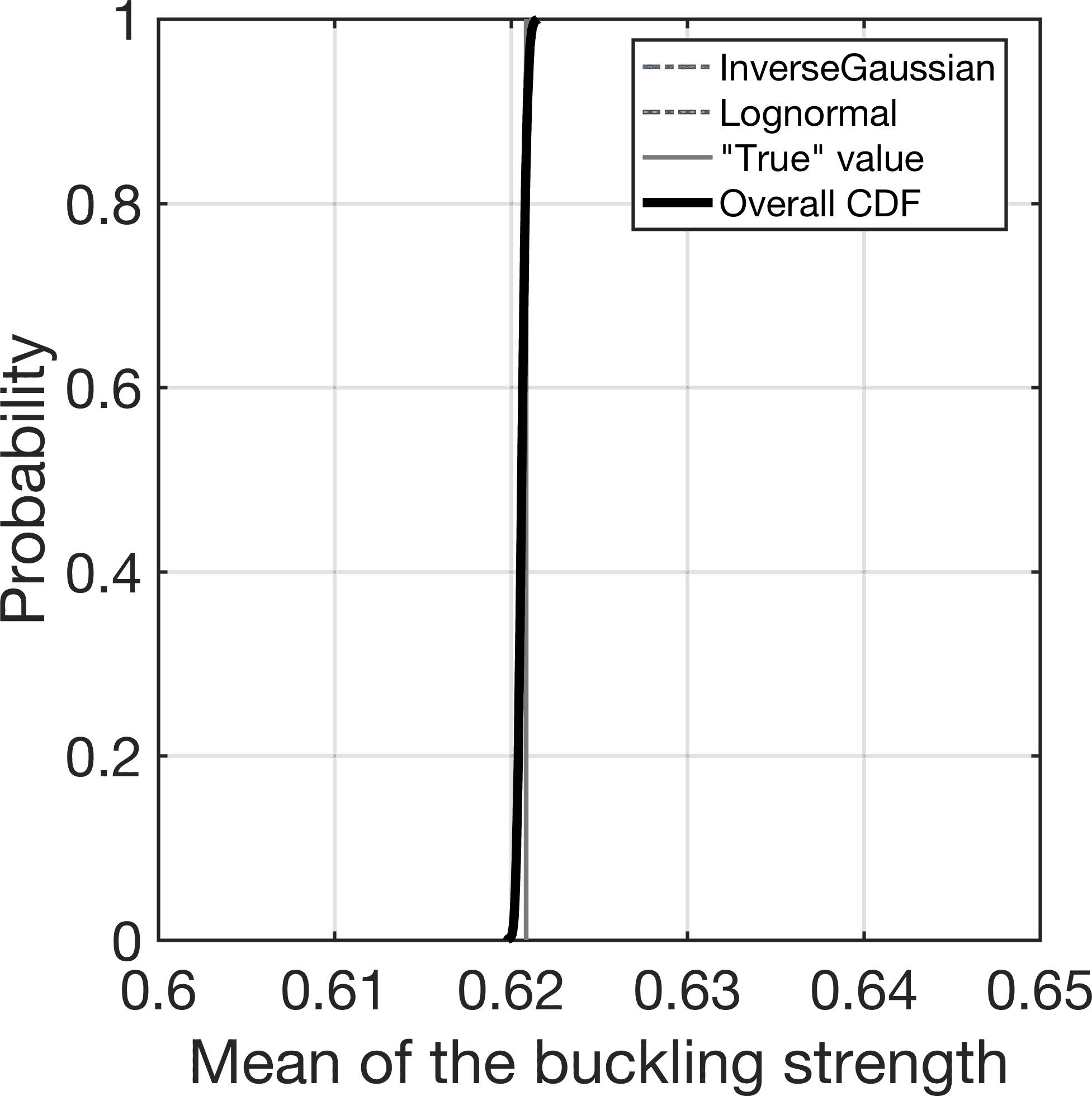}} 
&\parbox[c][1.35in]{1.3in}{\includegraphics[height=1.2in]{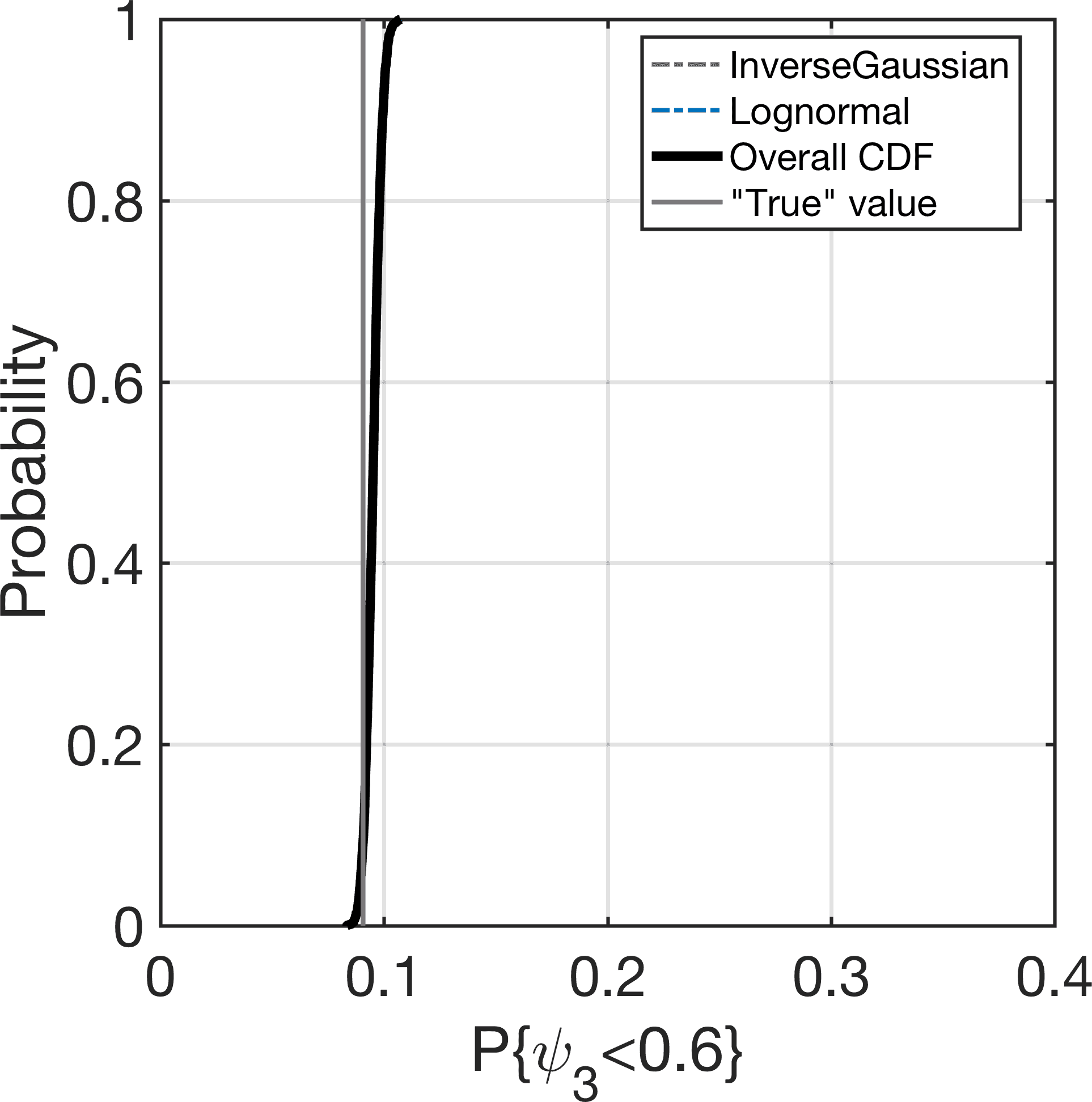}} \\
 \hline
\end{tabular}
\end{table}
The left column shows the set of 5000 probability densities identified from MC sampling of the quantified uncertainties in model form and parameters along with the optimal sampling density for propagation. The second column shows the set of CDFs for the buckling strength along with the true buckling strength CDF from propagating the true lognormal model. Notice that the true is fully encapsulated within the set of propagated distributions. The final columns show CDFs for the mean buckling strength and probability of failure ($P_f=P(\psi<0.6)$). The bold line gives the overall CDF considering all model forms (and their probabilities) while the colored CDFs are conditional CDFs for each plausible model form. Also shown in these figures are the true mean buckling strength and $P_f$ which, in all cases, fall within the range of the CDFs.

As expected, the uncertainty diminishes with increased dataset size. More specifically, the band of distributions in the input PDFs and output CDFs narrow toward their correct distributions. Also, the range of the CDFs for mean buckling strength and $P_f$ narrow toward the true values as the dataset size increases. 

These trends are clear and illustrate the method's performance for cases where a good prior (ABS-B) is selected. But, do these trends hold for other priors? To assess the effect of different priors, we quantify the convergence of the mean buckling strength, variance of buckling strength, and $P_f$ using two different metrics. The first is a simple quantile confidence metric which defines the 95\% confidence range for statistic $Y$ given $n$ data by:
\begin{equation}
{ \delta  }_{ { {Y} } }^{ (n) }=Q_{0.975}(Y^{(n)})-Q_{0.025}(Y^{(n)}) \label{eq:confidence_bounds}
\end{equation}
The ranges for the mean, variance, and $P_f$ are therefore denoted ${ \delta  }_{ { {\mu} } }^{ (n) }$, ${ \delta  }_{ { {\sigma^2} } }^{ (n) }$, and ${ \delta  }_{ { {(\psi<\psi^*)} } }^{ (n) }$.  The second is an accuracy metric, the ``area validation metric'' \cite{ferson2008, roy2011}, that measures the difference in area between the CDF and the true value for statistic $Y$ given $n$ data as:
\begin{equation}
d_Y^{(n)}(F, T) = \int_{-\infty}^{\infty} {\left| F(Y) - T(Y) \right|dy}
\label{eq:accuracy}
\end{equation}
Where $F(Y)$ is the CDF from the simulation and $T(Y)$ is the true value. For the mean, variance, and $P_f$ the accuracy metrics are therefore denoted by ${d}_{ { {\mu} } }^{ (n) }$, ${d}_{ { {\sigma^2} } }^{ (n) }$, and ${d}_{ { {(\psi<\psi^*)} } }^{ (n) }$, respectively.

\begin{figure}[!ht]	
	\centering
	\subfigure[]{\includegraphics[height=2in]{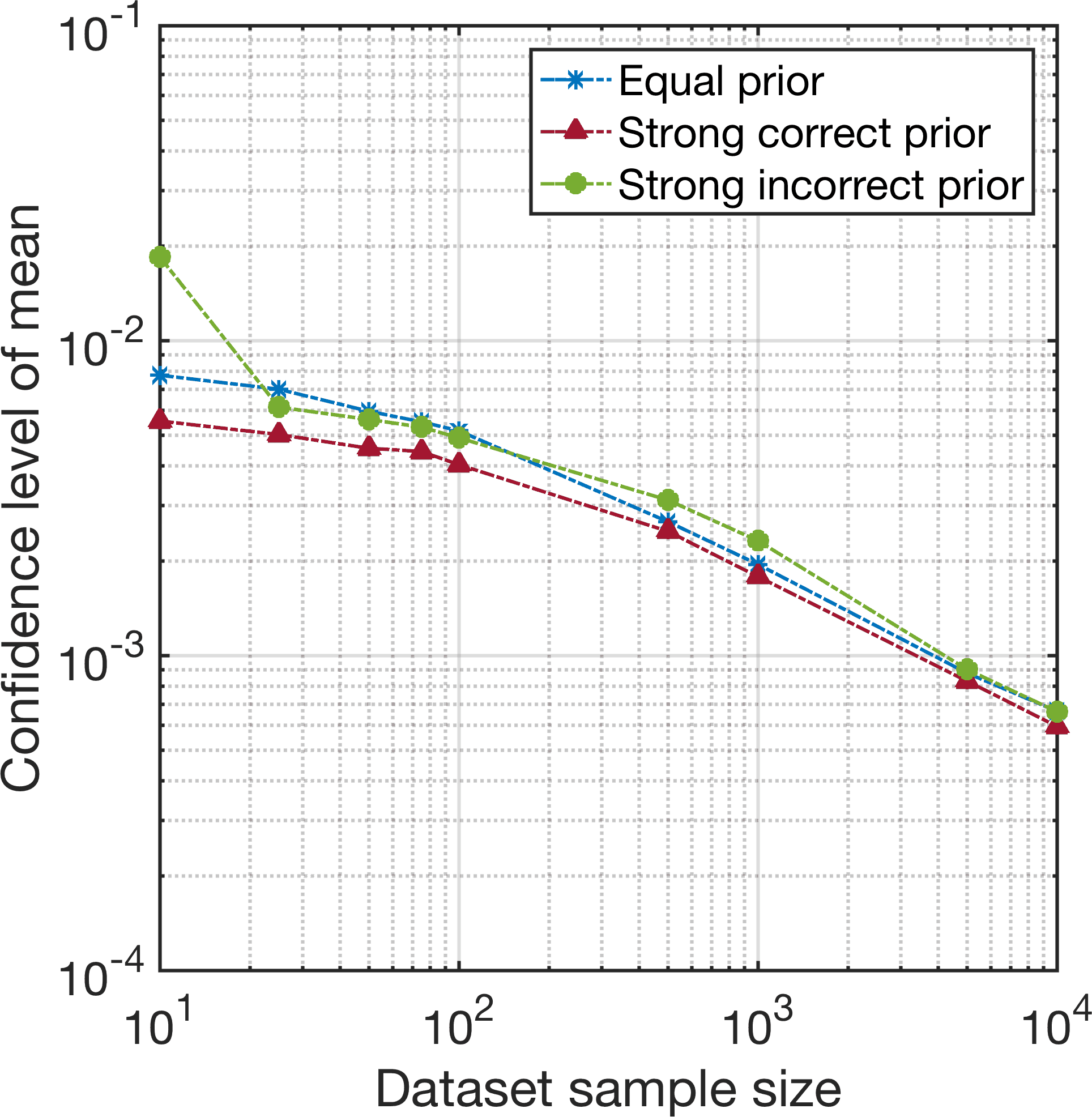}}
	\subfigure[]{\includegraphics[height=2in]{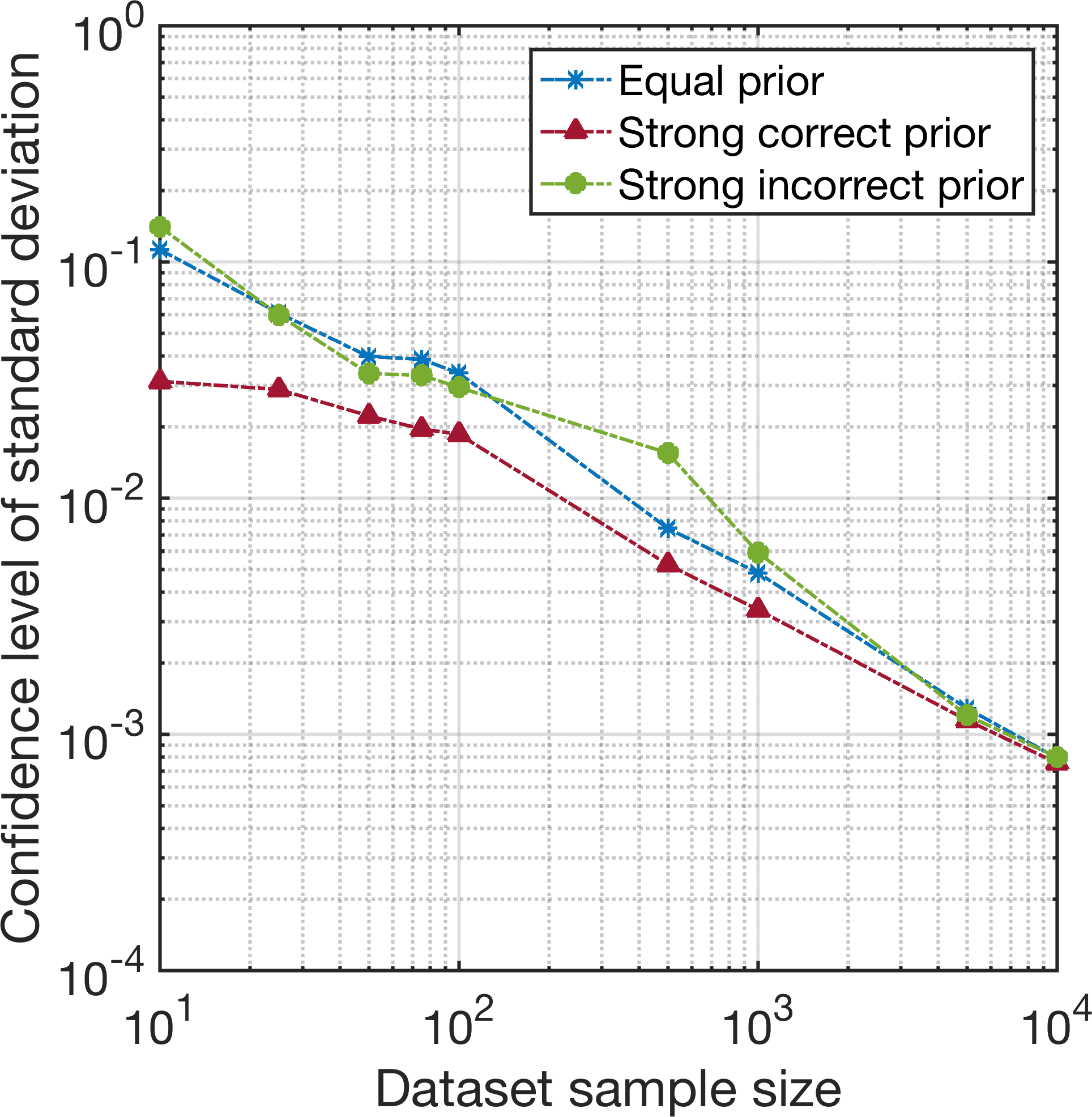}} 
	\subfigure[]{\includegraphics[height=2.1in]{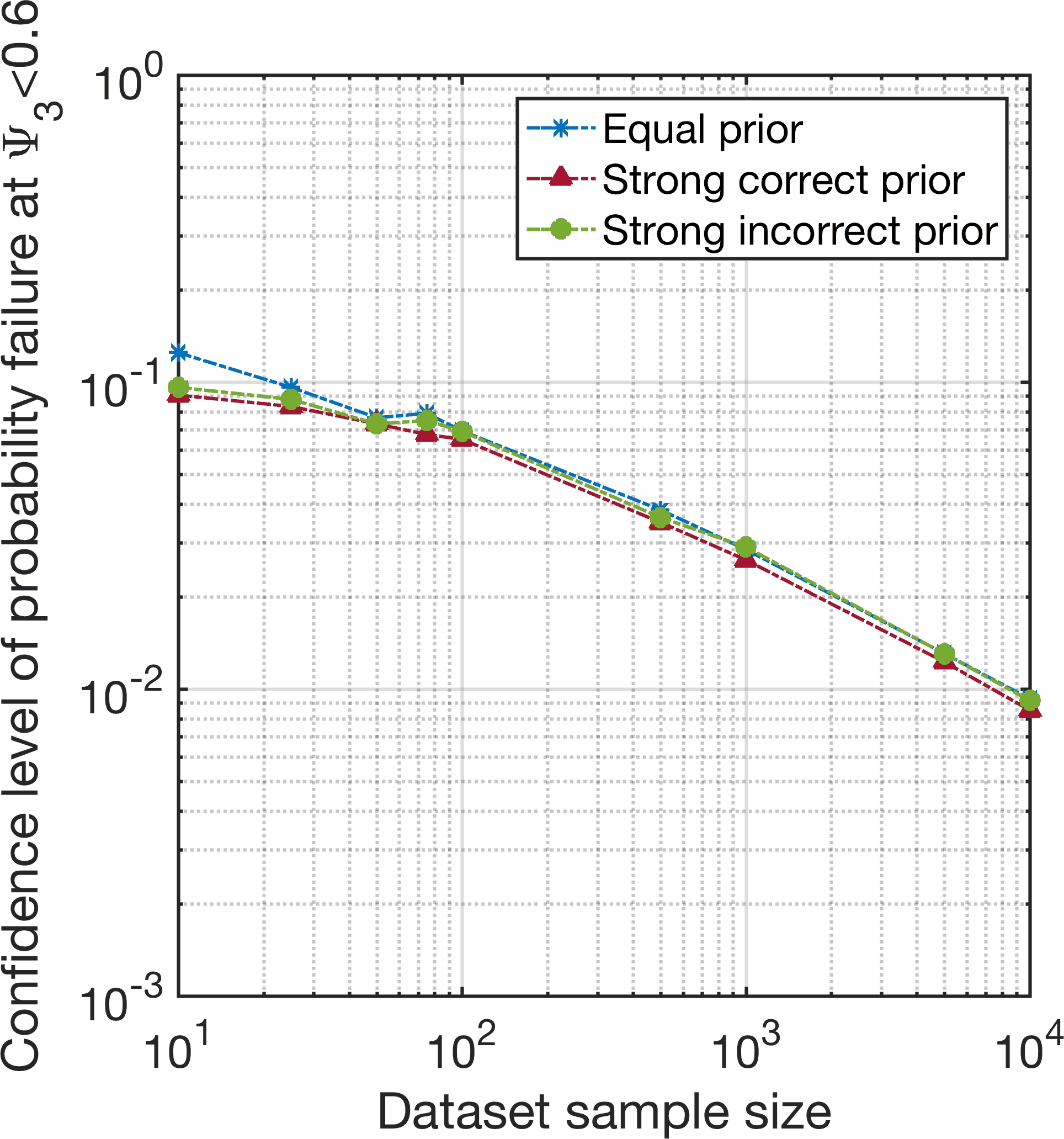}} \\
	\subfigure[]{\includegraphics[height=2in]{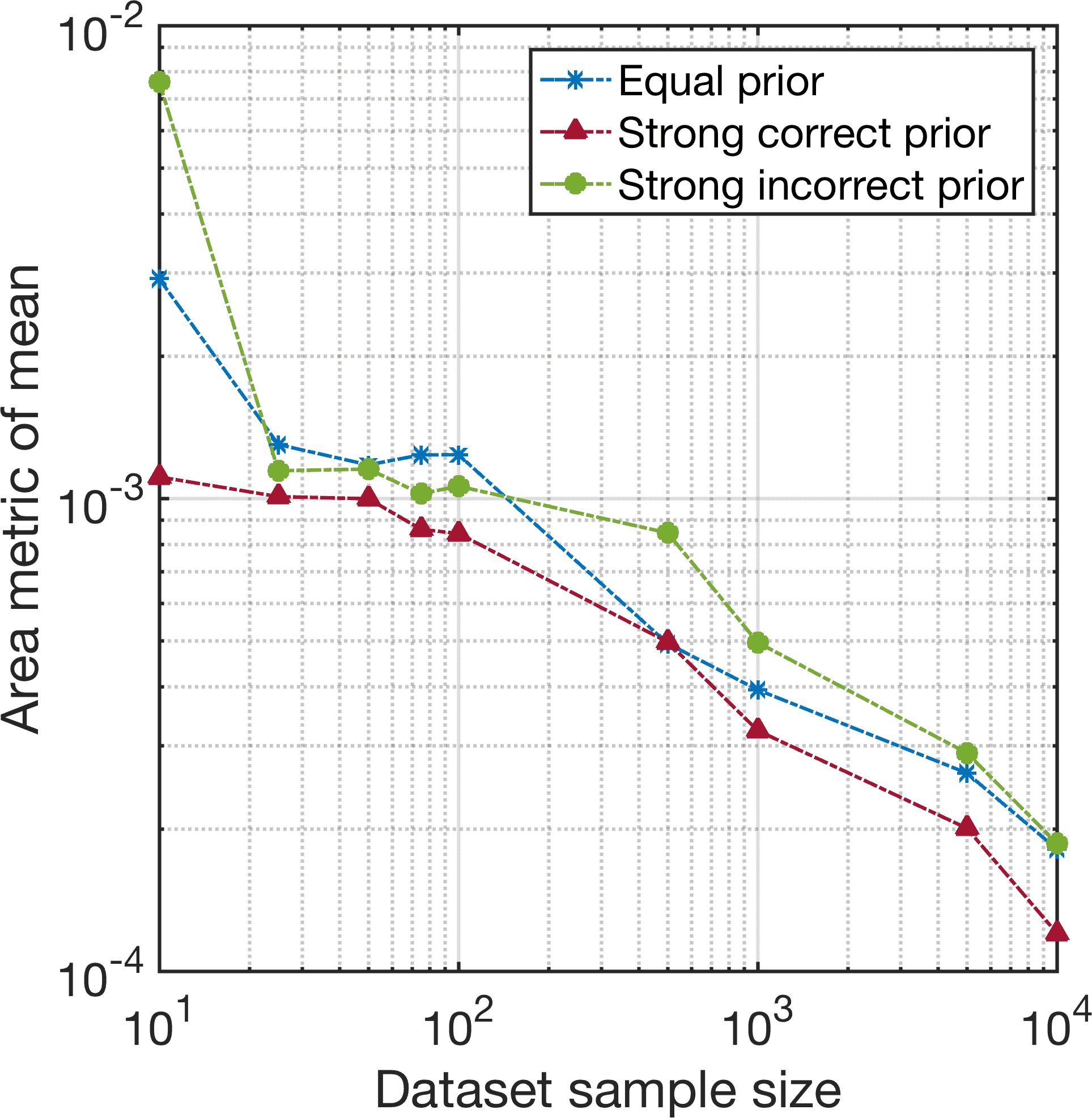}}
	\subfigure[]{\includegraphics[height=2in]{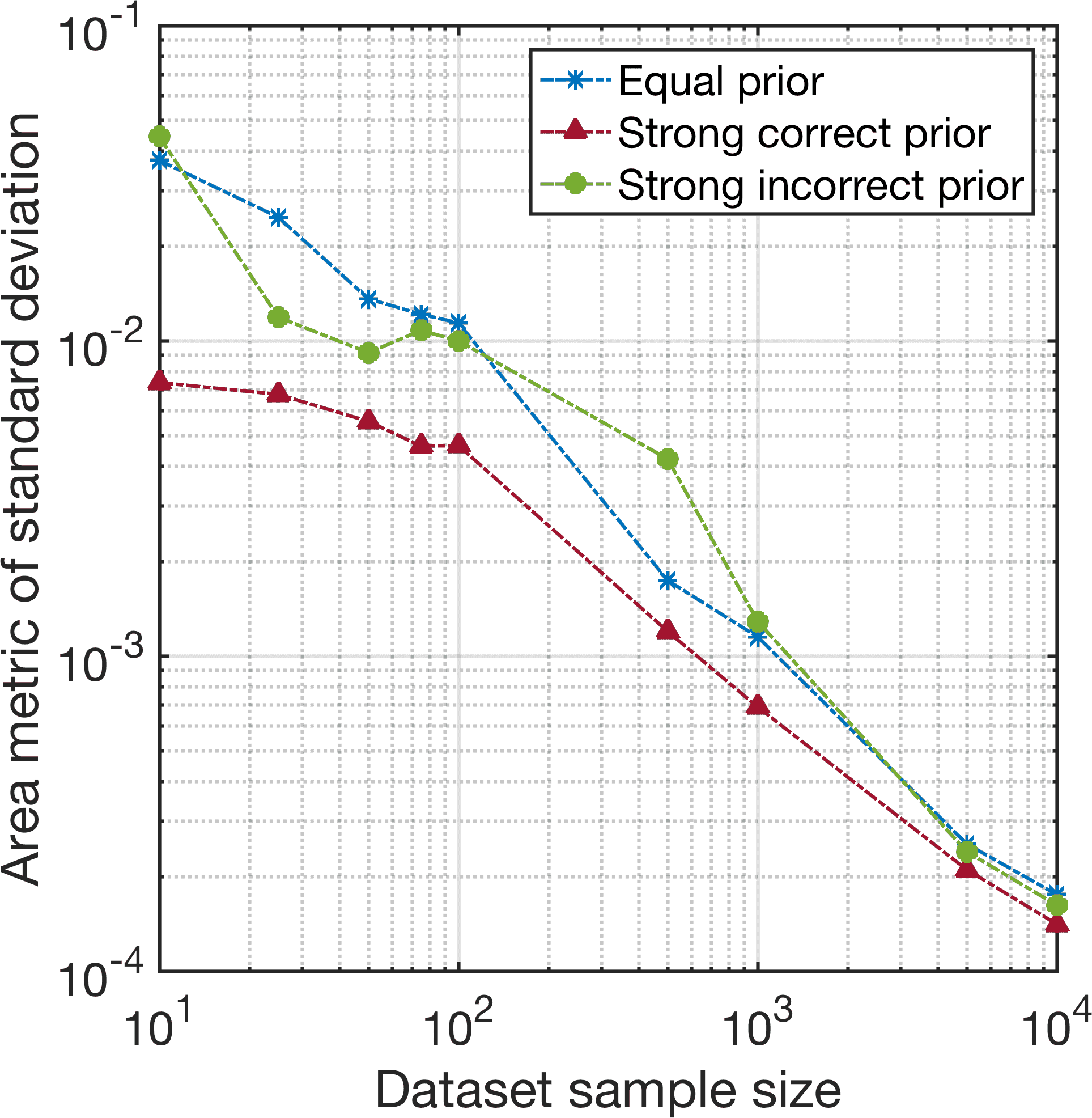}}
	\subfigure[]{\includegraphics[height=2in]{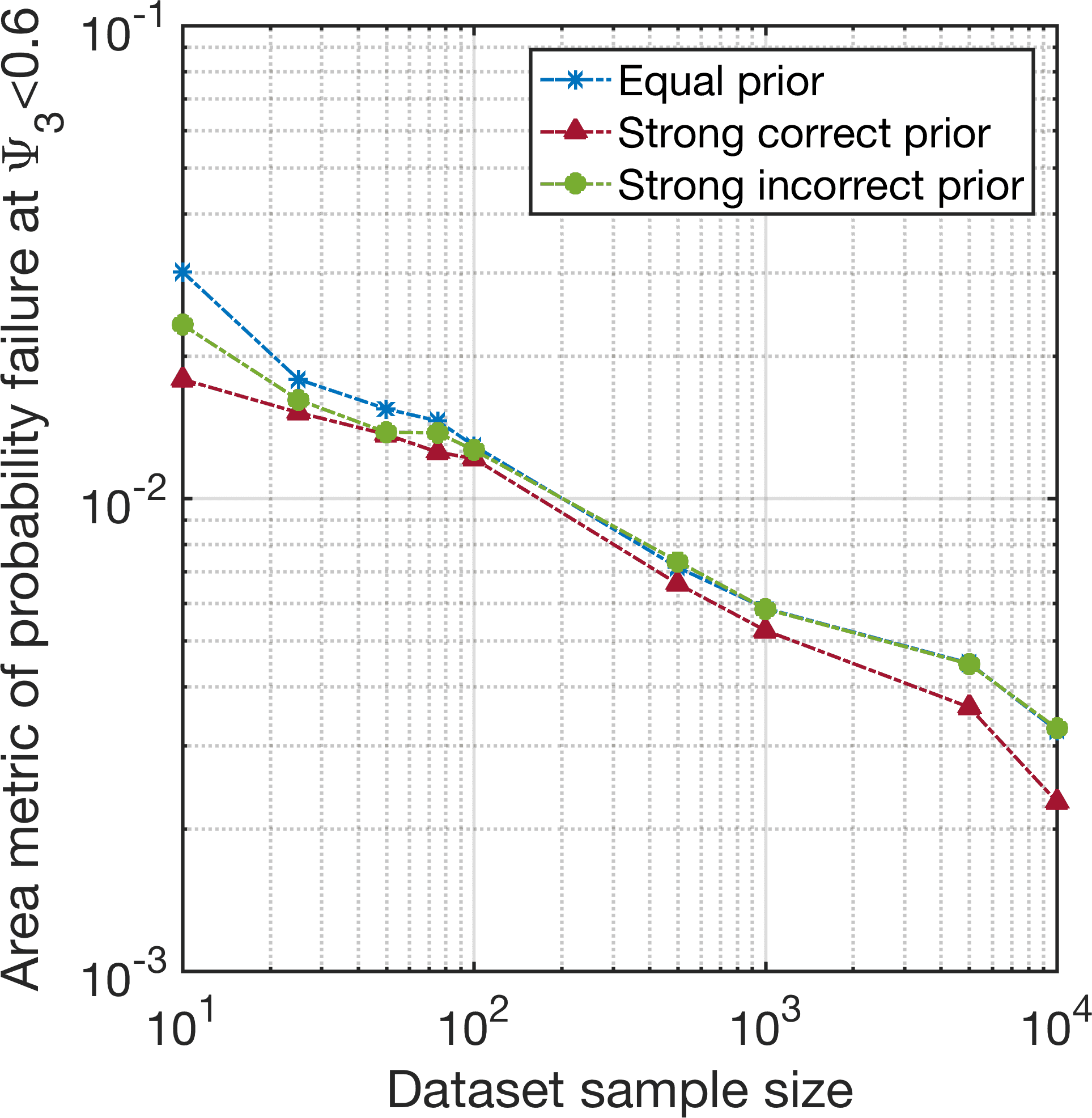}}
	\caption[]{Compare the effect of prior model probability for ABS-B prior - convergence of confidence level of (a) mean, (b) variance (c) probability of failure; and area validation metric for (d) mean, (e) variance and (f) probability of failure}  \label{fig:convergence1}
\end{figure}

Let us begin by investigating the effect of the prior model probability while retaining the correct ABS-B parameter prior. Figure \ref{fig:convergence1} shows convergence of the confidence metric (Eq.\ \eqref{eq:confidence_bounds} and the area accuracy metric (Eq.\ \eqref{eq:accuracy}).
These figures show that the strong correct model prior probabilities provide significant improvements in both confidence and accuracy for the mean buckling strength and variance of buckling strength for small datasets. The improvement diminishes as the dataset size increases. For $P_f$, on the other hand, the prior probabilities have relatively modest effect on convergence. 

\begin{figure}[!ht]	
	\centering
	\subfigure[]{\includegraphics[height=2in]{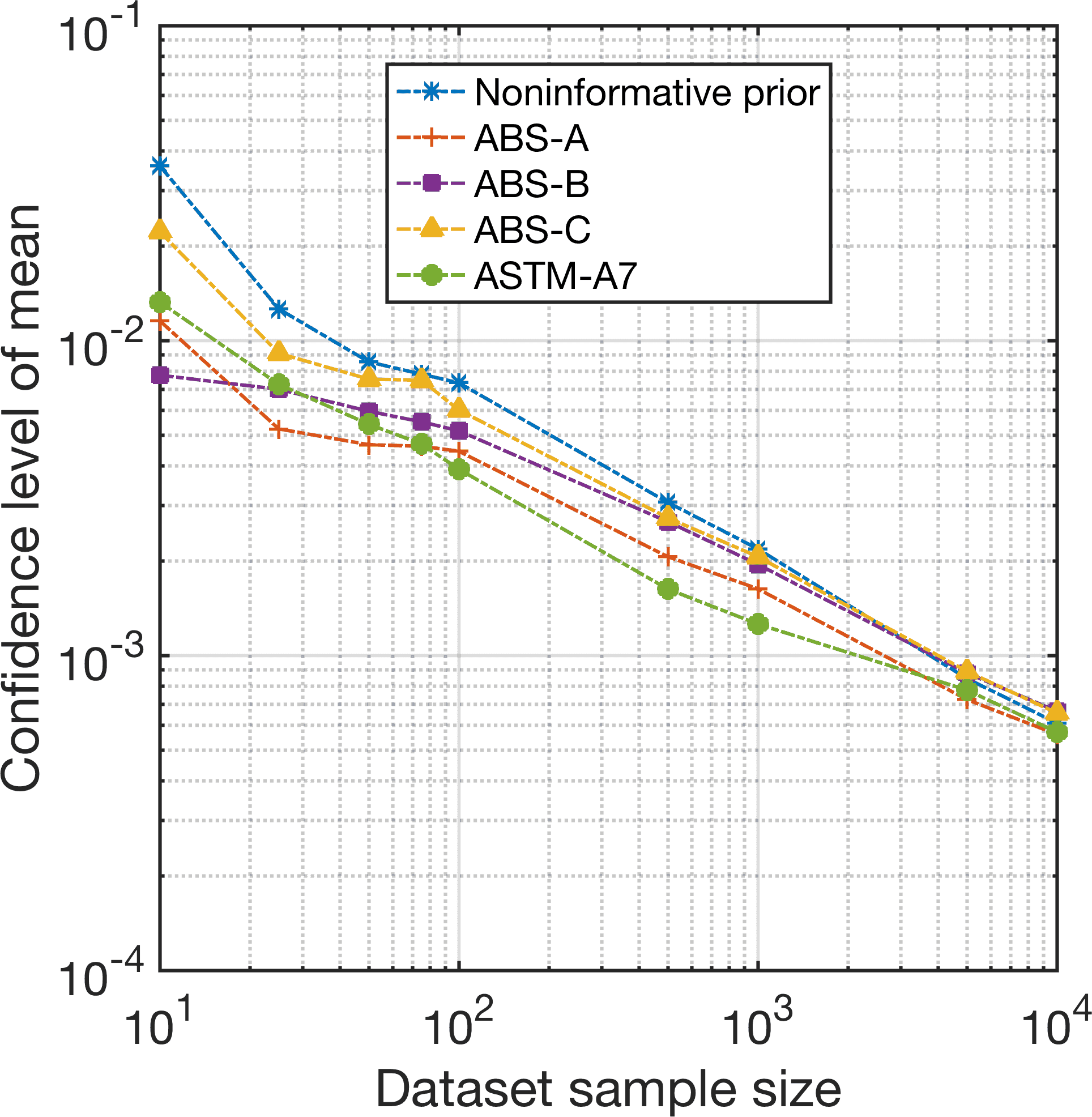}}
	\subfigure[]{\includegraphics[height=2in]{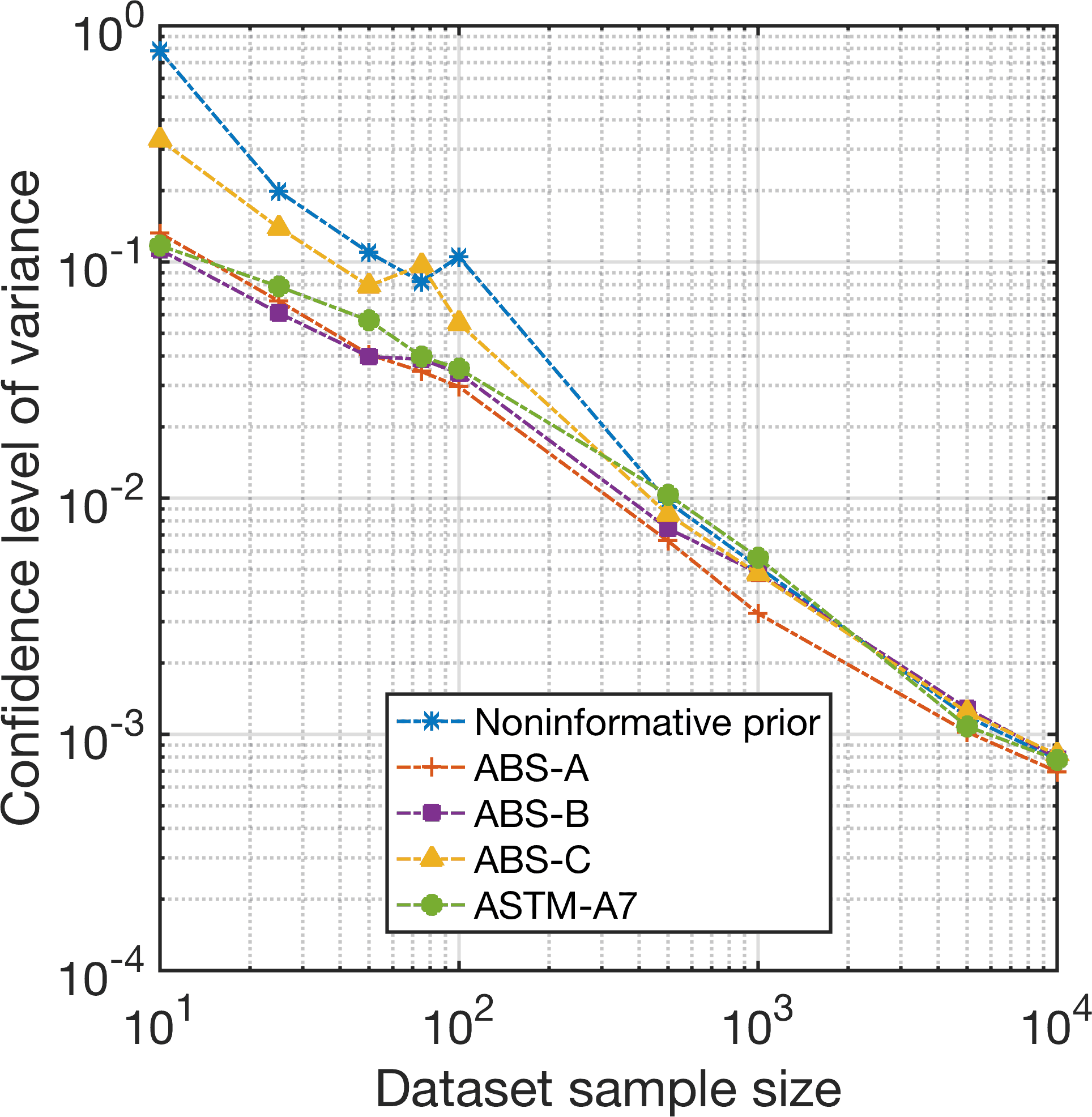}} 
	\subfigure[]{\includegraphics[height=2.1in]{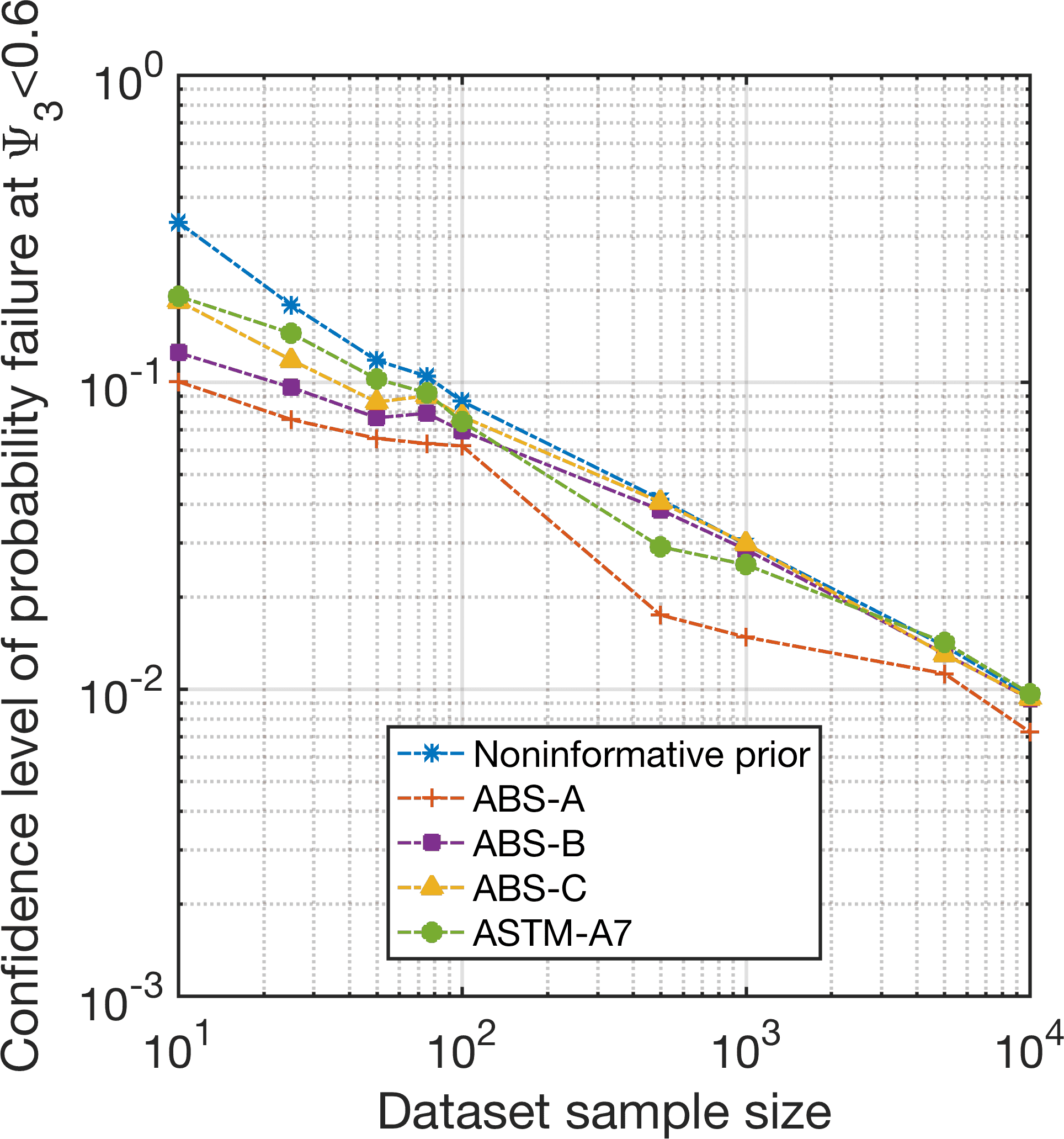}} \\
	\subfigure[]{\includegraphics[height=2in]{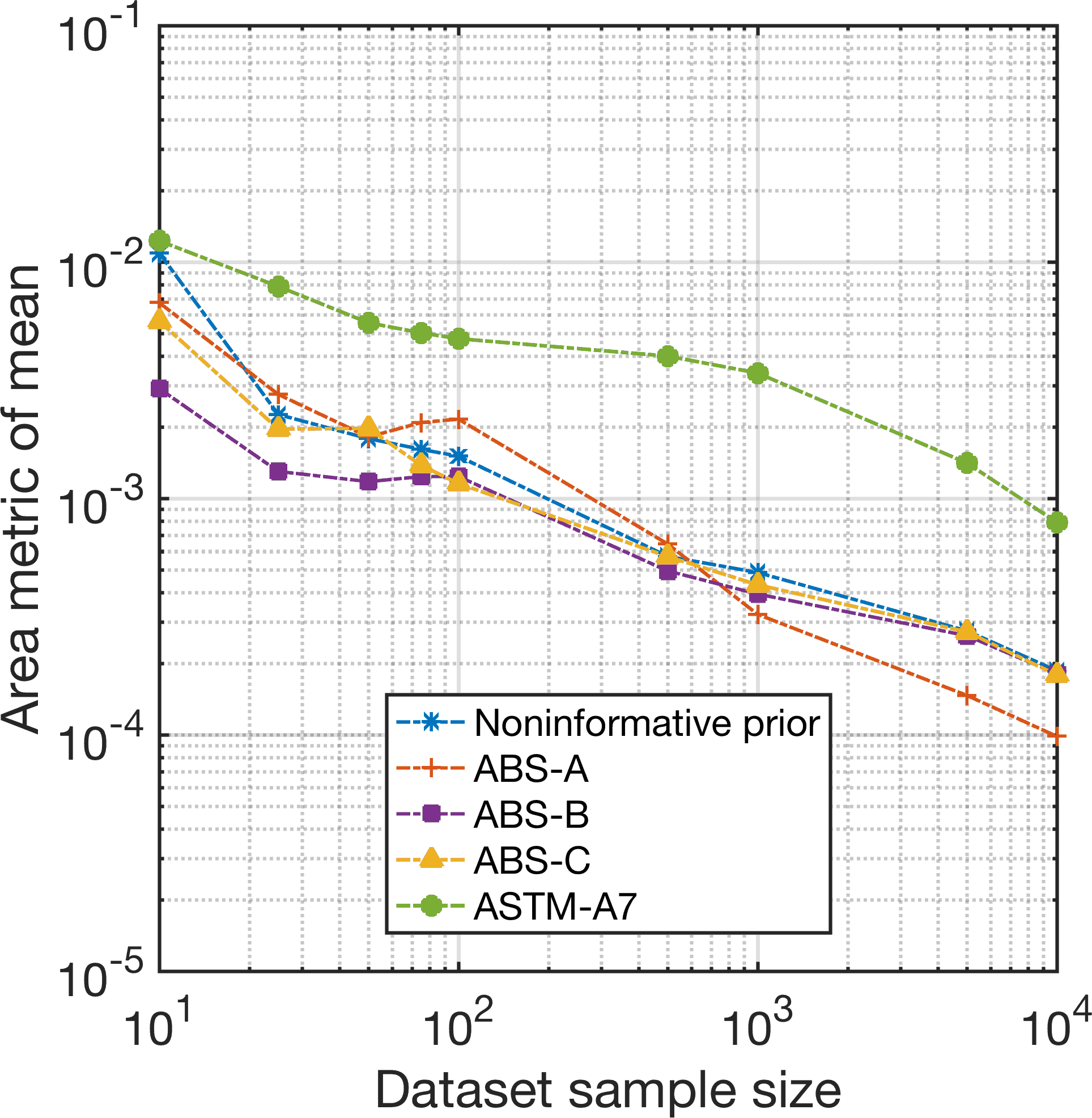}}
	\subfigure[]{\includegraphics[height=2in]{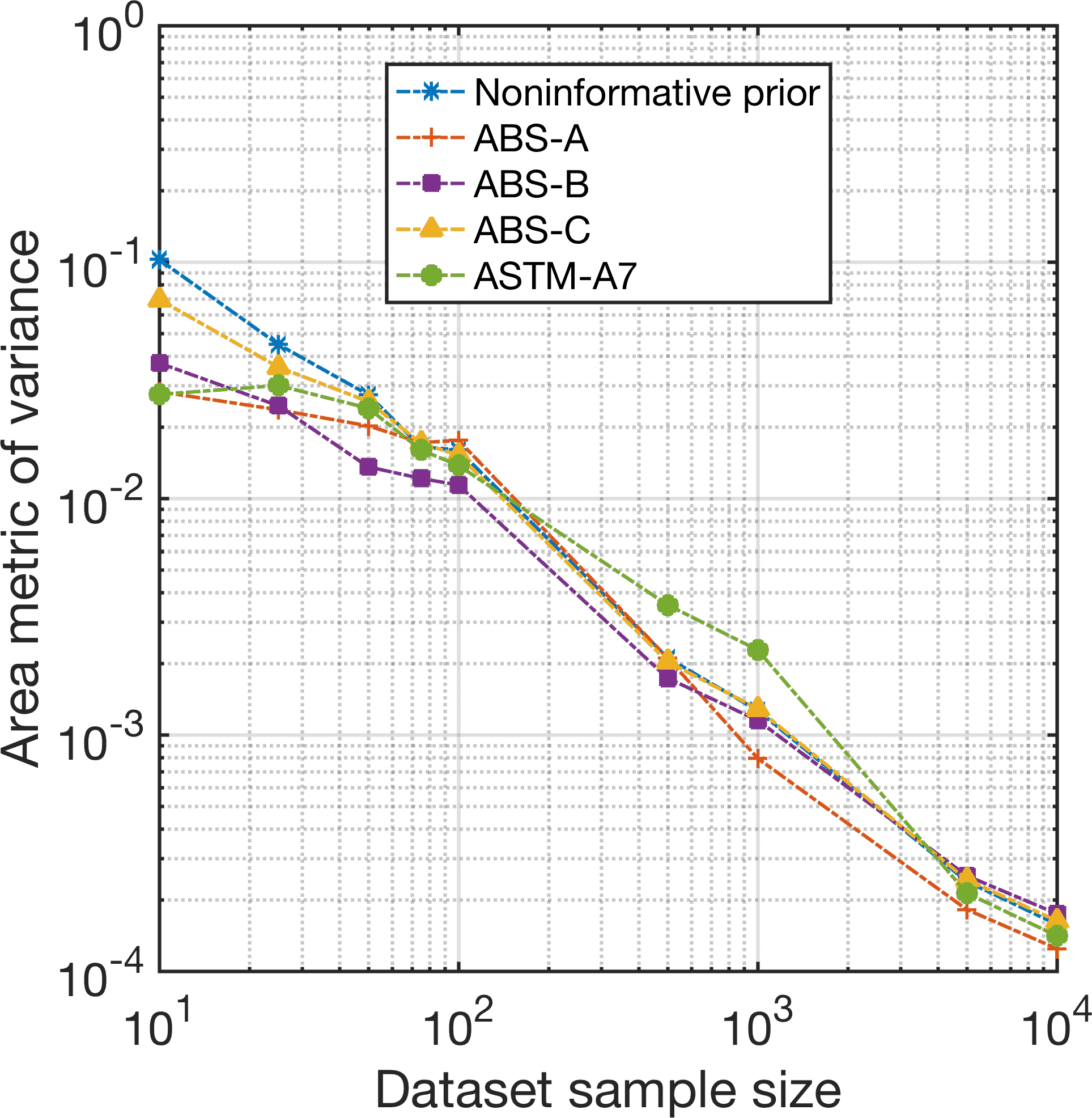}}
	\subfigure[]{\includegraphics[height=2in]{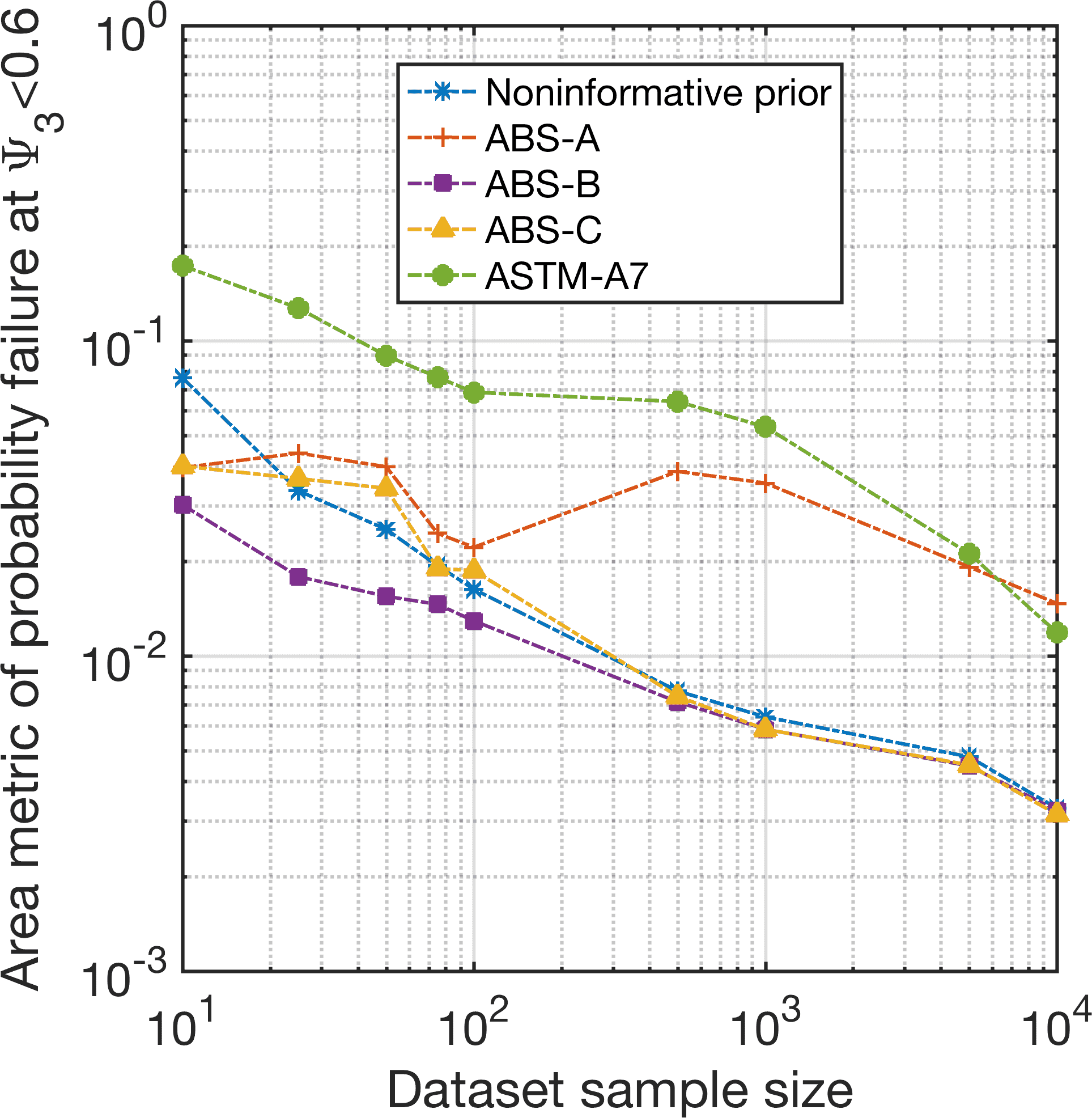}}
	\caption[]{Equal prior model probability and different parameter priors - convergence of confidence level of (a) mean, (b) variance and (c) probability of failure; and area validation metric for (d) mean, (e) variance and (f) probability of failure}  \label{fig:convergence2}
\end{figure}

Next, consider the effect of the parameter priors. Here, we employ equal prior model probabilities and vary the parameter prior. Figure \ref{fig:convergence2} shows convergence of the confidence and area metrics for the mean buckling strength, variance of buckling strength, and $P_f$ with dataset size. 
As expected, we see that the ABS-B prior shows consistently good performance in terms of both confidence and accuracy (it is a good prior). In fact, most of the other priors show reasonable performance as well and all converge in confidence at approximately the same rate. The problem lies in the accuracy convergence of the mean buckling strength and $P_f$ using the ASTM-A7 and ABS-A priors. Recall that these models did not accurately quantify input uncertainty. Consequently, the accuracy of response statistics is slow to converge. 

\begin{figure}[!ht]	
	\centering
	\subfigure[]{\includegraphics[height=2in]{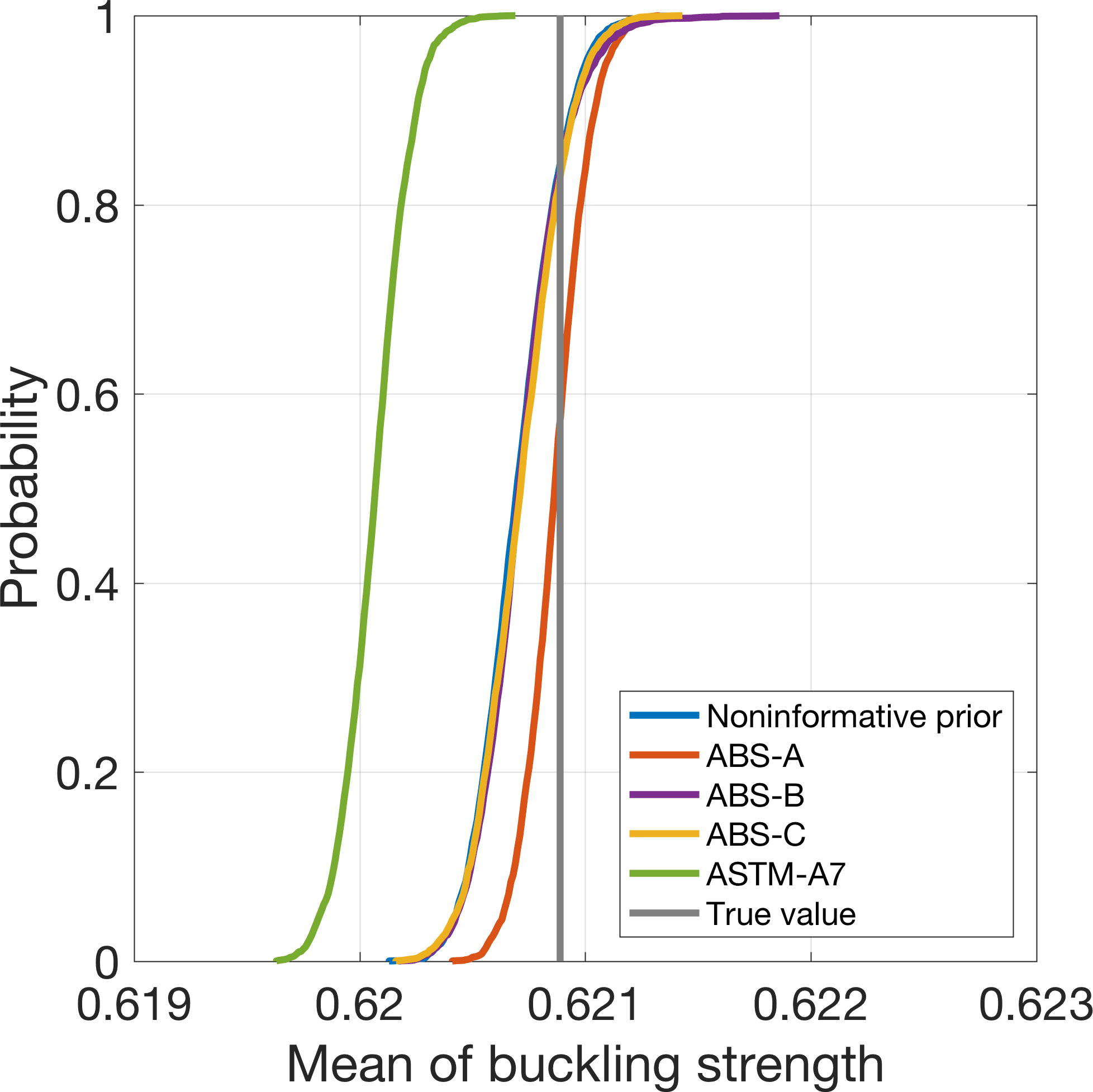}}
	\subfigure[]{\includegraphics[height=2in]{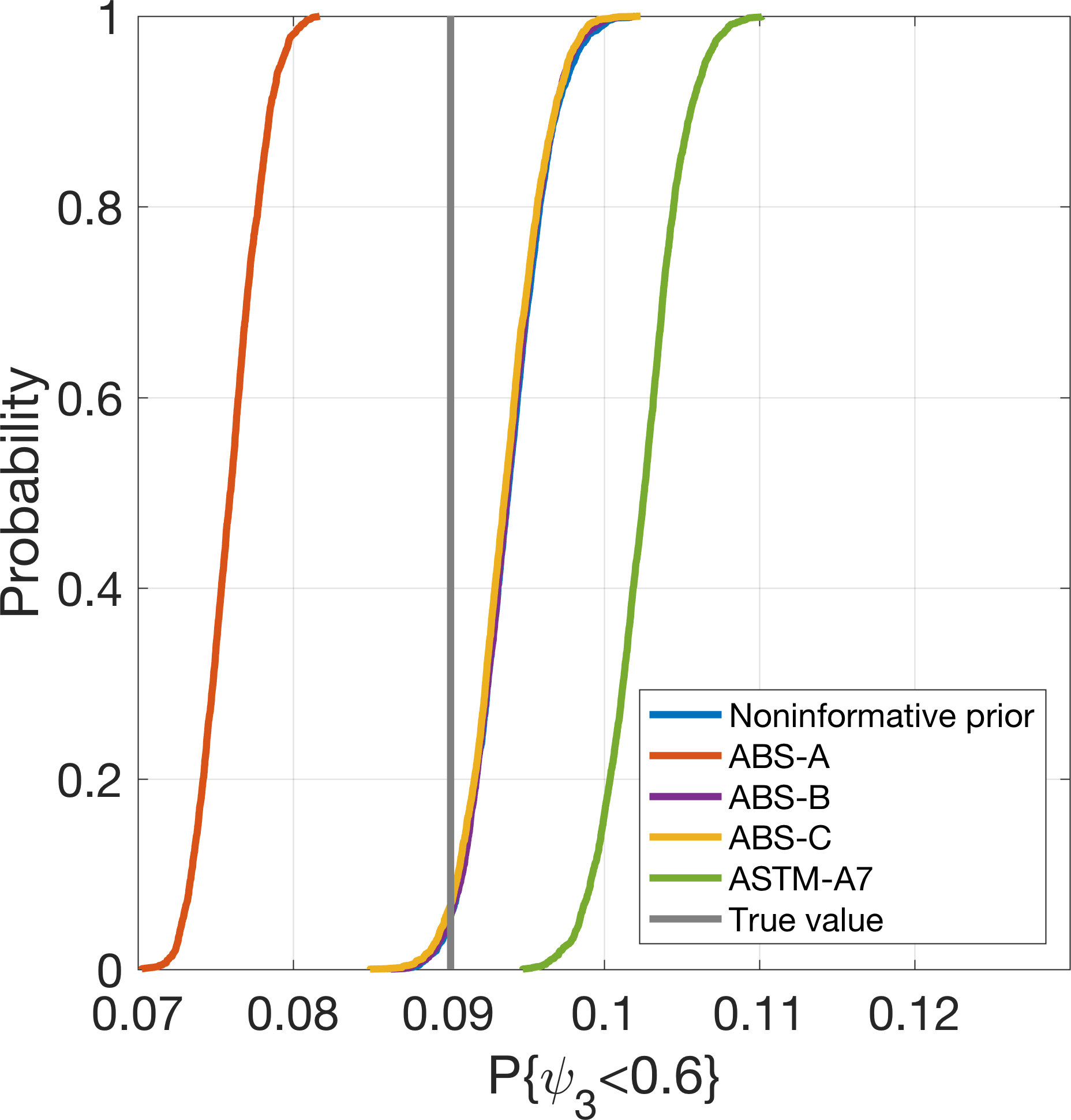}}
	\caption[]{ Empirical CDFs of (a) mean of buckling strength and (b) probability of failure at $\psi_3<0.6$ given 10000 data with equal prior model probability for different parameter priors}
    \label{fig:CDF_compare}
\end{figure}

This poses a significant problem because Figure \ref{fig:convergence2} a-c suggest a high level of confidence regardless of the prior but d-f suggest that accuracy depends on the prior. The result in these cases is high confidence in inaccurate statistics. This is more clearly illustrated in Figure \ref{fig:CDF_compare}, which shows CDFs for the mean buckling strength and $P_f$ for different priors given 10,000 data with equal prior model probabilities. 
In the mean value, the CDF for the ASTM-A7 prior is narrow but does not intersect the true value. Quantitatively, its confidence metric is small suggesting 95\% probability that the value lies in the range $[0.61979, 0.62036]$ but is inaccurate as the true value is $\mu_\psi=0.62089$. Similarly, the ASTM-A7 and ABS-A priors yield high confidence in incorrect $P_f$ estimates. Their values of 95\% probability lie in the range $[0.07249,0.07974]$ and $[0.09758,0.10725]$ respectively but are both inaccurate given that the true value is $P_f = 0.090132$. The result using these priors, even for large datasets, is high confidence in the wrong answer.

\subsection{Discussion}
The objective of imprecise probabilities in general, and the Bayesian multimodel UQ and propagation method proposed here more specifically, is to provide a near-complete picture of both epistemic and aleatory uncertainty in computational modeling. While the presented methodology is robust under noninformative priors, the work here has shown that it is not immune to biases introduced by improperly informed priors. Even with only slight mis-information in the prior (e.g.\ materials data that are from similar but not identical materials), the Bayesian approach can produce erroneous results that propagate through the computational modeling process yielding incorrect predictions of system performance and probability bounds that do not include the true response. Noninformative priors, while robust in bounding the real uncertainties, may be unnecessarily wide when compared to properly informed priors when datasets are small (for large datasets there is little benefit to informative priors). Consequently, it is the modelers responsibility the judge the relative ``safety'' of using noninformative priors with the risk and benefits of using informative priors.

\section{Conclusion}
In this work, the multimodel uncertainty quantification and propagation method previously proposed by the authors \cite{zhang2018} is recast in a fully Bayesian framework. This provides additional robustness in terms of quantifying uncertainties associated with probability model form in particular. Within this Bayesian framework, we are primarily interested in understanding the influence of prior probabilities in both probability model-form and probability model parameters on multimodel UQ and the propagation of these uncertainties. 

The paper deals primarily with the case where uncertainties are quantified from small datasets, which necessitates a multimodel approach and makes prior probabilities important. Through an example considering the analytical buckling analysis of a simply supported plate, we systematically explore the effect of various prior model-form and model parameter probabilities on multimodel uncertainty quantification and propagation. With regard to model-form uncertainties, it is shown that assumptions about prior probabilities have a significant influence on quantified uncertainties when datasets are small but incorrect prior probabilities can be overcome by large datasets if the parameter priors are appropriate. With regard to model parameter priors, it is shown that priors derived from \textcolor{black}{historical} datasets of varying suitability to the present analysis have a clear influence on uncertainties quantified from small datasets. Moreover, parameter priors derived from \textcolor{black}{historical} datasets that are similar to the presently collected data (but nonetheless different) can introduce biases in the multimodel inference that persist even as very large datasets are collected.

The combined effects of model-form and model parameter priors on uncertainty propagation are then investigated. Again, it is shown that uncertainties in response quantities depend strongly on both priors and biases introduced by incorrect priors persist yielding inaccurate probabilistic response quantities even in the large data limit.

% Based on the proposed methodology of our last paper, this work aims at studying the effect of informative, data-driven prior, including the parameter prior and model prior on the uncertainty estimation and propagation. We develop a nonparametric approach to formulate the informative prior knowledge and a novel adaptive updating approach to improve the optimality of sampling density for importance sampling. The study of convergence presents that the informative prior is critical and sensitive on the improvement of uncertainty quantification and propagation. We need to care about the choice of weak incorrect prior since it may improve the convergence of confidence level but it also leads to an incorrect estimate compared to the true value.  Therefore, it is necessary to make use of past data, experience to formulate an effective and informative prior as the pre-prior to improve the convergence and reduce the uncertainty for decision maker in engineering practice. 

\section{Acknowledgements}
The work presented herein has been supported by the Office of Naval Research under Award Number N00014-16-1-2582 with Dr. Paul Hess as program officer.
\\
\\
\textbf{References}
%\begin{spacing}{1}
%{\scriptsize
\bibliographystyle{elsarticle-num}
\bibliography{paper}
%}
%\end{spacing}

\end{spacing}
\end{document}